\documentclass[traditabstract]{aa}
\pdfoutput=1
\usepackage{txfonts,lscape,rotating,longtable,caption}
\usepackage{natbib}
\usepackage[dvips]{epsfig}
\usepackage{float}
\usepackage{graphicx}
\usepackage{color}

\def\S{Sect.\,}

\title{The VLT-FLAMES Tarantula Survey XVII. Physical and wind
  properties of massive stars at the top of the main sequence}

\author{Joachim\,M.\,Bestenlehner \inst{1}
\and G\"otz\,Gr\"afener \inst{1}
\and Jorick\,S.\,Vink \inst{1}
\and F.\,Najarro \inst{2}
\and A.\,de Koter \inst{3,4}
\and H.\,Sana \inst{5}
\and C.\,J.\,Evans \inst{6}
\and P.\,A.\,Crowther \inst{7}
\and V.\,H\'enault-Brunet \inst{8}
\and A.\,Herrero \inst{9,10}
\and N.\,Langer \inst{11}
\and F. R. N. Schneider \inst{11}
\and S. Sim\'{o}n-D\'{i}az \inst{9,10}
\and W.\,D.\,Taylor \inst{6}
\and N. R. Walborn \inst{12} 
 }

\institute{Armagh Observatory, College Hill,
  Armagh BT61 9DG, United Kingdom
\and Centro de Astrobiolog\'ia (CSIC-INTA), Ctra. de Torrej\'on a Ajalvir km-4, E-28850 Torrej\'on de Ardoz, Madrid, Spain
\and Astronomical Institute Anton Pannekoek, Amsterdam University, Science Park 904, 1098~XH, Amsterdam, The Netherlands
\and Instituut voor Sterrenkunde, Universiteit Leuven, Celestijnenlaan 200 D, 3001, Leuven, Belgium
\and ESA/STScI, 3700 San Martin Drive, Baltimore, MD21210, USA
\and UK Astronomy Technology Centre, Royal Observatory Edinburgh, Blackford Hill, Edinburgh, EH9 3HJ, UK
\and Dept. of Physics \& Astronomy, Hounsfield Road, University of Sheffield, S3 7RH, UK
\and Department of Physics, Faculty of Engineering and Physical Sciences, University of Surrey, Guildford, GU2 7XH, UK
\and Instituto de Astrof\'isica de Canarias, E-38200 La Laguna, Tenerife, Spain 
\and Departamento de Astrof\'isica, Universidad de La Laguna, E-38205 La Laguna, Tenerife, Spain
\and Argelander-Institut f\"ur Astronomie der Universit\"at Bonn, Auf dem H\"ugel 71, 53121 Bonn, Germany
\and Space Telescope Science Institute, 3700 San Martin Drive, Baltimore, MD 21218, USA
}

\date{Received /Accepted }

\begin{document}

\abstract{The evolution and fate of very massive stars (VMS) is
  tightly connected to their mass-loss properties. Their initial and
  final masses differ significantly as a result of mass loss. VMS have
  strong stellar winds and extremely high ionising fluxes, which are
  thought to be critical sources of both mechanical and radiative
  feedback in giant H\,{\sc ii} regions. However, how VMS mass-loss
  properties change during stellar evolution is poorly understood. In
  the framework of the VLT-Flames Tarantula Survey (VFTS), we explore
  the mass-loss transition region from optically thin O star winds to
  denser WNh Wolf-Rayet star winds, thereby testing theoretical
  predictions. To this purpose we select 62 O, Of, Of/WN, and WNh
  stars, an unprecedented sample of stars with the highest masses and
  luminosities known. We perform a spectral analysis of optical VFTS
  as well as near-infrared VLT/SINFONI data using the non-LTE
  radiative transfer code CMFGEN to obtain both stellar and wind
  parameters. For the first time, we
      observationally resolve the transition between optically thin
  O star winds and optically thick hydrogen-rich WNh Wolf-Rayet winds.  Our results suggest the
  existence of a ``kink'' between both mass-loss regimes, in agreement
  with recent Monte Carlo simulations. For the optically thick regime,
  we confirm the steep dependence on the classical Eddington factor
  $\Gamma_{\rm e}$ from previous theoretical and observational
  studies.  The transition occurs on the main sequence near a
  luminosity of $10^{6.1}\,L_\odot$, or a mass of 80...90\,$M_\odot$.
  Above this limit, we find that -- even when accounting for moderate
  wind clumping (with $f_{\rm v}$ = 0.1) -- wind mass-loss rates are {\it
    enhanced} with respect to standard prescriptions currently adopted
  in stellar evolution calculations. We also show that this results in
  substantial helium surface enrichment. Finally, based on our
  spectroscopic analyses, we are able to provide the most accurate
  ionising fluxes for VMS known to date, confirming the pivotal role
  of VMS in ionising and shaping their environments.}

\titlerunning{VFTS XVII. Physical and wind
  properties of massive stars at the top of the main sequence}

\maketitle 

\section{Introduction}
Very massive stars (VMS) are defined to be initially more massive than
100\,$M_{\odot}$ \citep[cf.][]{vink2013:IAU}.  Whilst a lot of
attention has been devoted to the role of VMS at extremely low
metallicity in the very early Universe \citep[e.g.][]{bromm1999,
  abel2002}, the role of VMS in our local Universe has largely been
ignored until recently.  \cite{hamann2006} found the
existence of a number of highly luminous Galactic objects (up to $10 ^{7} L_{\odot}$). 
\cite{crowther2010}
confirmed this finding, reporting the existence of VMS up to 320\,$M_{\odot}$ 
for the R136 core stars in the Tarantula nebula of the Large
Magellanic Cloud (LMC). Given the clustered nature of these objects
it was important that \cite{bestenlehner2011} identified an
almost identical twin of R136a3, VFTS 682, in the context of the VFTS. Its 
isolated nature increased confidence that VMS really exist above the
``canonical'' upper limit of about 150\,$M_{\odot}$ \citep{figer2005}.

Interestingly, the presence of VMS of up to 320\,$M_{\odot}$ 
would bring them into the predicted initial mass range of pair-instability supernovae of 140 - 260\,$M_{\odot}$ \citep{heger2002, langer2007}.  As such
explosions would disrupt the entire star, all metals would be returned
to the interstellar medium (ISM). One such object at the top of the
mass function of a massive-star cluster would produce more
metals than all the other lower mass stars together
\citep{langer2009}. It could significantly affect chemical evolution
modelling of galaxies.

However, the evolution and fate of VMS are highly uncertain. 
The main culprit is the unknown rate of wind mass loss, given that VMS are
thought to evolve close to chemically homogeneously
\citep[e.g.][]{graefener2011}.  Whilst recent evolution models have been
published by e.g.\  \cite{yungelson2008} and \cite{yusof2013}, the way
mass loss has been implemented in these models is entirely different
and it is not known a priori which way is more appropriate.  The key
issue to decide whether pair-instability supernovae are likely to
exist in the present day universe is that of stellar mass loss.

Fortunately, there has recently been significant progress in
our theoretical understanding of mass loss at the top of the
Hertzsprung-Russell diagram (HRD). In particular, \cite{vink2011}
computed mass-loss rates for VMS up to 300\,$M_{\odot}$ and discovered
a ``kink'' in the mass loss versus $\Gamma$ dependence, involving a
relatively shallow slope for optically thin O star winds, but changing
into a steeper slope for optically thick WNh Wolf-Rayet stars. This
latter result appears to be in qualitative agreement with theoretical
and observational studies of very massive WNh stars by
\cite{graefener2008} and \cite{graefener2011}. Furthermore,
\cite{vink2012} calibrated currently used mass-loss rates for VMS at
the transition point from Of to WNh. Alternative VMS wind models have
been proposed by \cite{pauldrach2012} where it is claimed that VMS
winds remain optically thin. In other words, there are not only
quantitative, but even qualitative differences in the behaviour of VMS
stellar wind models. For these reasons, it is imperative to determine
the VMS mass-loss behaviour empirically, thereby testing theoretical
concepts.

This has now become possible with the VFTS, providing a sufficiently 
large dataset at the top end of the HRD. This data-set
involves several groups of H-rich Wolf-Rayet stars (WNh), but also
transition Of/WN stars, as well as very luminous Of and other O stars.
These stars cover a large range in mass-loss rate (which we find to be 
of more than two orders of magnitude), which enables
us to study the transition from optically thin O star winds to 
optically thick WNh Wolf-Rayet winds.

In \S\,\ref{obs_data}, we give an overview of the observational data
and describe our target selection. We describe the details of our spectral analysis regarding theoretical model
computations and analysing techniques in \S\,\ref{spec}. On the basis of
the results we investigate in \S\,\ref{s:discussion} the mass-loss
properties, evolutionary stages, and ionising fluxes of our targets.
Our conclusions are summarised in \S\,\ref{s:conclusion}.

\section{Observational data and target selection\label{obs_data}}
The data used in this work have been obtained in the framework of the
VLT-FLAMES Tarantula Survey \citep[VFTS;][]{evans2011}, a large
spectroscopic survey of over 800 O-, B-, and Wolf-Rayet (WR) stars in the 30
Doradus region of the Large Magellanic Cloud. From this data-set we
selected 62 massive and very massive star candidates. The data have been augmented by
existing archival data and follow-up observations in the infrared (IR)
range. In the following we give an overview of the spectroscopic and
photometric data used (in \S\,\ref{s:so} and \S\,\ref{s:phot}), and
describe the target selection process (\S\,\ref{s:target}).

\subsection{Spectroscopic Observations \label{s:so}}
Within the VFTS project the entire 30 Doradus region, centred on the
dense cluster R136, has been covered with optical fibre spectroscopy
using the MEDUSA-GIRAFFE ($\lambda 4000-7000$\,\AA) mode of the FLAMES
instrument. Roughly 70\% of the stars with $V<17$\,mag have been
observed in this programme. For the densely populated region
near R\,136 additional integral-field observations were obtained using
FLAMES/UVES ($\lambda4200-6100$\,\AA) and ARGUS/GIRAFFE
  ($\lambda4000-4600$\,\AA).

Specifically for the analysis of VMS in the central region of
  30\,Dor, $K$-band data were obtained with SINFONI the
  Spectrograph for INtegral Field Observations in the Near Infrared on
  the VLT (Program ID\,084.D-0980, PI: Gr{\"a}fener). The spectra
  cover the central arcsec around R\,136 with a seeing constraint
  $\leq 0.8$\,arcsec and a spectral resolution of $R=4000$ over the
  wavelength range of $\lambda2.0-2.4\mathrm{\mu m}$.  Due to the
  increased IR wind-emission of VMS compared to lower-mass
    stars with weaker winds, these data provide a homogeneous set of
  wind and abundance diagnostics for these objects that is less prone
  to crowding issues than e.g.\ optical data.  The data were reduced
  using the standard ESO pipeline. The observed telluric
    standards were corrected for their intrinsic H and He absorption
    features with the aid of synthetic telluric spectra from
    \citet[][2014 submitted]{Smette2010}.

For some targets close to the centre of 30 Dor,
archival data were available from VLT/UVES (Program ID: 70.D-0164, PI: Crowther) and the Hubble Space Telescope (HST) satellite, specifically HST/FOS (Program ID: 6417, PI: Massey), HST/STIS
(Program ID: 7739, PI: Massey), HST/GHRS (Program ID: 3030, PI: Ebbets), and
HST/COS (Program ID: 11484, PI: Hartig). Some Of/WN and WNh stars had archival
data from the International Ultraviolet Explorer (IUE) available. 
  The additional optical archival observations are chosen in a way
  such that we can use the same diagnostic lines for all our targets,
  as to allow for a homogeneous spectroscopic analysis.  If {\it one}
  diagnostic line was not observed, the star was removed from the
  discussion (see \S\,\ref{s:discussion}). However,
  $\mathrm{H}_{\alpha}$ and He\,{\sc ii} $\lambda 4686$ could be
  substituted by the near-IR lines He\,{\sc ii} at
  $2.165$~$\mathrm{\mu m}$ and $\mathrm{Br}_{\gamma}$ if they show
  sufficient emission, or vice versa (see
  Appendix\,\ref{he_temp}). The UV-data are only used to constrain the
  terminal velocity. An overview of the data used for each target is given in the appendix
Table~\ref{t:spectra}. Aliases of the VFTS stars are given in Table~\ref{t:aliases}.

\begin{figure}
\begin{center}
\resizebox{\hsize}{!}{\includegraphics{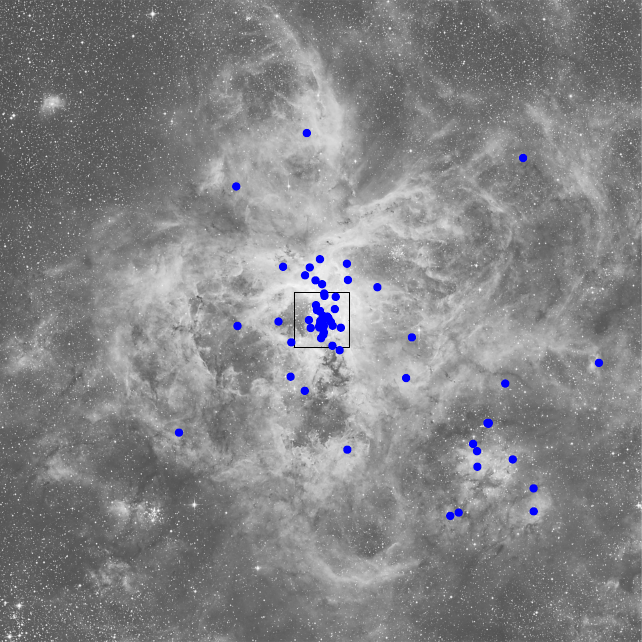}}
\end{center}
\caption{Location of our target stars. The physical dimension of the field is 280\,pc $\times$ 280\,pc, adopting a distance modulus of 18.49\,mag. See Fig.~\ref{f:pos2} for a close up on the core R136.}
\label{f:pos1}
\end{figure}
\begin{figure}
\begin{center}
\resizebox{\hsize}{!}{\includegraphics{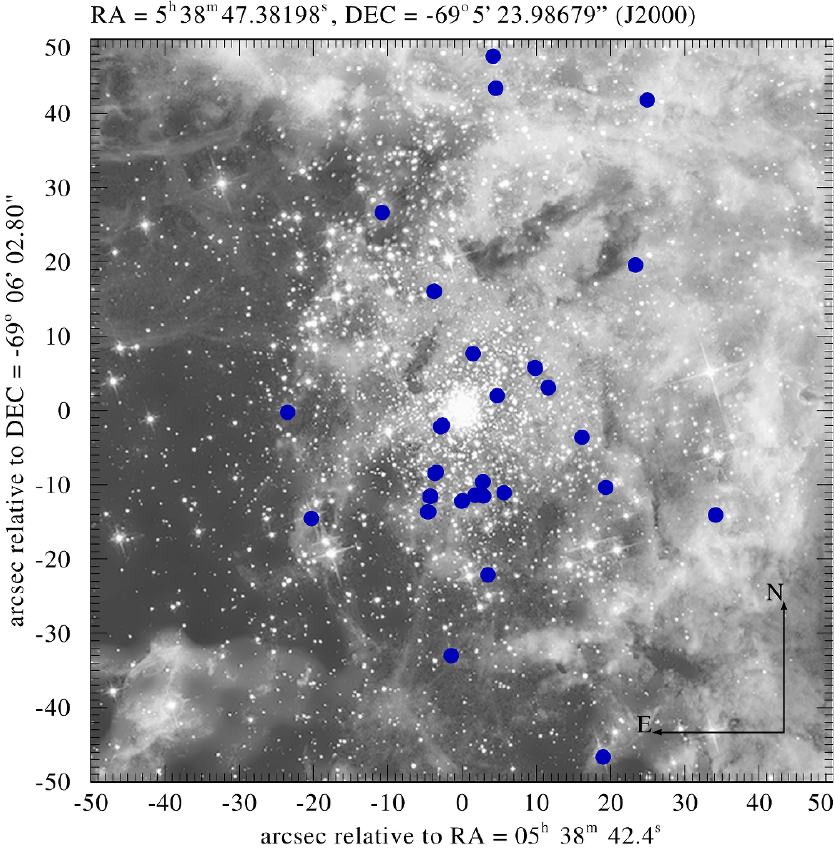}}
\end{center}
\caption{Location of our target stars around the cluster R136. The physical dimension is 24.24\,pc $\times$ 24.24\,pc.}
\label{f:pos2}
\end{figure}

\subsection{Supplementary photometry \label{s:phot}}

\begin{table}
  \caption{Photometry for stars which are not included in the catalogue of \cite{evans2011} or which have large uncertainties, e.g.\ due to the presence of strong emission lines or crowding.}
\label{t:phot}
\centering
\begin{tabular}{lcccccc}
\hline
\hline
\rule{0cm}{2.2ex}Name	& $B$	& $V$	& $H$	& $K_{\rm s}$\\
\hline
VFTS\,108		& 14.02$^1$		& 13.89$^1$		& --		& --		\\
VFTS\,147		& 12.22$^1$		& 12.09$^1$		& --		& --		\\
VFTS\,151 		& 13.08$^1$		& 12.98$^1$		& --		& --		\\
VFTS\,402 		& 12.33$^1$		& 12.61$^1$		& --		& --		\\
VFTS\,682		& 16.11$^2$		& 16.75$^2$		& --		& --		\\
VFTS\,1001	& 12.76$^1$		& 12.58$^1$			& --		& --			\\
VFTS\,1014	& 14.23$^3$		& 14.18$^3$		&	--	&	--	\\
VFTS\,1018	& 14.65$^3$		& 14.47$^3$		&	--	&	--	\\
VFTS\,1026	& 14.59$^3$		& 14.51$^3$		&	--	&	--	\\
VFTS\,1028	& 13.84$^3$		& 13.85$^3$		&	--	&	--	\\
VFTS\,1025 / R136c &	13.74$^3$& 13.43$^3$	& 11.64$^4$	 	& 11.38$^4$	 \\
Mk42	& 12.86$^3$		& 12.78$^3$		& 12.26$^4$	 	& 12.17$^4$	 \\
\hline

\end{tabular}
\tablefoot{
\tablefoottext{{\rm 1}}{\cite{massey2002:2},} 
\tablefoottext{\rm 2}{\cite{parker1993},} 
\tablefoottext{\rm 3}{\cite{deMarchi2011},}
\tablefoottext{\rm 4}{\cite{campbell2010}}
}
\end{table}

To constrain stellar luminosities it is necessary to model their
spectral energy distribution (SED) including interstellar extinction.
For this purpose, we use optical and near-IR photometry.  Optical
photometry has been obtained by \citet{evans2011} using the Wide Field Imager (WFI) at the
2.2\,m MPG/ESO telescope at La Silla. Near-IR photometry is provided by
the VISTA Magellanic Clouds (VMC) Survey \citep{cioni2011}.
Specifically, we use PSF-fitting photometry from \cite{rubele2012} and
cross-matches to the VFTS sources from Zaggia et al.~(in prep.). 
For objects that are not included in the VMC survey we
    use the InfraRed Survey Facility (IRSF) Magellanic Clouds
  catalogue \citep{kato2007} as listed in \cite{evans2011}. For
stars near the centre of the R\,136 cluster a higher spatial
resolution is required due to crowding.  In this case we utilise
optical HST-WFC3 photometry from \cite{deMarchi2011} and near-IR
$H/K_{\rm s}$-Band VLT/MAD photometry from \cite{campbell2010}.
The photometry from \citeauthor{evans2011} has been
  transformed from the original WFI system to Johnson colours.  In the
  presence of strong emission lines this can lead to large
  uncertainties. Furthermore uncertainties can arise from crowding in
  strongly populated fields. Alternative photometry for objects with
  such uncertainties or objects that are missing in the catalogue of
  \cite{evans2011} is given in Table~\ref{t:phot}.

\subsection{Target Selection \label{s:target}}
Our interest in this study is focussed on the most massive stars in
30\,Dor sampling the top of the main-sequence, including VMS with
masses in excess of $\sim 100\,M_\odot$. We are interested
  in objects that are suitable for an emission-line analysis and are
brighter than $\sim10^{5.5}L_{\odot}$, and we avoided further
selection criteria for the O stars to avoid a biassed sample. That said,
irrespective of their brightness, we included all of the most promising VMS
candidates \citep[namely the Of/WN and WNh stars from][]{evans2011}.

From the sample of 360 O stars \citep{evans2011, walborn2014} we
pre-selected stars showing emission-lines in He\,{\sc ii}
$\lambda4686$, N\,{\sc iii} $\lambda4634/4640$, N\,{\sc iv}
$\lambda4058$, or $\mathrm{H_{\alpha}}$ $\lambda6562$. Based on our theoretical models (\S\,\ref{cmfgen}) we divided the sample into three temperature classes to estimate their bolometric corrections (class 1: $T_{\mathrm{eff}} \lesssim 40$kK, class 2: $40\mathrm{kK} \lesssim T_{\mathrm{eff}} \lesssim 45$kK, class 3: $T_{\mathrm{eff}} \gtrsim 45$kK).
For each class intrinsic colours $B-V$ and $V-K_{\mathrm{s}}$ are
derived from our models. The $B$, $V$, and $K_{\mathrm{s}}$-band
magnitudes are determined by applying the filter function for the
MPG/ESO/WFI $B$ and $V$-band filters, and the 2MASS (2 Micron All Sky
Survey) $K_{\mathrm{s}}$-band filter. The obtained intrinsic colours
are then used to estimate the extinction parameters $E(B-V)$ and
$E(V-K_{\mathrm{s}})$ from the observed optical and near-IR
photometry. As outlined in \citet{bestenlehner2011} we estimate $A_V$
and $R_V$ for each object adopting the extinction law from
\cite{cardelli1989}. For the absolute visual magnitude $M_{\rm V}$ we
adopt a distance modulus of 18.49~mag \citep{pietrzynski2013}. Stellar
luminosities are then obtained using bolometric corrections from
\cite{martins2006} for each temperature class. Stars with
  estimated luminosities $\lesssim10^{5.5}L_{\odot}$ are excluded from
  our sample.

For the stars near R\,136 that are only covered by ARGUS
integral-field spectroscopy we further added objects based on the
presence of emission lines for He\,{\sc ii} at $2.165$ and
$2.189~\mathrm{\mu m}$ and $\mathrm{Br}_{\gamma}$ at
$2.165~\mathrm{\mu m}$ in our SINFONI data.  We expect that these
objects cover a similar mass-loss range as the ones selected on the
basis of optical emission lines from the MEDUSA and UVES observations.
Figures~\ref{f:pos1} and \ref{f:pos2} show the location of our targets in
the Tarantula Nebula.

In total we selected 62 stars combining the VFTS-MEDUSA, ARGUS and
UVES, the VLT-SINFONI, and archival HST and IUE observations. The
sample is deliberately heterogeneous covering a large range of emission-line
strengths, and including a variety of luminosity classes and spectral
types. A summary of the selected targets and the available
spectroscopy is given in Table~\ref{t:spectra}. The table also
indicates stars that are or may be spectroscopic binaries, higher order multiple systems,
or contaminated as a result of crowding.

\section{Spectral analyses\label{spec}}

In this section we describe the spectral analyses of the 62 objects in
our sample. In \S\,\ref{cmfgen} we describe the model computations performed to
produce a large grid of {\sc cmfgen} non-LTE atmosphere/wind models.
In \S\,\ref{spec_an} we introduce our method of analysis which
combines the classical, subjective approach to fit spectra by eye with
a more systematic $\chi^2$-approach, followed by a discussion of
important diagnostic spectral lines in Sect.\,\ref{s:lines}. In
Sect.\,\ref{s:lum_red} we describe our method to fit the observed
spectral energy distribution (SED), including the effects of
interstellar extinction. The uncertainties arising from our
  analysis are discussed in \S\,\ref{s:error}, and comparisons with
  previous results are performed in \S\,\ref{s:comp}.

\subsection{Model Grid\label{cmfgen}}

\begin{table}
  \caption{Main input parameters and parameter ranges of the computed model grids. Numbers in brackets indicate models that extend the standard grid in specific parameter ranges.}
\label{t:parameter}
\centering
\begin{tabular}{l@{~}|c@{~}c@{~}c}
\hline
\hline
Grid name	&Parameter	& Range	& \# of steps\\
\hline
main		&	$f_{\rm v}$	&	0.1, 1.0	& 2 \\
		&	$T_{\mathrm{eff}}$\,[kK]	&	35.5\,to\,56.2 &	9 \\
		&	$\log \dot{M_{\mathrm{t}}}$\,[$M_{\odot}$/yr]	&	$-6.28$\,to\,$-4.78$	&	10\\
		&	abundance	&	Table\,\ref{t:abundance}	&	8\\
		&	$\varv_{\infty}$\,[km\,s$^{-1}$]	&	2800	&	1\\
		&	$\log g$		&	4.0	&	1\\
\hline
$\log g$	&	$f_{\rm v}$	&	0.1, 1.0	& 2 \\
		&	$T_{\mathrm{eff}}$\,[kK]	&	35.5\,to\,44.7 &	5 \\
		&	$\log \dot{M_{\mathrm{t}}}$\,[$M_{\odot}$/yr]	&	$-6.28$\,to\,$-4.78$	&	10\\
		&	abundance	&	Table\,\ref{t:abundance}\,(row 1 to 3)	&	3\\
		&	$\varv_{\infty}$\,[km\,s$^{-1}$]	&	2800	&	1\\
		&	$\log g$		&	3.75, 4.25	&	2\\
\hline
WNh		&	$f_{\rm v}$	&	0.1	& 1 \\
		&	$T_{\mathrm{eff}}$\,[kK]	&	39.8\,to\,50.1\,(56.2) &	5\,(7) \\
		&	$\log \dot{M_{\mathrm{t}}}$\,[$M_{\odot}$/yr]	&	$-4.61$\,to\,$-4.11$	&	4\\
		&	$Y$ abundance	&	62.5\% to 92.5\%	&	5\\
		&	$\varv_{\infty}$\,[km\,s$^{-1}$]	&	2800	&	1\\
		&	$\log g$		&	4.0	&	1\\
\hline
late WNh	&	$f_{\rm v}$	&	0.1	& 1 \\
		&	$T_{\mathrm{eff}}$\,[kK]	&	35.5\,to\,42.2\,(50.1)	&	4\,(7) \\
		&	$\log \dot{M_{\mathrm{t}}}$\,[$M_{\odot}$/yr]	&	$-4.61$\,to\,$-3.61$	&	6\\
		&	$Y$ abundance	&	62.5\% to 99.0\%	&	6\\
		&	$\varv_{\infty}$\,[km\,s$^{-1}$]	&	1800	&	1\\
		&	$\log g$		&	4.0\,(3.75)	&	1\,(2)\\
\hline
\end{tabular}
\end{table}

The atmosphere models used in this work are computed with the {\sc
  cmfgen} code by \citet[][]{hillier1998}, and comprise complex model
atoms for H\,{\sc i}, He\,{\sc i-ii}, C\,{\sc iii-iv}, N\,{\sc iii-v},
O\,{\sc iii-vi}, Si\,{\sc iv}, P\,{\sc iv-v}, S\,{\sc iv-vi} and
Fe\,{\sc iv-vii} (cf.\,Table\,\ref{t:atom_model}). The models take
advantage of the concept of super-levels by \citet{anderson1989},
where the statistical equations of complex atoms are solved by
combining many close-by atomic energy levels into few super-levels.

The parameters of our main model grid are described in
Table\,\ref{t:parameter}. We varied three fundamental parameters,
effective temperature $T_{\rm eff}$, transformed mass-loss rate
$\dot{M}_{\rm t}$ and helium surface mass fraction $Y$ over their
relevant parameter range. Furthermore we took wind clumping
  with a volume-filling factor $f_{\rm v}$ into account by computing
  two identical model grids assuming a clumped wind with a volume-filling factor $f_{\rm v}=0.1$ starting at a velocity of
100\,km\,s$^{-1}$, and an unclumped wind with $f_{\rm
  v}=1$\footnote{We note that the adopted clumping structure is
  simplified, and that clumping might in reality be depth dependent 
   \citep{puls2006, najarro2011} and/or involve optically thick 
    structures \citep[e.g.][]{hillier2008, petrov2014}.
 Moreover, the effects of optically thick clumps may be different for O\,stars and WR\,stars \citep{hillier2008, sundqvist2011}.  
  Because of the
  generally poor constraints, wind clumping constitutes a likely source 
   of uncertainty in our quantitative analysis (cf.\ the discussion in Sects.\,\ref{mdot_gamma} and
  \ref{efficiency}).
However, the clumping implementation issue is an important limitation affecting
  {\it any} empirical mass-loss rate currently in literature. More
importantly, there is as yet no evidence that the clumping properties of O-type
stars would be altogether different from those of WR stars. There is thus
no a priori reason to suspect that our conclusions will be affected in a qualitative 
sense.}.
We computed one sub-grid for
  unprocessed material with $Y=0.25$ and metal abundances
  corresponding to half solar metallicity based on solar abundances
  from \cite{asplund2005}.  Furthermore we computed multiple sub-grids
  for material that has been processed in the CNO-cycle with a
  nitrogen-enriched composition and He mass fractions between 0.25 and
  0.7, (cf.\,Tab.\,\ref{t:abundance}).

In our model grids we keep the luminosity and terminal wind speed
fixed to values of $L=10^6L_\odot$ and
$\varv_\infty=2800$\,km\,s$^{-1}$. The use of the transformed
mass-loss rate $\dot{M}_{\rm t}$, as defined by \citet{graefener2013},
allows us to scale our results to other values of $L$ and
$\varv_\infty$, and to correct for wind clumping with an arbitrary
volume-filling factor $f_{\rm v}$.  For a synthetic emission-line spectrum with given
$\dot{M}_{\rm t}$ the true mass-loss rate $\dot{M}$ follows from the
relation
\begin{eqnarray}
  \log(\dot{M}) = \log(\dot{M}_{\rm t}) + 0.5\log(f_{\rm v}) + \log \left(\frac{\varv_{\infty}}{1000\,\mathrm{km\,s}^{-1}} \right) \nonumber \\ 
  +~0.75\log \left(\frac{L}{10^6L_{\odot}} \right)\,.
\label{trans}
\end{eqnarray}
We note that this scaling relation is equivalent to the approach using
the `transformed radius' by \cite{schmutz1989} and \cite{hamann2004}.

Furthermore, the model structure is computed assuming hydrostatic
equilibrium with a fixed surface gravity of $\log g = 4$ in the inner part
of the atmosphere, and a $\beta$-type velocity law with a fixed
$\beta$-parameter of 1.0 in the outer wind.  The reference radius
($R_{\rm ref}$) and effective temperature ($T_{\mathrm{eff}}$) of the
star are defined at an optical depth of $\tau = 2/3$. The stellar
temperature ($T_{\star}$) and radius ($R_{\star}$) of the inner
boundary of the atmosphere are defined at $\tau = 100$.

Additional grid computations were necessary to cover the parameter
range of WNh stars because of their higher mass-loss rates. The WNh grid
covers the range of $\log(\dot{M}_{\rm t}/ (M_{\odot}{\rm yr}^{-1}))$ from $-4.61$ to
$-4.11$ in steps of $1/6\,\mathrm{dex}$ with helium mass fractions of
 0.625, 0.7, 0.775, 0.85 and 0.925.  Furthermore, a small sub-grid was
computed for late WN6-8h stars, because their terminal velocities are
significantly lower. This sub-grid is computed for a terminal velocity
of 1800\,km\,s$^{-1}$ and covers the temperature $\log T[\mathrm{K}]$ from
$4.550$ to $4.625$ in $0.025\,\mathrm{dex}$ steps and the transformed
mass-loss range $\log(\dot{M}_{\rm t}/ (M_{\odot}{\rm yr}^{-1}))$ from $-4.61$ to
$-3.61$ in $1/6\,\mathrm{dex}$ steps with an additional helium abundance of 99.0\%
compared to the WNh grid. Because of convergence difficulties some of the WNh models required a lower $\log g$ of 3.75. As the winds of WNh stars are optically thick this choice has no effect on their emission-line spectra.
 
As the determination of temperatures and mass-loss rates for O stars
is sensitive to the surface gravity we computed sub-grids in the low
mass-loss and temperature range with $\log g = 3.75$ and 4.25.  We
used these models to quantify the impact of $\log g$ and estimate the
resulting uncertainties. The sub-grids cover a transformed mass-loss
range $\log(\dot{M}_{\rm t}/ (M_{\odot} {\rm yr}^{-1}))$ from $-6.28$ to $-4.78$,
temperature range from $\log T[\mathrm{K}]$ $4.55$ to $4.65$ and
helium mass fractions of 0.25 and 0.325. 

In addition, some models with $\log g = 3.5$ and
    with intermediate helium mass fractions were computed
  to investigate the influence of the assumed $\log g$ on the derived helium
  abundances.
  The results of these tests are discussed in Appendix\,\ref{a:logg}.

  Together with additional test models more
  than 2500 models were computed.

\subsection{Method of analysis\label{spec_an}}

The stellar sample analysed in this work covers a broad parameter
range, reaching from O stars with a combination of weak emission and
absorption lines to WR stars with their strong emission line spectra. 
Especially for the latter it is necessary to perform time intensive
non-LTE computations with complex numerical codes such as {\sc cmfgen}. 
As we are interested in a homogeneous analysis of the
transition between the O and WR stars we chose to analyse the complete
sample using the complex {\sc cmfgen} models. A drawback of this approach is that
the number of parameters that we are able to vary in our analysis is
limited. Our results will thus be less accurate than e.g.\ the ongoing
O star analyses in the VFTS collaboration that are based on the much
faster code {\sc fastwind} by \citet{puls2005}, but they are sufficient for our
investigation of the trends of the wind parameters. (The future works will be
based on tailored analyses for each object involving more
extensive grids and/or advanced fitting methods, however, they are
limited to the O star range.)

In this work we use the grid computations described in
Sect.\,\ref{cmfgen} and try to quantify the quality of the fit by 
introducing a tailored merit function ($\chi^2_{\rm tmf}$) that mimics a least-square fit method. This method is based
on measuring equivalent widths (EWs) and is thus independent of the
adopted terminal wind velocity ($\varv_{\infty}$) in our grid\footnote{A dependence of the EW on $\varv_{\infty}$ is only expected for P-Cygni type emission lines (i.e.\ scattering lines). The optical/IR emission lines used in this work are predominantly recombination lines whose EW is independent of $\varv_{\infty}$.}, the
projected rotational velocity ($\varv_{\rm e} \sin\,i$), and (micro) turbulent
velocity at the stellar surface (see \S\,\ref{cmfgen}). However, due
to the variety of spectral types in our sample and their different
line diagnostics we found it necessary to perform a final visual
assessment of the fit quality thus introducing a subjective
component in our approach.

The following continuum subtracted line profiles are taken into
account for the analysis: the N\,{\sc iii} doublet $\lambda4634/4640$,
N\,{\sc iv}\,$\lambda4058$, the N\,{\sc v} doublet $\lambda4604/4620$,
He\,{\sc i}\,$\lambda4471$, He\,{\sc ii}\,$\lambda4686$, He\,{\sc
  ii}\,$\lambda2.189\mathrm{\mu m}$ and the Balmer lines
$\mathrm{H_{\alpha - \delta}}$, and $\mathrm{Br}_{\gamma}$. 
 To ensure that the method is independent of the continuum
  normalisation, we determine the continuum locally and subtract it in
  exactly the same way from the observations and the synthetic
  spectra. The EWs are obtained by defining two continuum points in
the blue and red direction of the diagnostic line, fitting a straight
line through the points, subtracting it as the continuum, and
integrating over the spectra (consequently absorption lines
  have negative, and emission lines positive
  EWs). 
Complications can still arise from a contamination of the observed
spectra with nebular emission lines.  This is usually the case for
He\,{\sc i}\,$\lambda4471$ and the Balmer lines $\mathrm{H_{\alpha -
    \delta}}$. In these cases we remove the nebular features from the
observed spectrum and use a Gaussian fit to the line wings to
estimate the line profile shape in between. Again, this is done in
the same way for observations and models.  Lines that are strongly
affected by nebulosity are excluded from our analysis.  The fit
quality of the grid models can then be quantified for each object by
computing the sum
\begin{equation}
\chi^2_{\rm tmf} = \displaystyle\sum\limits_{\mathnormal{i}} \frac{1}{\sqrt{|ew^{\mathrm{obs.}}_{\mathnormal{i}}| + \cal C}} \left(ew^{\mathrm{obs.}}_{\mathnormal{i}} - ew^{\mathrm{mod.}}_{\mathnormal{i}}\right)^2~,
\end{equation}
where $i$ indicates the different diagnostic lines. Here the division
by the square root of the observed EW balances the weight between
stronger and weaker lines. The constant $\cal C$ (which is
  of the order $10^{-4}$) prevents divisions by zero and gives a lower
  relative weight to weak lines that are more affected by noise.

The final decision about the best-fitting model is done by eye,
considering the models with the smallest $\chi^2_{\rm tmf}$.  To this
purpose the model spectra are convolved with instrumental and rotational
profiles. This allows us to roughly constrain
  the projected rotational velocities $\varv_{\rm e} \sin\,i$ from the
  observed line profiles. Investigating the parameter range with
small $\chi^2_{\rm tmf}$ also helps to identify possible
  ambiguities in the parameters that can reproduce the observed
  spectra.

\subsection{Line diagnostics\label{s:lines}}

The determination of mass-loss rates and helium abundances is based on
strong H and He lines, mainly He\,{\sc ii}\,$\lambda4686$ and
$\mathrm{H_{\alpha}}$ in the optical and, if available, He\,{\sc
  ii}\,$\lambda2.189\mathrm{\mu m}$ and $\mathrm{Br}_{\gamma}$ in
the near-IR.  For high mass-loss rates other lines of the Balmer
series are also useful ($\mathrm{H_{\beta - \delta}}$). As the strength of these lines also
depends on $T_{\rm eff}$ the accuracy of our results depends on how
well $T_{\rm eff}$ can be constrained from other diagnostic
lines. The emission line strengths of $\mathrm{H_{\alpha}}$,
$\mathrm{Br}_{\gamma}$, and He\,{\sc ii}\,$\lambda2.189\mathrm{\mu
  m}$ roughly show a linear dependence on $T_{\rm eff}$. He\,{\sc
  ii}\,$\lambda4686$ shows a much stronger sensitivity at certain
temperatures (cf.\,Fig.\,\ref{f:temp}) and also depends on gravity 
(Fig.\,\ref{f:logg_highT}). In the absence of IR data large uncertainties in the derived He abundance may occur. In
Fig.\,\ref{f:he} we show the sensitivity of the diagnostic lines with
respect to the Helium mass fraction. The mass-loss rates determined
from pure emission lines scale with the square root of the adopted
volume-filling factor $\sqrt{f_{\rm v}}$ (cf.\ Eq.\,\ref{trans}). In contrast to
this, absorption lines, which form below the region where
we assume clumping to start, are affected by the actual mass-loss
rate and are thus independent of clumping.

To constrain $T_{\rm eff}$, we use He\,{\sc i}\,$\lambda4471$, as well
as N\,{\sc iii} $\lambda4634/4640$, N\,{\sc iv}\,$\lambda4058$,
N\,{\sc v} $\lambda4604/4620$. The He\,{\sc i}\,$\lambda4471$
absorption line is a classical temperature indicator for O stars,
however, it is only available for temperatures up to
$T_{\mathrm{eff}}\approx45$\,kK. For this reason we use the nitrogen
lines as an additional indicator. The ionisation stages of nitrogen cover 
N\,{\sc iii} to N\,{\sc v} and can thus be used
over the whole temperature range investigated here. However, lines from these ions 
are usually in emission, which means their analysis is
complicated by the effects of mass-loss and abundance variations
\citep[cf.][]{RPMN2012}. Moreover, their strength depends on detailed
atomic physics, such as dielectronic recombination channels
\citep{mihalas1973}. Finally, line overlaps between N\,{\sc iii}
$\lambda4634/4640$, N\,{\sc v} $\lambda4604/4620$ and He\,{\sc
  ii}\,$\lambda4686$ can lead to difficulties, particularly for high
mass-loss rates.

The terminal wind velocities ($\varv_{\infty}$) are best derived from
the P-Cygni profiles of UV resonance lines, in particular the C\,{\sc iv} $\lambda 1548/1551$ doublet. For objects without
available UV spectroscopy we estimate $\varv_{\infty}$ from the width
of the $\mathrm{H_{\alpha}}$ emission line by a detailed comparison
with synthetic line profiles. For VFTS\,016, 482, 1022, and
  MK42 we could use the traditional black absorption edge method
  \citep[e.g.][]{prinja1990, prinja1998} to test the
  $\varv_{\infty}$ measurements based on spectral modelling for
  $\mathrm{H_{\alpha}}$ and the C\,{\sc iv} doublet. For WNh stars of
late subtype (WN6-8h) the He\,{\sc i}\,$\lambda4471$ line has a
P-Cygni type profile and can be used to derive $\varv_{\infty}$ with
  similar accuracy as from the C\,{\sc iv}\,$\lambda 1548/1551$
  doublet.  For O and Of stars $\mathrm{H_{\alpha}}$ becomes too weak
to estimate $\varv_{\infty}$.  In these cases we adopt the values from
the relation by \cite{lamers1995} (cf.\,\S\,\ref{s:vinfty}).

  To constrain the surface nitrogen enrichment due to the
  CNO-cycle we performed a subjective comparison of the overall
  strength of the observed nitrogen features with our models.  We
  tentatively classified each object as N-enhanced ($e$), N-normal
  ($n$) or partially enhanced ($ne$). The results are listed in
  Tab.\,\ref{t:parameters}.  For He-enriched stars ($Y>0.25$) we found
  that the strength of the observed nitrogen features is generally in
  good agreement with an enhanced N abundance ($e$), as adopted in our
  models (cf.\,Tab.\,\ref{t:abundance}). For stars with normal He
  ($Y=0.25$), we computed N-normal and N-enhanced models
  (cf.\,Tab.\,\ref{t:abundance}) and found that the observed
    nitrogen features indeed reflect a variety of nitrogen
    abundances.

\subsection{Luminosity and interstellar extinction\label{s:lum_red}}

To determine the bolometric luminosities ($L_{\star}$) of our sample stars we use
the model flux as obtained from our grid computations (which is
computed for a luminosity of $10^6\,L_\odot$) and scale it to match
the observed SED in the $B$, $V$ and $K_{\mathrm s}$ bands
(cf.\,Sect.\,\ref{s:phot}).  Simultaneously we determine the
extinction parameters $R_\mathrm{V}$ and $A_\mathrm{V}$ as described
in Sect.\,\ref{s:target}, i.e., we compute
the intrinsic brightness in the $B$, $V$ and $K_{\mathrm s}$ bands
from the model flux using appropriate filter
functions\footnote{We approximate the filter functions for
  each filter system individually using Gaussian and boxcar functions
  with appropriate widths and wavelengths.} and match the resulting
values for $E(B-V)$ and $E(V-K_{\mathrm{s}})$ as described by
\citet{bestenlehner2011}. The results are given in
  Tables~\ref{t:parameters} and \ref{t:parameters_ex}.

As the extinction is the smallest in the near-IR range the resulting
luminosities rely chiefly on the $K_{\mathrm s}$-band photometry.
$R_\mathrm{V}$ and $A_\mathrm{V}$ mainly follow from $B$ and $V$ in
relation to $K_{\mathrm s}$.  This method thus leads to very reliable 
$L_\star$ as long as the $K_{\mathrm s}$ magnitude is not
contaminated by other sources, such as thermal dust emission or nearby
stars. To avoid such instances we generally compare the reddened model
flux for each object with the complete available photometry, in some
cases reaching from the UV to the far infrared
(cf.\,Appendix\,\ref{a:plots}).  A similar method has previously been
used by \citet{hamann2006} for the analysis of Galactic WR stars.
 As visual and UV fluxes are strongly affected by deviations
  from a standard extinction, i.e. $R_\mathrm{V} = 3.1$, this method
  helps to avoid large errors in the luminosity determination that can
  arise from using these wavelength ranges alone. A good example for
  such a case is VFTS\,682 for which \citet{bestenlehner2011} obtained
  $\log{L/L_\odot}=6.5$ based on the derived $R_{\rm V}=4.7$, while a
  standard extinction law with $R_{\rm V}=3.1$ would suggest a much lower
  luminosity of $\log{L/L_\odot}=5.7$.

Uncertainties in our analysis chiefly arise from contributions of
  unresolved neighbouring stars. Such occurrences are further complicated by
  differences in the spatial resolution of optical and near-IR
  photometry.  For emission-line stars the use of broad band
  photometry can cause problems
  if the line contribution is not correctly taken into account.
  However, as we know the line contribution from our models and take
  it into account using appropriate filter functions these
  uncertainties should be minimised. In some cases we still obtain
  unrealistic $R_\mathrm{V}$ from our analysis (e.g.\ VFTS\,402 and
  147). These problems are most likely caused by unresolved
  companions. Stars for which this happens are excluded from our
  discussion (cf.\ Sect.\,\ref{wind_momentum}) and their parameters
  are given in Table\,\ref{t:parameters_ex}.  As these
    problems only affect the optical extinction the derived
    luminosities may still give reasonable estimates for the main
    contributor in the IR, where the extinction is low.

  Furthermore, many of our sample stars show an apparent UV mismatch
  in the SED plots in Appendix\,\ref{a:plots}. In the majority of
  cases this is caused by our use of a Galactic extinction law over
  the full wavelength range from UV to IR. Specific LMC
    extinction laws \citep[as e.g. from][]{Howarth1983} are indeed
    much more appropriate for the UV range and lead to better UV fits.
    In particular the UV 2200\,\AA\ feature appears to
      depend strongly on metallicity and is much weaker in the LMC.
      However, deviations from the standard extinction with $R_V=3.1$
      have only been thoroughly investigated for the Galaxy
      \citep[e.g.][]{cardelli1989}. This particularly affects the
      transition from the IR to the optical range and the overall
      slope of the extinction law in the UV, which is why we use the
      extinction law by \citeauthor{cardelli1989} in this work.  The
      UV mismatch does not affect our results as these are only based
      on optical/IR photometry, and is thus a purely cosmetic issue
      following from the adopted extinction law.

\subsection{Results and error discussion \label{s:error}}

The results of our analyses are compiled in Tables~\ref{t:parameters} and \ref{t:parameters_ex}.
To discuss the uncertainties we need to consider that the analysis is
performed in two steps. In step one the line spectrum is analysed
using the method described in Sect.\,\ref{spec_an}. In this step the
following parameters are determined: the transformed mass-loss rate
($\dot{M}_{\rm t}$), the effective temperature ($T_{\mathrm{eff}}$), the
surface helium mass fraction ($Y$), and the terminal wind velocity
($\varv_{\infty}$). In step two the luminosity ($L_{\star}$) and the
extinction parameters $A_{\rm V}$ and $R_{\rm V}$ are determined by
fitting the observed SED (cf.\,Sect.\,\ref{s:lum_red}).  The actual
mass-loss rates ($\dot{M}$) then follow from the scaling relation
Eq.\,\ref{trans}, which additionally depends on the adopted volume-filling
factor $f_{\rm v}$. The uncertainties in $T_{\mathrm{eff}}$, $Y$ and $\dot{M}_{\rm t}$ are
dominated by the limited resolution of our model grid. We estimate
values of $\Delta \log T_{\mathrm{eff}} = \pm 0.02$\,dex (corresponding
to $\pm 2$\,to\,$2.5$kK), $\Delta Y = \pm5$\% and $\Delta \dot{M}_{\rm t} =
\pm0.1$\,dex. For the terminal wind speed the measurement error is of
the order of $\Delta \varv_{\infty}=\pm 200$ to $400$\,km\,s$^{-1}$.

Moreover, systematic errors are introduced in our grid analysis
because the wind acceleration parameter $\beta$ and the surface
gravity $\log g$ are fixed. By changing $\beta$ from 1.0 to 1.5
\citep[as suggested e.g.\ by model computations by][]{vink2011} the
wind acceleration zone is extended and the wind density is increased in
the line-forming region. To compensate for this effect $\dot{M}_{\rm t}$
needs to be reduced by 0.1 to 0.2\,dex. The slope of the SED is mainly
affected in the optical range, impacting $R_V$ but leaving the
luminosity almost unchanged ($\Delta \log L \sim$+0.005~dex).

Changes due to variations in $\log g$ are investigated in
  Appendix\,\ref{a:logg}. A change in $\log g$ can influence the
  density and ionisation balance in the hydrostatic part of the
  atmosphere. For stars with optically thin winds this may affect the
  derived helium and nitrogen abundances. The impact on the latter turns out to be
  negligible, mainly because we only coarsely estimate the nitrogen enrichment
   based on the overall strength of the nitrogen lines (see
  last paragraph in \S\,\ref{s:lines}).  For stars with optically
  thick winds changes of the hydrostatic structure have practically no
  effect.

In Fig.\,\ref{f:logg_lowT}, we show the effect of lowering $\log g$ by
0.5\,dex in a critical temperature range around $T_{\rm eff}=40$\,kK.
In this range lowering the gravity decreases the strength of the main
temperature diagnostic He\,{\sc i}\,$\lambda4471$. Consequently a
lower temperature by $\sim$\,2500\,K is needed to reproduce this line.
 While the transformed mass-loss rate remains almost unaffected
  the derived helium mass fraction is lowered by 2...4\%. This
  systematic uncertainty affects our given errors of $\Delta Y = \pm
  5\,\%$ only marginally. As a result of the lower $T_{\rm eff}$ we
derive a lower luminosity.  Moreover, the terminal wind speed
$\varv_\infty$ is affected for the cases where it is derived from the
relation by \cite{lamers1995}.  The resulting change of the
  absolute mass-loss rate $\dot{M}$ as derived from Eq.\,\ref{trans}
  is of the order of $0.15$\,dex and is taken into account in the given
  error range.  The resulting systematic changes are indicated by a
vector in Fig.\,\ref{f:wm_giants} (see \S\,\ref{s:discussion}). We
note that they only affect the O supergiants in our sample, as these
objects have significantly lower $\log g$ than 4.0. 

  As we are mainly interested in a homogeneous analysis of the
  mass-loss behaviour across the transition region between O stars,
  Of/WN, and WNh stars, and it is not possible to determine $\log g$
  for the WNh stars, we decided to use the same analysis method for all
  objects in our sample, i.e., to keep $\log g$ fixed.  A detailed
  determination of $\log g$ and N-abundances of all O stars in the
  VFTS sample will be performed in future works (Ramirez-Agudelo et al., Sab\'{i}n-Sanjuli\'{a}n et al.,~in prep.).

The derived luminosities depend on the photometry (mainly in the near-IR)
and the model SED, which depends on $T_{\rm eff}$ and $\dot{M}_{\rm
  t}$. The photometric errors are of the order of $\pm 0.05$\,dex.
Considering the resolution of our model grid, the resulting
uncertainties are of the order of $\Delta \log L_{\star} \approx \pm
0.1$\,dex.

In summary, we estimate uncertainties of $\Delta \log \dot{M}_{\rm t} =
\pm0.1$\,dex, $\Delta \log T_{\mathrm{eff}} \approx \pm0.02$\,dex, $\Delta
\varv_{\infty} \approx \pm 200$ to $400$\,kms$^{-1}$ (see also
\S\,\ref{s:vinfty}) and $\Delta \log L_{\star} \approx \pm 0.1$\,dex. The
resulting mass-loss rates $\dot{M}$ according to Eq.\,\ref{trans} are
uncertain by $\Delta \log \dot{M}\approx \pm0.2...0.3$\,dex considering the uncertainties in $\log g$ and $\beta$, and additionally
scale with the square root of adopted volume-filling factor $\sqrt{f_{\rm v}}$.

\subsection{Comparison with previous works}
\label{s:comp}

  In our present work we tried to follow a well-defined method
  of analysis as described in Sects.\,\ref{spec_an}, \ref{s:lines} and
  \ref{s:lum_red}. In this section we present comparisons with
  previous works to get a better picture of the systematic
  uncertainties that can arise from different analysis methods.

  A comprehensive study of WR stars in the LMC has recently been
  published by \citet{Hainich2014}, who have analysed 17 stars
  from our present sample. These authors used a similar grid approach
  as in our work, based on the Potsdam Wolf-Rayet (PoWR) model
  atmosphere code \citep{Koesterke2002,Graefener2002,Hamann2003}. An
  important difference with respect to our work is how the stellar
  luminosities and interstellar extinction are determined.  As
  described in Sect.\,\ref{s:lum_red} our results mainly rely on IR
  $K_{\rm s}$-band photometry. The optical flux in $B$ and $V$ is
  matched simultaneously with the IR by adapting $R_V$ and $E(B-V)$ in
  the extinction law.  Hainich et al.\ keep $R_V$ fixed and mainly use
  the optical to UV flux to determine $E(B-V)$ based on the LMC
  extinction law by \citet{Howarth1983}. As discussed in
  Sect.\,\ref{s:lum_red} this approach may lead to substantial errors
  if $R_V$ deviates from the adopted value. The analysis of Hainich et
  al.\ largely relies on the UV range where the uncertainties in the
  extinction are very high. Our results rely on the IR where the
  extinction is almost negligible, however, there is a danger of
  contamination by circumstellar dust or other IR sources.

  An illustrative case is VFTS\,482 (BAT99-99, Mk\,39). For this star
  Hainich et al.\ obtain a luminosity of $\log(L/L_\odot)=5.9$ while
  our value lies significantly higher ($\log(L/L_\odot)=6.4$). As
  described above Hainich et al.\ do indeed substantially
  underestimate the IR flux for this object. However, there is
  possible evidence for crowding in HST images of this star (cf.\
  Table~\ref{t:spectra}). Could the excess IR flux thus be due to
  other sources? Our SINFONI IR data of VFTS\,482 suggests that this
  is not the case as the IR spectrum shows the correct line strength,
  i.e., it is unlikely that the IR is contaminated by other sources.
  The same holds for the optical range for which a comparison between
  HST and UVES spectroscopy shows no sign of contamination.  For
  VFTS\,482 we are thus confident that our approach is correct.

  Furthermore, the different wavelength coverage and larger S/N ratio
  of the VFTS data can explain discrepancies arising from differences
  in the line diagnostics. This partly leads to different
  He-abundances (VFTS\,482 and 545) and temperatures ($\Delta T
  \gtrsim 10\,000$K, e.g.\ for VFTS\,545, 617, 1017, and 1025).  After
  correcting for clumping, there is also a systematic offset of 0.1 to
  0.2\,dex in the mass-loss rates noticeable. This difference lies
  within our error bars and is most likely caused by different
  assumptions in the model physics and atomic data.

  VFTS\,\,1025 (R\,136c) has previously been analysed by
  \citet{crowther2010}. This star may be affected by crowding as it is
  located near the core of R\,136.  Due to the high S/N in the VFTS
  data we detect a weak photospheric He\,{\sc i}\,$\lambda4471$
  absorption line in the optical spectrum which is reproduced by our
  models. Based on this line we determine a lower temperature for this
  object than \citet{crowther2010}. It is not clear whether this line
  is intrinsic or stems from a nearby star (see the detailed
  discussion in Appendix \ref{s:targets}). We obtain
  $T_\star=42\pm2$\,kK, $\log(L/L_\odot)=6.6\pm0.1$ while
  \citeauthor{crowther2010} obtain $T_\star=51\pm5$\,kK,
  $\log(L/L_\odot)=6.75\pm0.1$. The obtained mass-loss rates and
  He-abundances are very similar. The differences have practically no
  effect on our results in Sect.\,\ref{s:discussion}.

  VFTS\,016 has been analysed by \cite{evans2010}. Their model has
  a lower surface gravity which results in a lower $T_{\rm eff}$ and
  luminosity compared to our values. The mass loss and He-abundance
  are in good agreement.

  VFTS\,072 has been analysed by \cite{rivero2012}. They derived a
  higher $T_{\rm eff}$ that might be a result of the higher gravity in
  their models. Mass-loss rate, luminosity and He-abundance are in
  good agreement.

VFTS\,482 has also been analysed by \cite{massey2005}. We derived a
$\sim$\,4000K higher $T_{\rm eff}$. The luminosity is higher by about
0.6\,dex. \cite{massey2005} used a standard $R_{\rm V} = 3.1$
and estimated the extinction $E(B-V)$ by averaging the colour excesses
in $B-V$ and $U-B$ based on the spectral type. In this work we
calculate the $R_{\rm V}$ (see \S\,\ref{s:lum_red}). The mass loss 
is higher by 0.3\,dex while the He-abundance is in agreement.

Mk\,42 has been analysed by \cite{puls1996}. The derived $T_{\rm eff}$
is about 3000\,K higher while the He-abundance is a bit lower compared
to our values. The luminosity agrees and our mass-loss rate is 
about 0.1\,dex lower.

  We conclude that the stellar parameters derived in this
  work largely agree with previous studies with some important
  exceptions due to systematics.  The main systematic uncertainties
  arise from the uncertain interstellar extinction (in
    particular because of its dependence on $R_V$), and from problems
  with crowding due to the large distance to the LMC.

\section{Discussion\label{s:discussion}}

In Sect.\,\ref{s:subclass} we start our discussion with a division of
our sample into sub-classes based on the spectral classifications in
Table~\ref{t:parameters}. The method by which we estimate masses and
Eddington factors for each object is discussed in \S\,\ref{s:edd}. In
\S\,\ref{s:vinfty} we compare the observed terminal wind velocities
obtained with different methods within this work with commonly adopted
relations.  In the subsequent sections we discuss the mass-loss
properties of our sample stars, namely their wind momenta
(\S\,\ref{wind_momentum}), the dependence of the mass-loss rate on the
Eddington factor (\S\,\ref{mdot_gamma}), and their wind efficiencies
and clumping factors (\S\,\ref{efficiency}). A comparison with VMS
evolutionary models is performed in
\S\,\ref{s:evo_stages}.  The discussion closes with a brief section
of the ionising fluxes (\S\,\ref{s:fluxes}).

\subsection{Definition of stellar sub-classes\label{s:subclass}}

  In the following we divide our sample into sub-classes with
  different emission line strengths. Based on the spectral
  classifications in Table~\ref{t:parameters} we designate stars with
  spectral types WNh and WN(h) as ``WNh stars'', stars with mixed
  Of/WN spectral types as ``Of/WN stars'', O stars in categories f and
  (f) as ``Of stars", and O stars in categories ((f)) or without
  category as ``O stars". If we discuss O stars in general we
  designate them as ``O-type stars". Using these designations the
  group of O stars is dominated by dwarfs with low $\dot{M}_{\rm t}$,
  and the group of Of stars by giants and supergiants with
  considerably higher $\dot{M}_{\rm t}$.

\subsection{Stellar masses and Eddington factors}\label{s:edd}

To characterise the proximity of our sample stars to the Eddington
limit we use a method introduced by \citet{graefener2011}. The
Eddington factor ($\Gamma$) is defined as the ratio between radiative
acceleration ($g_{\mathrm{rad}}$) and gravitational acceleration
($g$), $\Gamma =
g_{\mathrm{rad}}/g$.  In spherical symmetry it can be
expressed as
\begin{equation}
\label{Gamma}
  \Gamma(r) = \frac{g_{\mathrm{rad}}}{g} = \frac{\chi(r)}{4 \pi cG} \frac{L_{\star}}{M_{\star}}.
\end{equation}
Here $\chi(r)$ denotes the total flux-mean opacity per gram of material. 
In general, $\chi$ is a complex function of density and temperature.
For this reason $\Gamma$ depends on radius, i.e.\ it is impossible to
find a single value of $\Gamma$ which characterises the proximity of a
star to the Eddington limit. For the hot atmospheres that we
  consider here H and He are largely ionised and electron scattering
  is a dominant contributor to the total opacity. As the electron
  scattering opacity $\chi_{\mathrm{e}}$ depends only on the density
  of free electrons it is almost independent of the radius. For this
  reason we use the {\it classical} Eddington factor
\begin{equation} \label{e:cEdd} 
  \Gamma_{\rm e} =
  \frac{g_{\mathrm{e}}}{g} =
  \frac{\chi_{\mathrm{e}}}{4 \pi cG} \frac{L_{\star}}{M_{\star}}
\end{equation}
in our discussion in Sect.\,\ref{mdot_gamma}. Assuming a fully ionised plasma $\chi_{\mathrm{e}}$ depends only on the hydrogen mass fraction $X_{\rm
    s}$ at the stellar surface, and $\Gamma_{\rm e}$ can be expressed by
  the relation
\begin{equation}\label{e:eddington_factor}
  \log \Gamma_{\rm e} = -4.813 + \log(1 + X_{\rm s}) + \log(L/L_{\odot}) - \log(M/M_{\odot}).
\end{equation}
In this equation $L$ and $X_{\rm s}$ are determined in our spectral
analyses, i.e.\ the only unknown is the stellar mass $M$.

To estimate stellar masses for given values of $L$ and $X_{\rm s}$
\cite{graefener2011} computed mass-luminosity relations for chemically
homogeneous stars. In this context it is important to know that hot
massive stars have large convective cores. In this case the luminosity
of a star with given mass $M$ is chiefly determined by the hydrogen
mass fraction $X_{\rm c}$ in the convective core. As $X_{\rm c}$ is a
priori unknown, \citeauthor{graefener2011} made two extreme
assumptions about $X_{\rm c}$. The first assumption is $X_{\rm
  c}=X_{\rm s}$. In this case the star is chemically homogeneous, and
the derived mass provides an upper limit for the real mass of the
star. The second assumption is $X_{\rm c}=0$. In this case the
hydrogen in the core is exhausted and the star is in the phase of core
He-burning. The mass derived in this way provides a lower limit to the
real mass of the star.

Chemical homogeneity is expected for stars with strong rotational
mixing and/or strong mass-loss, i.e.\ when the timescales of mixing or
mass-loss are shorter than the nuclear burning timescale.  E.g., for
the extremely massive WNh stars R136a1, a2, a3 and c,
\citet{crowther2010} estimate present day masses of 265, 195, 135 and
175\,$M_\odot$ using evolutionary models. The corresponding
homogeneous masses following \citet{graefener2011} are 286, 206, 154
and 185\,$M_\odot$. The agreement is better than the error-margins on
the stellar masses given by \citeauthor{crowther2010}. Mass estimates
following \citet{graefener2011} for our sample stars are given in
Table~\ref{t:parameters} and \ref{t:parameters_ex}. A comparison with
evolutionary models for very massive stars from \citet[][submitted]{koehler2014} gives values that are on average $\sim$\,30\% lower. This is to
be expected, in particular for O stars with H-rich surface
compositions. Many of these stars will have evolved away from the
zero-age main-sequence (ZAMS), i.e.\ they are not chemically
homogeneous. The Eddington factors estimated here for individual stars
are thus likely higher in reality, in particular for relatively low
stellar masses.

 The use of evolutionary masses for all stars in our sample turned
  out to be problematic, in particular for stars with helium-enriched
  surface compositions. In these cases the evolutionary models do not
  always match the observed abundances, leading to qualitatively wrong
  estimates of $\Gamma_{\rm e}$ following
  Eq.\,\ref{e:eddington_factor} (the resulting masses are partly
  higher than the upper limits obtained under the assumption of
  homogeneity).
  The strength of our method is therefore to obtain model-independent
  qualitative information about a $\Gamma$-dependence, based on
  observed luminosities and surface compositions of a large stellar
  sample. Nevertheless, we will test the effect of our assumptions on
  our results for the H-rich O stars in our sample in
  Sect.\,\ref{mdot_gamma}.

\subsection{Terminal wind velocities}\label{s:vinfty}
\begin{figure}
\begin{center}
\resizebox{\hsize}{!}{\includegraphics{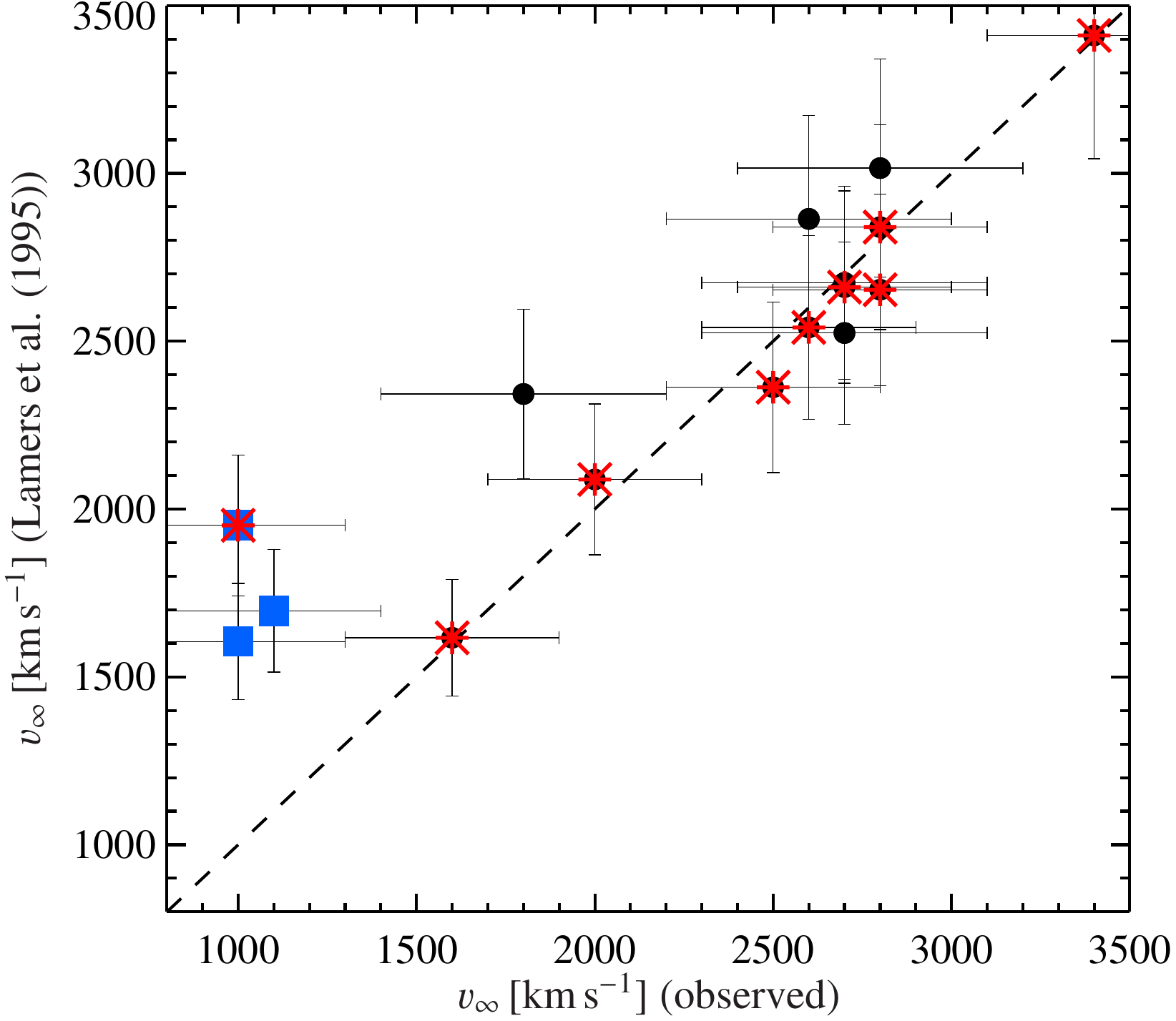}}
\end{center}
\caption{Observed vs.~predicted terminal velocity using the wind velocity relation by \cite{lamers1995}. Black dots represent measurements using  $\rm{H}_{\alpha}$; red asterisks are for measurements based on C\,{\sc iv} and blue squares are late WNh/(h)6-8-stars with He\,{\sc i} measurements.}
\label{f:vinf}
\end{figure}

The terminal velocities ($\varv_{\infty}$) given in
Table~\ref{t:parameters} and \ref{t:parameters_ex} are determined on the basis of different
lines. Dependent on their availability we use P-Cygni line profiles of
C{\sc iv}$\lambda1550$ or He\,{\sc i}\,$\lambda4471$, or the width of
$\mathrm{H_{\alpha}}$ (cf.\,Sect.\,\ref{spec}). If no diagnostic lines
are available we use the empirical relation for O stars with
$T_{\mathrm{eff}} > 25\,000\mathrm{K}$ from \cite{lamers1995}, i.e.\
$\varv_{\infty} / \varv_{\mathrm{esc}}^{\mathrm{eff}}=2.51 \pm 0.27$.
Here the effective escape velocity
($\varv_{\mathrm{esc}}^{\mathrm{eff}}$) is defined as
\begin{equation}\label{e:vesc}
  \varv_{\mathrm{esc}}^{\mathrm{eff}}=\sqrt{2GM_{\star}(1- \Gamma_{\rm e})/R_{\mathrm{ref}}}
\end{equation}
with $\Gamma_{\rm e}$ from Eq.\,\ref{e:eddington_factor}.
$R_{\mathrm{ref}}$ is defined as $L_{\star}=4\pi \sigma
R_{\mathrm{ref}}^2 T_{\mathrm{eff}}^4$. These values are also given in Tables~\ref{t:parameters} and \ref{t:parameters_ex}.

In Fig.\,\ref{f:vinf} we compare our results with this relation. The
comparison suggests that the relation by \cite{lamers1995} can be used
for the Of, Of/WN, and WN5h stars in our sample (Fig.\,\ref{f:vinf}).
For late WN6-8h stars we find lower values than predicted. That said, the
relation from \citeauthor{lamers1995} is most suitable for the O and Of stars in our sample for which direct
observational values are lacking.

The comparison indicates uncertainties in $\varv_{\infty}$ of $\pm$200
to 300\,km\,s$^{-1}$ using the C\,{\sc iv} P-Cygni profile.  Similar
errors are obtained by deriving $\varv_{\infty}$ from the He\,{\sc i}
line for late WNh stars. A comparison of C\,{\sc iv} and He\,{\sc i}
measurements shows a good agreement. The errors for measurements based
on the width of $\mathrm{H_{\alpha}}$ are of the order of $\pm$300 to
400\,km\,s$^{-1}$ depending on the quality of the model fit.

\subsection{Wind Momentum-Luminosity relation\label{wind_momentum}}

\begin{figure}[t]
\begin{center}
\resizebox{\hsize}{!}{\includegraphics{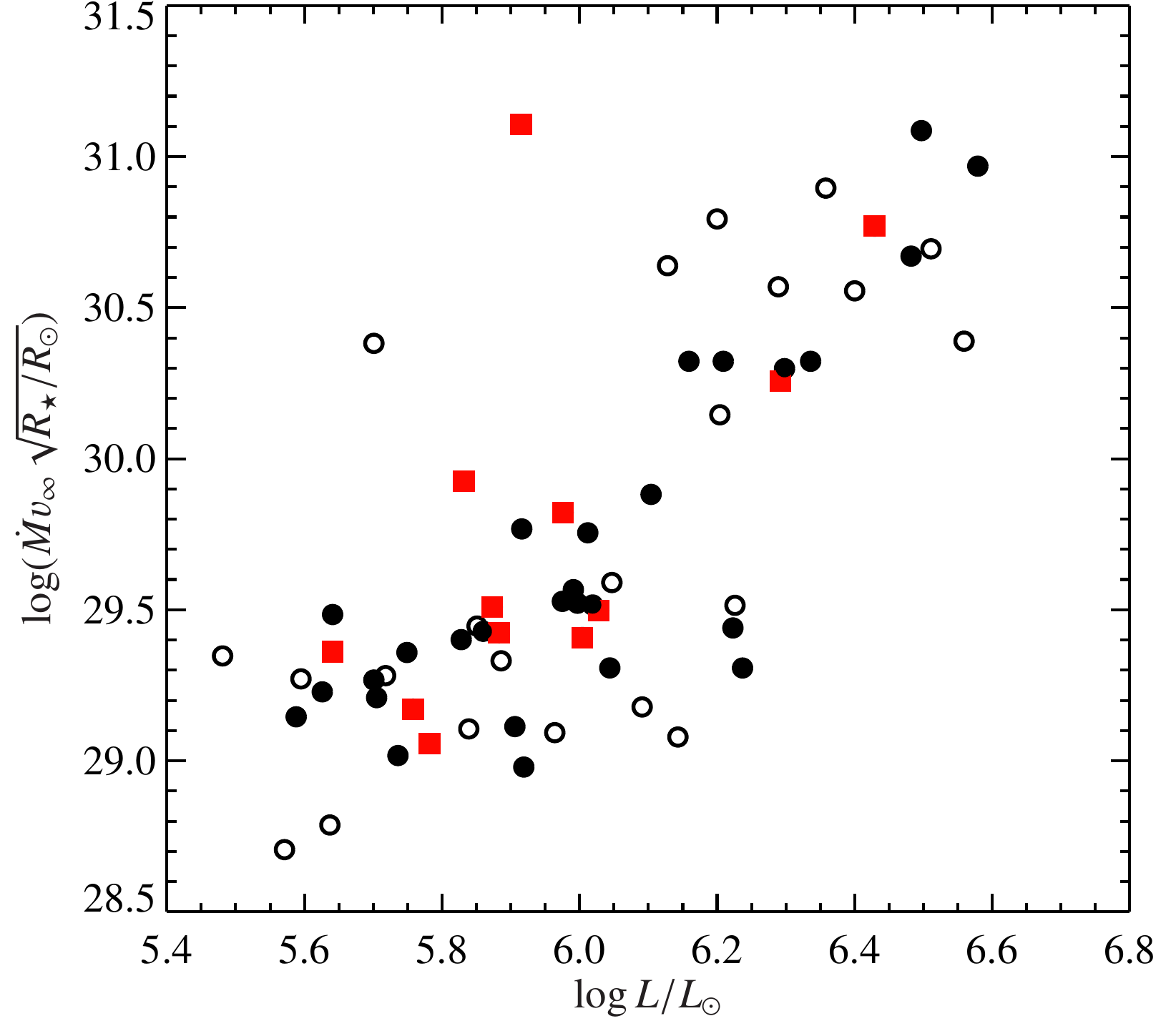}}
\end{center}
\caption{Unclumped wind momentum versus luminosity. Black open
  circles are single stars. Black filled circles are stars with weak
  radial velocity variations or are single lined spectroscopic binaries (SB1). The red squares indicate stars
  which have a spectroscopic binary (SB2) or for which the spectra are
  contaminated by nearby stars due to crowding.}
\label{f:windmomentum}
\end{figure}

\begin{figure}[t]
\begin{center}
\resizebox{\hsize}{!}{\includegraphics{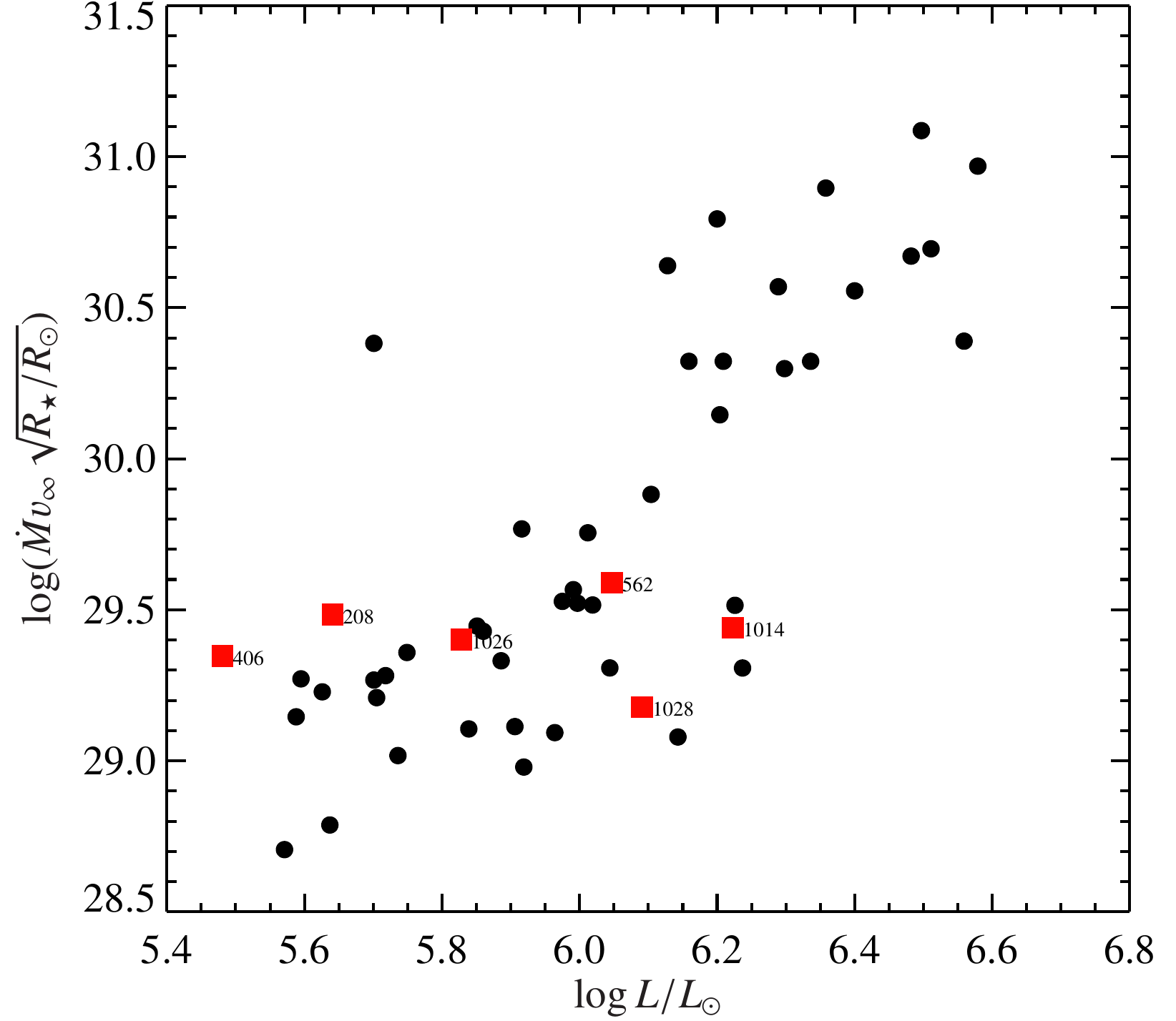}}
\end{center}
\caption{Unclumped wind momentum versus luminosity. Black filled
  circles: stars included in the analysis. The red squares indicate
  stars which have been removed as a result of an uncertain stellar
  parameter determination.}
\label{f:wm_up}
\end{figure}
\begin{figure}[t!]
\begin{center}
\resizebox{\hsize}{!}{\includegraphics{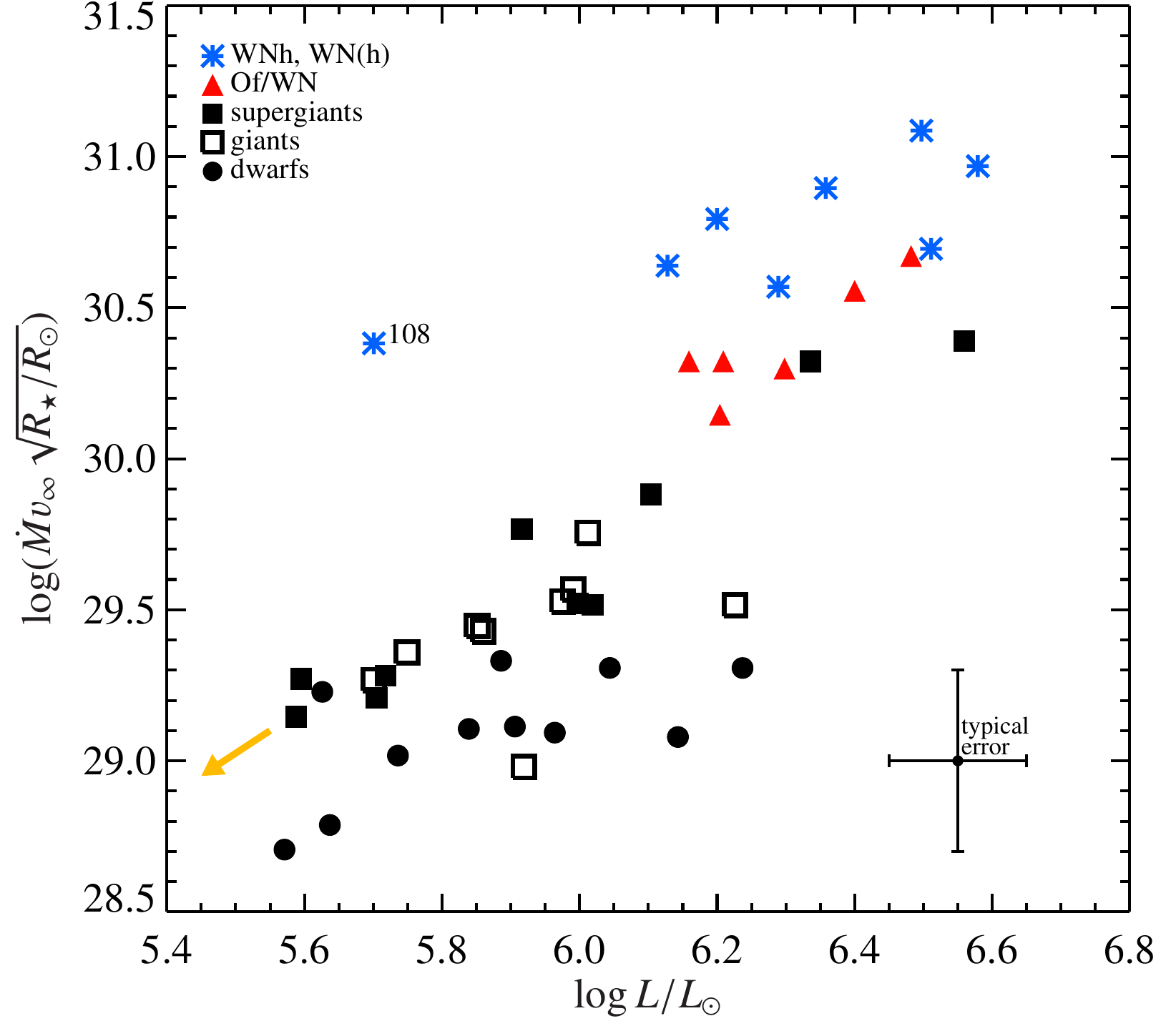}}
\end{center}
\caption{Unclumped wind momentum versus luminosity. Black filled
  circles are dwarfs (luminosity class {\sc V}), open black squares
  are giants ({\sc III-IV}), filled squares are supergiants ({\sc
    I-II}), red triangles are Of/WN stars, and blue asterisks are
  WNh and WN(h) stars. The yellow arrow indicates the shift of an O star, if a
  gravity lowered by 0.5\,dex would be used for the analysis of the
  (super)giants. The typical error $\Delta \log (\dot{M} \varv_{\infty} \sqrt{R_{\star}/R_{\odot}})$ is $\pm0.3$\,dex and $\Delta \log L$ is $\pm0.1$\,dex.}
\label{f:wm_giants}
\end{figure}

The wind momentum of a stellar wind is given by the product of its
mass-loss rate and terminal wind velocity ($\dot{M} \varv_{\infty}$).
\cite{kudritzki1999} introduced the modified wind momentum $\dot{M}
\varv_{\infty} \sqrt{R_{\star}/R_{\odot}}$, as this quantity is
expected to be almost independent of the stellar mass, i.e.\ it
depends chiefly on the luminosity of the star.  \cite{puls1996} and 
\cite{repolust2004} found a tight wind-momentum luminosity relation (WLR)
for O stars and supergiants. \cite{martins2008} studied the most
massive stars in the Arches cluster and found two WLRs, for O stars
and WNh stars. \cite{mokiem2007} investigated the WLR for a sample of
28 OB stars in the LMC based on the data of the VLT-FLAMES Survey of
Massive Stars \citep{evans2005}. Their sample included stars in the
luminosity range of $\log L/L_{\odot} = 4.5...6.0$. 

In this work we focus on LMC stars with even higher luminosities in
the range between $\log L/L_{\odot} = 5.5$ and 6.6, including Of/WN
and WNh stars. The mass-loss rates given here are for the case without
clumping, to facilitate a comparison with earlier works. This means
that even if our modelling indicates a clumped wind structure, e.g.\
based on the electron scattering wings of some emission lines, we
correct them for clumping according to Eq.\,\ref{trans}. For our
models with $f_{\rm v}=0.1$ this means that $\dot{M}$ has to be
multiplied with a factor of $\sqrt{10}$.

In Fig.\,\ref{f:windmomentum} we show the WLR for all stars in our
sample. Stars whose spectra are affected by multiplicity (i.e.\ SB2's
and stars in crowded fields) are indicated by squares. Although for the
following discussion these objects are removed from our sample, it is
remarkable that even though these objects are clearly contaminated by
other stars the resulting WLR is hardly affected. Black symbols
indicate SB1's or stars with weak RV variations. We assume that the
spectra of these objects are dominated by one star and we treat
  them as single stars in our analysis. The corresponding errors in
  the derived luminosities are most likely negligible compared to
  the impact of the photometric errors.

In Fig.\,\ref{f:wm_up} we indicate stars with highly uncertain stellar
parameters. Two of these stars have peculiar spectra with possible
indications of fast rotation (VFTS\,208 and 406).
For the rest some essential diagnostic features are missing or are of
poor quality (VFTS\,562, 1014, 1026, 1028). 
In total we excluded six objects from the following discussion (see Appendix\,\ref{a:plots} for more details).

In Fig.\,\ref{f:wm_giants} we show the WLR for the remaining sample
indicating dwarfs, giants, and supergiants. To investigate
the systematic effects of our choice of $\log\,g=4.0$ in our grid
computations we highlight the positions of O and Of (super)giants in
the WLR. The arrow indicates the estimated shift in the WLR if we
adopt a more realistic value of $\log(g)=3.5$ for these objects
(cf.\,Sect.\,\ref{s:error}). The resulting shift of
  -0.1\,dex in $\log(L)$ and -0.15\,dex in the modified wind momentum
  is also confirmed by an individual model for the O8\,Ib(f) star
  VFTS\,669 with $\log g = 3.5$ (cf.\ Figs.\,\ref{a:669} and
      \ref{a:669_test}). Therefore the corresponding objects would
move along the WLR towards lower luminosities and wind-momenta.  Their
qualitative position on the WLR is thus unaffected by the change in
$\log\,g$, i.e.\ the (super)giants still form a group along the WLR,
but with systematically lower wind momenta. For O dwarfs the
  effect may go exactly in the opposite direction as their gravities
  may be up to 0.2...0.3 dex higher than our adopted value.  Also
  those stars would remain on the {same} WLR.

\begin{figure}[t!]
\begin{center}
\resizebox{\hsize}{!}{\includegraphics{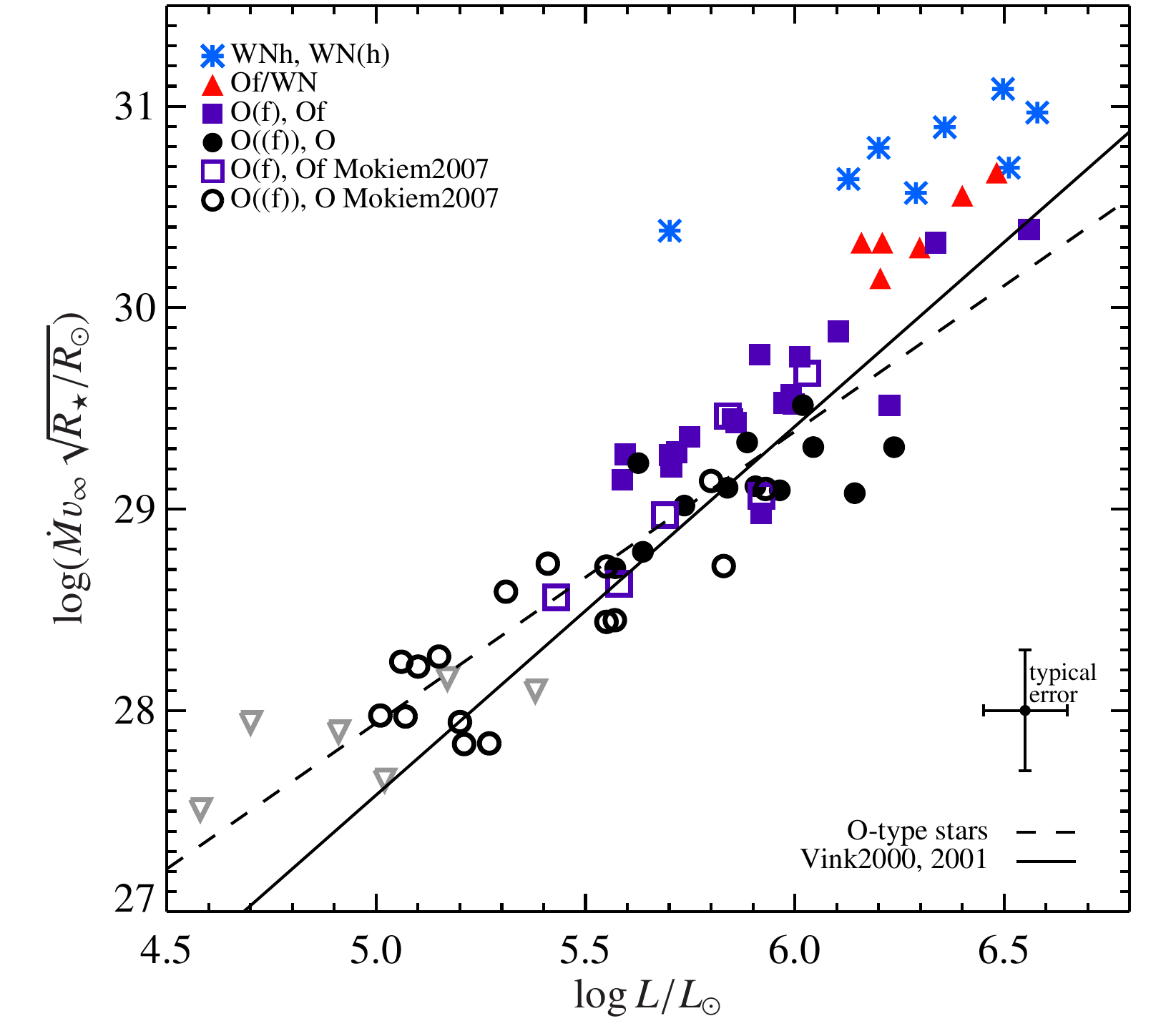}}
\end{center}
\caption{Unclumped wind momentum versus luminosity for the
    combined sample from this work and \cite{mokiem2007}. The
  different symbols indicate stellar sub-classes following
  Sect.\,\ref{s:subclass}. The solid line shows the theoretical
  prediction by \cite{vink2000, vink2001}.  The dashed line is a fit
  through the O-type stars including both samples. The grey triangles
  are stars from \cite{mokiem2007} that have only an upper limit in
  $\dot{M}$ and are excluded from the fit. }
\label{f:wm_unique}
\end{figure}

 \begin{table*}[t]
\caption{WLR coefficients.}
\label{t:wm}
\centering
\begin{tabular}{lccc}
\hline
\hline
				&	$m_0$	& $C_0$ 	& comments\\
\hline
O-type stars (observed)	&	$1.45\pm0.16$	&	$20.7\pm 0.88$	& Fig.\,\ref{f:wm_unique}\\
LMC predictions		&	1.83 &	18.43	& \cite{vink2000, vink2001}\\
\hline
\end{tabular}
\end{table*}

In Fig.\,\ref{f:wm_unique} we show the WLR for the combination of our
sample with the sample of \cite{mokiem2007}. We indicate the four groups of O stars, Of stars, Of/WN stars and
WNh stars (see last paragraph of \S\,\ref{s:target}) using
different symbols. The majority of the
O-type stars in both samples follow a clearly defined WLR. We note that the work
by \citeauthor{mokiem2007} is based on a different analysis method and
different numerical models than ours. However, the good agreement between both
samples in the overlapping region between $\log L/L_{\odot} = 5.5$ and
6.0 supports the reliability of the comparison. For the stars at the
highest luminosities, including the Of/WN, WNh and some Of stars from
our sample, the observed wind momenta are clearly enhanced with
respect to the overall WLR.  In Sect.\,\ref{mdot_gamma} we will show
that this mass-loss enhancement can be explained by the proximity of
these stars to the Eddington limit.

The only object that clearly stands out in
  Fig.\,\ref{f:wm_unique} is the WNh star VFTS\,108. This star has a
  high helium surface mass fraction $Y=0.775$ and may thus be in the
  phase of core He-burning.  Also the other four WNh stars with
  $Y>0.75$ (427, 695, 758, 1001) have increased modified wind-momenta
  and may already have reached the core He-burning stage.

Focussing only on the O-type stars we obtain an observed WLR of the form
\begin{equation}
\log (\dot{M} \varv_{\infty} \sqrt{R_{\star}/R_{\odot}}) = m_0 \log (L/L_{\odot}) + C_0,
\end{equation}
with coefficients $m_0$ and $C_0$ given in Table~\ref{t:wm}.
  The observed relation is derived in two steps. The first step is to
  fit a simple linear fit through the O-type stars ignoring the
  uncertainties in the luminosity and wind momentum. The result is
  used in the second step as the starting point for a
    Levenberg-Marquardt fit including error ellipses
  \citep{markwardt2009}. The coefficients for the resulting
    relation are given in Table\,\ref{t:wm}.

  In Fig.\,\ref{f:wm_unique} we compare this relation with the
  predictions by \citet{vink2001} in the form given by
  \citet[][coefficients are given in Table\,\ref{t:wm}]{mokiem2007}. As
  previously noted by \cite{mokiem2007} the observed WLR for O-type
  stars at low luminosities lies slightly above the theoretical
  predictions, which can be resolved by adopting moderate
  wind-clumping to reduce the observed mass-loss rates. For O-type
  stars with higher luminosities, around $\log{L/L_\odot}\sim 6$, the
  values derived here agree well with the predictions, so that a
  further downward correction would lead to values below the
  theoretical expectations.  While these effects are relatively small,
  and will be subject of future dedicated studies within the VFTS
  collaboration, we find a substantial increase of the wind momenta
  for the Of/WN and WNh stars at even higher luminosities, similar to
  the results of \citet{martins2008} for the luminous WNh stars in the
  Arches cluster. This increase has been interpreted by
  \citet{graefener2011} as being due to the proximity of these objects
  to the Eddington limit, and will be further discussed in
  Sect.\,\ref{mdot_gamma}.

  The average mass-loss relation for the complete sample,
    including WNh, Of/WN, and O-type stars, would lie    
$0.2...0.3$\,dex above the predictions. Moderate volume-filling
  factors around $f_{\rm v}=0.25$ would thus bring the observed
  average WLR in agreement with the predictions \citep[cf.\
  also][]{repolust2004, mokiem2007}.

\subsection{Mass-loss near the Eddington limit\label{mdot_gamma}}

\begin{figure}[t]
\begin{center}
\resizebox{\hsize}{!}{\includegraphics{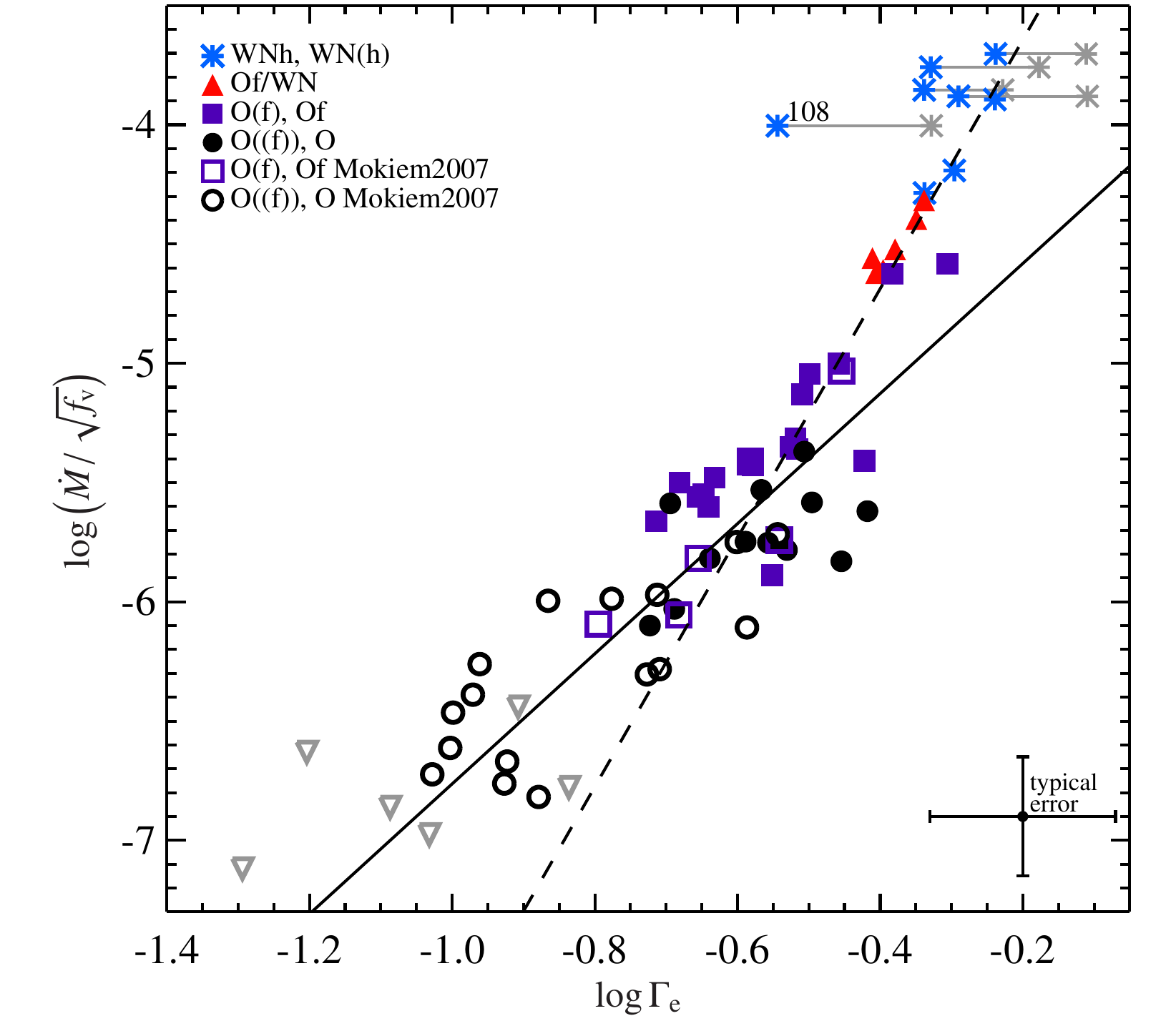}}
\end{center}
\caption{Unclumped $\log \dot{M}$ vs.~$\log \Gamma_{\rm e}$. Solid line: $\dot{M}-\Gamma_{\rm}$ relation for O stars. The
  different symbols indicate stellar sub-classes following
  Sect.\,\ref{s:subclass}. Dashed line: the steeper slope
  of the Of/WN and WNh stars. The {\it kink} occurs at $\log
  \Gamma_{\rm e} = -0.58$. The grey asterisks indicate the
    position of the stars with $Y > 0.75$ under the assumption of
    core He-burning. The grey upside down triangles are stars from
  \cite{mokiem2007} which have only an upper limit in $\dot{M}$ and
  are excluded from the fit.}
\label{f:mdot_gamma_obs}
\end{figure}
\begin{figure}[t]
\begin{center}
\resizebox{\hsize}{!}{\includegraphics{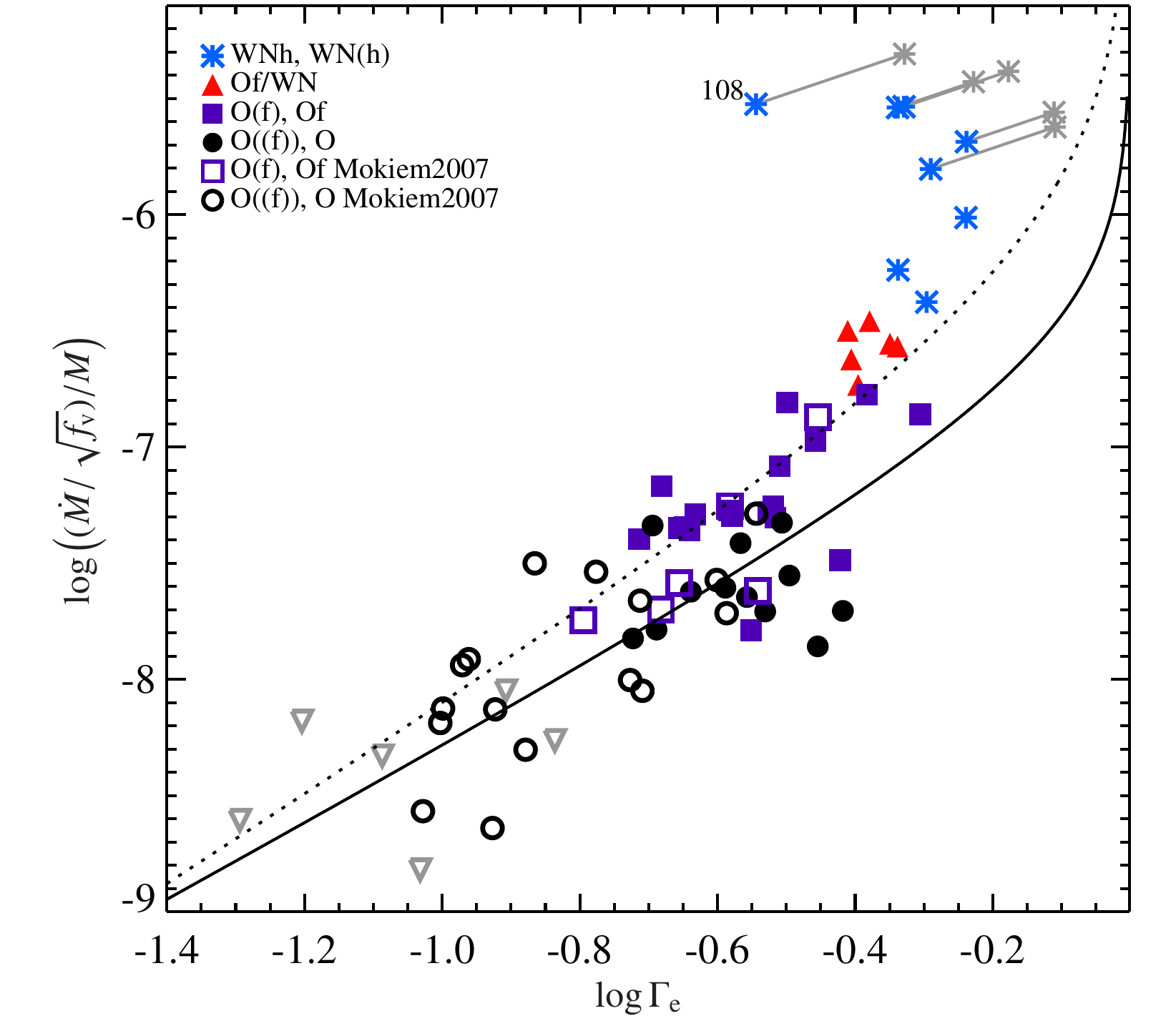}}
\end{center}
\caption{Unclumped $\log \dot{M}/M$ vs.~$\log \Gamma_{\rm e}$.
    The different symbols indicate stellar sub-classes following
    Sect.\,\ref{s:subclass}.  The black solid line is a fit using the
    CAK wind theory (with $\alpha = 0.62 \pm 0.08$).
    The dotted line indicates a tentative fit of the Of and
      Of/WN stars with $\alpha = 0.53$. The grey asterisks indicate
    the position of the stars with $Y > 0.75$ under the assumption of
    core He-burning. The grey triangles are stars from
    \cite{mokiem2007} with upper limits.  They are excluded from the
    fit.}
\label{f:mdot_gamma_cak}
\end{figure}

A primary goal of this work is to investigate the mass-loss properties of
VMS at the top of the main sequence, when they are still in
  the phase of core hydrogen-burning. \cite{graefener2008} and
\cite{vink2011} predicted a strong dependence on the classical
Eddington factor $\Gamma_{\rm e}$ in this regime. \cite{graefener2008}
also predicted a dependence on the environment metallicity $Z$, which
mainly decides around which values of $\Gamma_{\rm e}$ the strong
$\Gamma$-dependence sets in.  \citet{vink2011} investigated the
transition from the classical WLR for O stars to the
$\Gamma$-dependent mass-loss of VMS and found an abrupt change in the
form of a ``kink'' between both regimes.  \citet{graefener2011}
confirmed the strong $\Gamma$-dependence for the sample of VMS in the
Arches cluster near the Galactic Centre. They used the method outlined
in Sect.\,\ref{s:edd} to estimate Eddington factors which take
advantage of the characteristic dependence of the classical Eddington
factor on the hydrogen surface abundance with $\Gamma_{\rm
  e}\propto(1+X)$ (cf.\ Eq.\,\ref{e:eddington_factor}).

In Fig.\,\ref{f:mdot_gamma_obs} we show the resulting $\dot{M}$-
$\Gamma_{\rm e}$ relation for our sample and the sample of
\cite{mokiem2007}. To be consistent we also used masses based on the assumption of chemical homogeneity to calculate $\Gamma_{\rm e}$ for the sample of \citet{mokiem2007}. As predicted by \citet{vink2011} we find two
branches for high and low $\Gamma_{\rm e}$.  Separate fits for both
branches are shown using a relation of the form
\begin{equation}
\log \dot{M} = m_1 \log \Gamma_{\rm e} + C_1~,
\end{equation}
with $m_1$ and $C_1$ listed in Tab.\,\ref{t:mdot_gamma}. The fit
parameters and errors are obtained from a Levenberg-Marquardt analysis.

\begin{table}[t]
\caption{$\dot{M}-\Gamma_{\rm e}$ relations.}
\label{t:mdot_gamma}
\centering
\begin{tabular}{lcc}
\hline
\hline
				&	$m_1$	& $C_1$\\
\hline
Of/WN and WNh stars	&	$5.22\pm4.04$ 	&	$-2.6\pm1.46$ \\
O-type stars$^1$		&	$2.73\pm0.43$	&	$-4.04\pm0.30$ \\
\hline
\end{tabular}
\tablefoot{$^{(1)}$ including the stars from \cite{mokiem2007}}
\end{table}

The low-$\Gamma$ branch consists only of O-type stars and has a
relatively shallow slope with $\dot{M} \propto \Gamma_{\rm e}^{2.7}$.
As the hydrogen surface abundances of the stars on this branch do not
vary, their $\Gamma$-dependence translates into a dependence on
 $L/M$ (cf.\ Eq.\,\ref{e:eddington_factor}). As $M$ is computed
  from $L$ in our present approach (cf.\ Sect.\,\ref{s:edd}) this
  translates into a dependence on $L$ only.  It is thus not surprising
  that we find such a relation for the stars on this branch. The stars
  on the high-$\Gamma$ branch, on the other hand, are Of, Of/WN, and
  WNh stars with substantially varying hydrogen surface abundances.
  The $\Gamma$-dependence on this branch is thus a firm indication of
  a strong intrinsic dependence on the Eddington factor $\Gamma_{\rm
    e}$. On the high-$\Gamma$ branch we find a much steeper relation
  with $\dot{M} \propto \Gamma_{\rm e}^{5.2}$.  The five
  stars with $Y> 0.75$ have been excluded from the fit (see paragraph
  after next for details). The two branches intersect at $\log
\Gamma_{\rm e} = -0.58$.

  We further investigate whether the steep $\Gamma$-dependence
  can be explained within the standard wind theory for OB stars by
  \citet[][CAK]{castor1975}. Employing equation\,(46) from
  CAK in our fitting algorithm we obtain an (effective) force
  multiplier parameter $\alpha = 0.62 \pm 0.08$, in good agreement
  with typical values from the literature \citep{puls2008}. The
  resulting relation between $\dot{M}/M$ and $\Gamma_{\rm e}$ is shown
  in Fig.\,\ref{f:mdot_gamma_cak}. The relation matches the O-type
  stars in our sample fairly well, but the steep increase in $\dot{M}$
  occurs only for $\Gamma_{\rm e}$ {\it extremely} close to unity (as
  a result of the $1/(1-\Gamma)$ term in the CAK relation). In an
  attempt to match the steep observed increase for high $\Gamma_{\rm
    e}$ with a CAK-type relation we obtained better results for a
  reduced $\alpha= 0.53$. However, in this case the mass-loss rates of
  normal O-type stars are systematically over-estimated. To
  explain the observations in the framework of the CAK-theory it is
  thus necessary to adopt a varying $\alpha$-parameter, or two
  separate relations for O-type stars and Of/WN, WNh stars.

As outlined in Sect.\,\ref{s:edd} the $\Gamma_{\rm e}$ in
Fig.\,\ref{f:mdot_gamma_obs} are obtained under the assumption of
chemical homogeneity, i.e., they are lower limits.  For the five
  He-rich stars with $Y>0.75$ we also indicate upper limits for
  $\Gamma_{\rm e}$ based on the assumption that the stars are core
  He-burning.  The most significant change occurs for VFTS\,108. The
  assumption of core He-burning brings this star in agreement with the
  overall relation. 
  For the other He-rich stars the situation is less clear,
  i.e., they could be in a core H or core He-burning phase. For
VFTS\,108 we obtain $M=20.5\,M_\odot$ and $\Gamma_{\rm e}=0.46$ under
the assumption of core He-burning, and $M=33.1\,M_\odot$ and
$\Gamma_{\rm e}=0.29$ assuming chemical homogeneity. The lower mass is
also supported by the low terminal wind velocity of this star. Under
the assumption of chemical homogeneity the escape velocity of this
star would be higher by a factor 1.8 and the object would
significantly stand out in Fig.\,\ref{f:vinf}.  We thus conclude that
VFTS\,108 is most likely a core He-burning object in our sample.
 The situation is similar for VFTS\,427 and VFTS\,1001 which
  also have low terminal wind velocities ($\sim$1000 and
  $\sim$1100\,km\,s$^{-1}$).  The wind velocities of VFTS\,695 and VFTS\,758
  are higher ($\sim$1600 and $\sim$2000\,km\,s$^{-1}$) and in agreement with the
  relation by \cite{lamers1995} adopting chemical homogeneity.
  
As noted in Sect.\,\ref{s:edd} the masses derived under the assumption
of chemical homogeneity are upper limits, and the true masses of our
sample stars are likely smaller. Comparisons with evolutionary models
indicate an average difference of $\sim 30$\% (Schneider et al.~in prep.). This holds in particular for the stars with low
mass-loss rates and normal He or N abundances for which chemical
homogeneity is not expected from single star evolution. The true
Eddington factors for these stars may thus be significantly
higher than adopted in Fig.\,\ref{f:mdot_gamma_obs}.

  In \S\,\ref{s:error} and Appendix\,\ref{a:logg} we discuss that
  systematically lower $\log g$ for the O supergiants in our sample
  imply slightly lower luminosities for these objects. As this also
  results in a lower stellar mass the L/M ratio, and thus also
  $\Gamma_{\rm e}$, remain almost unaffected.  However, the absolute
  mass-loss rates would be affected, and the O supergiants would
  systematically shift downwards along the ordinate.  The effect on
  the slope of the O-type relation is rather small, as the O
  supergiants are evenly distributed along the abscissa.

  For the O-type relation we obtain under the assumption of
  chemical homogeneity $\log(\dot{M}/\sqrt{f_{\rm v}})=(2.73 \pm 0.43)
  \times \log \Gamma_{\rm e} - (4.04 \pm 0.3)$
  (Table\,\ref{t:mdot_gamma}). Using evolutionary masses from \citet[][submitted]{koehler2014}
  for the O-type stars with high hydrogen surface abundances the
  relation changes to $\log(\dot{M}/\sqrt{f_{\rm v}})=(2.68 \pm 0.44)
  \times \log \Gamma_{\rm e} - (4.29 \pm 0.27)$. The slope is only
  slightly shallower and well within the errors bars. The main
  difference is that the O-type relation is shifted towards higher
  $\Gamma_{\rm e}$ because of the systematically lower evolutionary
  masses.

  Furthermore, clumping could affect our results qualitatively if the
  clumping properties change systematically, e.g.\ between O and
  WR-type stars.  However, in \S\,\ref{efficiency} we will show that
  extreme volume-filling factors of the order of 0.01 or less
  \citep[as suggested for O stars e.g.\
  by][]{bouret2003,fullerton2006} are unlikely for our sample.
  Moderate volume-filling factors of the order of 0.1 may affect our
  results moderately if they change systematically, e.g., between
  different spectral types.

The main result of this work is that the observations {\em
  qualitatively} confirm the existence of a strong $\Gamma$-dependence
as predicted by \cite{graefener2008} and \cite{vink2011}. Furthermore
we find two branches with different slopes for low and high
$\Gamma_{\rm e}$ which appear to be connected by a kink as
predicted by \cite{vink2011}. The location of the kink in
Fig.\,\ref{f:mdot_gamma_obs} is at $\log \Gamma_{\rm e} = -0.58$ ($\Gamma_{\rm
e} = 0.26$).  For the
reasons outlined in Sect.\,\ref{s:edd} we believe that this is a lower
limit for the true value where the transition between normal O star
mass-loss and enhanced $\Gamma$-dependent mass-loss happens. E.g.\ the
models by \citet{graefener2008} suggest a higher value around
$\Gamma_{\rm e} \sim 0.5$ for LMC metallicity.

  The question whether the observed change in the mass-loss
  properties also marks a change in the underlying wind physics is
  presently difficult to answer. The models by \citet{graefener2008}
  suggest that the wind physics for the Of/WN and WNh stars on the
  steep branch is different from that for OB stars. The kink
  found by \citet{vink2011} also suggests that this is the case. In the
  present work we confirmed the steep $\Gamma$-dependence, but the
  nature of the transition remains unclear. In this context it is
noteworthy that, in terms of mass-loss, the kink in
Fig.\,\ref{f:mdot_gamma_obs} already appears in the regime of extreme
Of stars, well below the mass-loss range of Of/WN stars.  As we will
discuss in Sect.\,\ref{efficiency} the wind performance numbers $\eta$
in this regime are most likely still below unity (if we assume that
O star winds are clumped). The transition to a strong
$\Gamma$-dependence thus seems to happen for lower wind
efficiencies than the value of $\eta=1$ predicted by \cite{vink2011}.

The observed slope with $\dot{M} \propto \Gamma_{\rm e}^{5.2}$ on the
high-$\Gamma$ branch is similar to the prediction by \cite{vink2006} who
gives a relation with $\dot{M} \propto \Gamma_{\rm e}^{5}$ for VMS
stars approaching the Eddington limit.  \cite{graefener2008} and
\cite{vink2011} also predicted a steep increase but used more complex
descriptions than a simple power law.  Nevertheless, the relevant
result is the qualitative confirmation of a relation between
$\dot{M}$ and $\Gamma_{\rm e}$.

Finally there is an offset between the majority of the 
Of stars and the Of/WN stars in Fig.\,\ref{f:mdot_gamma_obs}. This
could indicate systematically lower volume-filling factors
($f_{\rm v}=0.1$) for the Of/WN and WNh stars, compared to more moderate
values ($f_{\rm v}=0.25$) for O-type stars \citep[see also][]{sundqvist2013}.
\citet{graefener2011} did not find a comparable gap in the Arches
cluster sample.

\subsection{Wind efficiency and clumping\label{efficiency}}

\begin{figure}[t]
\begin{center}
\resizebox{\hsize}{!}{\includegraphics{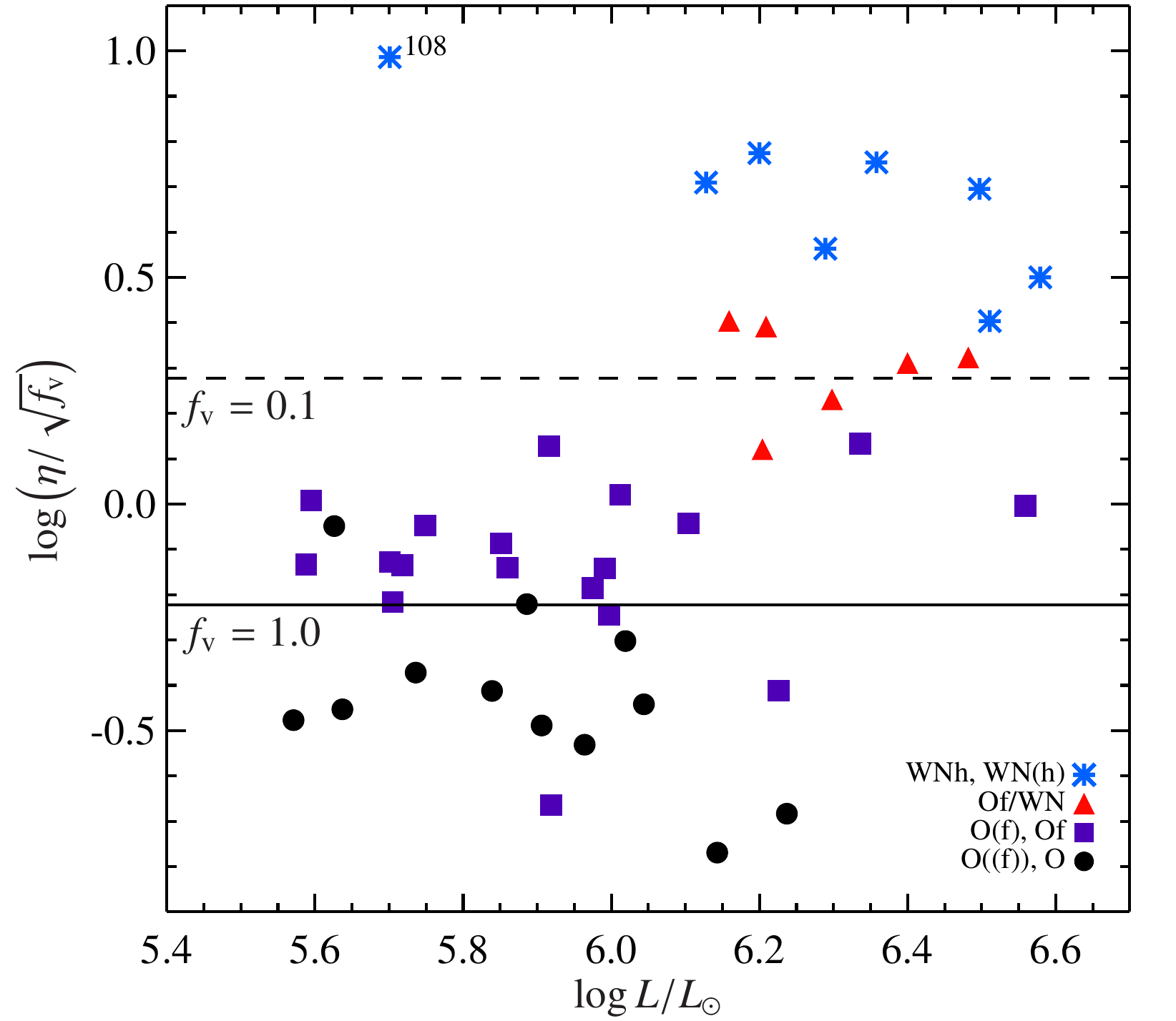}}
\end{center}
\caption{Wind efficiency $\eta$ vs.\ luminosity $L$.  The
    different symbols indicate stellar sub-classes following
    Sect.\,\ref{s:subclass}. The black solid line indicates
    $\tau = 1$ for unclumped winds ($f_{\rm v}=1.0$). The
  black dashed line indicates $\tau = 1$ for an assumed
  volume-filling factor of $f_{\rm v}=0.1$.}
\label{f:eta}
\end{figure}

The wind efficiency parameter $\eta = \dot{M} \varv_{\infty}/
(L_{\star}/c)$ gives the ratio between mechanical wind momentum
$\dot{M}\varv_{\infty}$ and radiative momentum $L_{\star}/c$ of a
star. $\eta=1$ denotes the single scattering limit, for which each
photon is scattered on average once to drive the stellar wind,
i.e. $\eta\approx\tau_{\rm wind}\approx 1$ where $\tau_{\rm
    wind}$ denotes the flux-mean optical depth at the sonic point
  \citep[cf.][Sect.\,7.2]{lamers1999}. \citet{vink2012} pointed out
that this relation is independent of wind clumping and used it to
calibrate the mass-loss rates of VMS at the single scattering limit.
Their model computations suggested that, more precisely, at the single
scattering limit $\eta=f \tau_{\rm wind}$ where $f$ depends
  on the ratio $\varv_\infty/\varv_{\rm esc}$ and is typically of the
  order of $f\approx 0.6$.  We thus expect stellar winds to become
optically thick for $\eta \approx 0.6$. At this point the
  majority of photons in the flux maximum are absorbed within the wind,
  with the consequence that the wind recombines and forms
  emission-line spectra.

 Our present stellar sample traces this transition
  observationally in unprecedented detail. In Fig.\,\ref{f:eta} we
show the wind efficiencies $\eta/\sqrt{f_{\rm v}}$ that follow from
our present analyses.  The dependence of $\eta$ on the adopted
volume-filling factor arises from Eq.\,\ref{trans}.  The point where
the winds become optically thick is indicated for the cases with
($f_{\rm v}=0.1$) and without ($f_{\rm v}=1.0$) clumping.
 Spectroscopically these points coincide with the domain of
  Of/WN stars (adopting $f_{\rm v}=0.1$) and the transition between O
  stars and Of stars (adopting $f_{\rm v}=1.0$), i.e., the regime
  where stars start to form emission lines. Clumping factors of this
  order of magnitude are thus plausible. Volume-filling factors of
this order of magnitude are commonly found for WR stars
\citep[e.g.][]{hamann1998} and have been proposed for O stars
\citep[e.g.][]{puls2006,surlan2013}.
Extreme volume-filling factors of the order of $f_{\rm v}=0.01$ as
proposed e.g.\ by \citet{bouret2003, fullerton2006} appear very
unlikely, as this would move the transition into the region of WNh
stars where strong emission line spectra are already fully developed.
We thus conclude that volume-filling factors at the transition between
optically thin and optically thick winds are likely moderate, i.e.,
they are likely of the order of $f_{\rm v}=0.1$ or higher.

\subsection{Stellar evolution in the upper HRD \label{s:evo_stages}}

\begin{figure}[t]
\begin{center}
\resizebox{\hsize}{!}{\includegraphics{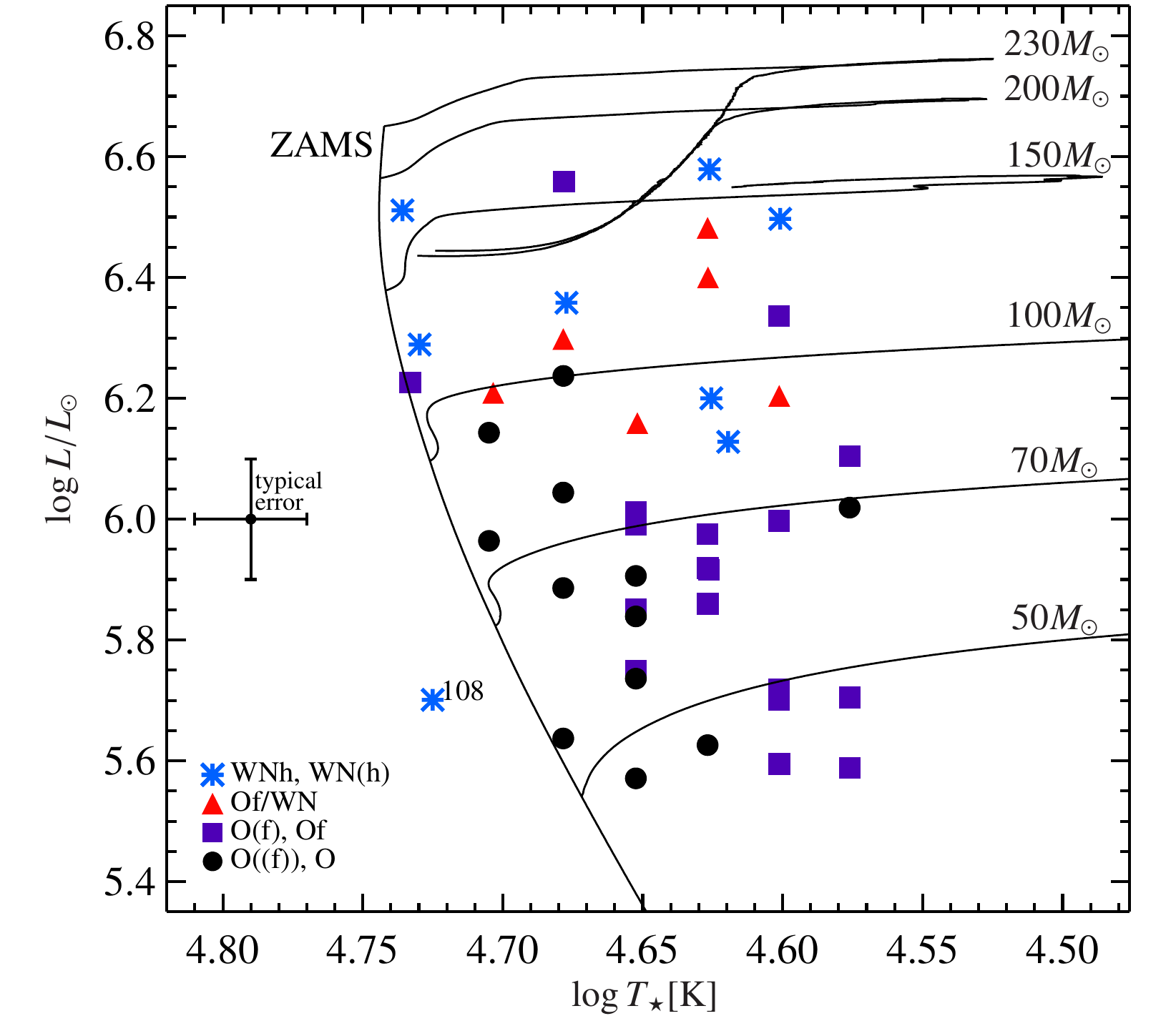}}
\end{center}
\caption{Distribution of spectral types of our sample in the HR-diagram.  The
  different symbols indicate stellar sub-classes following
  Sect.\,\ref{s:subclass}.  Black lines indicate evolutionary tracks
  from \citet[][submitted]{koehler2014} for an initial rotation rate of
  300\,km\,s$^{-1}$ and the location of the Zero-Age Main Sequence
  (ZAMS).}
\label{f:hrd_bonn_tstar}
\end{figure}
\begin{figure}[t]
\begin{center}
\resizebox{\hsize}{!}{\includegraphics{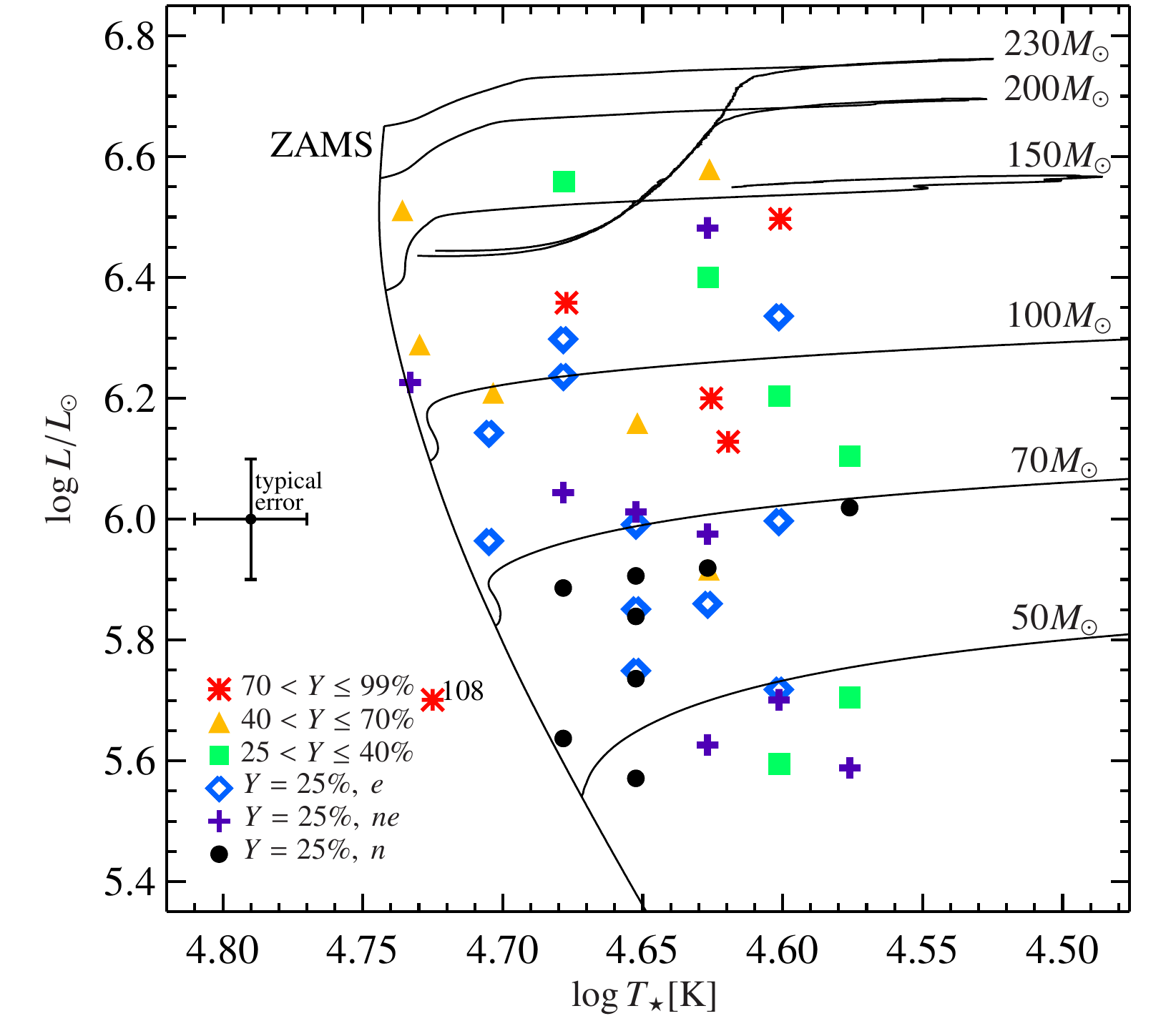}}
\end{center}
\caption{Chemical surface compositions throughout the
    HR-diagram. Different symbols indicate different surface helium
    mass fractions ($Y$) and different levels of nitrogen enhancement
    ($e$ enhanced; $ne$ partially enhanced; $n$ normal).}
\label{f:hrd_bonn_age}
\end{figure}

The data presented in the previous sections are unique as they provide
the largest stellar sample to date with luminosities above
$\sim$\,400\,000\,$L_{\odot}$.  They also provide a unique opportunity
to investigate the evolution of stars with masses above $\sim
40\,$M$_{\odot}$, for which constraints are otherwise sparse. In the
following we draw some straightforward conclusions from comparisons
with stellar evolution predictions.

   In addition to the effects of mass loss from the stellar wind,
   internal mixing processes (e.g., convective overshooting,
   rotationally-induced mixing) also constitute major uncertainties in
   evolutionary models.
  Furthermore, the stars considered here are very close to
  their Eddington limit, which may affect their envelope structure
  \citep[][submitted]{koehler2014} and give rise to instabilities
  \citep{Graefener2012a}, raising the question whether eruptive
  mass-loss, as observed e.g.\ for some LBVs and supernova
  progenitors, may also affect their evolution.  Finally, recent
    work by \citet{sana2012Sci}, \citet{Chini2012}, and \citet{sana2013} suggests that
    binary interaction greatly affects the evolution of massive
  stars, implying that several stars in our sample may have been
  subject to mass transfer or merger events
  \citep[see][]{deMink2014,schneider2014}.

  The observed HRD positions of our sample stars are indicated in
  Figs.\,\ref{f:hrd_bonn_tstar} and \ref{f:hrd_bonn_age},
  distinguishing between different spectral types and surface
  compositions. They are compared with single star evolutionary tracks
  from \citet[][submitted]{koehler2014} with an initial rotational velocity
  of $\varv_{\rm R, ini} = 300$\,km\,s$^{-1}$.  In agreement with our
  conclusions from Sect.\,\ref{mdot_gamma} the HRD positions of most
  of our sample stars agree with an evolutionary stage on the
  main-sequence. The only exception is VFTS\,108 which is found to the
  left of the Zero-Age Main-Sequence (ZAMS), supporting our previous
  conclusion that this star is most likely in the core He-burning
  phase.

  Considering the surface helium abundances of the main-sequence stars
  in our sample (Fig.\,\ref{f:hrd_bonn_age}), it appears convenient to
  divide the sample into two luminosity or mass ranges, with the
  dividing line roughly at $10^{6.1}\,$L$_{\odot}$ or
  80...90\,M$_{\odot}$.  The group of 24 stars below this luminosity
  limit consists of O-type stars of which only three have a somewhat
  enhanced helium abundance.  All other stars still show their
  original helium abundance. The group of 19 stars above the
  luminosity limit is dominated by Of/WN and WNh stars. The majority
  (13 out of 19) of stars in this group have enhanced helium surface
  abundances. Furthermore, there appears to exist a similar limit for
  the enrichment in nitrogen.  Stars without evidence for
  nitrogen-enrichment are only found in the low-luminosity group.

  Based on their evolutionary models \citet[][submitted]{koehler2014} identify similar luminosity limits as the ones described
    above.
    They find that the minimum luminosity for which the surfaces of
    main-sequence stars show considerable helium enrichment depends on
    the initial rotational velocity and varies between
    $\log(L/L_\odot) \approx 4.8$ for fast rotators and
    $\log(L/L_\odot) \approx 6.3$ for slow rotators.  Based on the
    models by K\"ohler et al.\ the empirical boundary reported here
    ($\log(L/L_\odot) \approx 6.1$) suggests that the bulk of our
    sample stars had initial rotational velocities that were slower than 
    $\sim$\,200\,km\,s$^{-1}$. 

\begin{figure}[t]
\begin{center}
\resizebox{\hsize}{!}{\includegraphics{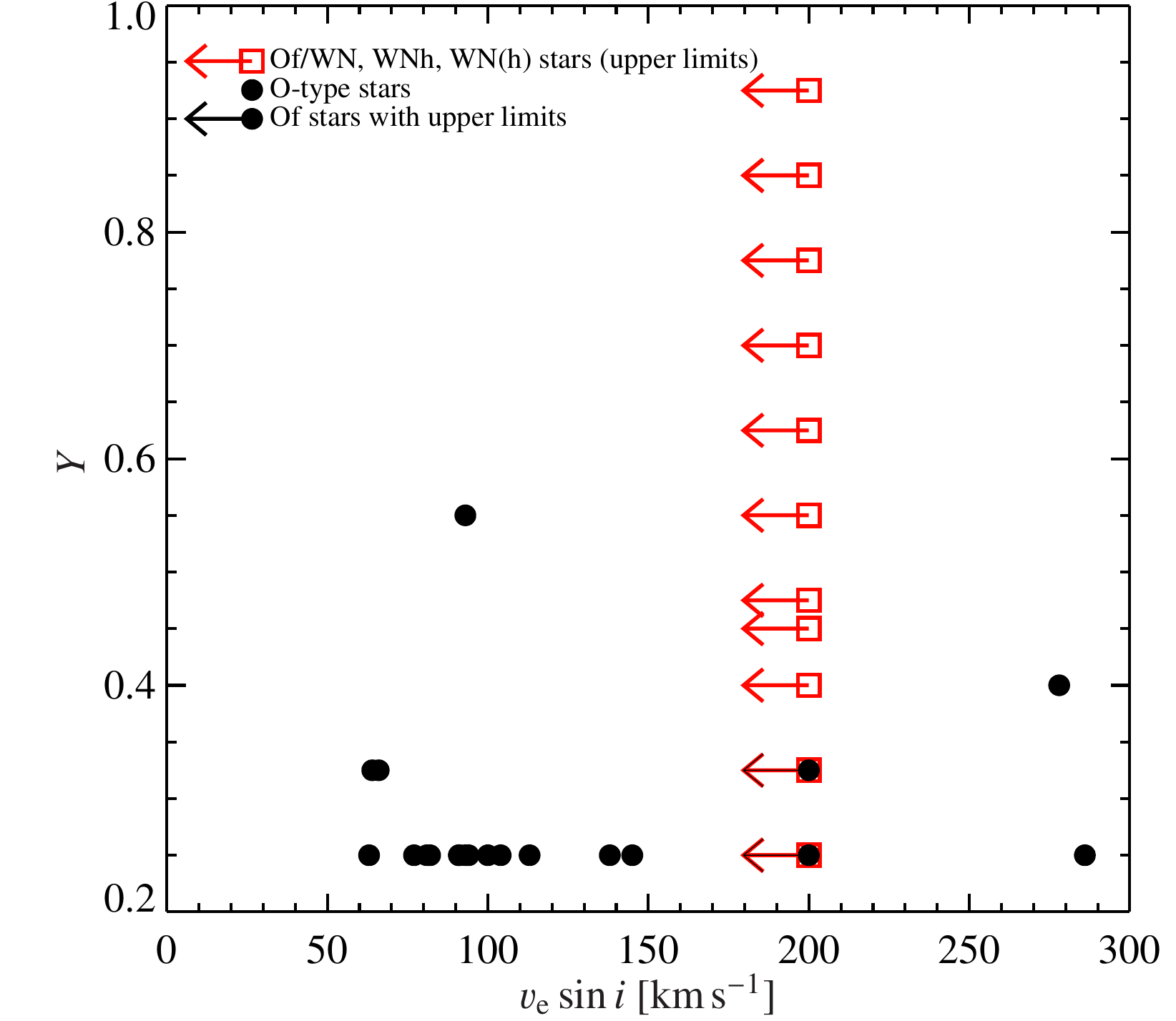}}
\end{center}
\caption{He surface mass fractions $Y$ vs.\ projected
  rotational velocities $\varv_{\rm e} \sin i$. For the Of/WN and WNh
  stars in this work we indicate conservative upper limits for
  $\varv_{\rm e} \sin i$ of 200\,km\,s$^{-1}$.}
\label{f:Yvsini}
\end{figure}
\begin{figure}[t]
\begin{center}
\resizebox{\hsize}{!}{\includegraphics{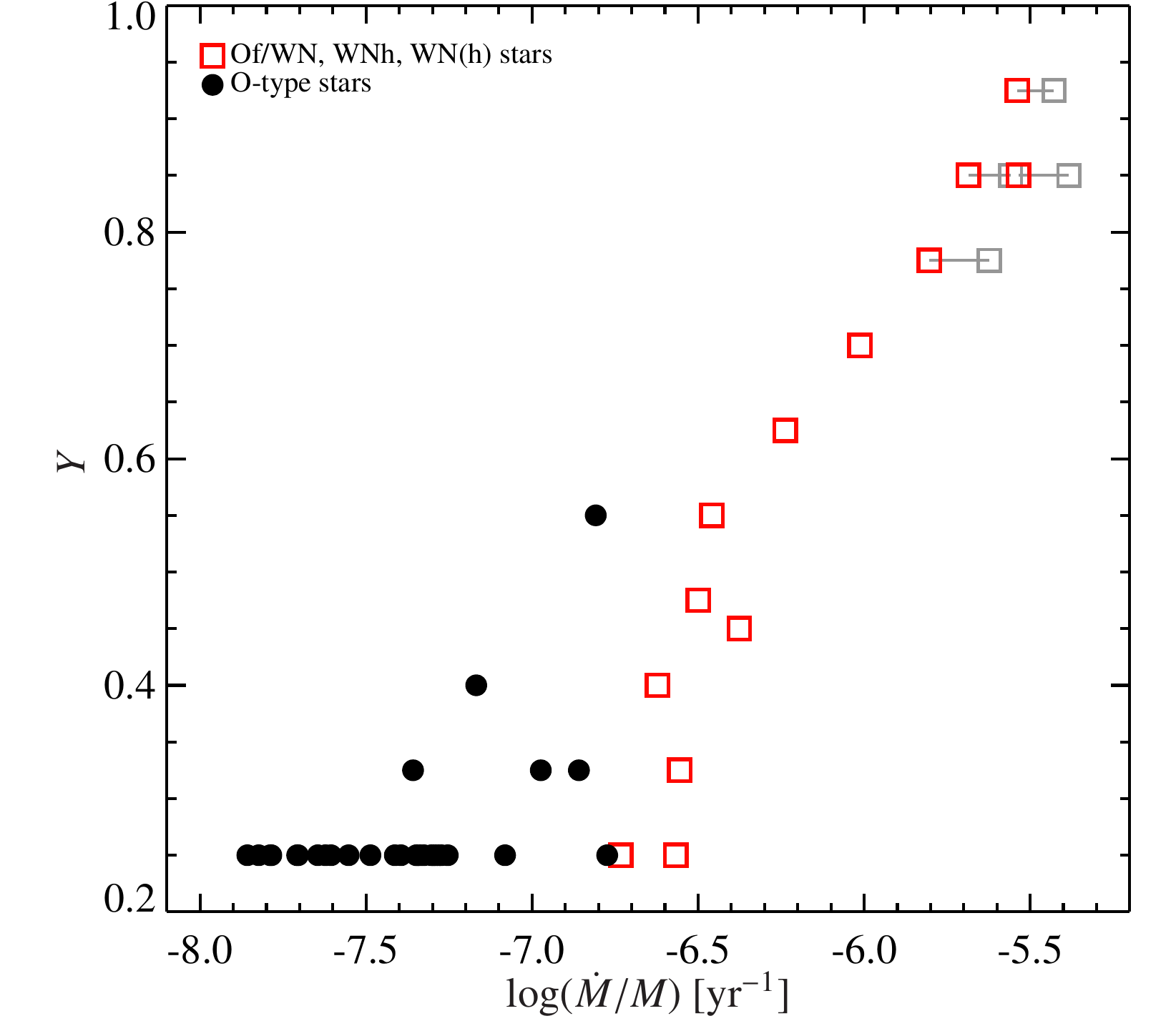}}
\end{center}
\caption{He surface mass fractions $Y$ vs.\ relative
  mass-loss rates $\log(\dot{M}/M)$ for the possible main-sequence
  stars in our sample. Masses have been estimated under the assumption
  of chemical homogeneity, i.e., they are upper limits. For potential
  He-burners we also indicate masses that would result from core
  He-burning in grey.}
\label{f:Ymdot}
\end{figure}

Projected equatorial rotational velocities $\varv_{\rm e} \sin i$ of
the presumed single O-type stars in the VFTS sample have been reported by
\citet{Ramirez2013}. After correcting for projection effects and macro-turbulence, they
find that the rotational velocity $\varv_{\rm e} \sin i$ distribution strongly peaks at 100
km\,s$^{-1}$, and that about 75\% of the sample has $\varv_{\rm e} \lesssim
200$ km\,s$^{-1}$. 

  Based on the synthetic emission line profiles from our models we
  are further able to constrain $\varv_{\rm e} \sin i$ for 15 Of/WN
  and WNh stars in our sample.  We use narrow emission lines (with an
  intrinsic width of 50 to 200\,km\,s$^{-1}$) that are formed near the
  stellar surface, and should thus well reflect the rotational speed
  at the surface. In contrast to the values for O-type stars from
  \citeauthor{Ramirez2013} these are not corrected for additional line broadening
  due to macro-turbulence. Because of this, and because the intrinsic
  line profiles are affected by the wind velocity, we are only able to
  give conservative upper limits for $\varv_{\rm e} \sin i$ of
  200\,km\,s$^{-1}$. The micro-turbulent velocities adopted in our
  models are much lower (10\,km\,s$^{-1}$) and have thus no effect on
  the given limits.

The results are presented in Fig.\,\ref{f:Yvsini} where we plot the observed helium surface mass
  fractions $Y$ vs.\ $\varv_{\rm e} \sin i$ for these objects and for
  the 20 O-type stars from \citet{Ramirez2013} and \citet{sabin2014} overlapping
  with our sample.  There is no evidence for a correlation between $Y$
  and $\varv_{\rm e} \sin i$ in this figure suggesting that
  rotationally induced mixing is not the dominant factor {responsible
    for} the helium surface enrichment.

In the whole sample there are only two fast rotators with $\varv_{\rm
  e} \sin i >$ 200\,km\,s$^{-1}$. These are VFTS\,626 with
278\,km\,s$^{-1}$ and VFTS\,755 with 286\,km\,s$^{-1}$.  While
VFTS\,626 belongs to our group of low-luminosity stars with
He-enrichment ($Y=0.4$), VFTS\,755 does not show any sign of surface
enrichment in He or N.  The main difference between both stars is
their mass-loss rate which is three times higher for VFTS\,626.
 
In Fig.\,\ref{f:Ymdot} we investigate the relation between
  surface helium enrichment and mass-loss. There appears to be a very
clear correlation between $Y$ and $\log(\dot{M}/M)$, indicating that
mass loss plays a dominant role in the helium enrichment for these objects. 
The X-axis
in Fig.\,\ref{f:Ymdot} can also be read as the inverse of the
mass-loss timescale $\tau_{\dot{M}}=M/\dot{M}$, or more precisely an
upper limit for $\tau_{\dot{M}}$ as the mass-estimates used are upper
limits (cf.\,Sect.\,\ref{s:edd}, some lower limits for potential
He-burners are also indicated in Fig.\,\ref{f:Ymdot}). For
$\log(\dot{M}/M)\gtrsim -6.5$, $\tau_{\dot{M}}$ becomes comparable to
the main-sequence lifetime\footnote{Note that $\dot{M}$ in
  Fig.\,\ref{f:Ymdot} is not corrected for clumping. For the case with
  clumping $\tau_{\dot{M}}$ would increase by a factor $1/\sqrt{f_{\rm
      v}}$ which is likely of the order of 3.} and the correlation
becomes very clear.

Our results thus suggest that the observed changes in the surface
abundances for luminosities $\gtrsim 10^{6.1}L_\odot$ are caused by
the change in the mass-loss properties discussed in
Sects.\,\ref{wind_momentum} and \ref{mdot_gamma}. The observed
luminosity threshold coincides with the luminosity above which we see
evidence for enhanced $\Gamma$-dependent mass-loss in our sample (cf.\
Fig.\,\ref{f:wm_unique}). In the high-luminosity group, with 
$L\gtrsim 10^{6.1}L_\odot$, we thus find predominantly Of/WN and WNh
stars whose hydrogen-rich layers have been stripped off by the
enhanced mass-loss during their main-sequence lifetime.

For the stars below the luminosity threshold our previous discussion
  suggests that mass loss and rotational mixing do not dominate the
  surface helium enrichment. In this case binary interaction may play
  a role, and the three He-enriched stars in this group may
  (predominantly) be binary interaction products. How does this
number compare with predictions from binary evolution models by
\citet[][]{deMink2013b,deMink2014}?  Assuming continuous star
formation and an initial binary fraction of 70\% \citet{deMink2014}
predict an incidence rate of $28$\,\% of binary interaction products,
50\,\% pre-interaction binaries and 22\,\% single stars. They suggest
that these numbers can be compared to the population in 30\,Dor where
\citet[][]{sana2013} observed a spectroscopic binary fraction of
35\,\%, and inferred that the intrinsic binary fraction is $\approx
50$\,\% which is comparable to the predicted fraction of
pre-interaction binaries from \citeauthor{deMink2014}

The spectroscopic binary fraction in our initial sample is 34\,\% (21
out of 62 stars), i.e., it is comparable to the results of
\citet[][]{sana2013} for 30\,Dor. As we removed 8 SB2
  systems in our previous discussion we expect
  a slightly higher incidence rate of binary products than predicted
by \citet{deMink2014}. Given the errors of roughly $50$\,\% from
\citeauthor{deMink2014}, and assuming that these objects would show an
enhanced He abundance, we thus expect 3...10 stars
of this kind in our low-luminosity sample. However, we have to be
careful as we also excluded several stars with uncertain parameters
from the initial sample.  Inspection of our provisional parameters for
these objects in Table\,\ref{t:parameters_ex} reveals that we excluded
three objects that would belong in the low-luminosity group and show
indications of an enhanced helium abundance (VFTS\,208, 406, and 562).
 Including these objects would bring our observed numbers in
agreement with the predictions. A thorough investigation of such
``problematic'' objects may thus be crucial to understand the role of
binary interactions in massive star evolution. Furthermore,
  a higher accuracy in $Y$ than achieved in this work ($\pm 5\,\%$;
  cf.\ Sect.\,\ref{s:error}) may be desirable.

Finally, we note the distribution of spectral types in the
upper HR-diagram (Fig.\,\ref{f:hrd_bonn_tstar}). Again, the
objects in this diagram can be divided in two groups, above and below
$10^{6.1}\,L_\odot$.  The low-luminosity group consists of O-type
stars with a clear separation between O stars and Of stars, indicating
that Of stars are cooler and more evolved than O stars. In contrast to
this the high-luminosity group, which consists of a mixture of O-type
stars, Of/WN and WNh stars, shows a much more erratic spectral type
distribution.

  A comparison with the empirical HRD of the VFTS-MEDUSA
  region in \citet[][their Fig.\,12a]{doran2013} suggests that
  our high-luminosity group covers the complete range of observed
  effective temperatures in the HRD, while our low-luminosity group
  appears to be truncated at $T_\star\approx 40$\,kK, omitting cooler
  temperatures and thus likely the most evolved stars on the main
  sequence.  This selection effect may also affect our previous
  discussion of the importance of binary interactions in the
  low-luminosity group.

    As the stars in the high-luminosity group cover the majority
    of the main-sequence phase this group will include relatively
    young objects evolving towards the red, but also more evolved
    objects evolving towards the blue. Their observed temperatures are
    generally higher than $\approx$\,40\,kK. The evolutionary tracks
    from K\"ohler et al.\ in Figs.\,\ref{f:hrd_bonn_tstar} and
    \ref{f:hrd_bonn_age} reach much lower temperatures than what is
    observed.  This is the case for all of their models with
  $\varv_{\rm R, ini} \le 300$\,km\,s$^{-1}$. One reason for this is the
  envelope inflation effect which has originally been predicted for
  massive stars near the Eddington limit by \citet{Ishii99}.  Recent
  theoretical studies of this effect suggest that it can substantially
  affect stellar radii and temperatures in the upper HRD and lead to
  LBV-type instabilities, depending on a variety of physical effects
  such as the detailed structure and dynamics of the sub-photospheric
  layers, the outer boundary conditions imposed by a stellar wind, or
  stellar rotation \citep[][submitted]{Petrovic2006,Graefener2012a,graefener2013,koehler2014}. The observed HRD
  positions in Fig.\,\ref{f:hrd_bonn_age} are likely affected by a
  combination of evolutionary effects and the complex envelope physics
  in this part of the HRD. Enhanced mass loss in the
    high-$\Gamma$ regime, as it has been discussed in the present
    work, is expected to reduce the inflation effect
    \citep[cf.][]{Petrovic2006}, and may thus help to resolve the
    apparent discrepancies.

\subsection{Ionising fluxes \label{s:fluxes}}
Recently \cite{doran2013} highlighted the pivotal role of VMS in the
ionising feedback from massive OB stars in 30 Doradus. In this section
we compare the number of ionising photons per second ($Q_0$) from our
work with the values from \cite{doran2013}. In contrast to our
grid analysis \cite{doran2013} used a template method. 
The ten stars with the highest ionising fluxes in 30 Doradus are
listed in Table 10 of \cite{doran2013}. Five stars are in our sample,
VFTS\,758, 402, 1001, 1025, and Mk42. VFTS\,758 (BAT99-122) and Mk42
agree within an uncertainty less than 50\%. VFTS\,1025 (R136c) has
half of the number of ionising photons, but still remains in the top
10. The value for VFTS\,1001 (R134) is a factor 4 lower.  VFTS\,402
(BAT99-095) has a factor 3 lower ionising flux. Both stars would drop
out from the 10 stars with the highest ionising flux.  However,
VFTS\,402 is a multiple system and our spectroscopic analysis is not
reliable.  On the other hand we determined higher values for VFTS\,682
and VFTS\,695 (BAT99-119) and they would move up into the top 10 with
the highest ionising fluxes. By adding up all ionising
photons of stars which are in both samples the uncertainties are
averaged out and we obtained similar total fluxes for both methods,
 so that the overall results of \cite{doran2013} are robust with
  respect to individual uncertainties.

\section{Conclusion \label{s:conclusion}}

In this work we analysed an unprecedented sample of 62 O, Of, Of/WN
and WNh stars with very high luminosities. The sample contains a
diversity of spectral types and luminosity classes and covers a range
of mass-loss rates of $\sim$2.5~dex. The analysis is based on
a large grid of complex stellar atmosphere models computed with the
non-LTE radiative transfer code CMFGEN \citep{hillier1998}.

  We investigated the mass-loss properties of our sample stars based
  on the classical approach for O stars using the wind-momentum
  luminosity relation (WLR) from \cite{kudritzki1999}, and the
  semi-empirical approach by \citet{graefener2011} to 
    investigate a dependence on the classical Eddington factor
  $\Gamma_{\rm e}$.  While the majority of O and Of stars in our
  sample follow the classical WLR, we find that the mass-loss rates of
  Of/WN, WNh, and some Of stars are enhanced with a steep dependence
  on $\Gamma_{\rm e}$, in qualitative agreement with theoretical
  mass-loss predictions for very massive stars (VMS) by
  \cite{graefener2008} and \cite{vink2011}. Furthermore, our
    results suggest that at the transition between both regimes
    the $\dot{M} - \Gamma_{\rm e}$ dependence shows a kink as
    predicted by \cite{vink2011}.

  The main uncertainties in our results are due to wind clumping
    which is poorly constrained, in particular for O-type stars.
    Assuming that the winds are moderately and uniformly clumped
  ($f_{\rm v}\approx0.1$) our results suggest that the mass-loss of
  VMS is enhanced with respect to standard mass-loss predictions from
  \citet{vink2000,vink2001}.  Furthermore, the dependence on
  $\Gamma_{\rm e}$ sets in earlier than expected, before the winds
  become optically thick. Its quantitative impact on the evolution of
  VMS may thus be larger than previously thought. Based on our present
  analysis the transition already occurs for luminous Of stars at
  $\Gamma_{\rm e} \approx 0.3$ with a steep mass-loss dependence
  $\dot{M} \propto \Gamma_{\rm e}^{5.2}$ above this value.  We note
  that these results are based on simplified mass estimates and the
  true Eddington factors are likely higher than adopted. This may at
  least form part of the reason why, quantitatively, theoretical
  models predict the steep $\Gamma$-dependence for higher values of
  $\Gamma_{\rm e} \approx 0.5$ \citep{graefener2008} and $\Gamma_{\rm
    e} \approx 0.7$ \citep{vink2011}.

  The observed HRD positions and mass-loss properties of our
    sample stars suggest that they are predominantly very massive
    main-sequence stars. Based on the observed stellar properties, the
    sample can be divided into two luminosity or mass ranges with a
    dividing line roughly at $10^{6.1}\,L_\odot$ or
    80...90\,$M_\odot$.  Above this limit we find predominantly Of/WN
    and WNh stars whose surfaces are enriched in helium due to their
    enhanced mass-loss.  Below this limit only a few stars are
    He-enriched.  These objects may be binary interaction products.
    Based on single star evolution models from \citet[][submitted]{koehler2014} the location of the observed luminosity limit suggests that
    rotational mixing is relatively unimportant for the
      He-enrichment in our sample.

  Finally, we confirm the important role of VMS in the ionising budget
  of 30\,Dor found by \cite{doran2013}, providing evidence that these
  objects play a pivotal role in ionising and shaping the interstellar
  environment in young starburst-like regions.

\begin{acknowledgements}
  We thank the anonymous referee for providing constructive comments, Joachim Puls and Jon Sundqvist for the fruitful discussion
  about mass-loss, and Danny Lennon and Selma de Mink for providing us with HST/F775W-band images (GO 12499) to check for visual companions. JMB, GG, and JSV thank the Science \& Technology
  Facilities Council (grant No.\ ST/J001082/1) and the Department of
  Culture, Arts and Leisure in Northern Ireland for financial support.
  This work is based on observations with the European Southern
  Observatory Very Large Telescope, programmes 182.D-0222, and
  084.D-0980.
\end{acknowledgements}

\bibliographystyle{aa}
\bibliography{reference}

\begin{thebibliography}{102}
\expandafter\ifx\csname natexlab\endcsname\relax\def\natexlab#1{#1}\fi

\bibitem[{{Abel} {et~al.}(2002){Abel}, {Bryan}, \& {Norman}}]{abel2002}
{Abel}, T., {Bryan}, G.~L., \& {Norman}, M.~L. 2002, Science, 295, 93

\bibitem[{{Anderson}(1989)}]{anderson1989}
{Anderson}, L.~S. 1989, \apj, 339, 558

\bibitem[{{Asplund} {et~al.}(2005){Asplund}, {Grevesse}, \&
  {Sauval}}]{asplund2005}
{Asplund}, M., {Grevesse}, N., \& {Sauval}, A.~J. 2005, in Astronomical Society
  of the Pacific Conference Series, Vol. 336, Cosmic Abundances as Records of
  Stellar Evolution and Nucleosynthesis, ed. {T.~G.~Barnes III \& F.~N.~Bash},
  25

\bibitem[{{Bestenlehner} {et~al.}(2011){Bestenlehner}, {Vink}, {Gr{\"a}fener},
  {Najarro}, {Evans}, {Bastian}, {Bonanos}, {Bressert}, {Crowther}, {Doran},
  {Friedrich}, {H{\'e}nault-Brunet}, {Herrero}, {de Koter}, {Langer}, {Lennon},
  {Ma{\'{\i}}z Apell{\'a}niz}, {Sana}, {Soszynski}, \&
  {Taylor}}]{bestenlehner2011}
{Bestenlehner}, J.~M., {Vink}, J.~S., {Gr{\"a}fener}, G., {et~al.} 2011, \aap,
  530, L14

\bibitem[{{Bouret} {et~al.}(2003){Bouret}, {Lanz}, {Hillier}, {Heap}, {Hubeny},
  {Lennon}, {Smith}, \& {Evans}}]{bouret2003}
{Bouret}, J., {Lanz}, T., {Hillier}, D.~J., {et~al.} 2003, \apj, 595, 1182

\bibitem[{{Bromm} {et~al.}(1999){Bromm}, {Coppi}, \& {Larson}}]{bromm1999}
{Bromm}, V., {Coppi}, P.~S., \& {Larson}, R.~B. 1999, \apjl, 527, L5

\bibitem[{{Campbell} {et~al.}(2010){Campbell}, {Evans}, {Mackey}, {Gieles},
  {Alves}, {Ascenso}, {Bastian}, \& {Longmore}}]{campbell2010}
{Campbell}, M.~A., {Evans}, C.~J., {Mackey}, A.~D., {et~al.} 2010, \mnras, 416

\bibitem[{{Cardelli} {et~al.}(1989){Cardelli}, {Clayton}, \&
  {Mathis}}]{cardelli1989}
{Cardelli}, J.~A., {Clayton}, G.~C., \& {Mathis}, J.~S. 1989, \apj, 345, 245

\bibitem[{{Castor} {et~al.}(1975){Castor}, {Abbott}, \& {Klein}}]{castor1975}
{Castor}, J.~I., {Abbott}, D.~C., \& {Klein}, R.~I. 1975, \apj, 195, 157

\bibitem[{{Chini} {et~al.}(2012){Chini}, {Hoffmeister}, {Nasseri}, {Stahl}, \&
  {Zinnecker}}]{Chini2012}
{Chini}, R., {Hoffmeister}, V.~H., {Nasseri}, A., {Stahl}, O., \& {Zinnecker},
  H. 2012, \mnras, 424, 1925

\bibitem[{{Cioni} {et~al.}(2011){Cioni}, {Clementini}, {Girardi}, {Guandalini},
  {Gullieuszik}, {Miszalski}, {Moretti}, {Ripepi}, {Rubele}, {Bagheri},
  {Bekki}, {Cross}, {de Blok}, {de Grijs}, {Emerson}, {Evans}, {Gibson},
  {Gonzales-Solares}, {Groenewegen}, {Irwin}, {Ivanov}, {Lewis}, {Marconi},
  {Marquette}, {Mastropietro}, {Moore}, {Napiwotzki}, {Naylor}, {Oliveira},
  {Read}, {Sutorius}, {van Loon}, {Wilkinson}, \& {Wood}}]{cioni2011}
{Cioni}, M., {Clementini}, G., {Girardi}, L., {et~al.} 2011, \aap, 527, A116

\bibitem[{{Crowther} \& {Dessart}(1998)}]{crowther1998}
{Crowther}, P.~A. \& {Dessart}, L. 1998, \mnras, 296, 622

\bibitem[{{Crowther} {et~al.}(2010){Crowther}, {Schnurr}, {Hirschi}, {Yusof},
  {Parker}, {Goodwin}, \& {Kassim}}]{crowther2010}
{Crowther}, P.~A., {Schnurr}, O., {Hirschi}, R., {et~al.} 2010, \mnras, 408,
  731

\bibitem[{{Crowther} \& {Smith}(1997)}]{crowther1997}
{Crowther}, P.~A. \& {Smith}, L.~J. 1997, \aap, 320, 500

\bibitem[{{Crowther} \& {Walborn}(2011)}]{crowther2011}
{Crowther}, P.~A. \& {Walborn}, N.~R. 2011, \mnras, 416, 1311

\bibitem[{{De Marchi} {et~al.}(2011){De Marchi}, {Paresce}, {Panagia},
  {Beccari}, {Spezzi}, {Sirianni}, {Andersen}, {Mutchler}, {Balick}, {Dopita},
  {Frogel}, {Whitmore}, {Bond}, {Calzetti}, {Carollo}, {Disney}, {Hall},
  {Holtzman}, {Kimble}, {McCarthy}, {O'Connell}, {Saha}, {Silk}, {Trauger},
  {Walker}, {Windhorst}, \& {Young}}]{deMarchi2011}
{De Marchi}, G., {Paresce}, F., {Panagia}, N., {et~al.} 2011, \apj, 739, 27

\bibitem[{{de Mink} {et~al.}(2013){de Mink}, {Langer}, {Izzard}, {Sana}, \& {de
  Koter}}]{deMink2013b}
{de Mink}, S.~E., {Langer}, N., {Izzard}, R.~G., {Sana}, H., \& {de Koter}, A.
  2013, \apj, 764, 166

\bibitem[{{de Mink} {et~al.}(2014){de Mink}, {Sana}, {Langer}, {Izzard}, \&
  {Schneider}}]{deMink2014}
{de Mink}, S.~E., {Sana}, H., {Langer}, N., {Izzard}, R.~G., \& {Schneider},
  F.~R.~N. 2014, \apj, 782, 7

\bibitem[{{Doran} {et~al.}(2013){Doran}, {Crowther}, {de Koter}, {Evans},
  {McEvoy}, {Walborn}, {Bastian}, {Bestenlehner}, {Gr{\"a}fener}, {Herrero},
  {K{\"o}hler}, {Ma{\'{\i}}z Apell{\'a}niz}, {Najarro}, {Puls}, {Sana},
  {Schneider}, {Taylor}, {van Loon}, \& {Vink}}]{doran2013}
{Doran}, E.~I., {Crowther}, P.~A., {de Koter}, A., {et~al.} 2013, \aap, 558,
  A134

\bibitem[{{Evans} {et~al.}(2005){Evans}, {Smartt}, {Lee}, {Lennon}, {Kaufer},
  {Dufton}, {Trundle}, {Herrero}, {Sim{\'o}n-D{\'{\i}}az}, {de Koter},
  {Hamann}, {Hendry}, {Hunter}, {Irwin}, {Korn}, {Kudritzki}, {Langer},
  {Mokiem}, {Najarro}, {Pauldrach}, {Przybilla}, {Puls}, {Ryans}, {Urbaneja},
  {Venn}, \& {Villamariz}}]{evans2005}
{Evans}, C.~J., {Smartt}, S.~J., {Lee}, J.-K., {et~al.} 2005, \aap, 437, 467

\bibitem[{{Evans} {et~al.}(2011){Evans}, {Taylor}, {H{\'e}nault-Brunet},
  {Sana}, {de Koter}, {Sim{\'o}n-D{\'{\i}}az}, {Carraro}, {Bagnoli}, {Bastian},
  {Bestenlehner}, {Bonanos}, {Bressert}, {Brott}, {Campbell}, {Cantiello},
  {Clark}, {Costa}, {Crowther}, {de Mink}, {Doran}, {Dufton}, {Dunstall},
  {Friedrich}, {Garcia}, {Gieles}, {Gr{\"a}fener}, {Herrero}, {Howarth},
  {Izzard}, {Langer}, {Lennon}, {Ma{\'{\i}}z Apell{\'a}niz}, {Markova},
  {Najarro}, {Puls}, {Ramirez}, {Sab{\'{\i}}n-Sanjuli{\'a}n}, {Smartt},
  {Stroud}, {van Loon}, {Vink}, \& {Walborn}}]{evans2011}
{Evans}, C.~J., {Taylor}, W.~D., {H{\'e}nault-Brunet}, V., {et~al.} 2011, \aap,
  530, A108

\bibitem[{{Evans} {et~al.}(2010){Evans}, {Walborn}, {Crowther},
  {H{\'e}nault-Brunet}, {Massa}, {Taylor}, {Howarth}, {Sana}, {Lennon}, \& {van
  Loon}}]{evans2010}
{Evans}, C.~J., {Walborn}, N.~R., {Crowther}, P.~A., {et~al.} 2010, \apjl, 715,
  L74

\bibitem[{{Figer}(2005)}]{figer2005}
{Figer}, D.~F. 2005, \nat, 434, 192

\bibitem[{{Foellmi} {et~al.}(2003){Foellmi}, {Moffat}, \&
  {Guerrero}}]{foellmi2003}
{Foellmi}, C., {Moffat}, A.~F.~J., \& {Guerrero}, M.~A. 2003, \mnras, 338, 1025

\bibitem[{{Fullerton} {et~al.}(2006){Fullerton}, {Massa}, \&
  {Prinja}}]{fullerton2006}
{Fullerton}, A.~W., {Massa}, D.~L., \& {Prinja}, R.~K. 2006, \apj, 637, 1025

\bibitem[{{Gr{\"a}fener} \& {Hamann}(2008)}]{graefener2008}
{Gr{\"a}fener}, G. \& {Hamann}, W. 2008, \aap, 482, 945

\bibitem[{Gr\"afener {et~al.}(2002)Gr\"afener, Koesterke, \&
  Hamann}]{Graefener2002}
Gr\"afener, G., Koesterke, L., \& Hamann, W.-R. 2002, \aap, 387, 244

\bibitem[{{Gr{\"a}fener} {et~al.}(2012){Gr{\"a}fener}, {Owocki}, \&
  {Vink}}]{Graefener2012a}
{Gr{\"a}fener}, G., {Owocki}, S.~P., \& {Vink}, J.~S. 2012, \aap, 538, A40

\bibitem[{{Gr{\"a}fener} \& {Vink}(2013)}]{graefener2013}
{Gr{\"a}fener}, G. \& {Vink}, J.~S. 2013, \aap, 560, A6

\bibitem[{{Gr{\"a}fener} {et~al.}(2011){Gr{\"a}fener}, {Vink}, {de Koter}, \&
  {Langer}}]{graefener2011}
{Gr{\"a}fener}, G., {Vink}, J.~S., {de Koter}, A., \& {Langer}, N. 2011, \aap,
  535, A56

\bibitem[{{Guerrero} \& {Chu}(2008)}]{guerrero2008}
{Guerrero}, M.~A. \& {Chu}, Y.-H. 2008, \apjs, 177, 216

\bibitem[{{Hainich} {et~al.}(2014){Hainich}, {R{\"u}hling}, {Todt}, {Oskinova},
  {Liermann}, {Gr{\"a}fener}, {Foellmi}, {Schnurr}, \& {Hamann}}]{Hainich2014}
{Hainich}, R., {R{\"u}hling}, U., {Todt}, H., {et~al.} 2014, ArXiv e-prints

\bibitem[{{Hamann} \& {Gr{\"a}fener}(2004)}]{hamann2004}
{Hamann}, W. \& {Gr{\"a}fener}, G. 2004, \aap, 427, 697

\bibitem[{{Hamann} \& {Gr{\"a}fener}(2003)}]{Hamann2003}
{Hamann}, W.-R. \& {Gr{\"a}fener}, G. 2003, \aap, 410, 993

\bibitem[{{Hamann} {et~al.}(2006){Hamann}, {Gr{\"a}fener}, \&
  {Liermann}}]{hamann2006}
{Hamann}, W.-R., {Gr{\"a}fener}, G., \& {Liermann}, A. 2006, \aap, 457, 1015

\bibitem[{{Hamann} \& {Koesterke}(1998)}]{hamann1998}
{Hamann}, W.-R. \& {Koesterke}, L. 1998, \aap, 333, 251

\bibitem[{{Heger} \& {Woosley}(2002)}]{heger2002}
{Heger}, A. \& {Woosley}, S.~E. 2002, \apj, 567, 532

\bibitem[{{H{\'e}nault-Brunet} {et~al.}(2012){H{\'e}nault-Brunet}, {Evans},
  {Sana}, {Gieles}, {Bastian}, {Ma{\'{\i}}z Apell{\'a}niz}, {Markova},
  {Taylor}, {Bressert}, {Crowther}, \& {van Loon}}]{henault2012}
{H{\'e}nault-Brunet}, V., {Evans}, C.~J., {Sana}, H., {et~al.} 2012, \aap, 546,
  A73

\bibitem[{{Hillier}(2008)}]{hillier2008}
{Hillier}, D.~J. 2008, in Clumping in Hot-Star Winds, ed. W.-R. {Hamann},
  A.~{Feldmeier}, \& L.~M. {Oskinova}, 93

\bibitem[{{Hillier} {et~al.}(2012){Hillier}, {Bouret}, {Lanz}, \&
  {Busche}}]{hillier2012}
{Hillier}, D.~J., {Bouret}, J.-C., {Lanz}, T., \& {Busche}, J.~R. 2012, \mnras,
  426, 1043

\bibitem[{{Hillier} \& {Miller}(1998)}]{hillier1998}
{Hillier}, D.~J. \& {Miller}, D.~L. 1998, \apj, 496, 407

\bibitem[{{Howarth}(1983)}]{Howarth1983}
{Howarth}, I.~D. 1983, \mnras, 203, 301

\bibitem[{{Ishii} {et~al.}(1999){Ishii}, {Ueno}, \& {Kato}}]{Ishii99}
{Ishii}, M., {Ueno}, M., \& {Kato}, M. 1999, \pasj, 51, 417

\bibitem[{{Kato} {et~al.}(2007){Kato}, {Nagashima}, {Nagayama}, {Kurita},
  {Koerwer}, {Kawai}, {Yamamuro}, {Zenno}, {Nishiyama}, {Baba}, {Kadowaki},
  {Haba}, {Hatano}, {Shimizu}, {Nishimura}, {Nagata}, {Sato}, {Murai},
  {Kawazu}, {Nakajima}, {Nakaya}, {Kandori}, {Kusakabe}, {Ishihara},
  {Kaneyasu}, {Hashimoto}, {Tamura}, {Tanab{\'e}}, {Ita}, {Matsunaga},
  {Nakada}, {Sugitani}, {Wakamatsu}, {Glass}, {Feast}, {Menzies}, {Whitelock},
  {Fourie}, {Stoffels}, {Evans}, \& {Hasegawa}}]{kato2007}
{Kato}, D., {Nagashima}, C., {Nagayama}, T., {et~al.} 2007, \pasj, 59, 615

\bibitem[{{Koesterke} {et~al.}(2002){Koesterke}, {Hamann}, \& {Gr{\"
  a}fener}}]{Koesterke2002}
{Koesterke}, L., {Hamann}, W.-R., \& {Gr{\" a}fener}, G. 2002, \aap, 384, 562

\bibitem[{{K{\"o}hler} {et~al.}(2014){K{\"o}hler}, {Langer}, {de Koter}, {de
  Mink}, {Crowther}, {Evans}, {Gr{\"a}fener}, {Sana}, {Sanyal}, {Schneider}, \&
  {Vink}}]{koehler2014}
{K{\"o}hler}, K., {Langer}, N., {de Koter}, A., {et~al.} 2014, \aap, submitted

\bibitem[{{Kudritzki} {et~al.}(1999){Kudritzki}, {Puls}, {Lennon}, {Venn},
  {Reetz}, {Najarro}, {McCarthy}, \& {Herrero}}]{kudritzki1999}
{Kudritzki}, R.~P., {Puls}, J., {Lennon}, D.~J., {et~al.} 1999, \aap, 350, 970

\bibitem[{{Lamers} \& {Cassinelli}(1999)}]{lamers1999}
{Lamers}, H.~J.~G.~L.~M. \& {Cassinelli}, J.~P. 1999, {Introduction to Stellar
  Winds}

\bibitem[{{Lamers} {et~al.}(1995){Lamers}, {Snow}, \& {Lindholm}}]{lamers1995}
{Lamers}, H.~J.~G.~L.~M., {Snow}, T.~P., \& {Lindholm}, D.~M. 1995, \apj, 455,
  269

\bibitem[{{Langer}(2009)}]{langer2009}
{Langer}, N. 2009, \nat, 462, 579

\bibitem[{{Langer} {et~al.}(2007){Langer}, {Norman}, {de Koter}, {Vink},
  {Cantiello}, \& {Yoon}}]{langer2007}
{Langer}, N., {Norman}, C.~A., {de Koter}, A., {et~al.} 2007, \aap, 475, L19

\bibitem[{{Markwardt}(2009)}]{markwardt2009}
{Markwardt}, C.~B. 2009, in Astronomical Society of the Pacific Conference
  Series, Vol. 411, Astronomical Data Analysis Software and Systems XVIII, ed.
  D.~A. {Bohlender}, D.~{Durand}, \& P.~{Dowler}, 251

\bibitem[{{Martins} \& {Hillier}(2012)}]{martins2012}
{Martins}, F. \& {Hillier}, D.~J. 2012, \aap, 545, A95

\bibitem[{{Martins} {et~al.}(2008){Martins}, {Hillier}, {Paumard},
  {Eisenhauer}, {Ott}, \& {Genzel}}]{martins2008}
{Martins}, F., {Hillier}, D.~J., {Paumard}, T., {et~al.} 2008, \aap, 478, 219

\bibitem[{{Martins} \& {Plez}(2006)}]{martins2006}
{Martins}, F. \& {Plez}, B. 2006, \aap, 457, 637

\bibitem[{{Massey}(2002)}]{massey2002:2}
{Massey}, P. 2002, \apjs, 141, 81

\bibitem[{{Massey} \& {Hunter}(1998)}]{massey1998}
{Massey}, P. \& {Hunter}, D.~A. 1998, \apj, 493, 180

\bibitem[{{Massey} {et~al.}(2002){Massey}, {Penny}, \& {Vukovich}}]{massey2002}
{Massey}, P., {Penny}, L.~R., \& {Vukovich}, J. 2002, \apj, 565, 982

\bibitem[{{Massey} {et~al.}(2005){Massey}, {Puls}, {Pauldrach}, {Bresolin},
  {Kudritzki}, \& {Simon}}]{massey2005}
{Massey}, P., {Puls}, J., {Pauldrach}, A.~W.~A., {et~al.} 2005, \apj, 627, 477

\bibitem[{{Melnick}(1985)}]{melnick1985}
{Melnick}, J. 1985, \aap, 153, 235

\bibitem[{{Mihalas}(1973)}]{mihalas1973}
{Mihalas}, D. 1973, \pasp, 85, 593

\bibitem[{{Mokiem} {et~al.}(2007){Mokiem}, {de Koter}, {Evans}, {Puls},
  {Smartt}, {Crowther}, {Herrero}, {Langer}, {Lennon}, {Najarro}, {Villamariz},
  \& {Vink}}]{mokiem2007}
{Mokiem}, M.~R., {de Koter}, A., {Evans}, C.~J., {et~al.} 2007, \aap, 465, 1003

\bibitem[{{Najarro} {et~al.}(2011){Najarro}, {Hanson}, \& {Puls}}]{najarro2011}
{Najarro}, F., {Hanson}, M.~M., \& {Puls}, J. 2011, \aap, 535, A32

\bibitem[{{Parker}(1993)}]{parker1993}
{Parker}, J.~W. 1993, \aj, 106, 560

\bibitem[{{Pauldrach} {et~al.}(2012){Pauldrach}, {Vanbeveren}, \&
  {Hoffmann}}]{pauldrach2012}
{Pauldrach}, A.~W.~A., {Vanbeveren}, D., \& {Hoffmann}, T.~L. 2012, \aap, 538,
  A75

\bibitem[{{Petrov} {et~al.}(2014){Petrov}, {Vink}, \&
  {Gr{\"a}fener}}]{petrov2014}
{Petrov}, B., {Vink}, J.~S., \& {Gr{\"a}fener}, G. 2014, ArXiv e-prints

\bibitem[{{Petrovic} {et~al.}(2006){Petrovic}, {Pols}, \&
  {Langer}}]{Petrovic2006}
{Petrovic}, J., {Pols}, O., \& {Langer}, N. 2006, \aap, 450, 219

\bibitem[{{Pietrzy{\'n}ski} {et~al.}(2013){Pietrzy{\'n}ski}, {Graczyk},
  {Gieren}, {Thompson}, {Pilecki}, {Udalski}, {Soszy{\'n}ski}, {Koz{\l}owski},
  {Konorski}, {Suchomska}, {Bono}, {Moroni}, {Villanova}, {Nardetto},
  {Bresolin}, {Kudritzki}, {Storm}, {Gallenne}, {Smolec}, {Minniti}, {Kubiak},
  {Szyma{\'n}ski}, {Poleski}, {Wyrzykowski}, {Ulaczyk}, {Pietrukowicz},
  {G{\'o}rski}, \& {Karczmarek}}]{pietrzynski2013}
{Pietrzy{\'n}ski}, G., {Graczyk}, D., {Gieren}, W., {et~al.} 2013, \nat, 495,
  76

\bibitem[{{Portegies Zwart} {et~al.}(2002){Portegies Zwart}, {Pooley}, \&
  {Lewin}}]{portegies2002}
{Portegies Zwart}, S.~F., {Pooley}, D., \& {Lewin}, W.~H.~G. 2002, \apj, 574,
  762

\bibitem[{{Prinja} {et~al.}(1990){Prinja}, {Barlow}, \& {Howarth}}]{prinja1990}
{Prinja}, R.~K., {Barlow}, M.~J., \& {Howarth}, I.~D. 1990, \apj, 361, 607

\bibitem[{{Prinja} \& {Crowther}(1998)}]{prinja1998}
{Prinja}, R.~K. \& {Crowther}, P.~A. 1998, \mnras, 300, 828

\bibitem[{{Puls} {et~al.}(1996){Puls}, {Kudritzki}, {Herrero}, {Pauldrach},
  {Haser}, {Lennon}, {Gabler}, {Voels}, {Vilchez}, {Wachter}, \&
  {Feldmeier}}]{puls1996}
{Puls}, J., {Kudritzki}, R.-P., {Herrero}, A., {et~al.} 1996, \aap, 305, 171

\bibitem[{{Puls} {et~al.}(2006){Puls}, {Markova}, {Scuderi}, {Stanghellini},
  {Taranova}, {Burnley}, \& {Howarth}}]{puls2006}
{Puls}, J., {Markova}, N., {Scuderi}, S., {et~al.} 2006, \aap, 454, 625

\bibitem[{{Puls} {et~al.}(2005){Puls}, {Urbaneja}, {Venero}, {Repolust},
  {Springmann}, {Jokuthy}, \& {Mokiem}}]{puls2005}
{Puls}, J., {Urbaneja}, M.~A., {Venero}, R., {et~al.} 2005, \aap, 435, 669

\bibitem[{{Puls} {et~al.}(2008){Puls}, {Vink}, \& {Najarro}}]{puls2008}
{Puls}, J., {Vink}, J.~S., \& {Najarro}, F. 2008, \aapr, 16, 209

\bibitem[{{Ram{\'{\i}}rez-Agudelo} {et~al.}(2013){Ram{\'{\i}}rez-Agudelo},
  {Sim{\'o}n-D{\'{\i}}az}, {Sana}, {de Koter}, {Sab{\'{\i}}n-Sanjul{\'{\i}}an},
  {de Mink}, {Dufton}, {Gr{\"a}fener}, {Evans}, {Herrero}, {Langer}, {Lennon},
  {Ma{\'{\i}}z Apell{\'a}niz}, {Markova}, {Najarro}, {Puls}, {Taylor}, \&
  {Vink}}]{Ramirez2013}
{Ram{\'{\i}}rez-Agudelo}, O.~H., {Sim{\'o}n-D{\'{\i}}az}, S., {Sana}, H.,
  {et~al.} 2013, \aap, 560, A29

\bibitem[{{Repolust} {et~al.}(2004){Repolust}, {Puls}, \&
  {Herrero}}]{repolust2004}
{Repolust}, T., {Puls}, J., \& {Herrero}, A. 2004, \aap, 415, 349

\bibitem[{{Rivero Gonz{\'a}lez} {et~al.}(2012{\natexlab{a}}){Rivero
  Gonz{\'a}lez}, {Puls}, {Massey}, \& {Najarro}}]{RPMN2012}
{Rivero Gonz{\'a}lez}, J.~G., {Puls}, J., {Massey}, P., \& {Najarro}, F.
  2012{\natexlab{a}}, \aap, 543, A95

\bibitem[{{Rivero Gonz{\'a}lez} {et~al.}(2012{\natexlab{b}}){Rivero
  Gonz{\'a}lez}, {Puls}, {Najarro}, \& {Brott}}]{rivero2012}
{Rivero Gonz{\'a}lez}, J.~G., {Puls}, J., {Najarro}, F., \& {Brott}, I.
  2012{\natexlab{b}}, \aap, 537, A79

\bibitem[{{Rubele} {et~al.}(2012){Rubele}, {Kerber}, {Girardi}, {Cioni},
  {Marigo}, {Zaggia}, {Bekki}, {de Grijs}, {Emerson}, {Groenewegen},
  {Gullieuszik}, {Ivanov}, {Miszalski}, {Oliveira}, {Tatton}, \& {van
  Loon}}]{rubele2012}
{Rubele}, S., {Kerber}, L., {Girardi}, L., {et~al.} 2012, \aap, 537, A106

\bibitem[{{Sab{\'{\i}}n-Sanjuli{\'a}n}
  {et~al.}(2014){Sab{\'{\i}}n-Sanjuli{\'a}n}, {Sim{\'o}n-D{\'{\i}}az},
  {Herrero}, {Walborn}, {Puls}, {Ma{\'{\i}}z Apell{\'a}niz}, {Evans}, {Brott},
  {de Koter}, {Garcia}, {Markova}, {Najarro}, {Ram{\'{\i}}rez-Agudelo}, {Sana},
  {Taylor}, \& {Vink}}]{sabin2014}
{Sab{\'{\i}}n-Sanjuli{\'a}n}, C., {Sim{\'o}n-D{\'{\i}}az}, S., {Herrero}, A.,
  {et~al.} 2014, \aap, 564, A39

\bibitem[{{Sana} {et~al.}(2013){Sana}, {de Koter}, {de Mink}, {Dunstall},
  {Evans}, {H{\'e}nault-Brunet}, {Ma{\'{\i}}z Apell{\'a}niz},
  {Ram{\'{\i}}rez-Agudelo}, {Taylor}, {Walborn}, {Clark}, {Crowther},
  {Herrero}, {Gieles}, {Langer}, {Lennon}, \& {Vink}}]{sana2013}
{Sana}, H., {de Koter}, A., {de Mink}, S.~E., {et~al.} 2013, \aap, 550, A107

\bibitem[{{Sana} {et~al.}(2012){Sana}, {de Mink}, {de Koter}, {Langer},
  {Evans}, {Gieles}, {Gosset}, {Izzard}, {Le Bouquin}, \&
  {Schneider}}]{sana2012Sci}
{Sana}, H., {de Mink}, S.~E., {de Koter}, A., {et~al.} 2012, Science, 337, 444

\bibitem[{{Schmutz} {et~al.}(1989){Schmutz}, {Hamann}, \&
  {Wessolowski}}]{schmutz1989}
{Schmutz}, W., {Hamann}, W., \& {Wessolowski}, U. 1989, \aap, 210, 236

\bibitem[{{Schneider} {et~al.}(2014){Schneider}, {Izzard}, {de Mink}, {Langer},
  {Stolte}, {de Koter}, {Gvaramadze}, {Hu{\ss}mann}, {Liermann}, \&
  {Sana}}]{schneider2014}
{Schneider}, F.~R.~N., {Izzard}, R.~G., {de Mink}, S.~E., {et~al.} 2014, \apj,
  780, 117

\bibitem[{{Schnurr} {et~al.}(2009){Schnurr}, {Chen{\'e}}, {Casoli}, {Moffat},
  \& {St-Louis}}]{schnurr2009}
{Schnurr}, O., {Chen{\'e}}, A.-N., {Casoli}, J., {Moffat}, A.~F.~J., \&
  {St-Louis}, N. 2009, \mnras, 397, 2049

\bibitem[{{Schnurr} {et~al.}(2008){Schnurr}, {Moffat}, {St-Louis}, {Morrell},
  \& {Guerrero}}]{schnurr2008}
{Schnurr}, O., {Moffat}, A.~F.~J., {St-Louis}, N., {Morrell}, N.~I., \&
  {Guerrero}, M.~A. 2008, \mnras, 389, 806

\bibitem[{{Smette} {et~al.}(2010){Smette}, {Sana}, \& {Horst}}]{Smette2010}
{Smette}, A., {Sana}, H., \& {Horst}, H. 2010, Highlights of Astronomy, 15, 533

\bibitem[{{Sundqvist} \& {Owocki}(2013)}]{sundqvist2013}
{Sundqvist}, J.~O. \& {Owocki}, S.~P. 2013, \mnras, 428, 1837

\bibitem[{{Sundqvist} {et~al.}(2011){Sundqvist}, {Puls}, {Feldmeier}, \&
  {Owocki}}]{sundqvist2011}
{Sundqvist}, J.~O., {Puls}, J., {Feldmeier}, A., \& {Owocki}, S.~P. 2011, \aap,
  528, A64

\bibitem[{{{\v S}urlan} {et~al.}(2013){{\v S}urlan}, {Hamann}, {Aret},
  {Kub{\'a}t}, {Oskinova}, \& {Torres}}]{surlan2013}
{{\v S}urlan}, B., {Hamann}, W.-R., {Aret}, A., {et~al.} 2013, ArXiv e-prints

\bibitem[{{Vink}(2006)}]{vink2006}
{Vink}, J.~S. 2006, in Astronomical Society of the Pacific Conference Series,
  Vol. 353, Stellar Evolution at Low Metallicity: Mass Loss, Explosions,
  Cosmology, ed. H.~J.~G.~L.~M. {Lamers}, N.~{Langer}, T.~{Nugis}, \&
  K.~{Annuk}, 113

\bibitem[{{Vink} {et~al.}(2000){Vink}, {de Koter}, \& {Lamers}}]{vink2000}
{Vink}, J.~S., {de Koter}, A., \& {Lamers}, H.~J.~G.~L.~M. 2000, \aap, 362, 295

\bibitem[{{Vink} {et~al.}(2001){Vink}, {de Koter}, \& {Lamers}}]{vink2001}
{Vink}, J.~S., {de Koter}, A., \& {Lamers}, H.~J.~G.~L.~M. 2001, \aap, 369, 574

\bibitem[{{Vink} \& {Gr{\"a}fener}(2012)}]{vink2012}
{Vink}, J.~S. \& {Gr{\"a}fener}, G. 2012, \apjl, 751, L34

\bibitem[{{Vink} {et~al.}(2013){Vink}, {Heger}, {Krumholz}, {Puls}, {Banerjee},
  {Castro}, {Chen}, {Chene}, {Crowther}, {Daminelli}, {Grafener}, {Groh},
  {Hamann}, {Heap}, {Herrero}, {Kaper}, {Najarro}, {Oskinova}, {Roman-Lopes},
  {Rosen}, {Sander}, {Shirazi}, {Sugawara}, {Tramper}, {Vanbeveren}, {Voss},
  {Wofford}, \& {Zhang}}]{vink2013:IAU}
{Vink}, J.~S., {Heger}, A., {Krumholz}, M.~R., {et~al.} 2013, ArXiv e-prints

\bibitem[{{Vink} {et~al.}(2011){Vink}, {Muijres}, {Anthonisse}, {de Koter},
  {Gr{\"a}fener}, \& {Langer}}]{vink2011}
{Vink}, J.~S., {Muijres}, L.~E., {Anthonisse}, B., {et~al.} 2011, \aap, 531,
  A132

\bibitem[{{Walborn} \& {Blades}(1997)}]{walborn1997}
{Walborn}, N.~R. \& {Blades}, J.~C. 1997, \apjs, 112, 457

\bibitem[{{Walborn} {et~al.}(2010){Walborn}, {Howarth}, {Evans}, {Crowther},
  {Moffat}, {St-Louis}, {Fari{\~n}a}, {Bosch}, {Morrell}, {Barb{\'a}}, \& {van
  Loon}}]{walborn2010a}
{Walborn}, N.~R., {Howarth}, I.~D., {Evans}, C.~J., {et~al.} 2010, \aj, 139,
  1283

\bibitem[{{Walborn} {et~al.}(2014){Walborn}, {Sana}, {Sim{\'o}n-D{\'{\i}}az},
  {Ma{\'{\i}}z Apell{\'a}niz}, {Taylor}, {Evans}, {Markova}, {Lennon}, \& {de
  Koter}}]{walborn2014}
{Walborn}, N.~R., {Sana}, H., {Sim{\'o}n-D{\'{\i}}az}, S., {et~al.} 2014, \aap,
  564, A40

\bibitem[{{Yungelson} {et~al.}(2008){Yungelson}, {van den Heuvel}, {Vink},
  {Portegies Zwart}, \& {de Koter}}]{yungelson2008}
{Yungelson}, L.~R., {van den Heuvel}, E.~P.~J., {Vink}, J.~S., {Portegies
  Zwart}, S.~F., \& {de Koter}, A. 2008, \aap, 477, 223

\bibitem[{{Yusof} {et~al.}(2013){Yusof}, {Hirschi}, {Meynet}, {Crowther},
  {Ekstr{\"o}m}, {Frischknecht}, {Georgy}, {Abu Kassim}, \&
  {Schnurr}}]{yusof2013}
{Yusof}, N., {Hirschi}, R., {Meynet}, G., {et~al.} 2013, \mnras, 433, 1114

\end{thebibliography}

\onecolumn
\begin{landscape}
\begin{longtable}{l@{\,}l@{~}c@{~~}c@{~}c@{~~~~}c@{~}c@{~}c@{~}c@{~}c@{~}c@{~}c@{~}c@{~}c@{~}c@{~}c@{~}c@{~~~}c@{~}c@{~}c@{~}c}
\caption{\label{t:parameters} Stellar parameters of stars that are considered in the discussion.}\\
\hline\hline
Star	&Spectral type	&source& $\log L_{\star}/L_{\odot}$	& $T_{\mathrm{eff}}$	&$T_{\star}$	& $\log \dot{M}/\sqrt{f}$	& $\varv_{\infty}$$^{\ddagger}$	& $\varv_{\infty}$$^{\blacktriangle}$	& $Y^{\dagger}$ & $N^{\bigstar}$	& $\log Q_0$	& $M_{\star}/M_{\odot}^{\blacklozenge}$	& $\Gamma_{\mathrm{e}}^{\blacklozenge}$	& $\varv \sin i$ &source& $M_{\mathrm{V}}$ &	$R_{\mathrm{V}}$ &	$E(B-V)~$	&	$E(V-K_{\mathrm{s}})$\\
VFTS&			&&						& [kK]					& [kK]			& [$M_{\odot}$/yr]						& [km\,s$^{-1}$] & [km\,s$^{-1}$] & && [ph\,s$^{-1}$]	& &&[km\,s$^{-1}$]&&[mag]\\
\hline
\endfirsthead
\caption{Stellar parameters of stars that are considered in the discussion.}\\
\hline\hline
Star	&Spectral type	&source& $\log L_{\star}/L_{\odot}$	& $T_{\mathrm{eff}}$	&$T_{\star}$	& $\log (\dot{M}/\sqrt{f_{\rm v}})$	& $\varv_{\infty}$$^{\ddagger}$	& $\varv_{\infty}$$^{\blacktriangle}$	& $Y^{\dagger}$ & $N^{\bigstar}$	& $\log Q_0$	& $M_{\star}/M_{\odot}^{\blacklozenge}$	& $\Gamma_{\mathrm{e}}^{\blacklozenge}$	& $\varv \sin i$ &source& $M_{\mathrm{V}}$ &	$R_{\mathrm{V}}$ &	$E(B-V)~$	&	$E(V-K_{\mathrm{s}})$\\
VFTS&			&&						& [kK]					& [kK]			& [$M_{\odot}$/yr]						& [km\,s$^{-1}$] & [km\,s$^{-1}$] & && [ph\,s$^{-1}$]	& &&[km\,s$^{-1}$]&&[mag]\\
\hline
\endhead
\hline
\endfoot
016	&	O2 III-If*	&	1 		& 6.23	& 53.1	& 54.1	& -5.4	& 	3400	& 3410	&	0.25 	& $ne$	& 50.08		& 120	& 0.38 & 94	&{\sc i}	& -6.1	& 3.34	& 0.356	& 1.046\\
064	&	O7.5 II(f)	&	1		& 6.0	& 39.8	& 39.9	& -5.4	&	--	&2640	&	0.25 	& $e$ & 49.68		& 88	& 0.3 & 104	&{\sc i}	&-6.5	& 5.85	& 0.45	& 2.266	\\
072	&	O2 V-III(n)((f*))	&	1	 	& 5.96	& 50.1	& 50.7	& -5.8	& 	--	& 3340 &	0.25 	& $e$ 	& 49.75		& 84	& 0.29 & 200	&{\sc i}	& -5.7	& 4.84	& 0.18	& 0.755\\
108	&	WN7h	&	3 			& 5.7	& 39.8	& 53.1	& -4.0	& 	1000	& 1950 &	0.775 	& $e$ 	& 49.50		& 33$^{\#}$	& 0.29 & $<200$	& {\sc iii}		& -5.9	& 4.28	& 0.311	& 1.157\\
169	&	O2.5 V(n)((f*))	&	1		& 5.89	& 47.3	& 47.7	& -5.5	&	--	& 3200 &	0.25	& $n$ &49.70		& 76	& 0.27 & 200	&{\sc ii}	& -5.7	& 5.07	& 0.35	& 1.534\\
171	&	O8 II-III(f)	&	1		& 5.59	& 37.6	& 37.7	& -5.7	&	--	& 2670 &	0.25	& $ne$& 49.18		& 54	& 0.19	& 81	&{\sc i}	& -5.7 & 4.63	& 0.268	& 1.077\\
180	&	O3 If*	&	4			& 5.92	& 42.2	& 42.3	& -5.0	& 	2500	& 2360 &	0.55	& $e$ & 49.65		& 58	& 0.32 & 93	&{\sc i}	& -6.1	& 4.89	& 0.237	& 1.004\\
216	&	O4 V((fc))	&	1		& 5.84	& 44.7	& 44.9	& -5.7	&	--	& 3040 &	0.25	& $n$ & 49.62		& 72	& 0.26 & 100	&{\sc ii}	& -5.7	& 2.91	& 0.571	& 1.474\\
259	&	O6 Iaf	&	1			& 6.1	& 37.6	& 37.7	& -5.0	&	--	& 2350 &	0.325	& $e$ 	& 49.71		& 94	& 0.35 & 66	&{\sc i}	& -7.0	& 4.0	& 0.529	& 1.844\\
267	&	O3 III-I(n)f*	&	1		& 6.01	& 44.7	& 44.9	& -5.1	&	--	& 2960 &	0.25	& $ne$& 49.79		& 89	& 0.31 & 145	&{\sc i}	& -6.2	& 4.35	& 0.27	& 1.02\\
333	&	O8 II-III((f))	&	1		& 6.02	& 37.6	& 37.7	& -5.4	&	--	& 2490 &	0.25 	& $n$	& 49.61		& 90	& 0.31 & 77	&{\sc i}	& -6.8	& 2.91	& 0.258	& 0.666\\
404	&	O3.5 V(n)((fc))	&	1		& 5.91	& 44.7	& 44.9	& -5.8	&	--	& 3010 &	0.25	& $n$ & 49.69		& 78	& 0.28 & --	& --		& -5.9	& 4.57	& 0.341	& 1.352\\
422	&	O4 III(f)	&	1			& 5.7	& 39.8	& 39.9	& -5.6	&	--	& 2770 &	 0.25	& $ne$ & 49.38		& 61	& 0.22 & 356	&{\sc i}	& -5.8	& 4.3	& 0.56	&2.093\\
427	&	WN8(h)	&	3			& 6.13	& 39.8	& 41.6	& -3.9	& 	1000	& 1610 &	0.925	& $e$ 	& 49.90		& 48$^{\#}$	& 0.46 & $<200$	&{\sc iii}		& -7.0	& 4.16	& 0.534	& 1.934\\
455	&	O5: V:n	&	1			& 5.63	& 42.2	& 42.3	& -5.6	& 	--	& 2970 &	0.25	& $ne$	& 49.36		& 56	& 0.2  & --	& --		& -5.4	& 3.7  	& 0.47	& 1.525\\
457	&	O3.5 If*/WN7	&	4		& 6.2	& 39.8	& 39.9	& -4.6	& 	1800	& 2340 &	0.4	& $e$ 	& 49.89		& 100	& 0.39 & $<200$	& {\sc iii}		& -7.0	& 2.83	& 0.806	& 2.015\\
482	&	O2.5 If*/WN6	&	4		& 6.4	& 42.2	& 42.3	& -4.4 	& 	2600	& 2540 &	0.325	& $e$ 	& 50.14		& 145	& 0.45 & $<200$	& {\sc iii}		& -7.3	& 3.85	& 0.436	& 1.467\\
506	&	ON2 V((n))((f*))	&	1		& 6.24	& 47.3	& 47.7	& -5.6	&	--	& 3040 &	0.25	& $e$ & 50.05		& 122	& 0.38 & 100	&{\sc ii}	& -6.5	& 3.99	& 0.341	& 1.187\\
512	&	O2 V-III((f*))	&	1		& 6.04	& 47.3	& 47.7	& -5.6	& 	--	& 3120 &	0.25	& $ne$& 49.86		& 93	& 0.32 & --	& --		& -6.1	& 3.57	& 0.52	& 1.628\\
518	&	O3.5 III(f*)	&	1		& 5.75	& 44.7	& 44.9	& -5.5	&	--	& 3090 &	0.25	& $e$ 	& 49.53		& 65	& 0.23 & 91	&{\sc i}	& -5.5	& 3.6  	& 0.591	& 1.864\\
532	&	O3 V(n)((f*))z + OB	&	1	& 5.74	& 44.7	& 44.9	& -5.8	&	--	& 3090 &	0.25 	& $n$	& 49.52		& 64	& 0.23 & --	& --		& -5.5	& 3.36	& 0.521	& 1.542\\
542	&	O2 If*/WN5	&	4		& 6.16	& 44.7	& 44.9	& -4.6	& 	2700	& 2520 &	0.475	& $e$ 	& 49.94		& 87	& 0.39 & $<200$	& {\sc iii}		& -6.5	& 3.71	& 0.408	& 1.324\\
545	&	O2 If*/WN5	&	4		& 6.3	& 47.3	& 47.7	& -4.6	& 	2800	& 3020 &	0.25	& $e$ & 50.11		& 133	& 0.4  & $<200$	& {\sc iii}		& -6.7	& 3.67	& 0.436	& 1.421\\
566	&	O3 III(f*)	&	1		& 5.85	& 44.7	& 44.9	& -5.4	&	--	& 3030 &	0.25	& $e$ & 49.63		& 73	& 0.26 & 91	&{\sc i}	& -5.8	& 3.48	& 0.381	& 1.162\\
599	&	O3 III(f*)	&	1		& 5.99	& 44.7	& 44.9	& -5.3	&	--	& 2970 &	0.25	& $e$ & 49.77		& 87	& 0.3  & 113	&{\sc i}	& -6.1	& 3.55	& 0.401	& 1.248\\
603	&	O4 III(fc)	&	1		& 5.98	& 42.2	& 42.3	& -5.4	& 	--	& 2810 &	0.25	& $ne$& 49.71		& 85	& 0.3  & --	& --		& -6.3	& 4.9  	& 0.360	& 1.528\\
608	&	O4 III(f)	&	1			& 5.86	& 42.2	& 42.3	& -5.4	&	--	& 2860 &	0.25	& $e$ & 49.60		& 74	& 0.26 & --	& --		& -6.0	& 3.54	& 0.481	& 1.49\\
617	&	WN5ha	&	8			& 6.29	& 53.1	& 53.7	& -4.3	& 	2800	& 2650 &	0.625	& $e$ 	& 50.14		& 90	& 0.46 & $<200$	& {\sc iii}		& -6.3	& 4.5  	& 0.212	& 0.848\\
621	&	O2 V((f*))z	&	1		& 6.14	& 50.1	& 50.7	& -5.8	&	--	& 3260 &	0.25	& $e$ 	& 49.97		& 107	& 0.35 & 80	&{\sc ii}	& -6.1	& 5.09 	& 0.591	& 2.595\\
626	&	O5-6 n(f)p	&	1		& 5.6	& 39.8	& 39.9	& -5.5	&	--	& 2590 &	0.4	& $e$ & 49.27		& 46	& 0.21 & 278	&{\sc i}	& -5.5	& 3.28	& 0.581	& 1.679\\
664	&	O7 II(f)	&	1			& 5.72	& 39.8	& 39.9	& -5.5	&	--	& 2760 &	0.25	& $e$ & 49.40		& 62	& 0.23 & 63	&{\sc i}	& -5.8	& 3.39	& 0.46	& 1.371\\
669	&	O8 Ib(f)	&	1			& 5.71	& 37.6	& 37.7	& -5.6	&	--	& 2510 &	0.325	& $e$ 	& 49.31		& 57	& 0.23 & 64	&{\sc i}	& -6.0	& 2.8	& 0.590	& 1.466\\
682	&	WN5h	&	2			& 6.51	& 52.1	& 54.4	& -4.2	& 	2600	& 2860 &	0.45	& $e$ 	& 50.35		& 153	& 0.51 & $<200$	& {\sc iii}		& -6.9	& 4.74	& 0.941	& 3.863\\
695	&	WN6h + ?	&	2			& 6.5	& 39.8	& 39.9	& -3.7 	& 	1600	& 1620 &	0.85	& $e$ 	& 50.22		& 96$^{\#}$	& 0.58 & $<200$	& {\sc iii}		& -7.8	& 4.42	& 0.315	& 1.211\\
755	&	O3 Vn((f*))	&	1		& 5.64	& 47.3	& 47.7	& -6.0	&	--	& 3330 &	0.25	& $n$& 49.45		& 57   	& 0.2  & 285	&{\sc ii}	& -5.1	& 2.86	& 0.561	& 1.421\\
758	&	WN5h	&	2			& 6.36	& 47.3	& 47.6	& -3.9	& 	2000	& 2090 &	0.775	& $e$ 	& 50.17		& 84$^{\#}$	& 0.51 & $<200$	& {\sc iii}		& -7.0	& 3.63	& 0.428	& 1.362\\
797	&	O3.5 V((n))((fc))	&	1		& 5.57	& 44.7	& 44.9	& -6.1 	&	--	& 3180 &	0.25	& $n$& 49.35		& 53	& 0.19 & 140	&{\sc ii}	& -5.1	& 3.38	& 0.372	& 1.106\\
1001	&	WN6(h)	&	5			& 6.2	& 39.8	& 42.2	& -3.8	& 	1100	& 1700 &	0.85	& $e$ 	& 49.96		& 60$^{\#}$	& 0.47 & $<200$	& {\sc iii}		& -7.4	& 2.93	& 0.523	& 1.36\\
1017	&	O2 If*/WN5	&	4		& 6.21	& 50.1	& 50.5	& -4.5	& 	2700	& 2670 &	0.55	& $e$ 	& 50.04		& 86	& 0.42 & $<200$	& {\sc iii}		& -6.3	& 3.87	& 0.597	& 2.015\\
1018	&	O3 III(f*) + mid/late O	&	5,6	& 5.92	& 42.2	& 42.3	& -5.9	&	--	& 2830 &	0.25 	& $n$	& 49.65		& 79	& 0.28 & --	& --		& -6.1	& 4.15	& 0.507	& 1.83\\
1021	&	O4 If+	&	5			& 6.34	& 39.8	& 39.9	& -4.6	& 	--	& 2530 &	0.25	& $e$ & 50.02		& 141	& 0.41 & $<200$	& {\sc iii}		& -7.3	& 4.16	& 0.519	& 1.881\\
1022	&	O3.5 If*/WN7	&	4		& 6.48	& 42.2	& 42.3	& -4.3	& 	2700	& 2670 &	0.25	& $ne$ & 50.22		& 178	& 0.46 & $<200$	& {\sc iii}		& -7.5	& 4.36	& 0.567	& 2.149\\
1025	&	WN5h	&	7			& 6.58	& 42.2	& 42.3	& -3.9	&	--	& 1910 &	0.7	& $e$ 	& 50.32		& 132	& 0.58 & $<200$	& {\sc iii}		& -7.9	& 4.5	& 0.608	& 2.373\\
Mk42	&	O2 If*	&	4			& 6.56	& 47.3	& 47.7	& -4.6	& 	2800	& 2840 &	0.325	& $e$ 	& 50.37		& 189	& 0.49 & $<200$	& {\sc iii}		& -7.4	& 4.09	& 0.4  	& 1.426\\
\hline
\end{longtable}
\tablefoot{$^{(\dagger)}$Y is the helium mass fraction, $^{(\ddagger)}$ measured $\varv_{\infty}$, $^{(\blacktriangle)}$ calculated $\varv_{\infty}$ using \cite{lamers1995} with Eq.\,\ref{e:vesc}, nitrogen abundance: $n$ for normal, $ne$ for between normal and enhanced, and $e$ for enhanced, $^{(\blacklozenge)}$using the mass-luminosity relation for chemical homogeneity by \cite{graefener2011}.\\
Sources: $^{(1)}$\cite{walborn2014}, $^{(2)}$\cite{evans2011}, $^{(3)}$\cite{crowther1997}, $^{(4)}$\cite{crowther2011}, $^{(5)}$\cite{massey1998}, $^{(6)}$\cite{henault2012}, $^{(7)}$\cite{crowther1998}, $^{(8)}$\cite{foellmi2003}, $^{(9)}$\cite{melnick1985}, $^{(10)}$\cite{walborn1997}.\\
($\#$) Stellar masses and Eddington factors under the assumption of He-core burning using the relation by \cite{graefener2011}: VFTS\,108 (20\,$M_{\odot}$, 0.29), 427 (38\,$M_{\odot}$, 0.46), 695 (72\,$M_{\odot}$, 0.58), 758 (55\,$M_{\odot}$, 0.51), and 1001 (42\,$M_{\odot}$, 0.47).\\
Sources for $\varv \sin i$: ({\sc i}) \cite{Ramirez2013}, ({\sc ii}) \cite{sabin2014}, ({\sc iii}) conservative upper limits from this work.}

\begin{table*}[h]
\centering
\caption{\label{t:parameters_ex} Stars with uncertain stellar parameters that are excluded from the discussion.}
\begin{tabular}{l@{~}l@{}c@{}@{}c@{\,}c@{~~~}c@{~}c@{~}c@{~}c@{~}c@{~}c@{~}c@{~}c@{~}c@{~}c@{\,}c@{~}c@{~}c@{\,}c@{\,}c}
\hline\hline
Star	&Spectral type	&src.& $\log L_{\star}/L_{\odot}$	& $T_{\mathrm{eff}}$	&$T_{\star}$	& $\log (\dot{M}/\sqrt{f_{\rm v}})$	& $\varv_{\infty}$$^{\ddagger}$	& $\varv_{\infty}$$^{\blacktriangle}$	& $Y^{\dagger}$ & $N^{\bigstar}$	& $\log Q_0$	& $M_{\star}/M_{\odot}^{\blacklozenge}$	& $\Gamma_{\mathrm{e}}^{\blacklozenge}$	& $\varv \sin i$ &src.& $M_{\mathrm{V}}$ &	$R_{\mathrm{V}}$ &	$E(B-V)~$	&	$E(V-K_{\mathrm{s}})$\\
VFTS&			&						&& [kK]					& [kK]			& [$M_{\odot}$\,yr$^{-1}$]						& [km\,s$^{-1}$] & [km\,s$^{-1}$] &&& [ph\,s$^{-1}$]	& &&[km\,s$^{-1}$]&&[mag]\\
\hline
063	&	O5 III(n)(fc)+ sec	&1	& 5.76	& 42.2& 42.3& -5.7&	--	& 2910 &	0.25 		& $ne$	& 49.49	& 65	& 0.24 & --  & -- 	& -5.7 & 4.56& 0.321& 1.27\\
094	&	O3.5 Inf*p + sec		&1	& 5.98	& 42.2& 42.3& -5.0&	--	& 2690&		0.325 		& $e$ 	& 49.71	& 79	& 0.31 & --  & -- 	& -6.3 & 4.39& 0.43 & 1.641\\
145	&	O8fp			&1	& 5.87	& 39.8& 39.9& -5.3&	--	& 2470 &	0.4		& $e$ 	& 49.55	& 65	& 0.28 & 93  & {\sc i}	& -6.2 & 3.92& 0.511& 1.748\\
147	&	WN6(h)			&2	& 5.83	& 39.8& 42.5& -4.5& 	1100	& 2030	& 	0.7		& $e$ 	& 49.61	& 43	& 0.31 & $<200$ & {\sc iii} 	& -- & --	& --	& --\\
151	&	O6.5 II(f)p		&1	& 6.03	& 42.2& 42.3& -5.4&	--	& 2550 &	0.4		& $e$ 	& 49.77	& 79	& 0.33 & 89  & {\sc i}	& -6.4 & 2.09& 0.423& 0.802\\
208	&	O6(n)fp			&1	& 5.64	& 37.6& 37.7& -5.3&	--	& 2530 &	0.325		& $e$ 	& 49.25	& 53	& 0.21 & 271 & {\sc i}	& -5.8 & 3.02& 0.649& 1.731\\
402	&	WN7(h) + OB		&2	& 5.92	& 39.8& 46.4& -3.6&	1800	& 1770 &	0.85 		& $e$	& 49.69	& 40	& 0.37 & $<200$ & {\sc iii}	& -6.7 & 0.85& 0.575& 0.495\\
406	&	O6 Vnn		&1	& 5.48	& 37.6& 37.7& -5.4&	--	& 2600	&	0.325		& $e$ 	& 49.09	& 44	& 0.18 & 356 & {\sc i}	& -5.4 & 3.67& 0.329& 1.058\\
440	&	O6-6.5 II(f)		&1	& 5.88	& 39.8& 39.9& -5.4&	--	& 2690 &	0.25 		& $e$ 	& 49.56	& 76	& 0.27 & 94  & {\sc i}	& -6.2 & 4.06& 0.34 & 1.205\\
445	&	O3-4 V:((fc)): + O4-7 V:((fc)):&1	& 5.78	& 44.7& 44.9& -5.8& 	--	& 3070 &	0.25		& $ne$	& 49.56	& 67	& 0.24 & --  & --	& -5.6 & 4.39& 0.422& 1.607\\
468	&	O2 V((f*)) + OB		&1	& 6.0	& 44.7& 44.9& -5.5& 	--	& 2960 &	0.25		& $e$ 	& 49.79	& 89	& 0.31 & 80  & {\sc ii}	& -6.1 & 6.22& 0.361& 1.928\\
509	&	WN5(h) + early O	&2	& 6.43	& 42.2& 42.3& -4.1& 	2200	& 1840 &	0.775		& $e$ 	& 50.17	& 94	& 0.54 & $<200$ & {\sc iii}	& -7.4 & 5.37& 0.295& 1.364\\
527	&	O6.5 Iafc + O6 Iaf	&1	& 6.29	& 34.0& 34.9& -5.5&	1200	& -- 	&	0.25		& $e$ 	& 49.99	& --	& --   & --	& --	& -7.6 & 2.96& 0.359& 1.293\\
538	&	ON9 Ia: + O7.5: I:(f):	&1	& 5.64	& 37.6& 37.7& -5.5&	--	& 2640 &	0.25		& $e$ 	& 49.24	& 57	& 0.21 & --	& --	& -5.8 & 4.49& 0.289& 1.124\\
562	&	O4V			&9	& 6.05	& 42.2& 42.3& -5.3&	--	& 2780 &	$\geqq0.25$	& $e$ 	& 49.79	& 94  	& 0.32 & --	& --	& -6.4 & 3.91& 0.411& 1.401\\
1014	&	O3 V + mid/late O	&5,6	& 6.22	& 44.7& 44.9& -5.5&	--	& 2870 &	$\geqq0.25$ 	& $ne$	& 50.00	& 119	& 0.38 & --	& --	& -6.7 & 6.52& 0.366& 2.044\\
1026	&	O3 III(f*) + mid/late O	&5,6	& 5.83	& 42.2& 42.3& -5.5&	--	& 2880 &	0.25		& $e$ 	& 49.57	& 71 	& 0.25 & --	& --	& -5.9 & 4.74& 0.403& 1.653\\
1028	&	O3 III(f*) or O4-5V&5,6,10&6.09& 47.3& 47.7& -5.7&	--	& 3100 &	0.25		& $e$ 	& 49.91	& 99 	& 0.33 & --	& --	& -6.2 & 4.54& 0.336& 1.322\\
\hline
\end{tabular}
\tablefoot{$^{(\dagger)}$Y is the helium mass fraction, $^{(\ddagger)}$ measured $\varv_{\infty}$, $^{(\blacktriangle)}$ calculated $\varv_{\infty}$ using \cite{lamers1995} with Eq.\,\ref{e:vesc}, nitrogen abundance: $n$ for normal, $ne$ for between normal and enhanced, and $e$ for enhanced, $^{(\blacklozenge)}$using the mass-luminosity relation chemical homogeneity by \cite{graefener2011}.\\
Sources: $^{(1)}$\cite{walborn2014}, $^{(2)}$\cite{evans2011}, $^{(3)}$\cite{crowther1997}, $^{(4)}$\cite{crowther2011}, $^{(5)}$\cite{massey1998}, $^{(6)}$\cite{henault2012}, $^{(7)}$\cite{crowther1998}, $^{(8)}$\cite{foellmi2003}, $^{(9)}$\cite{melnick1985}, $^{(10)}$\cite{walborn1997}.\\
Sources for $\varv \sin i$: ({\sc i}) \cite{Ramirez2013}, ({\sc ii}) \cite{sabin2014}, ({\sc iii}) conservative upper limits from this work.}
\end{table*}
\end{landscape}

\clearpage
\begin{appendix}

\twocolumn

\section{Comments on individual objects}

\label{s:targets}
Here we describe the individual properties of the Of/WN and WNh stars
in our sample (\S\,\ref{s:slash_wn}). This is followed by a discussion of two peculiarities that occurred during the spectroscopic analysis (\S\,\ref{s:pot}).

\subsection{Of/WN, WNh and WN(h) stars}\label{s:slash_wn}
We now give a short summary of individual stars and their
characteristics.

\begin{figure}
\begin{center}
\includegraphics[width=17cm]{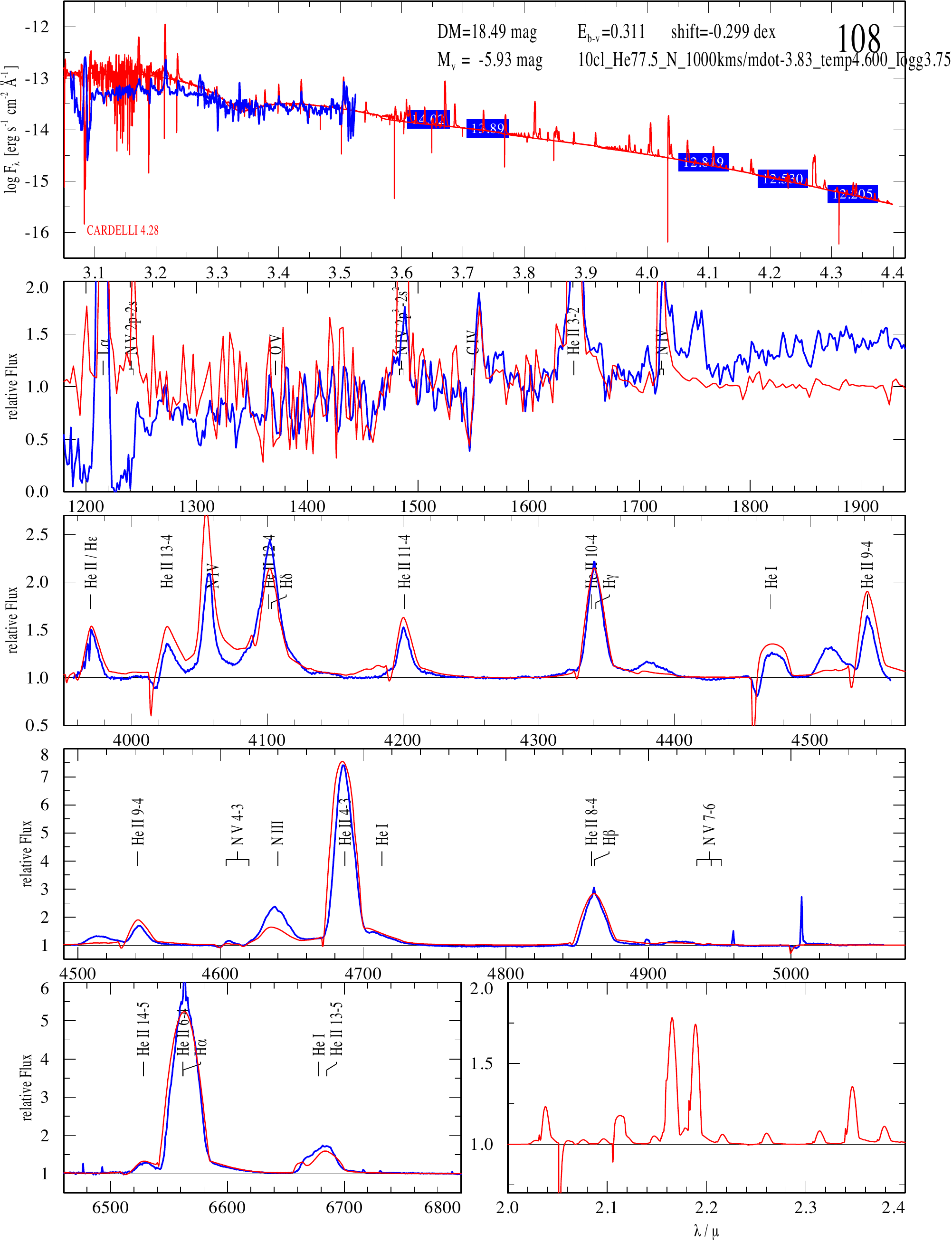}
\end{center}
\caption{The temperature of VFTS\,108 (WN7h) is based on the lines He\,{\sc i}\,$\lambda 4471$,  N\,{\sc iii}\,$\lambda 4634/4640$, N\,{\sc iv}\,$\lambda 4058$,  N\,{\sc v} $\lambda4604/4620$. The He mass fraction of the star is slightly lower than that in the $77.5$\% model. The N\,{\sc iv}\,$\lambda 4058$ and $\mathrm{H_{\delta}}$ lines do not fit because of the normalisation. Even though the star is still relatively H-rich its position in the HRD suggests that the star is He-burning. Because of convergence difficulties the model had to be calculated with a lower $\log g$. For emission lines $\log g$ does not affect the results.}
\label{a:108}
\end{figure}

VFTS\,108 (BAT99-089/Brey71): A single WN7h-star with no noticeable
variation in the peak intensity of the emission lines between the
observed epochs. There are no absorption lines suitable for RV
measurements to investigate the presence of a companion or the stars's
runaway status. The star is not located in a cluster and lies in the
field surrounded by a few fainter objects. The quality of the model is
reasonably good and the derived stellar parameters are well
constrained (Fig.\,\ref{a:108}).

\begin{figure}
\begin{center}
\includegraphics[width=17cm]{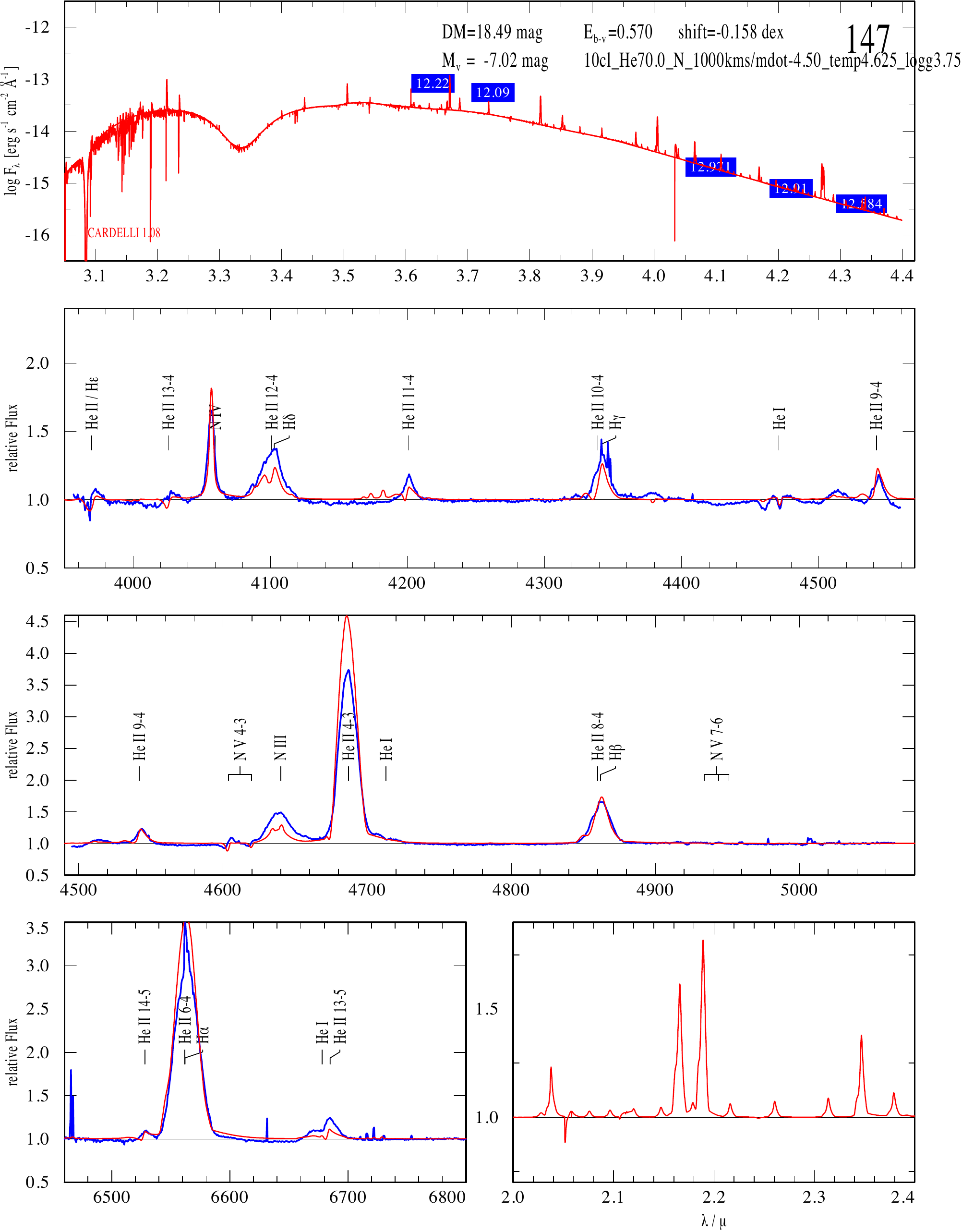}
\end{center}
\caption{The temperature of VFTS\,147 (WN6(h)) is based on the lines He\,{\sc i}\,$\lambda 4471$,  N\,{\sc iii}\,$\lambda 4634/4640$, N\,{\sc iv}\,$\lambda 4058$,  N\,{\sc v} $\lambda4604/4620$. The spectrum is contaminated by multiple nearby stars so as a result the obtained parameters are unreliable. The photometry is uncertain, too. Because of convergence difficulties the model has to be calculated with a lower $\log g$. For emission lines $\log g$ does not affect the results.}
\label{a:147}
\end{figure}

VFTS\,147 (BAT99-091/Brey73): The star has the spectral type WN6(h)
and is not a known binary or multiple component system. According to
\cite{schnurr2008} it does not show any variability. However, the star
is located in the Brey~73 cluster with a bright object nearby. In
several epochs absorption features from a possible OB star are visible
in the VFTS spectra.  The presence/absence of these features is due to
slightly different offsets of the MEDUSA fibre position and/or
variations in the atmospheric seeing. The mass-loss rate and
temperature estimation are uncertain. The low spatial resolution of
the available optical photometry 
unrealistically high flux in the $B$ and $V$ bands. The continuum
estimated from the near-IR VISTA photometry is less affected (or
unaffected), which results in a negative extinction parameter
$E(V-K_{\mathrm{s}})$. The estimated luminosity is therefore based on
the $K_{\mathrm{s}}$ magnitude. 
photometry (\S\,\ref{s:lum_red}), but The luminosity might be up to
$\sim$0.2~dex too high depending on the uncertainties in the near-IR
photometry and the near-IR flux of the nearby star. The star is
excluded from the discussion (Fig.\,\ref{a:147}).

\begin{figure}
\begin{center}
\includegraphics[width=17cm]{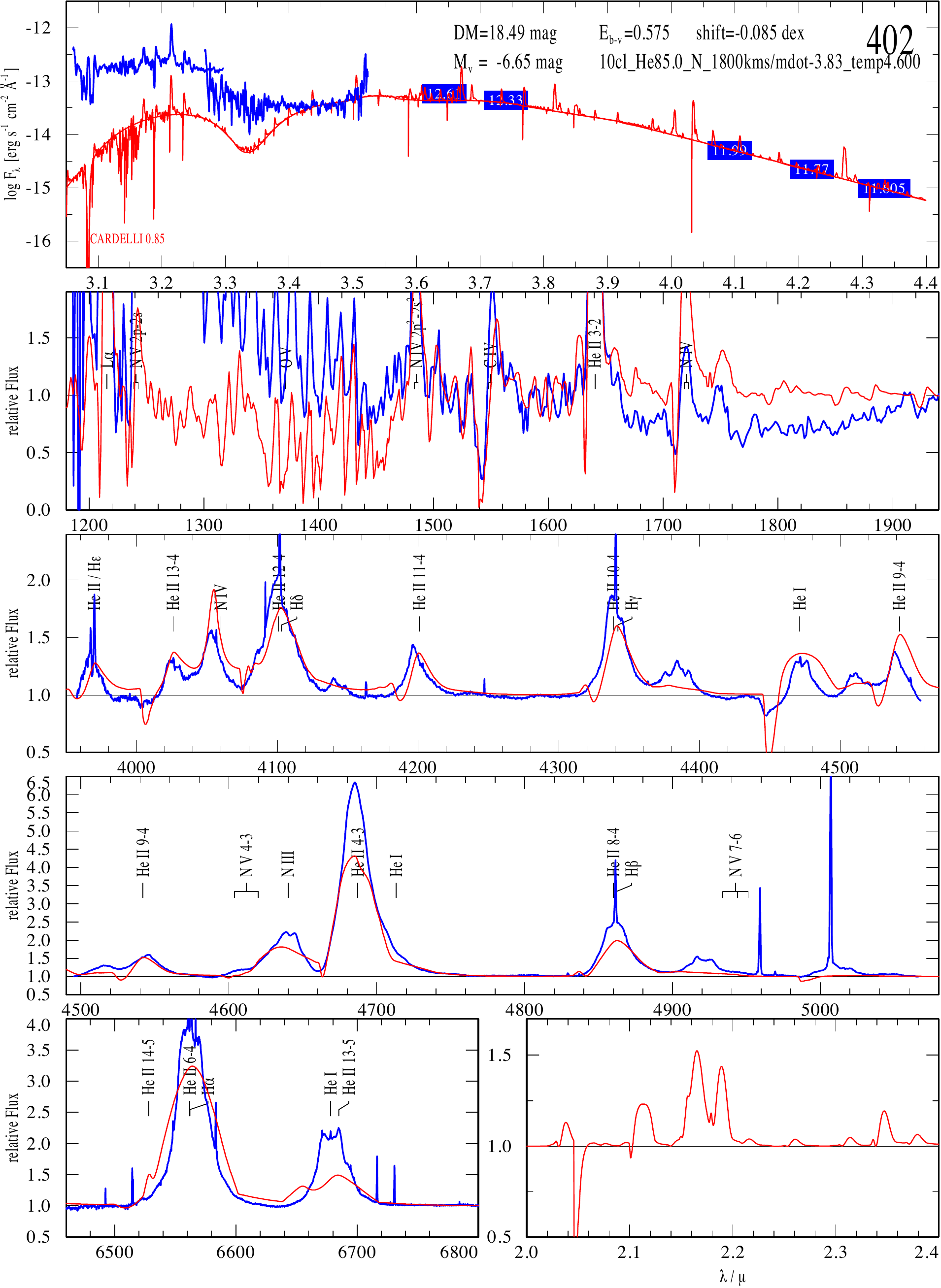}
\end{center}
\caption{The spectrum of VFTS\,402 (WN7(h) + OB) shows the characteristic of a SB2. The temperature is based on the lines He\,{\sc i}\,$\lambda 4471$, N\,{\sc iii}\,$\lambda 4634/4640$, and N\,{\sc iv}\,$\lambda 4058$. $\dot{M}$ is based on He\,{\sc ii}\,$\lambda 4686$ and $\mathrm{H}_{\alpha}$. The fit quality is very poor as result of the secondary. The He mass fraction is between 85.0 and 92.5\%. The optical photometry is contaminated by nearby stars which results in an unusually low $R_{\rm V}=0.85$.}
\label{a:402}
\end{figure}

VFTS\,402 (BAT99-095/Brey80): This WN7(h) star has a probable OB
companion. \cite{schnurr2008} classified it as binary system with a
short orbital period of $\approx$2 days. The model fit quality is very
poor as a result of the presence of the secondary. The temperature,
mass-loss rate and helium abundance determinations are highly uncertain
(Fig.\,\ref{a:402}). The impact of the secondary on the total
continuum is not known, which leads to an incorrect luminosity. The
star is excluded from the discussion.

\begin{figure}
\begin{center}
\includegraphics[width=17cm]{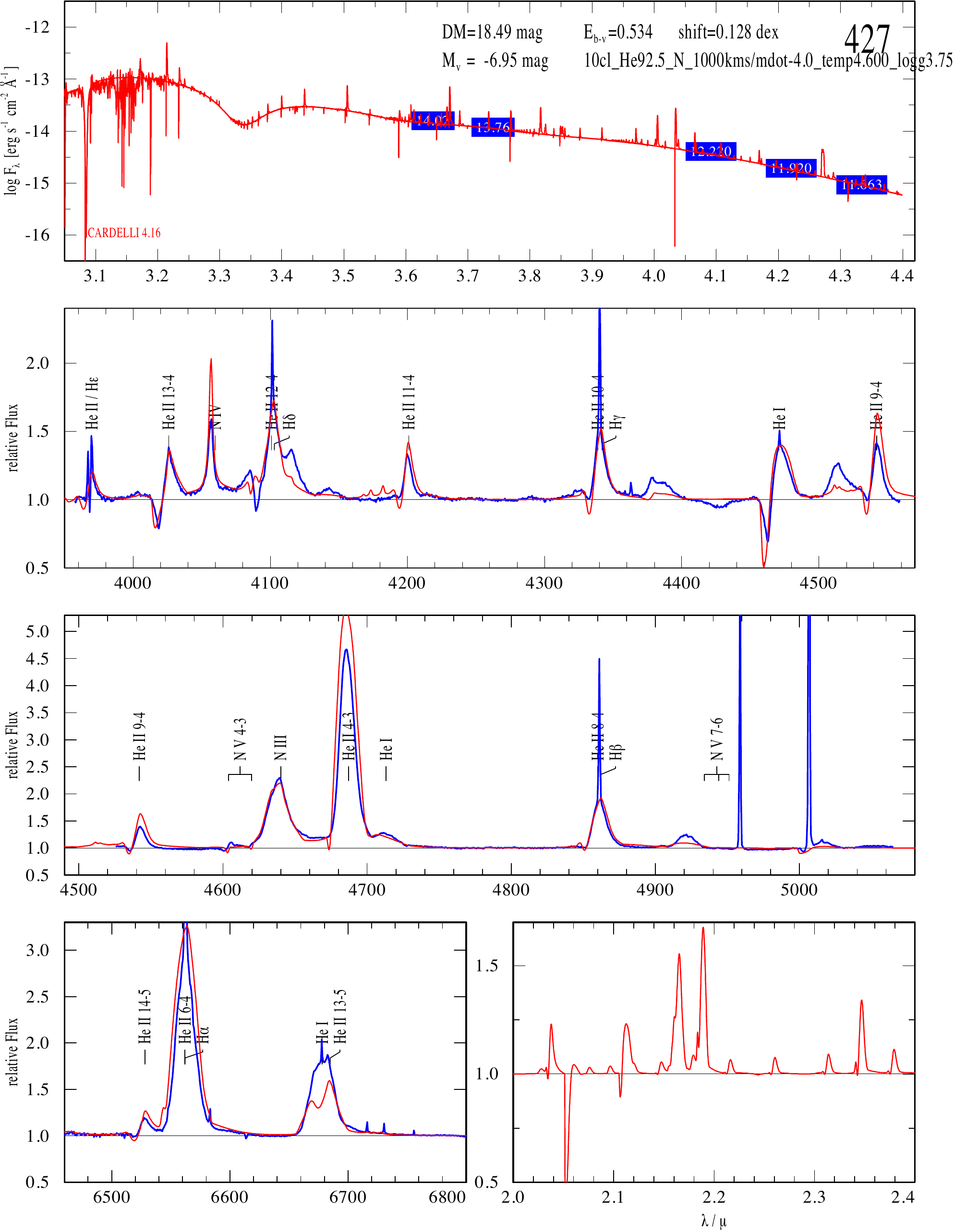}
\end{center}
\caption{The temperature of VFTS\,427 (WN8(h)) is based on the lines He\,{\sc i}\,$\lambda 4471$, N\,{\sc iii}\,$\lambda 4634/4640$, N\,{\sc iv}\,$\lambda 4058$, and N\,{\sc v} $\lambda4604/4620$. $\dot{M}$ and He abundance are based on He\,{\sc ii}\,$\lambda 4686$, $\mathrm{H}_{\alpha}$. The fit quality is reasonably good for a late WNh star.}
\label{a:427}
\end{figure}

VFTS\,427 (BAT99-096/Brey81):  A single star with no noticeable change
in the peak intensity of the emission lines between the observed
epochs. The WN8(h)-star is surrounded by a few nearby fainter objects,
which do not considerably influence the spectra or the continuum.
There are no photospheric absorption lines to study RV variations due to
any possible secondary. The quality of the model is good and the
derived stellar parameters are reasonably accurate
(Fig.\,\ref{a:427}).

VFTS\,457 (BAT99-097): The star is classified as O3.5{\sc I}f*/WN7 and
is not a known binary. Small LPVs are observed in the spectra which
might be a result of the normalisation process. The fit to the
nitrogen lines is poor, in particular for N\,{\sc iii}. However, test
calculations show that an increase of the nitrogen abundance, moving
the starting point of clumping closer to the stellar surface to a wind
velocity of 10~km\,s$^{-1}$, and a higher value of the wind parameter
$\beta$ improved the fit quality, but does not fundamentally change
the result (see Fig.\,\ref{f:457} and \ref{f:457_test}). These
different properties might be a result of an extended or asymmetric
atmosphere, as a result of rotation. The terminal velocity is between
900 and 1100~km\,s$^{-1}$ lower than the other Of/WN stars.
Nevertheless, the derived stellar properties are robust.

\begin{figure}
\begin{center}
\vspace{-0.2cm}
\includegraphics[width=17cm]{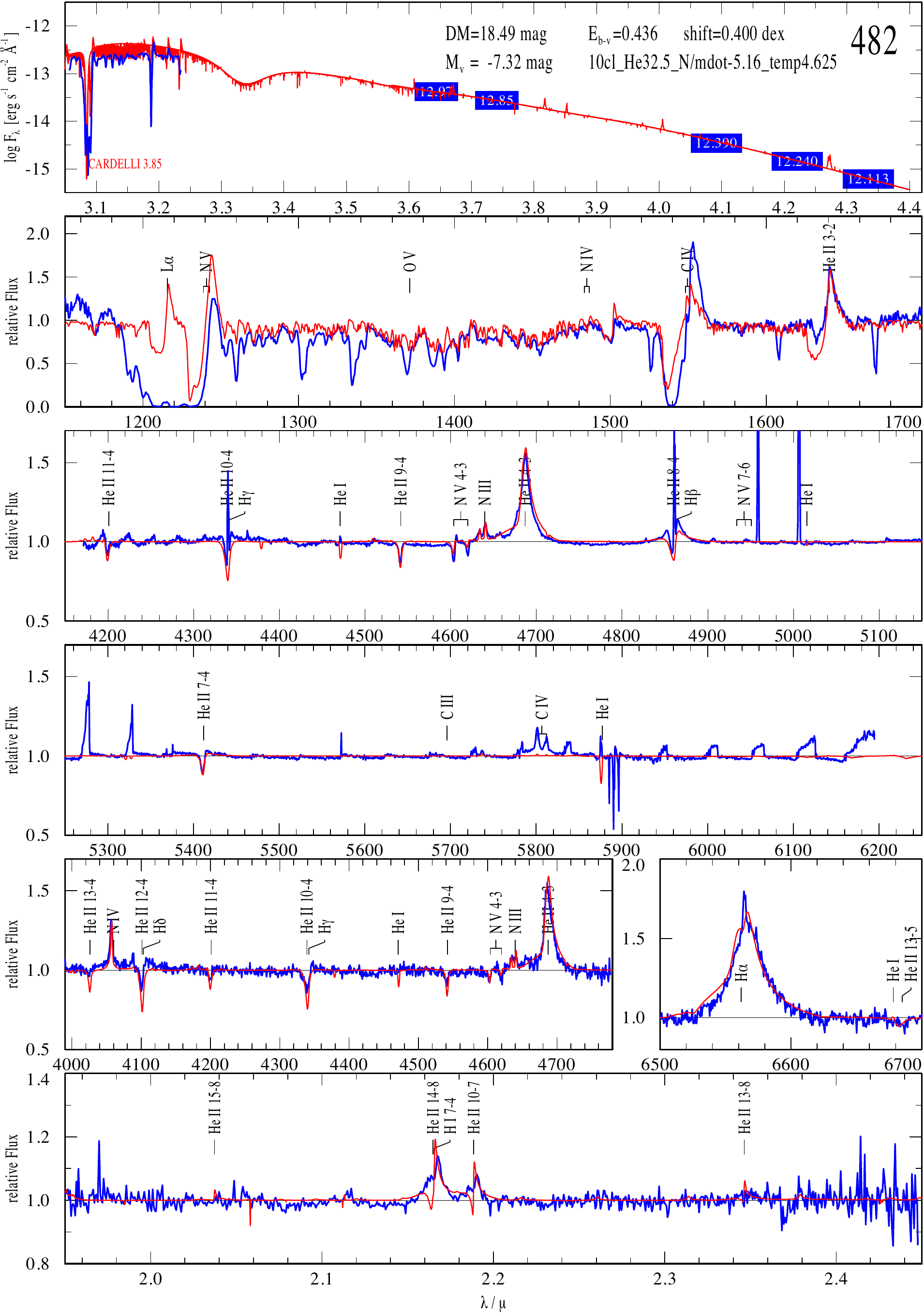}
\end{center}
\vspace{-0.6cm}
\caption{VFTS\,482 (O2.5 If*/WN6) is surrounded by nearby stars, which could have an impact on the spectroscopy and photometry. The temperature is based on the lines N\,{\sc iii}\,$\lambda 4634/4640$, N\,{\sc iv}\,$\lambda 4058$, and N\,{\sc v} $\lambda4604/4620$. $\dot{M}$ and He-abundance are based on the lines He\,{\sc ii}\,$\lambda 4686$, $\mathrm{H}_{\alpha}$, He\,{\sc ii}\,$2.19\mu m$, and $\mathrm{H}^{\rm Br}_{\gamma}$.}
\label{a:482}
\end{figure}

VFTS\,482 (BAT99-099/Brey78/Mk39): VFTS\,482 is classified as O2.5{\sc
I}f*/WN6. \cite{massey2002} identified spectral variability while
\cite{massey2005} detected a composite in the spectra, but provided no
classification of the secondary. According to \cite{schnurr2008} it is
a elliptical system with a period of 92.6~days (close to their
detection limit). The quality of the fit is good. The temperature,
mass-loss rate and He-abundance can be well determined
(Fig\,\ref{a:482}). The star is located in a relatively crowded field.
The luminosity might be over estimated because of its unclear binary
nature.

\begin{figure}
\begin{center}
\includegraphics[width=17cm]{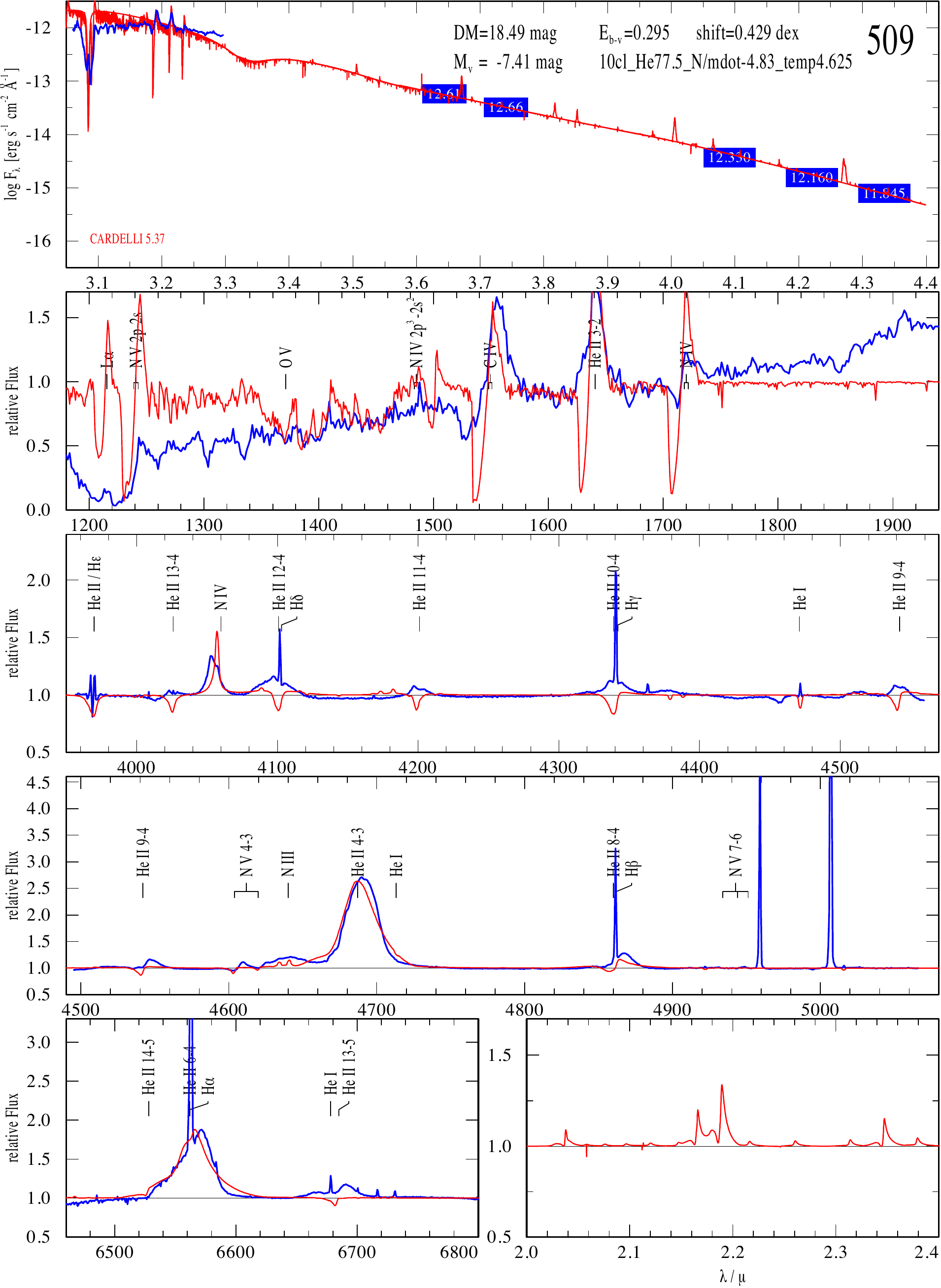}
\end{center}
\caption{The spectrum of VFTS\,509 (WN5(h) + early O) shows the characteristic of a SB2. The temperature is based on the lines N\,{\sc iii}\,$\lambda 4634/4640$, and N\,{\sc iv}\,$\lambda 4058$. $\dot{M}$  and He-abundance are based on He\,{\sc ii}\,$\lambda 4686$ and $\mathrm{H}_{\alpha}$. The fit quality is poor as result of the secondary.}
\label{a:509}
\end{figure}

VFTS\,509 (BAT99-103/Brey87): This WN5(h) star is part of a binary
system with a separated early O star companion \citep{evans2011,
doran2013}. \cite{schnurr2008} give a period of 2.76 days. The fit
quality is poor, but gives a fair hint of the actual stellar
parameters (Fig.\,\ref{a:509}). Still, the star is excluded from the
discussion.

\begin{figure}
\begin{center}
\includegraphics[width=17cm]{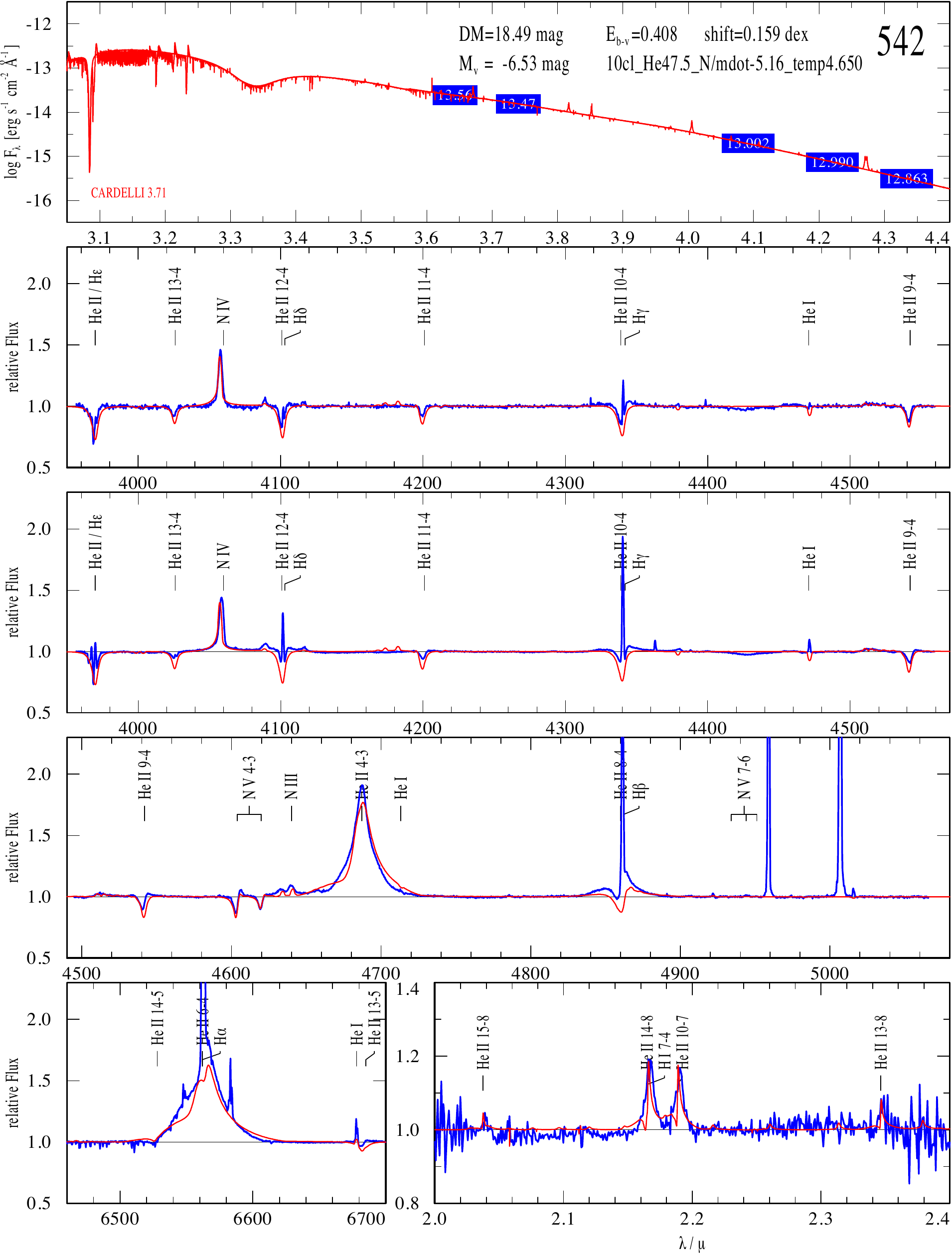}
\end{center}
\caption{VFTS\,542 (O2 If*/WN5): The second (ARGUS) and third (MEDUSA) panel cover the same wavelength range and the spectra are similar (except for the degree of the nebular contamination). The temperature is based on the lines N\,{\sc iii}\,$\lambda 4634/4640$, N\,{\sc iv}\,$\lambda 4058$, and N\,{\sc v} $\lambda4604/4620$. $\dot{M}$ and He-abundance are based on the lines He\,{\sc ii}\,$\lambda 4686$, $\mathrm{H}_{\alpha}$, He\,{\sc ii}\,$2.19\mu m$, and $\mathrm{H}^{\rm Br}_{\gamma}$.}
\label{a:542}
\end{figure}

VFTS\,542 (BAT99-113): This O2{\sc I}f*/WN5 star is a binary system
with a period of 4.7 days \citep{schnurr2008}. It is located in a
crowded field. The quality of the model fit is good. The temperature,
mass-loss and helium abundance can be determined (Fig.\,\ref{a:542}).
The luminosity is uncertain, because it is unclear how the continuum
is affected by the secondary.

\begin{figure}
\begin{center}
\includegraphics[width=17cm]{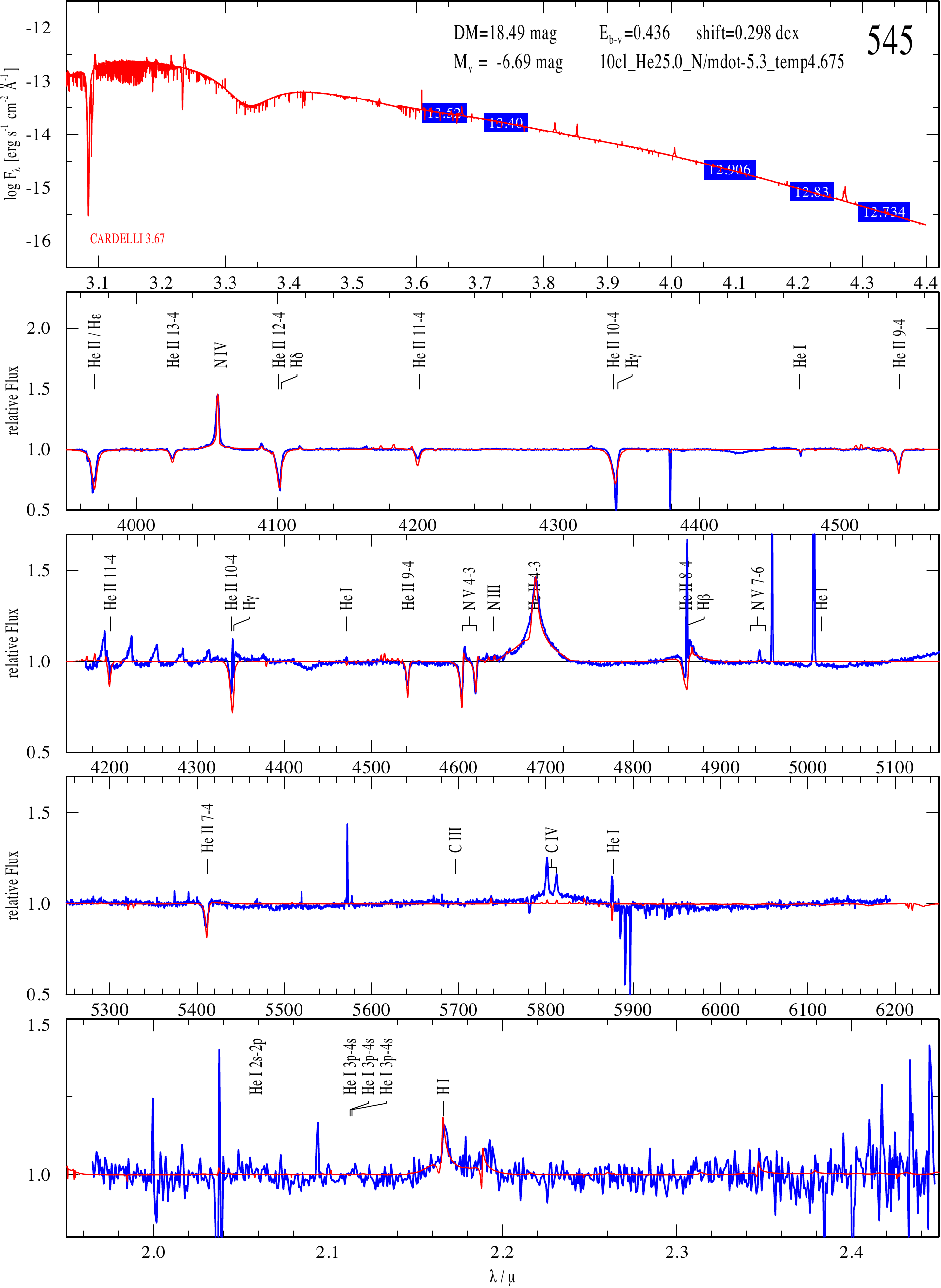}
\end{center}
\caption{The temperature of VFTS\,545 (O2 If*/WN5) is based on the lines N\,{\sc iii}\,$\lambda 4634/4640$, N\,{\sc iv}\,$\lambda 4058$, and N\,{\sc v} $\lambda4604/4620$. $\dot{M}$ and He-abundance are based on He\,{\sc ii}\,$\lambda 4686$, $\mathrm{H}_{\alpha}$, He\,{\sc ii}\,$2.19\mu m$, and $\mathrm{H}^{\rm Br}_{\gamma}$. N is enriched.}
\label{a:545}
\end{figure}

VFTS\,545 (BAT99-114): As reported by \cite{henault2012} there are
weak RV variations in the spectra of this O2{\sc I}f*/WN5 star. There
is no contamination of the stellar spectra; the fit is good and the
results are robust (Fig.\,\ref{a:545}).

\begin{figure}
\begin{center}
\includegraphics[width=17cm]{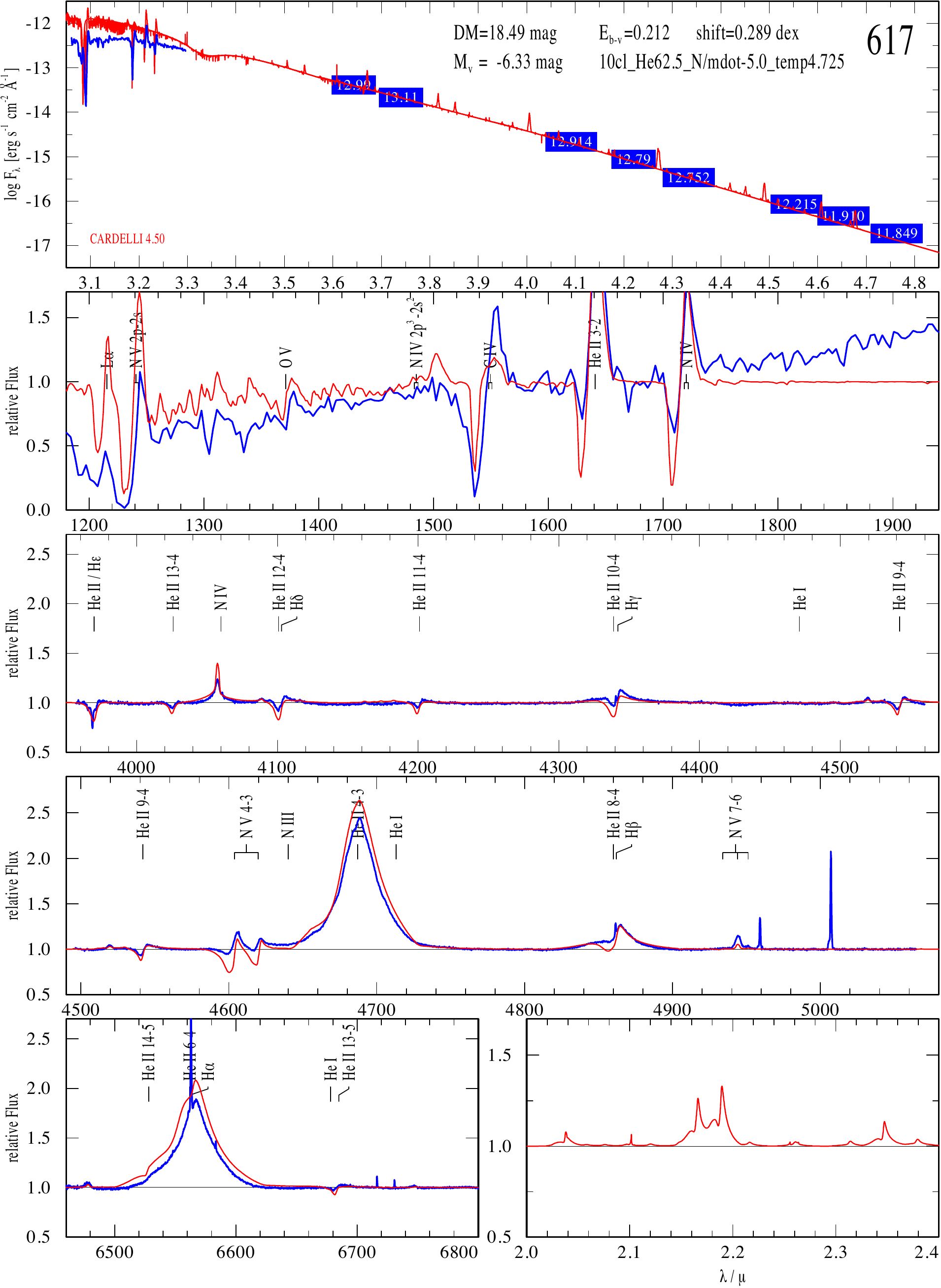}
\end{center}
\caption{The temperature of VFTS\,617 (WN5ha) is based on the lines N\,{\sc iv}\,$\lambda 4058$, N\,{\sc v} $\lambda4604/4620$, and N\,{\sc v} $\lambda4945$. $\dot{M}$ and He-abundance are based on He\,{\sc ii}\,$\lambda 4686$ and $\mathrm{H}_{\alpha}$.}
\label{a:617}
\end{figure}

VFTS\,617 (BAT99-117/Brey88): This single star is classified as a
WN5ha. The spectra show LPVs particularly in the Balmer series. The
quality of the model is good. $\dot{M}$ varies as a result of the
LPVs, but only within the error bars. (Fig.\,\ref{a:617}).

\begin{figure}
\begin{center}
\includegraphics[width=17cm]{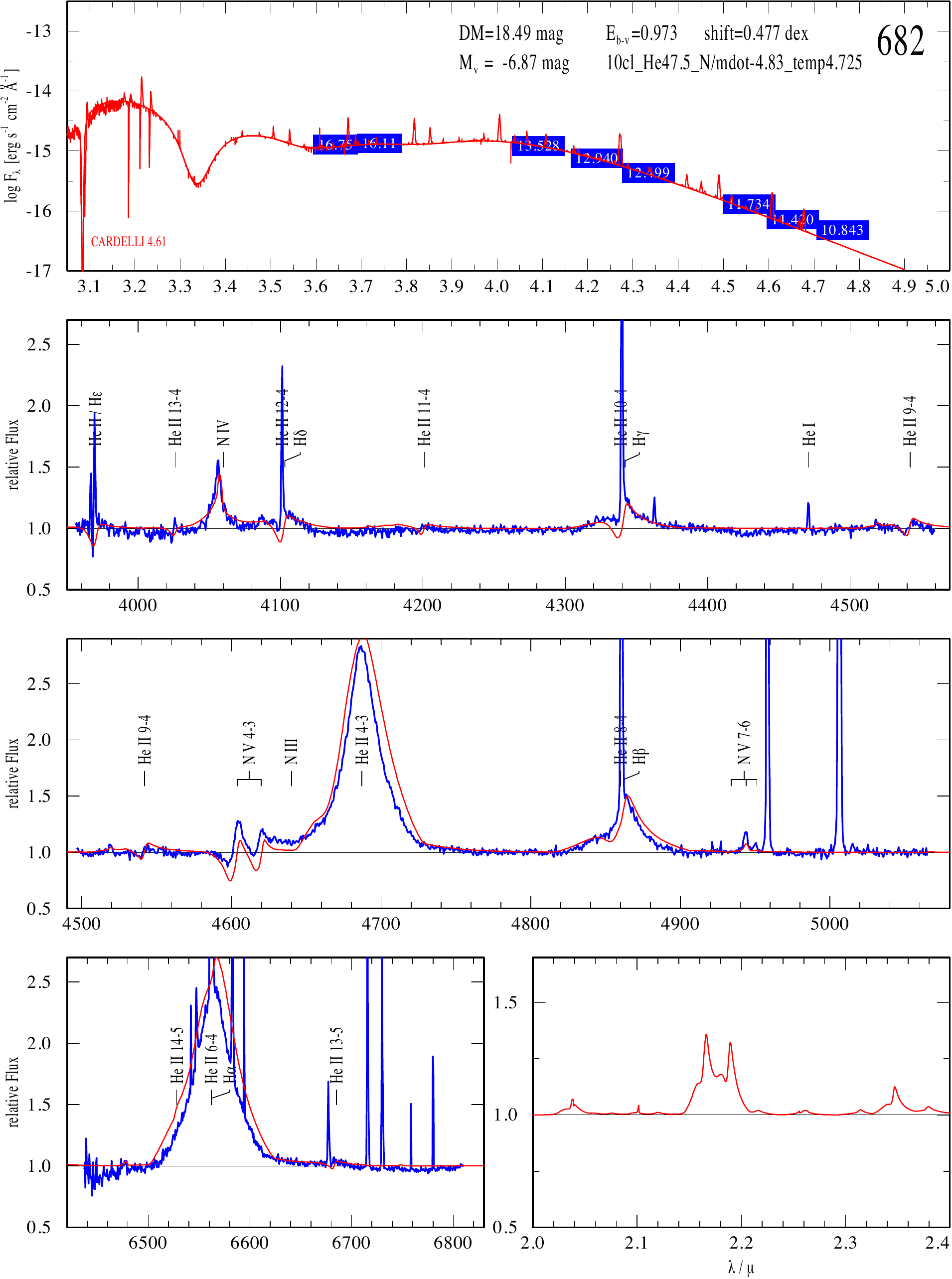}
\end{center}
\caption{The temperature of VFTS\,682 (WN5h) is based on the lines N\,{\sc iv}\,$\lambda 4058$, N\,{\sc v} $\lambda4604/4620$, and on the absence of the He\,{\sc i}\,$\lambda 4471$ line. $\dot{M}$ and He-abundance are based on He\,{\sc ii}\,$\lambda 4686$ and $\mathrm{H}_{\alpha}$. The fit quality is reasonably good for a WNh stars. Even though the best fitting grid model has a slightly too high $\dot{M}$. The results are similar to those from \cite{bestenlehner2011}.}
\label{a:682}
\end{figure}

VFTS\,682: The WN5h-star is a newly discovered WR-star by the VFTS
\citep{evans2011, bestenlehner2011}. The star appears to be single.
The model fit is good and the results based on the grid are comparable
with the parameters derived from \cite{bestenlehner2011}. The grid fit
is shown in Fig.\,\ref{a:682}.

\begin{figure}
\begin{center}
\includegraphics[width=17cm]{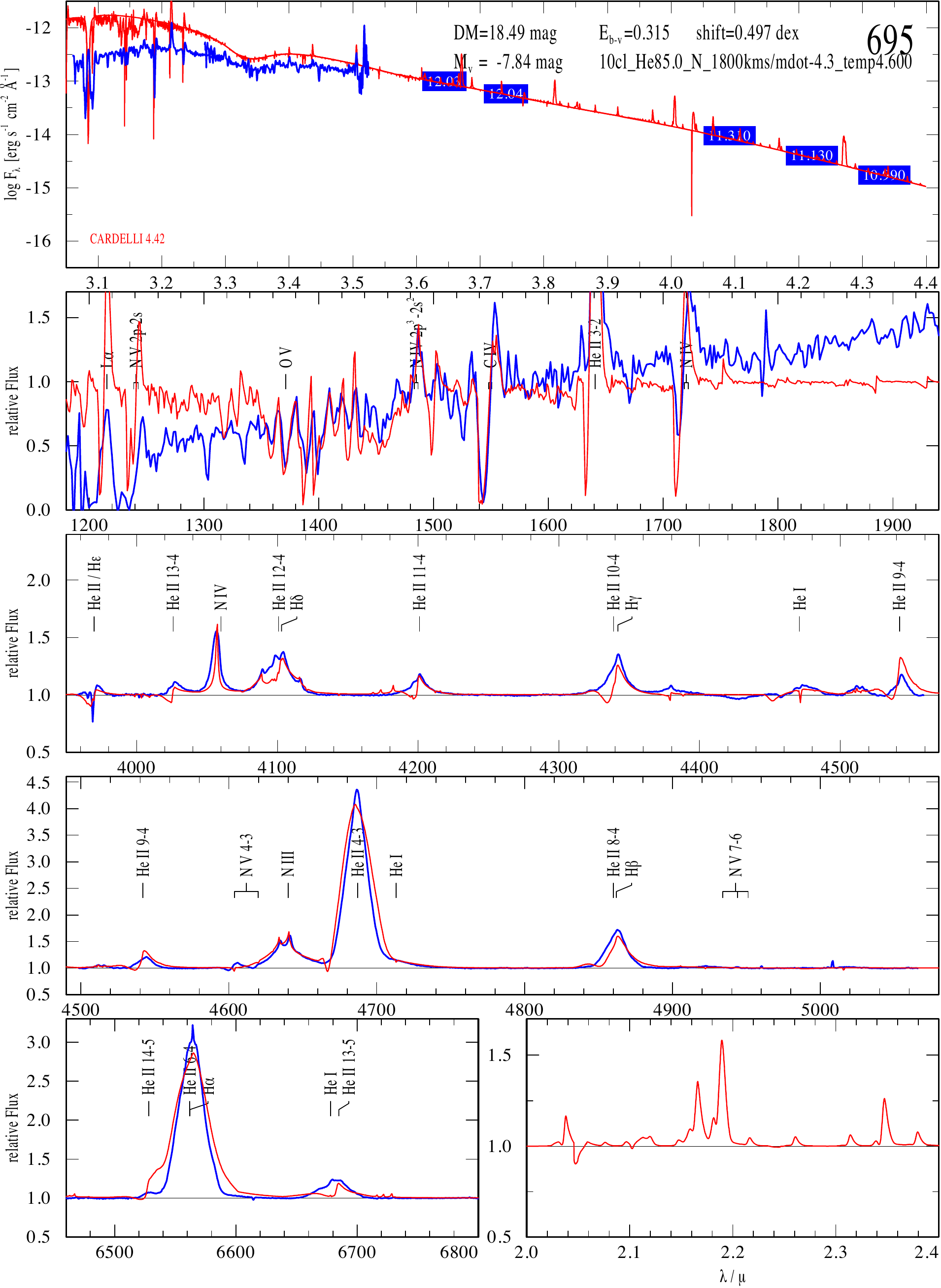}
\end{center}
\caption{The spectrum of VFTS\,695 (WN6h + ?) shows the characteristic of a SB1. The temperature is based on the lines He\,{\sc i}\,$\lambda 4471$, N\,{\sc iii}\,$\lambda 4634/4640$, and N\,{\sc iv}\,$\lambda 4058$. $\dot{M}$ and He-abundance are based on He\,{\sc ii}\,$\lambda 4686$ and $\mathrm{H}_{\alpha}$. The He-abundance is between 85.0 and 92.5\%.}
\label{a:695}
\end{figure}

VFTS\,695 (BAT99-119/Brey90): This is a WR-star with spectral type
of WN6h plus a companion. The binary system has a period of 158.8d
\citep{schnurr2008}. The spectra show LPVs in the Balmer series, in
particular for $\mathrm{H}_{\alpha}$. The observations can be
reasonably well reproduced by our model, however $L_{\star}$ might be
over estimated (Fig.\,\ref{a:695}).

\begin{figure}
\begin{center}
\includegraphics[width=17cm]{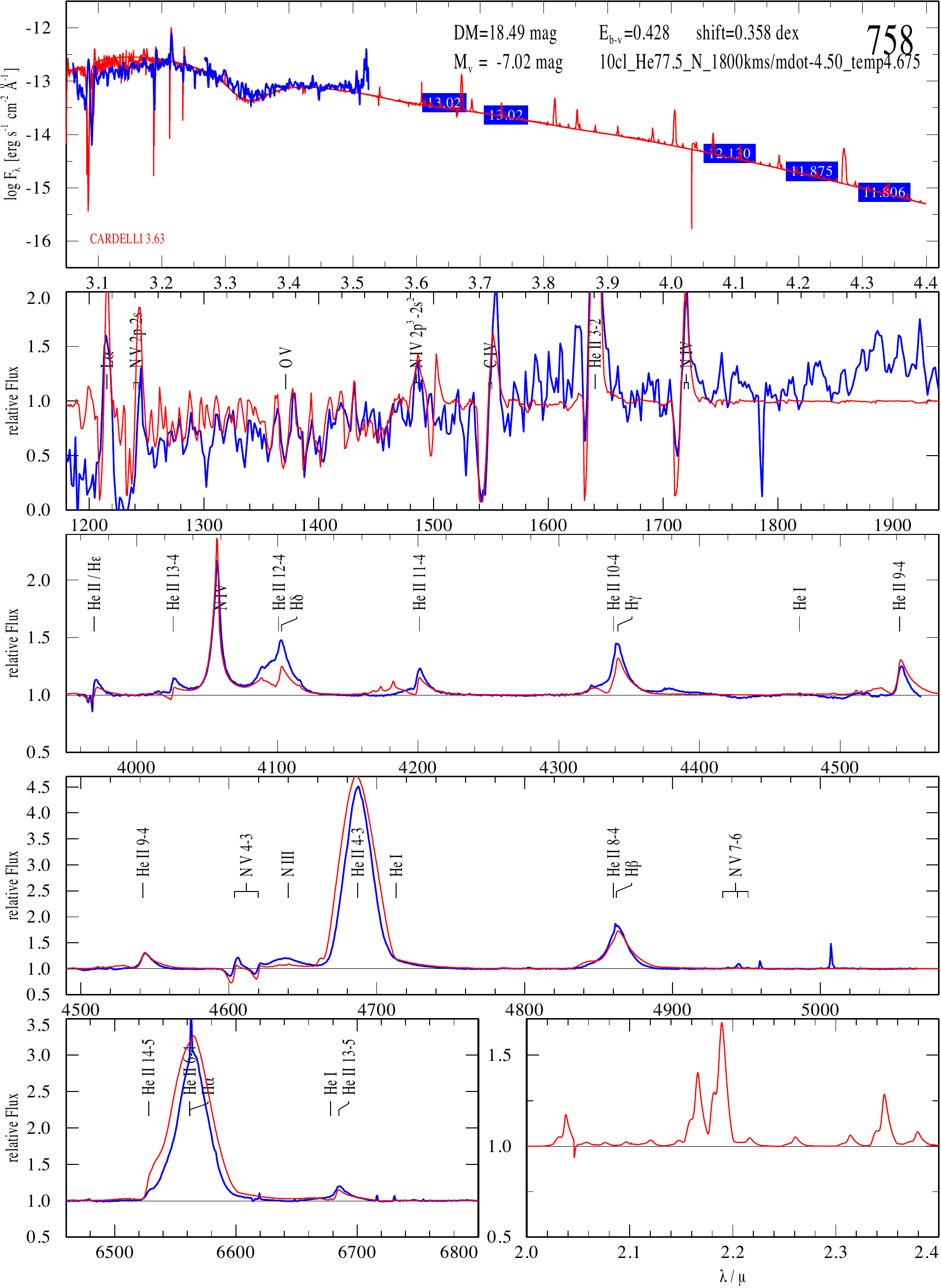}
\end{center}
\caption{The temperature of VFTS\,758 (WN5h) is based on the lines N\,{\sc iv}\,$\lambda 4058$, N\,{\sc v} $\lambda4604/4620$, and on the absence of the He\,{\sc i}\,$\lambda 4471$ line. $\dot{M}$ and He-abundance are based on He\,{\sc ii}\,$\lambda 4686$ and $\mathrm{H}_{\alpha}$. The fit quality is reasonably good for a WNh stars, although $\mathrm{H}_{\delta}$ could not be properly reproduced.}
\label{a:758}
\end{figure}

VFTS\,758 (BAT99-122/Brey91): This is a single WN5h-star. No
noticeable line-profile and RV variations were found in the spectra.
The fit quality regarding the grid resolution is good and the results
are reasonably good (Fig.\,\ref{a:758}).

\begin{figure}
\begin{center}
\includegraphics[width=17cm]{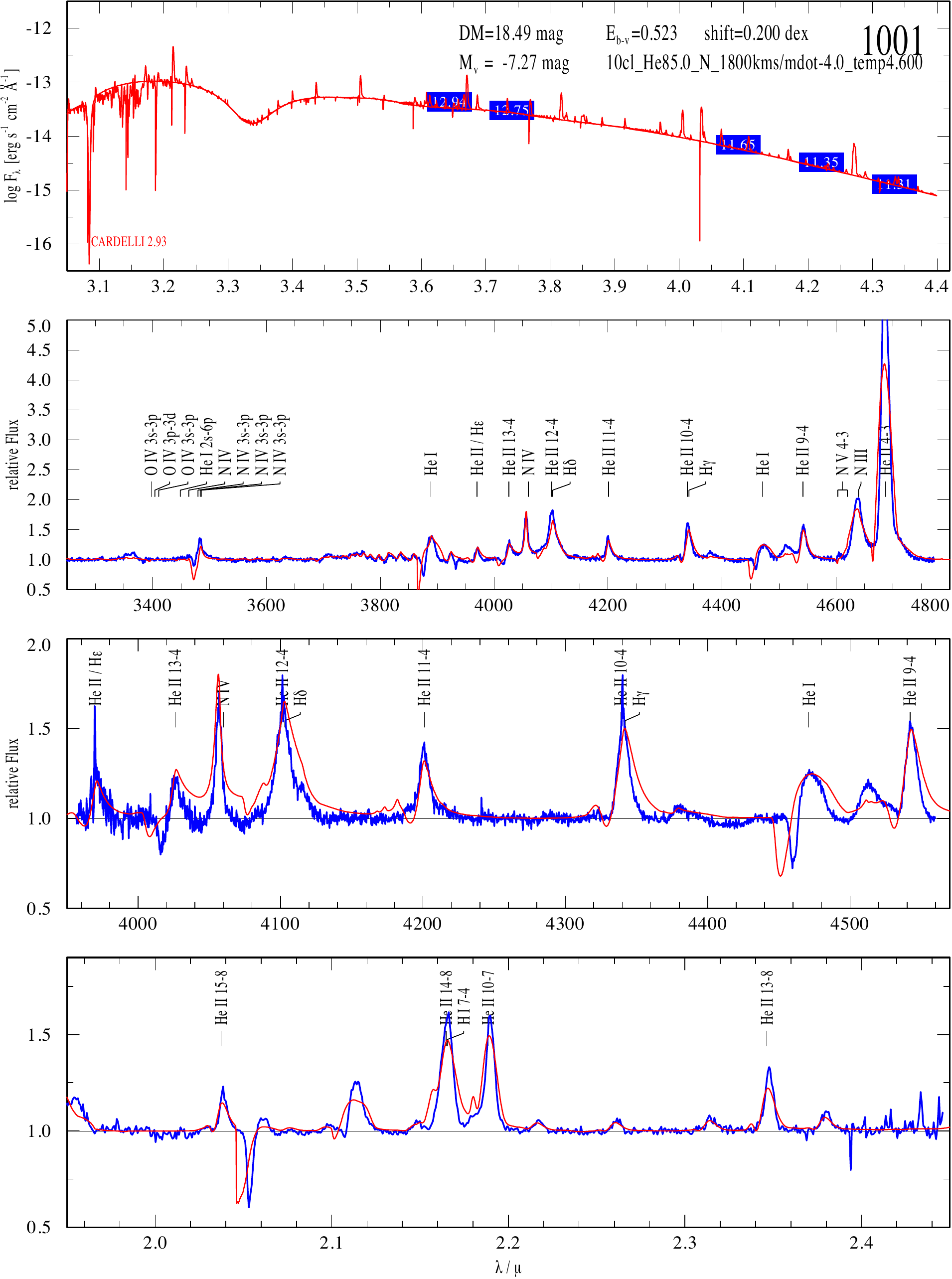}
\end{center}
\caption{VFTS\,1001 (WN6(h)): The temperature is based on the lines He\,{\sc i}\,$\lambda 4471$,  N\,{\sc iii}\,$\lambda 4634/4640$, N\,{\sc iv}\,$\lambda 4058$,  N\,{\sc v} $\lambda4604/4620$. $\dot{M}$ and He-abundance are based on the lines He\,{\sc ii}\,$\lambda 4686$, $\mathrm{H}_{\alpha}$, He\,{\sc ii}\,$2.19\mu m$, and $\mathrm{H}^{\rm Br}_{\gamma}$. $\dot{M}$ is slightly too low for the best fitting model. The model suggests a He-abundance of 85\%. (Second panel: HST/FOS, third panel: ARGUS. The ARGUS data were only approximately normalised as there prime use was the investigations of RV/LPV variations.)}
\label{a:1001}
\end{figure}

VFTS\,1001 (BAT99-100/Brey75): The WN6(h)-star is not a known binary.
\cite{henault2012} found possible LPVs in the He\,{\sc ii}\,$\lambda
4542$ line. The star is associated with an X-ray source.
\cite{portegies2002} suggested that these X-rays are potentially due
to a wind-wind collision in a binary system. However, our model fits
the observation well and we treat the object as if it is a single star
(Fig.\,\ref{a:1001}).

\begin{figure}
\begin{center}
\includegraphics[width=17cm]{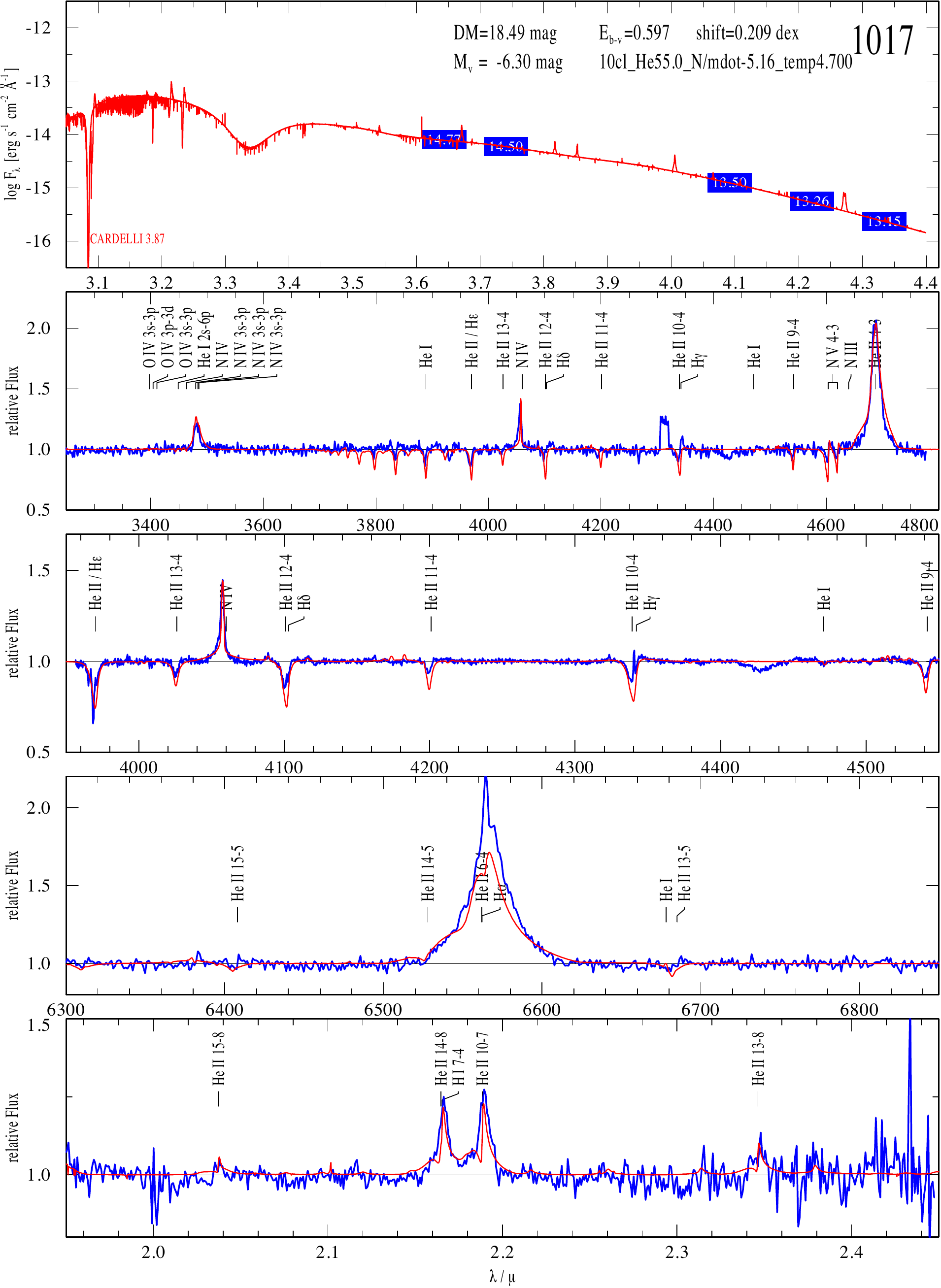}
\end{center}
\caption{The temperature of VFTS\,1017 (O2 If*/WN5) is based on the lines N\,{\sc iv}\,$\lambda 4058$, N\,{\sc v} $\lambda4604/4620$, and He\,{\sc i}\,$\lambda 4471$. $\dot{M}$ and He-abundance are based on the lines He\,{\sc ii}\,$\lambda 4686$, $\mathrm{H}_{\alpha}$, He\,{\sc ii}\,$2.19\mu m$, and $\mathrm{H}^{\rm Br}_{\gamma}$. The optical HST observation suggests a lower He-abundance while the near-IR suggests a slightly higher abundance than 55\%. (Second panel: HST/FOS, third panel: ARGUS, fourth panel: HST/STIS, fifth panel: SINFONI.)}
\label{a:1017}
\end{figure}

VFTS\,1017 (BAT99-104/Brey76): The O2 {\sc I}f*/WN5 spectra of this
star show LPVs and weak RV variations \citep{henault2012}. The
possible companion does not impact the fit quality. The results are
reliable with slightly larger uncertainties than most well fit stars
in this study (Fig.\,\ref{a:1017}).

\begin{figure}
\begin{center}
\includegraphics[width=17cm]{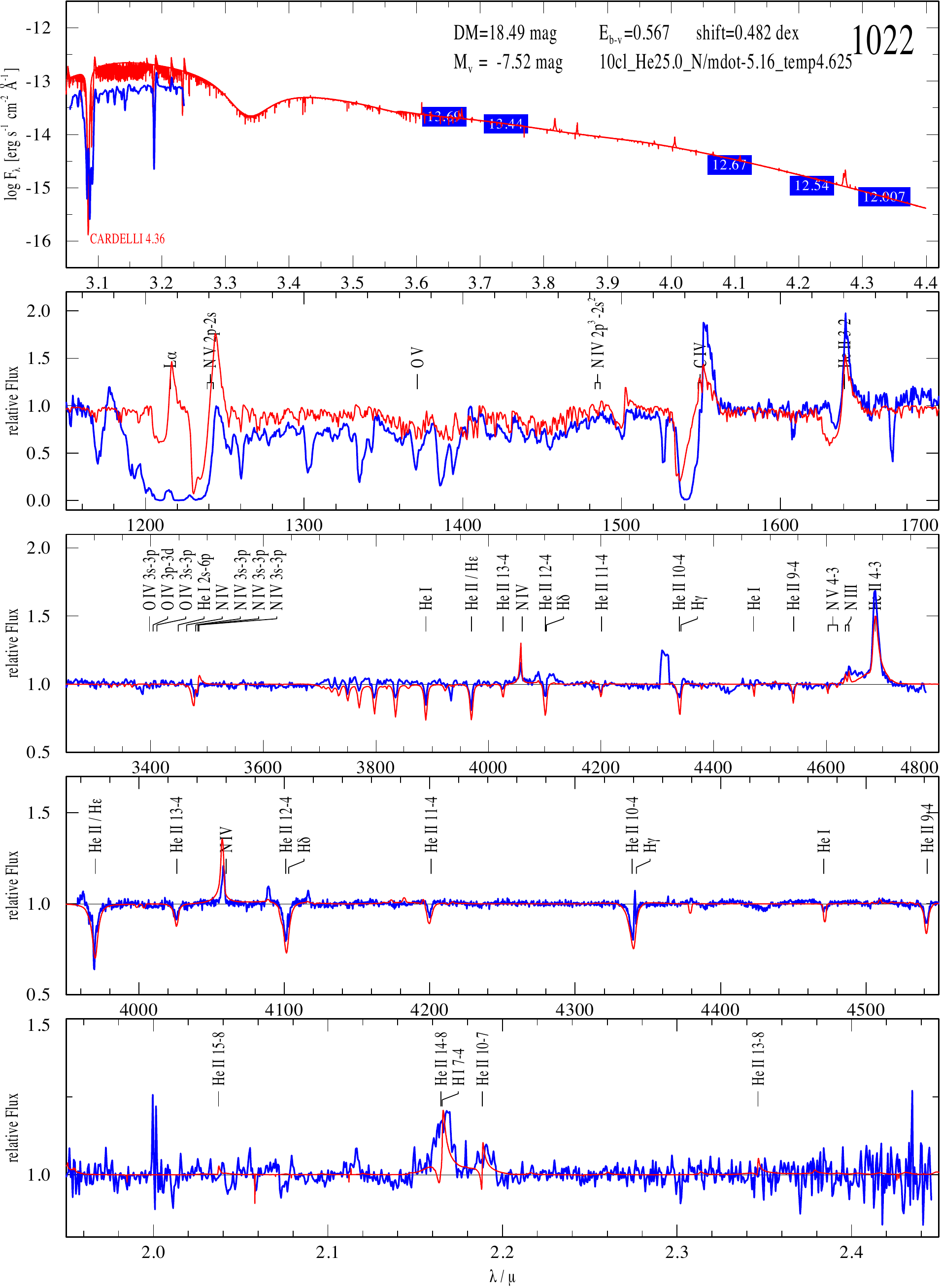}
\end{center}
\caption{The temperature of VFTS\,1022 (O3.5 If*/WN7) is based on the lines N\,{\sc iii}\,$\lambda 4634/4640$, N\,{\sc iv}\,$\lambda 4058$, and He\,{\sc i}\,$\lambda 4471$. $\dot{M}$ and He-abundance are based on the lines He\,{\sc ii}\,$\lambda 4686$, He\,{\sc ii}\,$2.19\mu m$, and $\mathrm{H}^{\rm Br}_{\gamma}$. $\dot{M}$ is too weak and N\,{\sc iv} is too strong. N-abundance is between normal and enriched. (Second panel: HST/STIS, third panel: HST/FOS, fourth panel: ARGUS.)}
\label{a:1022}
\end{figure}

VFTS\,1022: This star has been reclassified from O4 If+
\citep{massey1998} to O3.5 {\sc I}f*/WN7 by \cite{crowther2011}. The
star shows small RV variations. Even though it is likely a binary, the
derived stellar parameter are treated as if it were a single star
(Fig.\,\ref{a:1022}).

VFTS\,1025 (BAT99-112/Brey82/R136c): This WN5h-star is located in a
crowded field and surrounded by multiple sources in HST observations
by \citet{massey1998}. In their catalogue VFTS\,1025 is designated as
source No.\,10 with $M_V=-6.8$. One neighbouring HST source at an
angular distance $\lesssim$\,0.5\,arcsec is bright enough to affect
our optical spectrum directly (source No.\,57 with $M_V=-5.5$,
classified as O3\,III(f*)). Furthermore the bright O star R136b
(source No.\,9, $M_V=-6.9$, O4\,If+) is relatively close at an angular
distance of $\approx$\,1.3\,arcsec. VFTS\,1025 itself shows
indications of binarity based on its strong X-ray emission
\citep{guerrero2008} and possible low-amplitude RV variations
\citep{schnurr2009}.

In our analysis we determine a significantly lower temperature than
\cite{crowther2010}, mainly based on the He\,{\sc i} $\lambda
4471\,\AA$ absorption line in the optical range. Our cooler
temperature is further supported by the low terminal wind speed of
VFTS\,1025 compared to the other WN5(h) stars in R\,136 analysed by
\citeauthor{crowther2010} Their hotter temperature is based on the
N\,{\sc v} $\lambda 2.10\,\mu$m emission line in the IR. It is not
clear if the He\,{\sc i} line in our optical spectrum is intrinsic to
VFTS\,1025 or originates from a nearby star. We can exclude star
No.\,57 from \citeauthor{massey1998} as the origin of the He\,{\sc i}
line, based on the non-detection of strong enough He\,{\sc i} in its
HST spectrum. R136b has a weak He\,{\sc i} absorption line and could
possibly affect our spectrum. \cite{henault2012} found LPVs for the
prominent H and He\,{\sc ii} emission lines but not for the He\,{\sc
i} absorption line in the spectrum of VFTS\,1025. This is in line with
a photospheric nature of the absorption line and wind variability in
the emission lines as expected from our cool model, but also with a
contribution from a nearby star.

\subsection{Pecuilarities}\label{s:pot}
In this section we discuss three peculiarities that arose during our
spectroscopic analysis. These peculiarities occurred as discrepancies
between the observations and the model predictions. The plots of the
model fits are shown in Appendix\,\ref{a:plots}.

Firstly, our models predict the C\,{\sc iii} lines at $\lambda
4647/4650$\,{\AA} in emission. \citet{martins2012} discussed that
these lines may appear in absorption or emission, depending on the
detailed model assumptions. In our observations they appear in
absorption for VFTS\,171, 333, 532, and 669 and in emission for
VFTS\,216, 404, 422, 603, 608, and 797. Furthermore, VFTS\,532 has a
cooler companion that could contribute to the C\,{\sc iii} absorption.
In our analysis these lines are ignored.

Secondly, VFTS\,208, 406, and 626 show He\,{\sc ii}\,$\lambda
4686$\,{\AA} emission profiles with a central absorption, which are
characteristic for the fast rotating stars in the Onfp subclass
\citep[see][]{walborn2010a}. This type of line profile has been
modelled in 2D by \cite{hillier2012}. VFTS\,406 has been classified as
OVnn by \cite{walborn2014} who find that its composite
spectrum in the VFTS data is likely due to contamination by an
adjacent WN spectrum on the detector.  For VFTS\,208 and 406 we only
achieve a bad fit quality and the stars are excluded from our
discussion.

\begin{figure}
\begin{center}
\includegraphics[width=17cm]{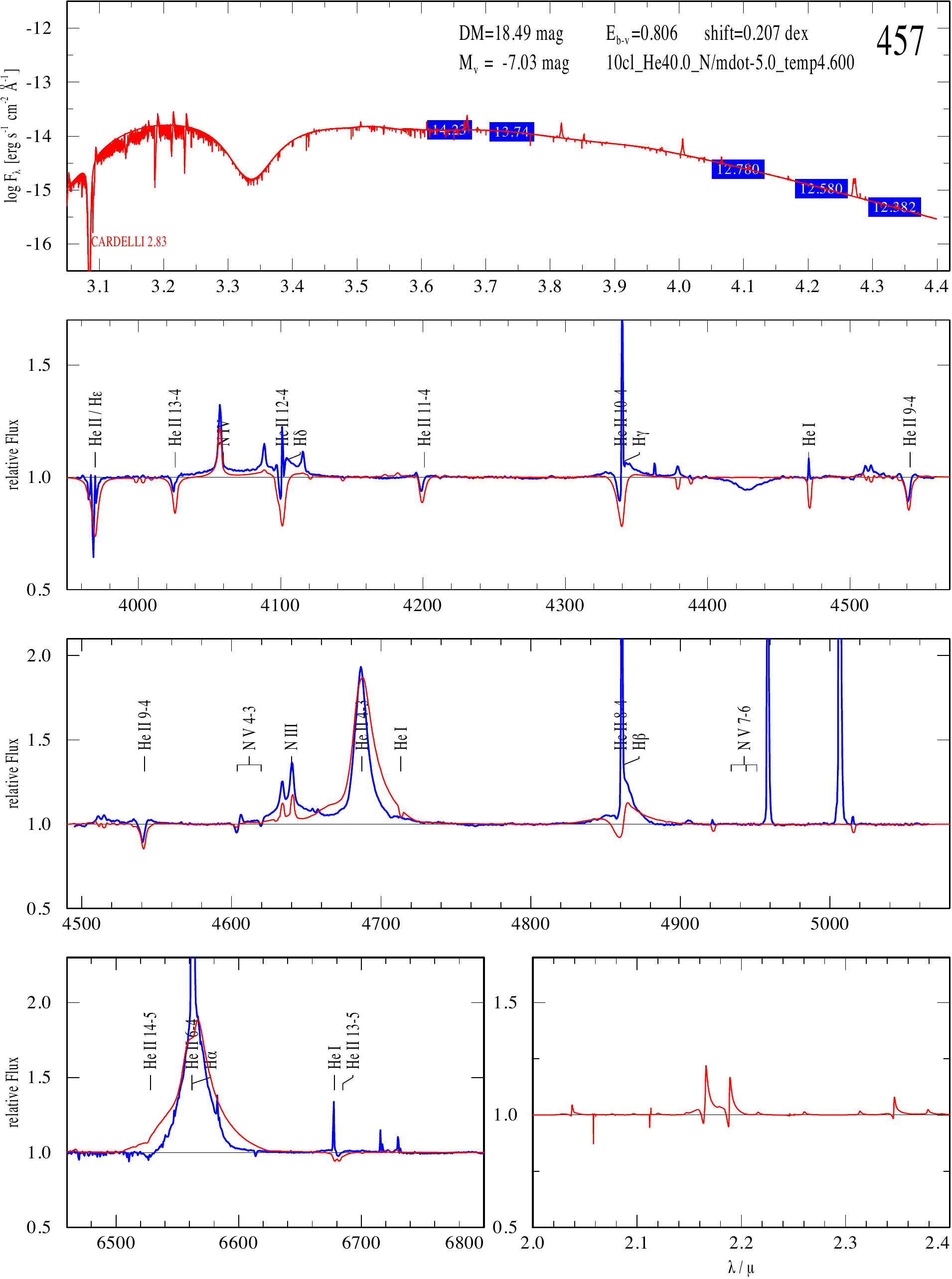}
\end{center}
\caption{The temperature of VFTS\,457 (O3.5 If*/WN7) is based on the lines N\,{\sc iii}\,$\lambda 4634/4640$, N\,{\sc iv}\,$\lambda 4058$, and N\,{\sc v} $\lambda4604/4620$. $\dot{M}$ is based on He\,{\sc ii}\,$\lambda 4686$ and $\mathrm{H}_{\alpha}$. The best fitting model $\dot{M}$ is slightly too high. The star suggests an unusually high N abundance. The quality of the fit is not good, but can be easily improved (see Fig.\,\ref{f:457_test}).}
\label{f:457}
\end{figure}

\begin{figure}
\begin{center}
\includegraphics[width=17cm]{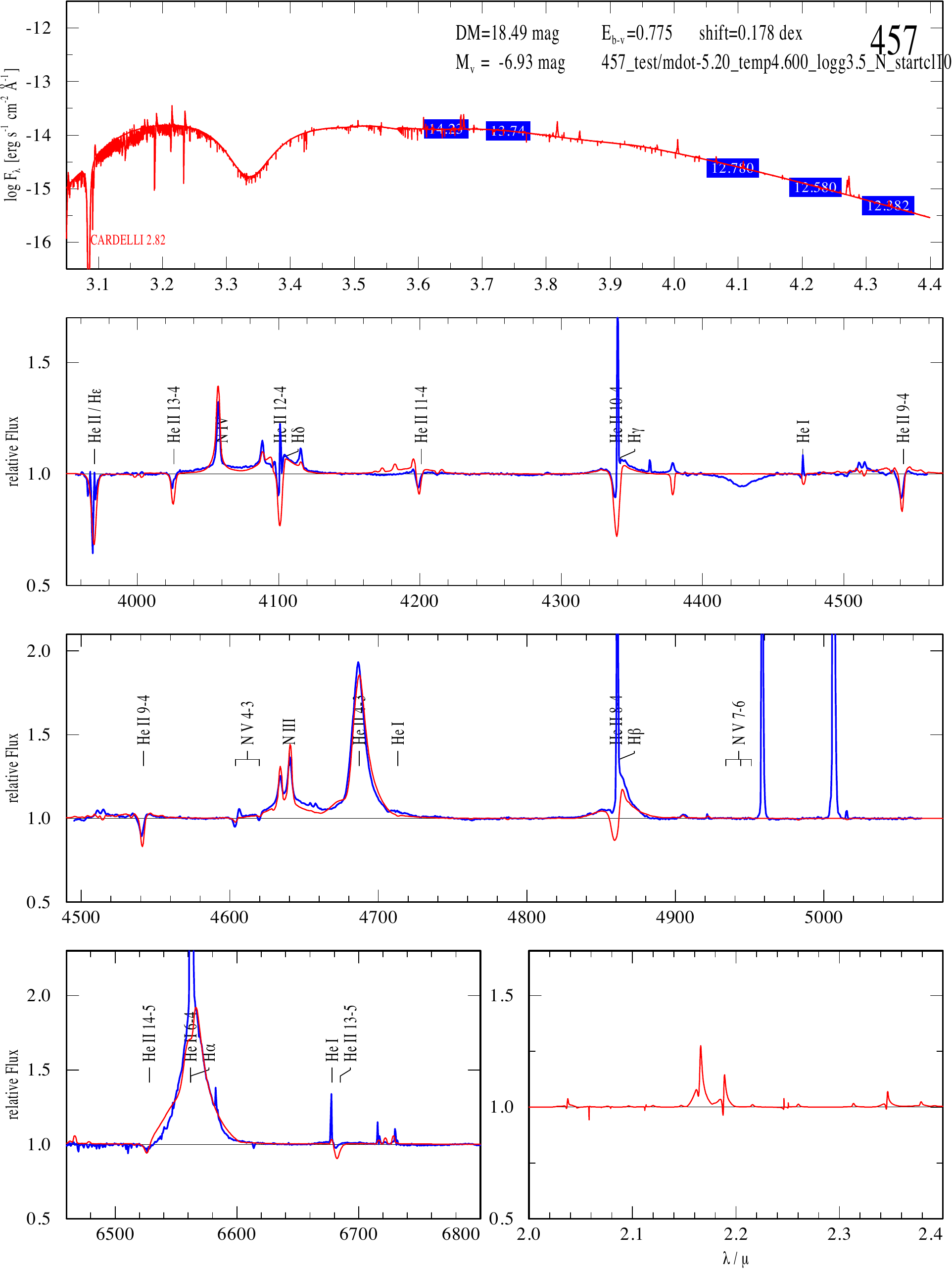}
\end{center}
\caption{Improved test model for VFTS\,457 (O3.5 If*/WN7) relative to Fig.\,\ref{f:457}. The N abundance is twice that of our enriched models. In addition, the model differs by having a lower $\log g$ (3.5\,dex) and by starting the clumping in the wind velocity law at 10 km\,s$^{-1}$. $\dot{M}$ was reduced as well to compensate the line strength increases as a result of the lower $\log g$. The difference in $\dot{M}$ is less than 0.1 dex.}
\label{f:457_test}
\end{figure}

Thirdly, for VFTS\,259, 457, and 1021 the observed N\,{\sc
iii}\,$\lambda 4634/4640$\,{\AA} lines are significantly stronger than
in our theoretical models.  Furthermore, there is a discrepancy
between the diagnostic He\,{\sc i}\,$\lambda 4471$\,{\AA} line, and
the N\,{\sc iii}\,$\lambda 4634/4640$\,{\AA} and N\,{\sc iv}\,$\lambda
4058$\,{\AA} lines for the effective temperature.  To estimate the
uncertainties we carried out test calculations for the O star 259 and
Of/WN star 457. We found that the discrepancies can be significantly
improved by simultaneously lowering $\log g$ by 0.5\,dex, increasing
the N-abundance by a factor 2, and moving the starting point of
clumping closer to the surface (see Fig.\,\ref{f:457} and
\ref{f:457_test} for the case of VFTS\,457). The resulting parameters
agree with those given in Tab.\,\ref{t:parameters} within the given
uncertainties.

\onecolumn
\section{Additional Tables}
\begin{center}
\begin{longtable}{l@{\,}c@{\,}c@{\,}c@{\,}c@{\,}c@{\,}c@{\,}c@{\,}c@{\,}c@{\,}|@{\,}c@{\,}c@{\,}cc}
\caption{\label{t:spectra} Sources of spectroscopic data, line profile variation (LPV), and comments about binarity/multiplicity.}\\
\hline\hline
Star	& Medusa	& UVES	& ARGUS	& SINFONI	& HST/FOS	& HST/STIS	& HST/COS & HST/GHRS		& IUE	& LPV & comments & source	&	$\dagger$\\
\hline
\endfirsthead
\caption{Spectroscopic data, line profile variation (LPV), and comments about binarity/multiplicity.}\\
\hline\hline
Star	& Medusa	& UVES	& ARGUS	& SINFONI	& HST/FOS	& HST/STIS	& HST/COS &HST/GHRS		& IUE	& LPV & comments & source	&	$\dagger$\\
\hline
\endhead
\hline
\endfoot
016	&	X$^{\rm a}$	& --		& --		& --		& --		& --		& X$^{\rm g}$	& --		& --		&no	& 	--	&	1	&	x\\
063	&	X$^{\rm a}$	& --		& --		& --		& --		& --		& --	& --		& --		&yes	& SB2	&	1&	--\\
064	&	X$^{\rm a}$	& --		& --		& --		& --		& --		& --	& --		& --		&yes	& weak RV	&	1	&	x\\
072	&	X$^{\rm a}$	& --		& --		& --		& --		& --		& --	& --		& --		&no	&	--	&	1	&	x\\
094	&	X$^{\rm a}$	& --		& --		& --		& --		& --		& --	& --		& --		&yes	& SB2	&	1	&	--\\
108	&	X$^{\rm a}$	& --		& --		& --		& --		& --		& --	& --		& X		&no	&	--	&	7	&	x\\
145	&	X$^{\rm a}$	& --		& --		& --		& --		& --		& --	& --		& --		&yes	& weak RV, MULT	&	1	&	--\\
147	&	X$^{\rm a}$	& --		& --		& --		& --		& --		& --	& --		& --		&yes?& MULT	&	7	&	--\\
151	&	X$^{\rm a}$	& --		& --		& --		& --		& --		& --	& --		& --		&yes	& weak RV, MULT	&	1	&	--\\
169	&	X$^{\rm a}$	& --		& --		& --		& --		& --		& --	& --		& --		&no	&	--	&	1	&	x\\
171	&	X$^{\rm a}$	& --		& --		& --		& --		& --		& --	& --		& --		&yes	& weak RV	&	1	&	x\\
180	&	X$^{\rm a}$	& --		& --		& --		& --		& --		& --	& --		& X		&yes	& weak RV	&	1	&	x\\
208	&	X$^{\rm a}$	& --		& --		& --		& --		& --		& --	& --		& --		&no	& weak RV	&	1	&	--\\
216	&	X$^{\rm a}$	& --		& --		& --		& --		& --		& --	& --		& --		&no	&	--	&	1	&	x\\
259	&	X$^{\rm a}$	& --		& --		& --		& --		& --		& --	& --		& --		&no	& weak RV	&	1	&	x\\
267	&	X$^{\rm a}$	& --		& --		& --		& --		& --		& --	& --		& --		&?	& weak RV	&	1	&	x\\
333	&	X$^{\rm a}$	& --		& --		& --		& --		& --		& --	& --		& --		&yes	& weak RV	&	1	&	x\\
402	&	X$^{\rm a}$	& --		& --		& --		& --		& --		& --	& --		& X		&yes	& SB2? 	&	3,4,7	&	--\\
404	&	X$^{\rm a}$	& --		& --		& --		& --		& --		& --	& --		& --		&no	& SB1	&	1	&	x\\
406	&	X$^{\rm a}$	& --		& --		& --		& --		& --		& --	& --		& --		&no	& ASYM?	&	1	&	--\\
422	&	X$^{\rm a}$	& --		& --		& --		& --		& --		& --	& --		& --		&?	& SB1	&	1	&	x\\
427	&	X$^{\rm a}$	& --		& --		& --		& --		& --		& --	& --		& --		&no	&	--	&	4	&	x\\
440	&	X$^{\rm a}$	& --		& --		& --		& --		& --		& --	& --		& --		&yes	& SB2?	&	1	&	--\\
445	&	X$^{\rm a}$	& --		& --		& --		& --		& --		& --	& --		& --		&no	& SB2	&	1	&	--\\
455	&	X$^{\rm a}$	& --		& --		& --		& --		& --		& --	& --		& --		&yes	& SB1, ASYM	&	1	&	x\\
457	&	X$^{\rm a}$	& --		& --		& --		& --		& --		& --	& --		& --		&no?	& 	--	& 	7	&	x\\
468	&	X$^{\rm a}$	& --		& --		& --		& --		& --		& --	& --		& --		&no	& MULT	&	1	&	--\\
482	&	--	& X$^{\rm a}$		& --		& X$^{\rm b}$		& X$^{\rm c}$		& X$^{\rm d,e}$		& --		& --	& --	&yes?	& SB1,MULT?	&	3,5	&	x\\
506	&	X$^{\rm a}$	& --		& --		& --		& --		& --		& --	& --		& --		&yes	& weak RV	&	1	&	x\\
509	&	X$^{\rm a}$	& --		& --		& --		& --		& --		& --	& --		& X		&yes	& SB2		&	3,7	&	--\\
512	&	X$^{\rm a}$	& --		& --		& --		& --		& --		& --	& --		& --		&yes	& SB1	&	1	&	x\\
518	&	X$^{\rm a}$	& --		& --		& --		& --		& --		& --	& --		& --		&no	& weak RV	&	1	&	x\\
527	&	X$^{\rm a}$	& --		& --		& --		& --		& --		& --	& --		& X		&yes	& SB2	&	1	&	--\\
532	&	X$^{\rm a}$	& --		& --		& --		& --		& --		& --	& --		& --		&no	& SB1	&	1	&	x\\
538	&	X$^{\rm a}$	& --		& --		& --		& --		& --		& --	& --		& --		&yes	& SB2	&	1	&	--\\
542	&	X$^{\rm a}$	& --		& X$^{\rm a}$		& X$^{\rm b}$		& --	& --		& --		& --		& --		&yes	& SB1	&	2	&	x\\
545	&	--	& X$^{\rm a}$		& X$^{\rm a}$		& X$^{\rm b}$		& --	& --		& --		& --		& --		&no	& weak RV	&	2	&	x\\
562	&	--	& X$^{\rm a}$		& --		& --		& --		& --		& --	& --		& --		&? 	& ?		&	--	&	--\\
566	&	X$^{\rm a}$	& --		& --		& --		& --		& --		& --	& --		& --		&no	&	--	&	1	&	x\\
599	&	X$^{\rm a}$	& --		& --		& --		& --		& --		& --	& --		& --		&no	& weak RV	&	1	&	x\\
603	&	X$^{\rm a}$	& --		& --		& --		& --		& --		& --	& --		& --		&yes	& SB1	&	1	&	x\\
608	&	X$^{\rm a}$	& --		& --		& --		& --		& --		& --	& --		& --		&yes	& SB1	&	1	&	x\\
617	&	X$^{\rm a}$	& --		& --		& --		& --		& --		& --	& --		& X		&yes	&	--	&	7	&	x\\
621	&	X$^{\rm a}$	& --		& --		& --		& --		& --		& --	& --		& --		&no	& DBL	&	1	&	x\\
626	&	X$^{\rm a}$	& --		& --		& --		& --		& --		& --	& --		& --		&no	&	--	&	1	&	x\\
664	&	X$^{\rm a}$	& --		& --		& --		& --		& --		& --	& --		& --		&no	&	--	&	1	&	x\\
669	&	X$^{\rm a}$	& --		& --		& --		& --		& --		& --	& --		& --		&no	& weak RV	&	1	&	x\\
682	&	X$^{\rm a}$	& --		& --		& --		& --		& --		& --	& --		& --		&no	&	--	&	6	&	x\\
695	&	X$^{\rm a}$	& --		& --		& --		& --		& --		& --	& --		& X		&yes	& SB1	& 3,4,7	&	x\\
755	&	X$^{\rm a}$	& --		& --		& --		& --		& --		& --	& --		& --		&no	&	--	&	1	&	x\\
758	&	X$^{\rm a}$	& --		& --		& --		& --		& --		& --	& --		& X		&no?	&	--	&	7	&	x\\
797	&	X$^{\rm a}$	& --		& --		& --		& --		& --		& --	& --		& --		&no	&	--	&	1	&	x\\
1001	&	--	& --		& X$^{\rm a}$		& X$^{\rm b}$		& X$^{\rm c}$	& --		& --		& --		& --		&no?	&	--	&	2	&	x\\
1014	&	--	& --		& X$^{\rm a}$		& --		& X$^{\rm c}$		& --	& --		& --		& --		&no	& SB1	&	2	&	--\\
1017	&	--	& --		& X$^{\rm a}$		& X$^{\rm b}$		& X$^{\rm c}$		& X$^{\rm d}$	& --		& --		& --		&yes	& weak RV	&	2	&	x\\
1018	&	--	& --		& X$^{\rm a}$		& X$^{\rm b}$		& X$^{\rm c}$		& --		& --	& --		& --		&no	& SB1	&	2	&	x\\
1021	&	--	& --		& X$^{\rm a}$		& X$^{\rm b}$		& X$^{\rm c}$		& --		& --	& --		& --		&no	& weak RV	&	2	&	x\\
1022	&	--	& --		& X$^{\rm a}$		& X$^{\rm b}$		& X$^{\rm c}$		& X $^{\rm e}$	& --		& --		& --		&no	& weak RV	&	2	&	x\\
1025	&	--	& --		& X$^{\rm a}$		& X$^{\rm b}$		& X$^{\rm c}$		& --	& --		& --	& --		&yes	& MULT?	&	2	&	x\\
1026	&	--	& --		& X$^{\rm a}$		& X$^{\rm b}$		& X$^{\rm c}$		& --		& --	& --		& --		&no	& SB1	&	2	&	--\\
1028	&	--	& --		& X$^{\rm a}$		& X$^{\rm b}$		& X$^{\rm c}$		& --		& --	& --		& --		&no	&	--	&	2	&	--\\
Mk42	&	--	& X$^{\rm f}$		& --		& X$^{\rm b}$		& --		& --		& --	& X$^{\rm h}$		& --		&?	&MULT?	&	--	&	x\\
\end{longtable}
\tablefoot{Comments: single-lined (SB1), double-lined (SB2)
spectroscopic binary, asymmetric profile (ASYM), and double (DBL) or
multiple (MULT) sources in HST images.\\
Sources: (1) \cite{sana2013}, (2) \cite{henault2012}, (3) \cite{schnurr2008}, (4) \cite{doran2013}, (5) \cite{massey2005}, (6) \cite{bestenlehner2011}, (7) this work.\\
Observations: (a) program ID:  182.D-0222 (PI:\,Evans), (b) program ID: 084.D-0980 (PI:\,Gr\"afener), (c) program ID: 6417 (PI:\,Massey), (d) program ID: 7739 (PI:\,Massey), (e) program ID: 9412 (PI:\,Massey), (f) program ID: 70.D-0164 (PI:\,Crowther), (g) program ID: 11484 (PI:\,Hartig), (h) program ID: 3030 (PI:\,Ebbets).\\
($\dagger$): Stars mark with x are considered in the discussion.}

\end{center}

\begin{table}[h]
\caption{Aliases.}
\label{t:aliases} 
\centering
\begin{tabular}{ll|cl}
\hline
\hline
VFTS & aliases & VFTS & aliases\\
\hline
016	&	30 Dor 016					&	518	&	P93 901\\
063	&	ST92 1-11					&	527	&	BAT99 107, Brey 86, R139, P93 952\\
064	&	ST92 1-12					&	532	&	P93 974\\
072	&	BI 253						&	538	&	Mk 22, P93 1024\\
094	&	ST92 1-28, OGLE LMC-LPV-75429			&	542	&	BAT99 113, Mk 30, P93 1018, R136-015\\	
108	&	BAT99 089, Brey 71, HD 269883, SK -69 233	&	545	&	BAT99 114, Mk 35, P93 1029, R136-012\\
145	&	T88-3						&	562	&	Mk 26, P93 1150\\	
147	&	BAT99 091, Brey 73, T88-1			&	566	&	Mk 23, P93 1163\\	
151	&	T88-2						&	599	&	P93 1311\\
169	&	ST92 1-71					&	603	&	Mk 10, P93 1341\\
171	&	ST92 1-72					&	608	&	Mk 14, P93 1350\\
180	&	BAT99 093, Brey 74a, ST92 1-78                  &	617	&	BAT99 117, Brey 88, R146, HD 269926, SK -69 245\\	
208	&	ST92 1-93                                       &	621	&	P93 1429\\
216	&	ST92 1-97                                       &	626	&	P93 1423\\
259	&	--                                              &	664	&	Mk 4, P93 1607\\
267	&	Dor IRS 2                                       &	669	&	P93 1619\\
333	&	R133, P93 42                                    &	682	&	P93 1732\\	
402	&	BAT99 095, Brey 80, R135, P93 355               &	695	&	BAT99 119, Brey 90, R145, HD 269928,	SK -69 248\\	
404	&	--                                              &	755	&	P93 2041\\
406	&	Mk 55, P93 370                                  &	758	&	BAT99 122, Brey 92, R147, HD 38344 , SK -69 251\\	
422	&	--                                              &	797	&	--\\
427	&	BAT99 096, Brey 81, Mk53                        &	1001	&	BAT99 100, Brey75, R134, R136-004\\
440	&	Mk47, P93 607                                   &	1014	&	R136-029, P93 863\\	
445	&	P93 621                                         &	1017	&	BAT99 104, Brey 76, R136-044, P93 897\\
455	&	P93 661                                         &	1018	&	R136-037, P93 900\\	
457	&	BAT99 097, Mk51, P93 666                        &	1021	&	R136-011, P93 917, Mk 37W\\
468	&	Mk 36, P93 706                                  &	1022	&	Mk 37a, R136-014\\
482	&	BAT99 99, Brey 78, Mk 39, P93 767, R136-007     &	1025	&	BAT99 112, Brey 82, R136c, R136-010, P93 998 \\	
506	&	Mk 25, P93 871                                  &	1026	&	R136-041, P93 1013, Mk 35N\\	
509	&	BAT99 103, Brey 87                              &	1028	&	Mk 35S, R136-023, P93 1036\\
512	&	P93 885                                         &	Mk42	&	BAT99 105, Brey 77, R136-002, P93 922\\
\hline
\end{tabular}			
\end{table}

\begin{table}[h]
\caption{Adopted atomic model for the grid calculation.}
\label{t:atom_model}
\centering
\begin{tabular}{lcc}
\hline
\hline
Ion	& super levels	& atomic levels\\
\hline
H\,{\sc i}		&	20	&	30	\\
He\,{\sc i}		&	45	&	69	\\
He\,{\sc ii}	&	22	&	30	\\
C\,{\sc iii}		&	51	&	84	\\
C\,{\sc iv}		&	64	&	64	\\
N\,{\sc iii}		&	41	&	82	\\
N\,{\sc iv}		&	44	&	76	\\
N\,{\sc v}		&	41	&	49	\\
O\,{\sc iii}		&	88	&	170	\\
O\,{\sc iv}		&	78	&	154	\\
O\,{\sc v}		&	32	&	56	\\
O\,{\sc vi}		&	25	&	31	\\
Si\,{\sc iv}		&	22	&	33	\\
P\,{\sc iv}		&	30	&	90	\\
P\,{\sc v}		&	16	&	62	\\
S\,{\sc iv}		&	51	&	142	\\
S\,{\sc v}		&	31	&	98	\\
S\,{\sc vi}		&	28	&	58	\\
Fe\,{\sc iv}	&	74	&	540	\\
Fe\,{\sc v}		&	50	&	220	\\
Fe\,{\sc vi}	&	44	&	433	\\
Fe\,{\sc vii}	&	29	&	153	\\
\hline
Total			&	926	&	2724	\\
\hline
\end{tabular}
\end{table}

\begin{table*}[h]
\caption{The eight different abundances of our main grid given in mass fraction. The first row corresponds to half-solar metallicity \citep{asplund2005}. In the following rows C and O are converted into N as a result of the CNO-cycle and the He mass fraction increases.}
\label{t:abundance}
\centering
\begin{tabular}{lccccccc}
\hline
\hline
He		& C	& N	& O	& Si	& P	& S	& Fe\\
\hline
$0.25$	& $1.08\,10^{-3}$	&$ 3.08\,10^{-4}$	& $2.67\,10^{-3}$	& $3.31\,10^{-4}$	& $5.19\,10^{-6}$	& $1.6\,10^{-4}$	& $5.76\,10^{-4}$\\
$0.25$	 & $1.08~10^{-6}$	&$ 4.05~10^{-3}$	& $2.67~10^{-6}$	& $3.31\,10^{-4}$	& $5.19\,10^{-6}$	& $1.6\,10^{-4}$	& $5.76\,10^{-4}$\\
$0.325$	 & $1.08~10^{-6}$	&$ 4.05~10^{-3}$	& $2.67~10^{-6}$	& $3.31\,10^{-4}$	& $5.19\,10^{-6}$	& $1.6\,10^{-4}$	& $5.76\,10^{-4}$\\
$0.4$	 & $1.08~10^{-6}$	&$ 4.05~10^{-3}$	& $2.67~10^{-6}$	& $3.31\,10^{-4}$	& $5.19\,10^{-6}$	& $1.6\,10^{-4}$	& $5.76\,10^{-4}$\\
$0.475$	 & $1.08~10^{-6}$	&$ 4.05~10^{-3}$	& $2.67~10^{-6}$	& $3.31\,10^{-4}$	& $5.19\,10^{-6}$	& $1.6\,10^{-4}$	& $5.76\,10^{-4}$\\
$0.55$	 & $1.08~10^{-6}$	&$ 4.05~10^{-3}$	& $2.67~10^{-6}$	& $3.31\,10^{-4}$	& $5.19\,10^{-6}$	& $1.6\,10^{-4}$	& $5.76\,10^{-4}$\\
$0.625$	 & $1.08~10^{-6}$	&$ 4.05~10^{-3}$	& $2.67~10^{-6}$	& $3.31\,10^{-4}$	& $5.19\,10^{-6}$	& $1.6\,10^{-4}$	& $5.76\,10^{-4}$\\
$0.7$	 & $1.08~10^{-6}$	&$ 4.05~10^{-3}$	& $2.67~10^{-6}$	& $3.31\,10^{-4}$	& $5.19\,10^{-6}$	& $1.6\,10^{-4}$	& $5.76\,10^{-4}$\\
\hline
\end{tabular}
\end{table*}

\clearpage

\newpage

\section{Temperature sensitivity and He-abundance\label{he_temp}}
\begin{figure*}[ht]
\begin{center}
\includegraphics[width=17cm]{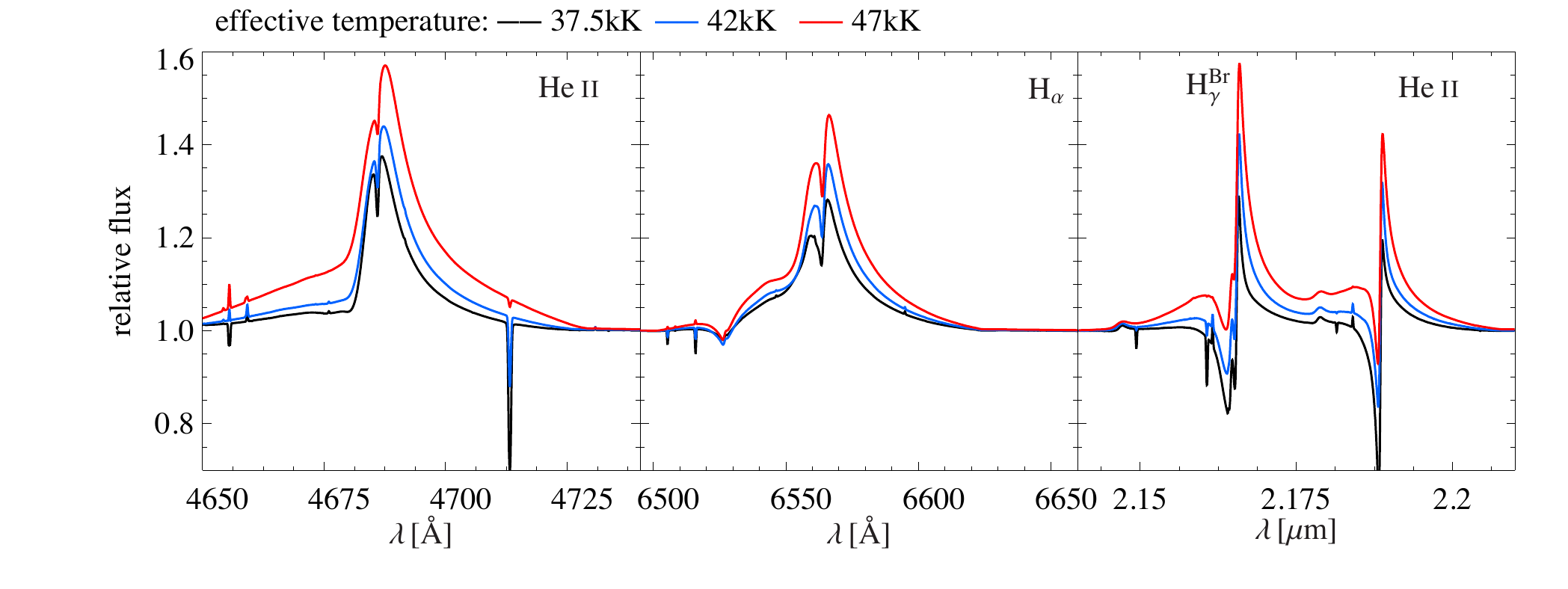}
\end{center}
\caption{Change of the line strengths of the helium and hydrogen emission lines with increasing temperature 
but constant luminosity and mass-loss rate. He\,{\sc ii} shows a strong sensitivity 
at certain temperatures. The near-IR is more homogeneous, but slightly clumping dependent. 
For reasons of clarity the relative near-IR flux ($F$) is scaled to the power 3 ($F^3$).} 
\label{f:temp}
\end{figure*}

\begin{figure*}[ht]
\begin{center}
\includegraphics[width=17cm]{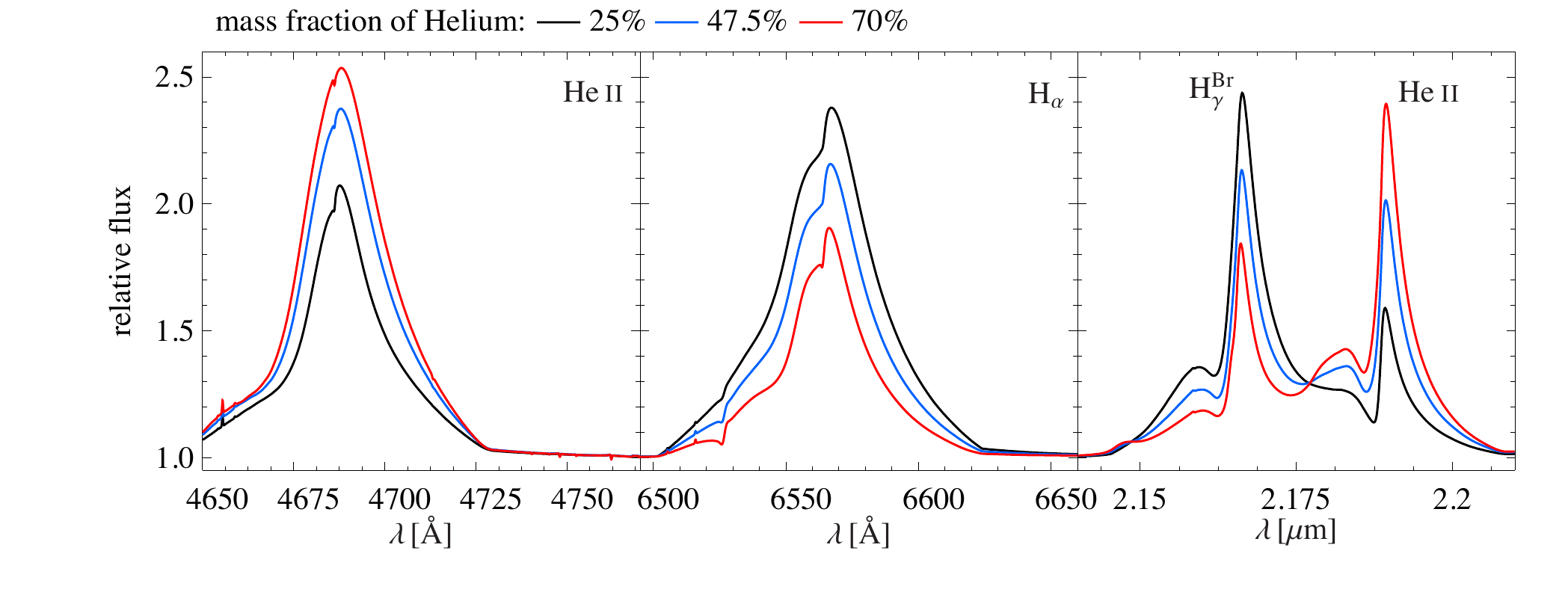}
\end{center}
\caption{Change of the ratio of the helium and hydrogen emission lines
with increasing helium abundance. 
The mass-loss rate of the models is $10^{-5}M_{\odot}yr^{-1}$ and the temperature is $\sim$\,$50\,000\mathrm{K}$. For reasons of clarity, 
the relative near-IR flux ($F$) is scaled to the power 3 ($F^3$).}
\label{f:he}
\end{figure*}
\section{The influence of $\log g$ on the stellar parameters \label{a:logg}}
A variation of $\log g$ changes the density and ionisation balance in
the stellar atmosphere. This can directly affect the derived
temperature and helium abundance and with it the luminosity and the
mass-loss rate. Here, we investigate the effect of varying surface
gravity on the result for the O-type stars. For this purpose, we used
our $\log g$ sub-grid with some additional smaller steps in helium
abundance, and analysed a small sub-sample of O stars in that
parameter space. More accurate parameters using a different stellar
atmosphere code including a variable $\log g$ in the analysis will be
provided by Ramirez-Agudelo et al. (in prep.) and
Sab\'{i}n-Sanjuli\'{a}n et al. (in prep.).

As shown in Fig.\,\ref{f:logg_lowT} and \ref{f:logg_highT} a lower
value of the surface gravity ($\Delta \log g =0.5$\,dex) requires a
lower model temperature ($\Delta T_{\rm eff} = -0.025...-0.03$\,dex)
to match the main O star temperature diagnostic He{\sc i} \,$\lambda
4471$. The transformed mass-loss rate is almost unchanged, within the
given error bars. The lower $T_{\rm eff}$ results in a lower
luminosity ($\sim -0.1$\,dex) and lower stellar mass. This gives a
lower terminal velocity for the O stars using the relation by
\cite{lamers1995}. These changes affect the mass-loss rates derived
from Eq.\,\ref{trans} by about $\sim 0.15$\,dex. The derived
He-abundance decreases by 2...4\%. The differences are larger for
increasing temperature.

  A variation of $\log g$ to values as low as 3.5 is only
  possible for the lowest $T_{\rm eff}$ in our model grid.  The reason
  is that for the hotter models the radiative flux ($F_{\rm
    rad}=4\pi\sigma T_{\rm eff}^4$) in the photosphere becomes so high
  that $\log g_{\rm rad}$ exceeds 3.5, leading to a situation where
  $\Gamma>1$ (cf.\ Eq.\,\ref{Gamma}) and no hydrostatic solution is
  possible. As $\Gamma$ is generally high for the hotter models, and
  the spectroscopically relevant quantity is $g_{\rm eff} = g
  (1-\Gamma)$, the $\log g$ needed to match the $\mathrm{H}_{\beta -
    \delta}$ line profiles of hot supergiants are not significantly
  lower than the $\log g$ of 4.0 adopted in our grid models. E.g.,
  \citet{evans2010} determined $\log g = 3.75$ for VFTS\,016
  (O2\,III-If*) in agreement with our own test computations.

\newpage
\begin{landscape}

\begin{figure}[ht]
\centering
\begin{minipage}[b]{12.3cm}
\centering
\includegraphics[width=12.3cm]{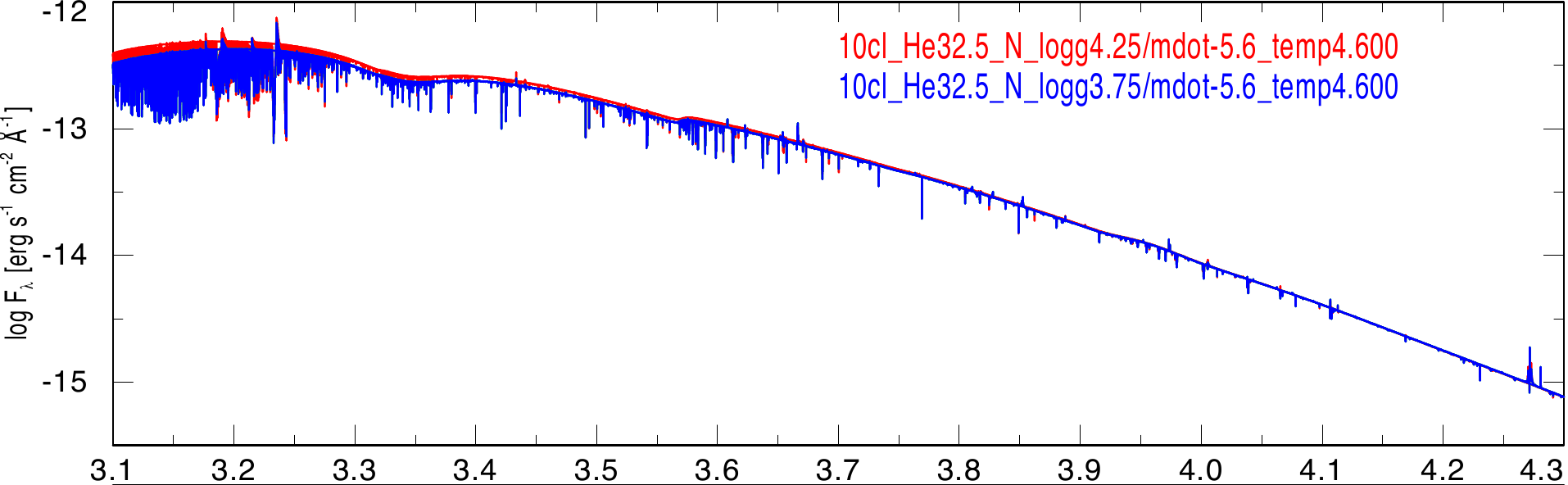}
\includegraphics[width=12.3cm]{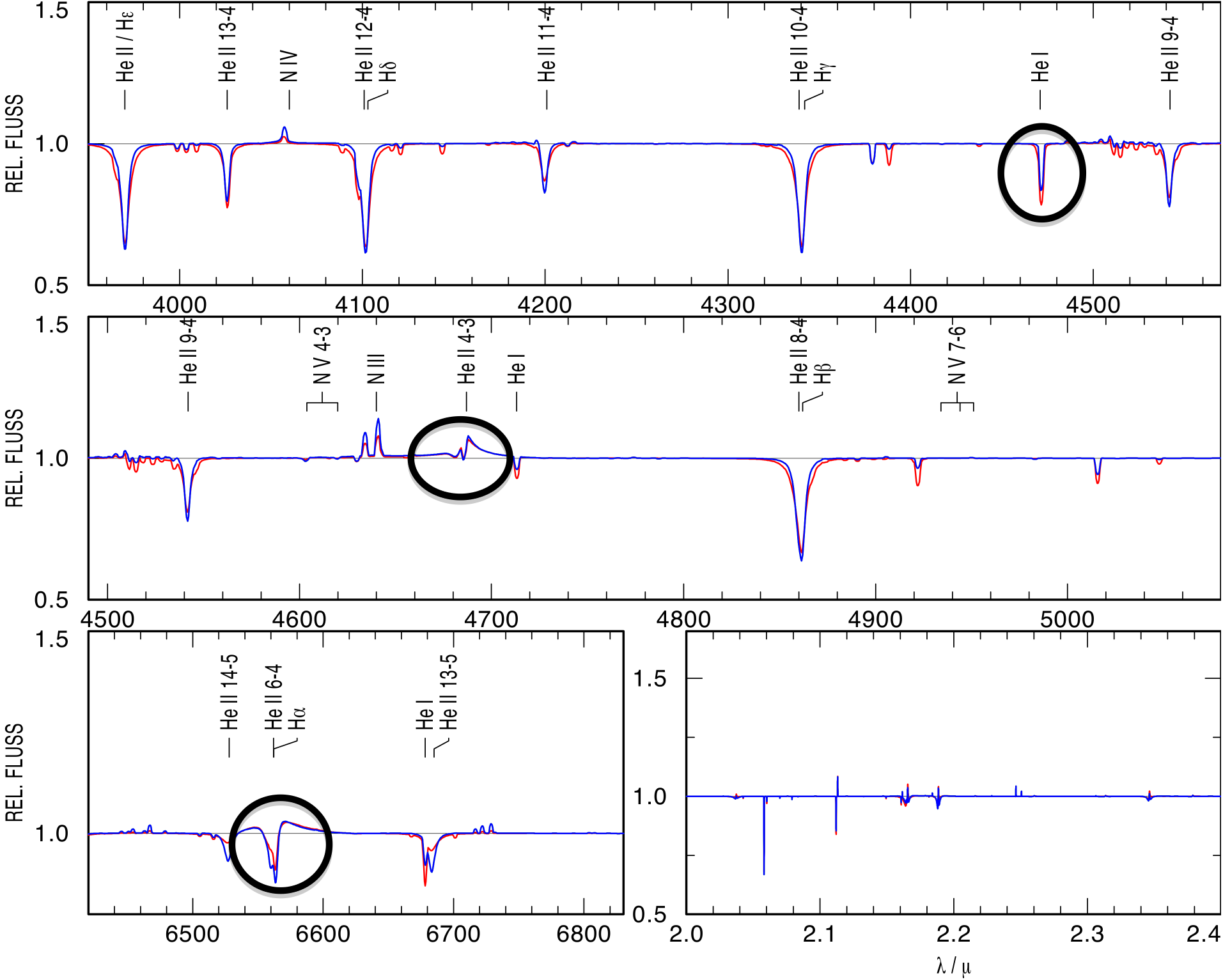}
\end{minipage}
\begin{minipage}[b]{12.3cm}
\centering
\includegraphics[width=12.3cm]{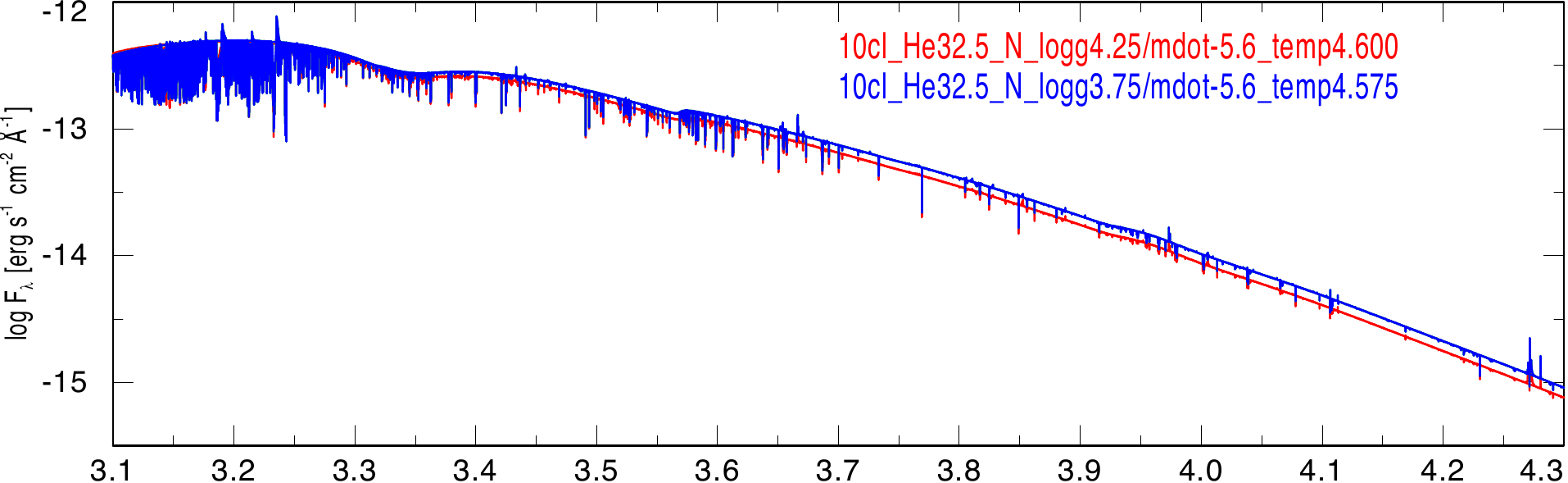}
\includegraphics[width=12.3cm]{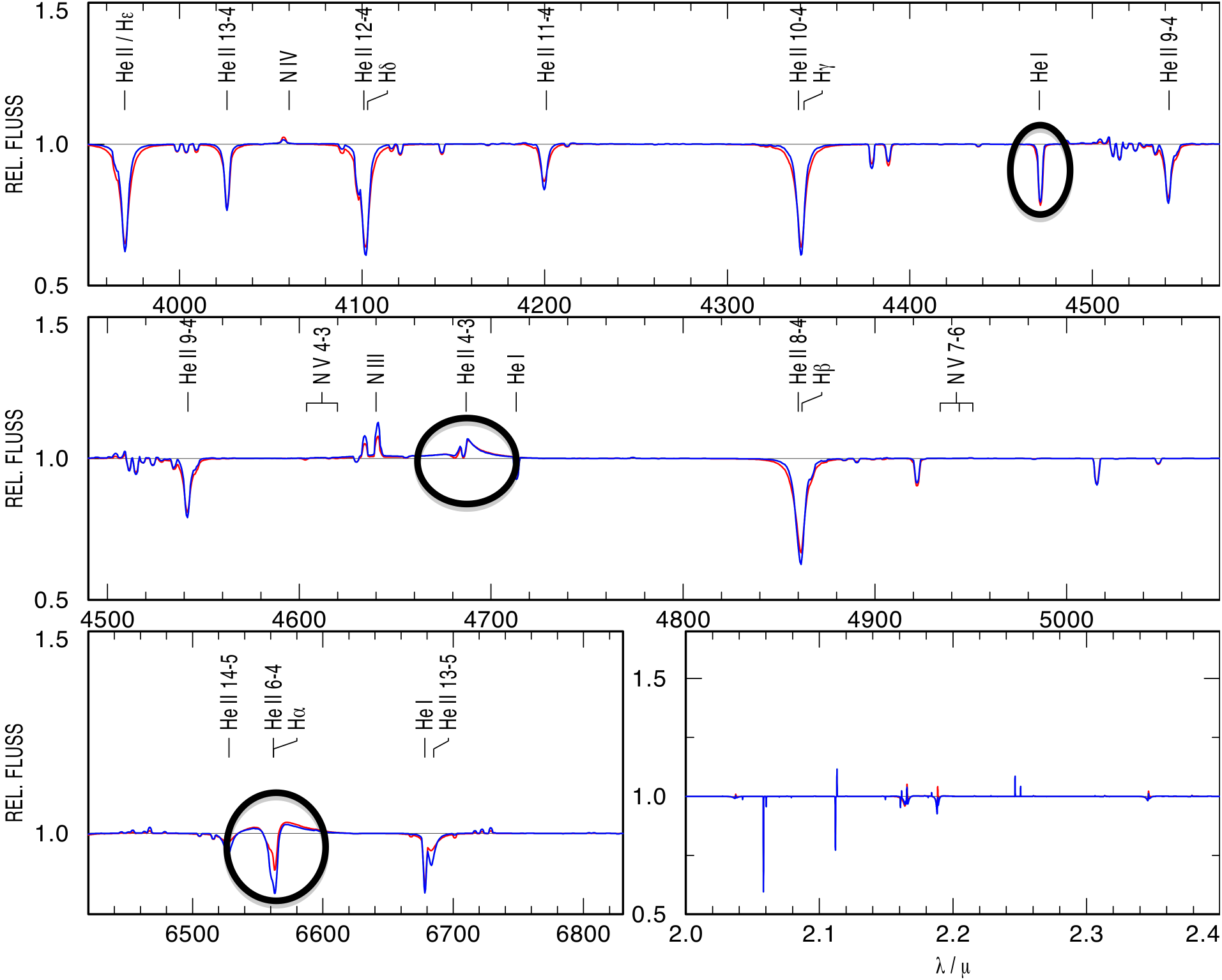}
\end{minipage}
\caption{Left: The blue model has a lower $\log g$ of 0.5~dex compare
to the red. Right: The temperature of the blue model is lower by
0.025~dex so that the line strength of the temperature diagnostic
He\,{\sc i}\,$\lambda4471$ matches the red model again. The
differences in luminosity are around 0.1\,dex, but the transformed mass-loss rate is unchanged.}
\label{f:logg_lowT}
\end{figure}
\newpage
\begin{figure}[ht]
\centering
\begin{minipage}[b]{12.3cm}
\centering
\includegraphics[width=12.3cm]{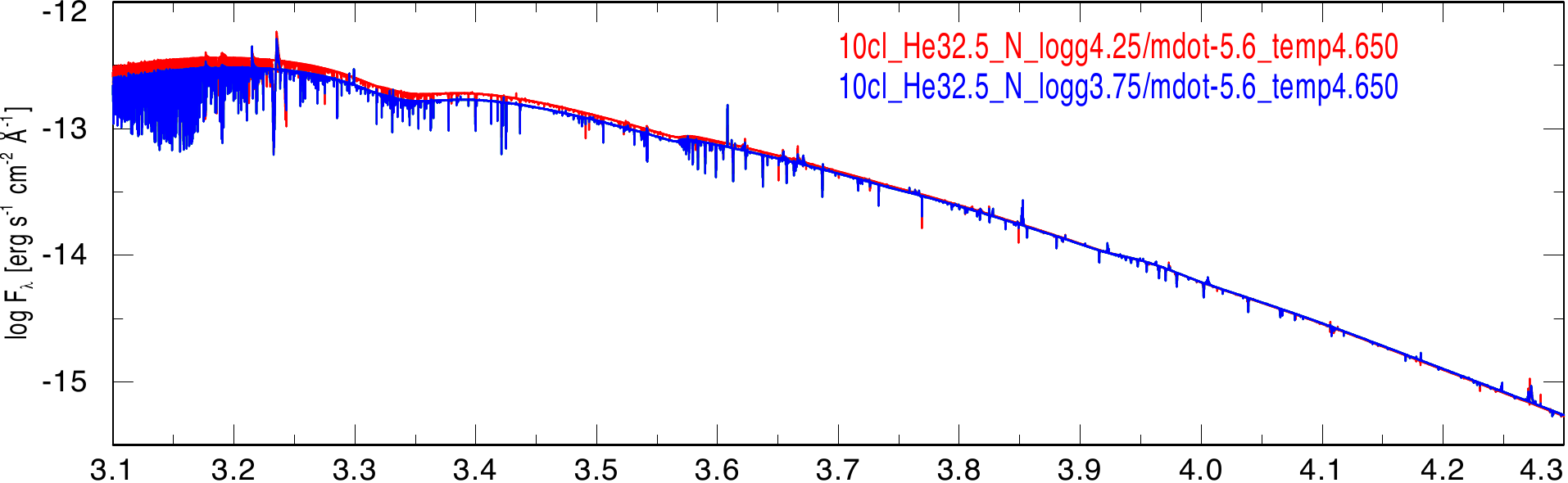}
\includegraphics[width=12.3cm]{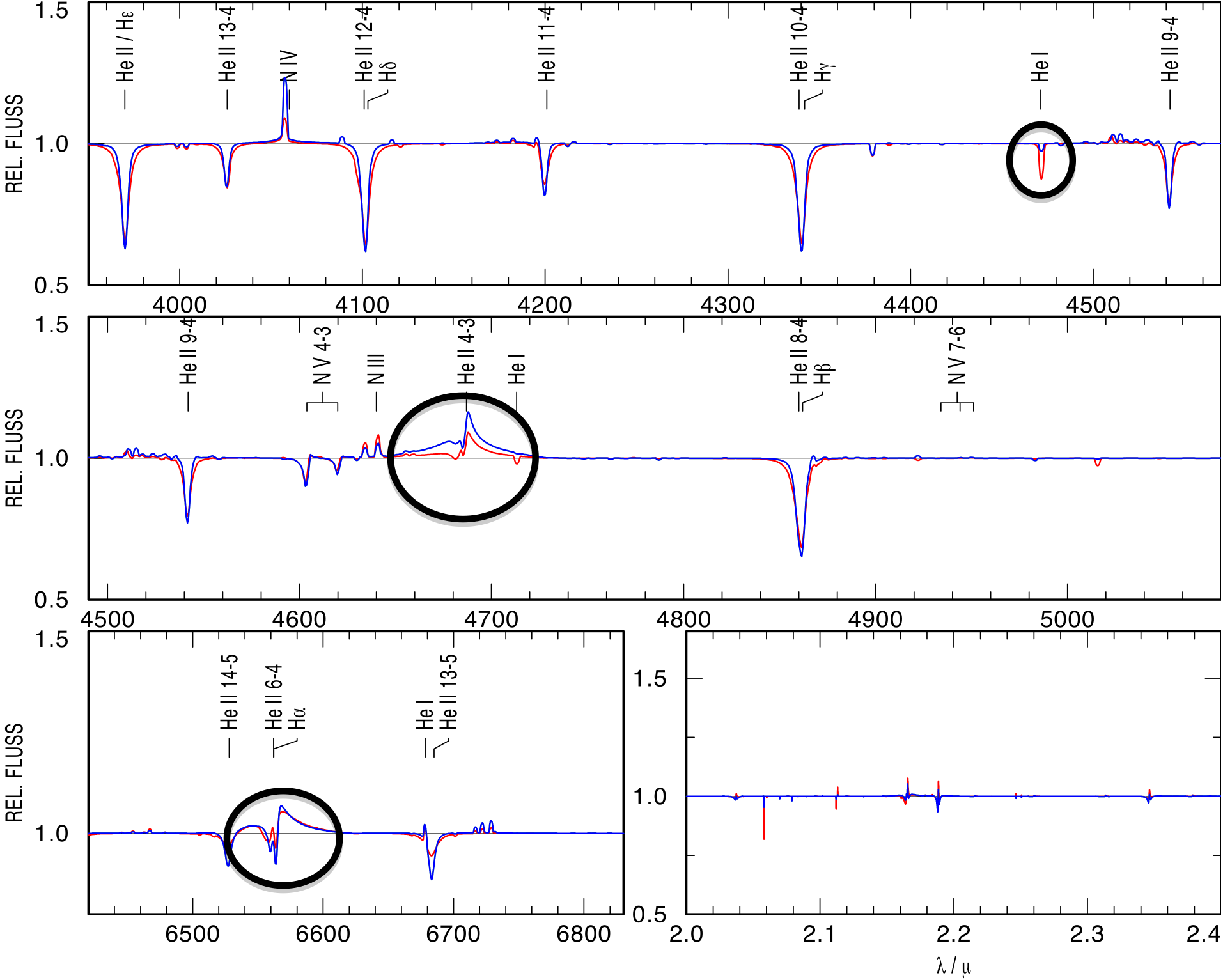}
\end{minipage}
\begin{minipage}[b]{12.3cm}
\centering
\includegraphics[width=12.3cm]{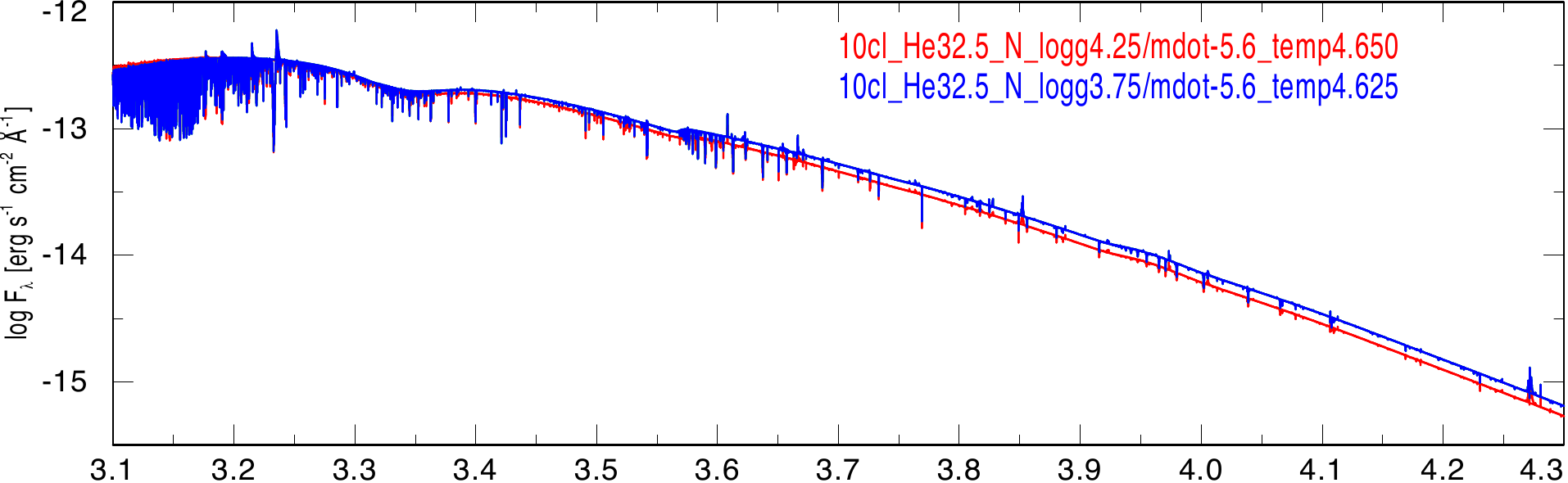}
\includegraphics[width=12.3cm]{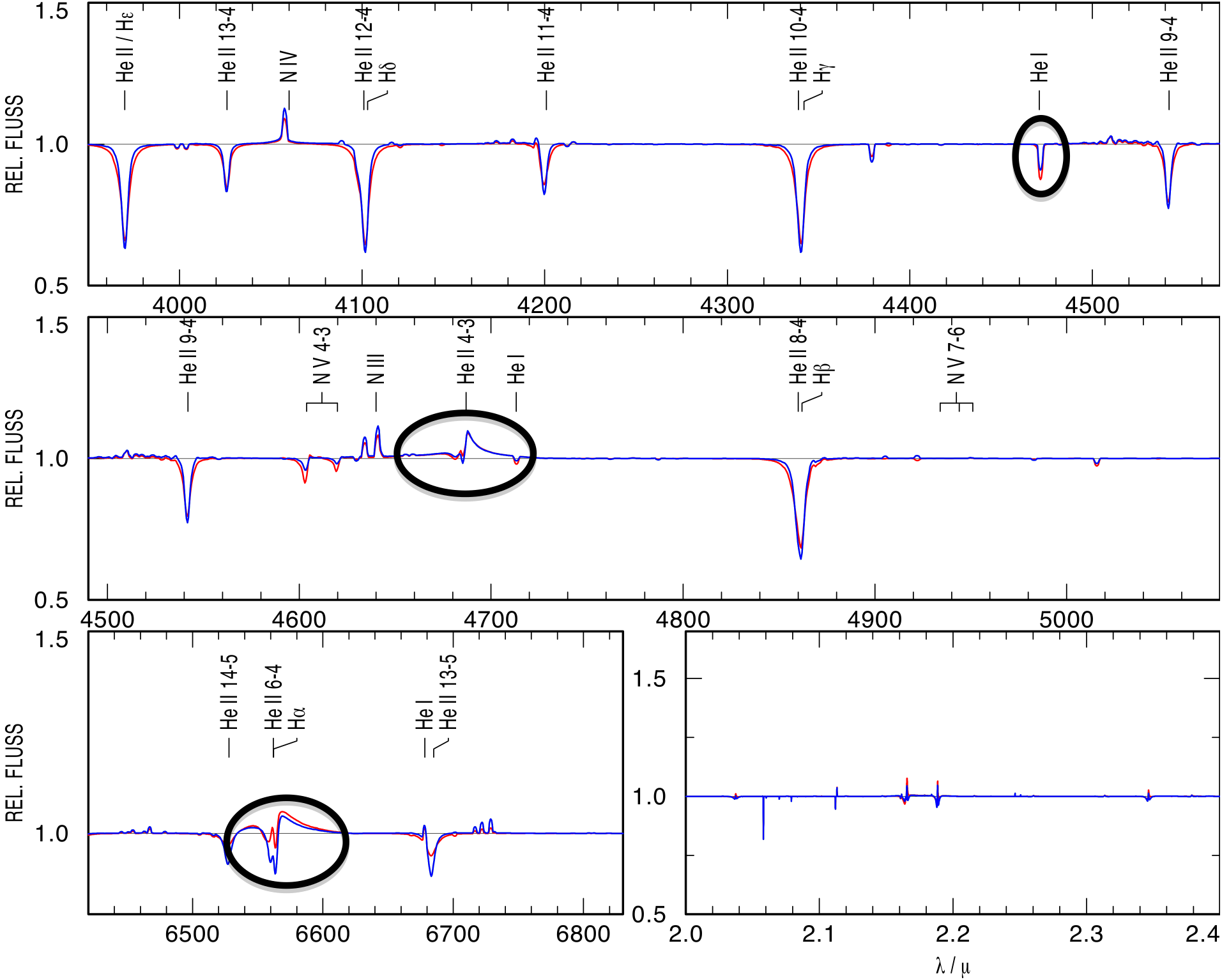}
\end{minipage}
\caption{Similar to Fig\,\ref{f:logg_lowT}, but at slightly higher temperature. He\,{\sc ii}\,$\lambda4686$ increases significantly after lowering $\log g$. The line strength is preserved after adjusting the temperature again.}
\label{f:logg_highT}
\end{figure}
\end{landscape}

\clearpage
\newpage
\Online
\section{Spectral modelling for each target star\label{a:plots}}

In the appendix we show model fits for all our targets and describe how we obtained the stellar parameters. The first panel in the figures is the model SED (red solid line) fit to the optical and near-IR photometry (blue boxes). The reddening law and the $R_{\rm V}$ is given in the bottom left corner. The following panels show the fits of the model spectra (red solid line) to the observations (blue solid line). 

\clearpage
\begin{figure}
\begin{center}
\includegraphics[width=17cm]{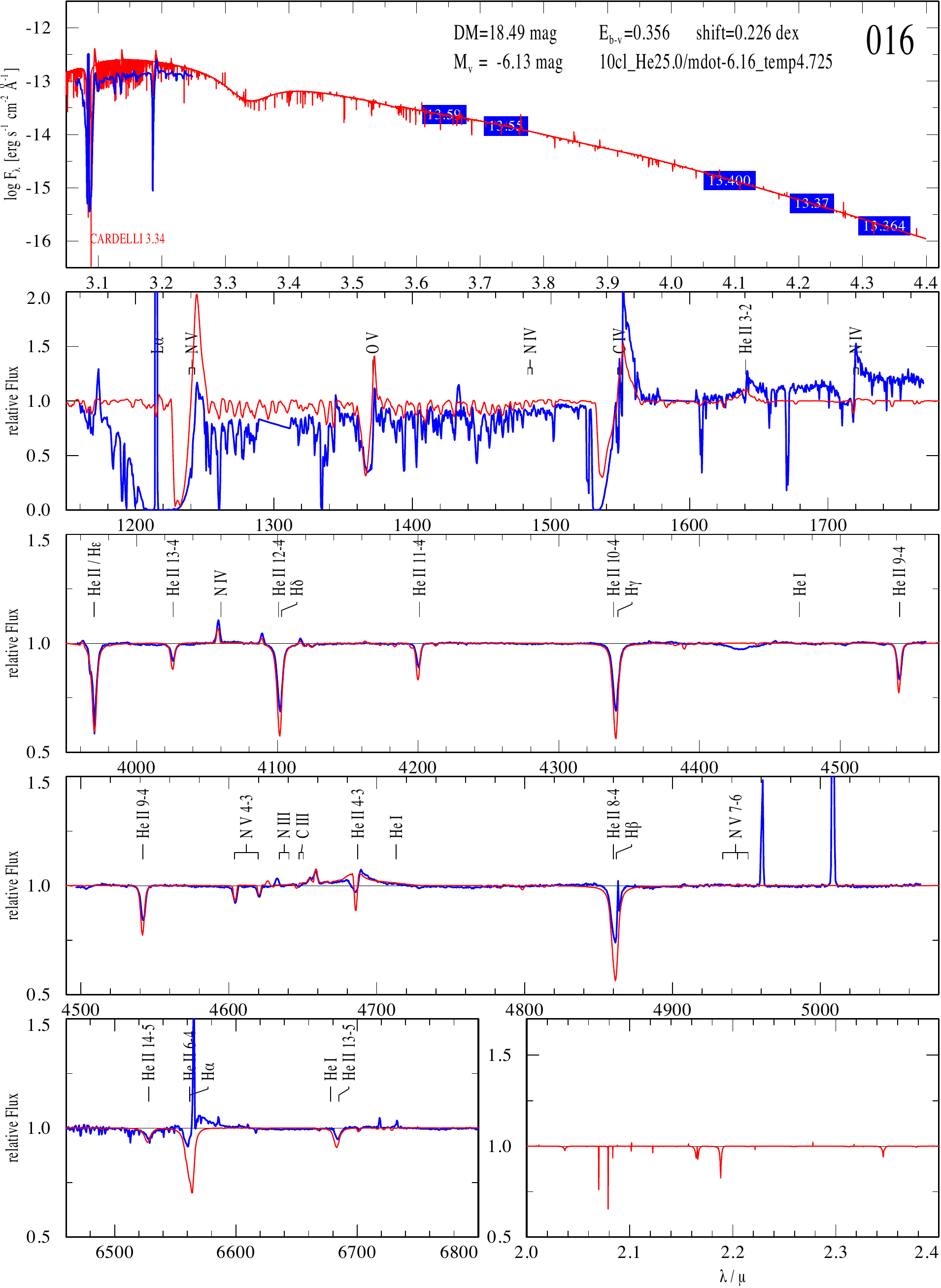}
\end{center}
\caption{VFTS\,016 (O2 III-If*) is a known runaway \citep{evans2010}. The temperature is based on the lines N\,{\sc iv}\,$\lambda 4058$, N\,{\sc v} $\lambda4604/4620$, and on the absence of He\,{\sc i}\,$\lambda 4471$. $\dot{M}$ is based on the line shape of He\,{\sc ii}\,$\lambda 4686$. The blue wing of He\,{\sc ii}\,$\lambda 4686$ and the red wing of $\mathrm{H_{\alpha}}$ is not properly reproduced by the model. However, the stellar parameters are in agreement with \cite{evans2010}.}
\label{f:016}
\end{figure}
\clearpage
\begin{figure}
\begin{center}
\includegraphics[width=17cm]{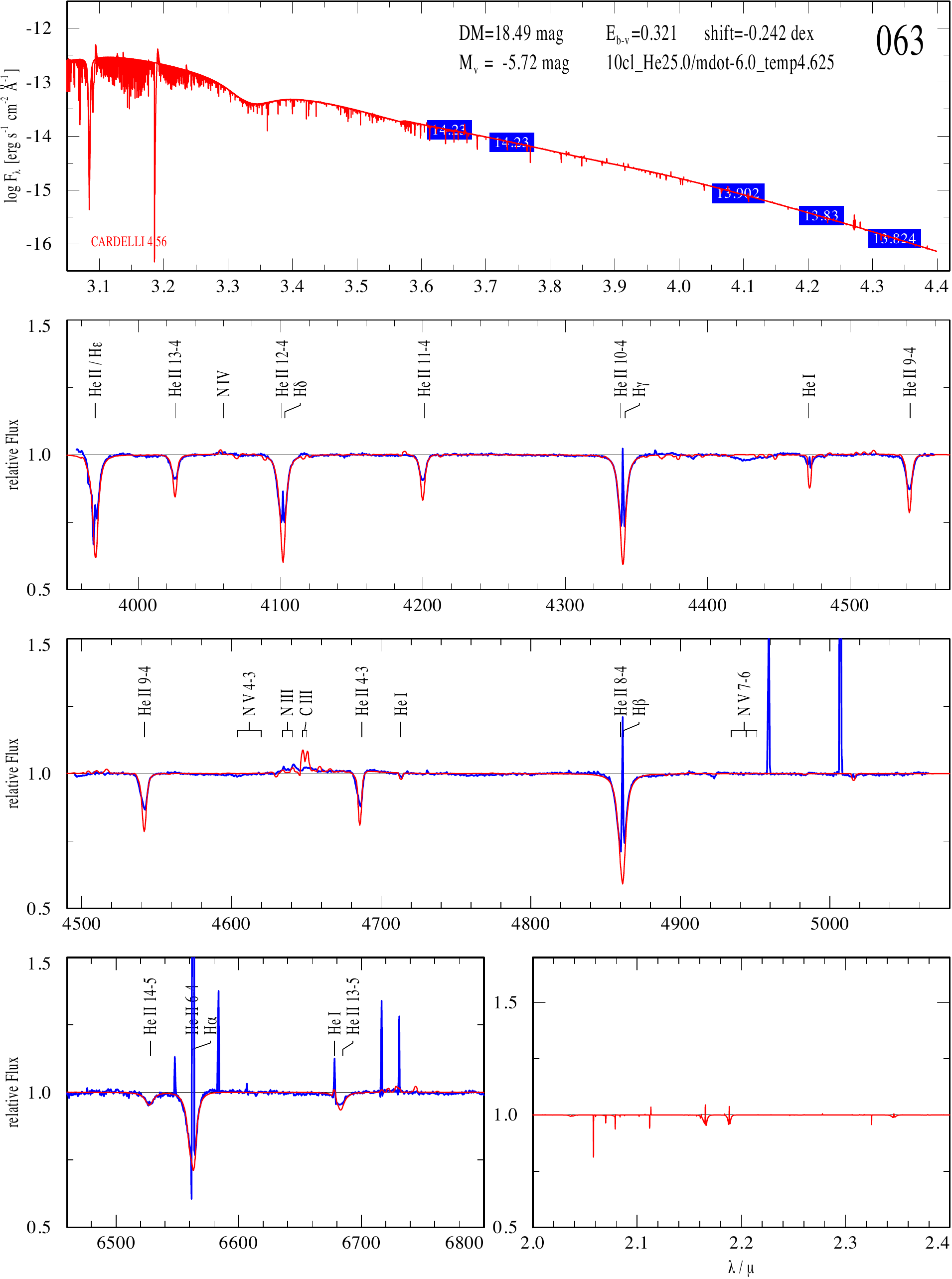}
\end{center}
\caption{VFTS\,063 (O5 III(n)(fc)+ sec): the He\,{\sc i}\,$\lambda 4471$ line is a bit too broad possibly as a result of the star being a SB2. The N\,{\sc iii}\,$\lambda 4634/4640$ lines are a bit too weak whilst the C\,{\sc iii}\,$\lambda 4647/4650$ lines are too strong. N-abundance is between normal and enriched.}
\end{figure}
\clearpage
\begin{figure}
\begin{center}
\includegraphics[width=17cm]{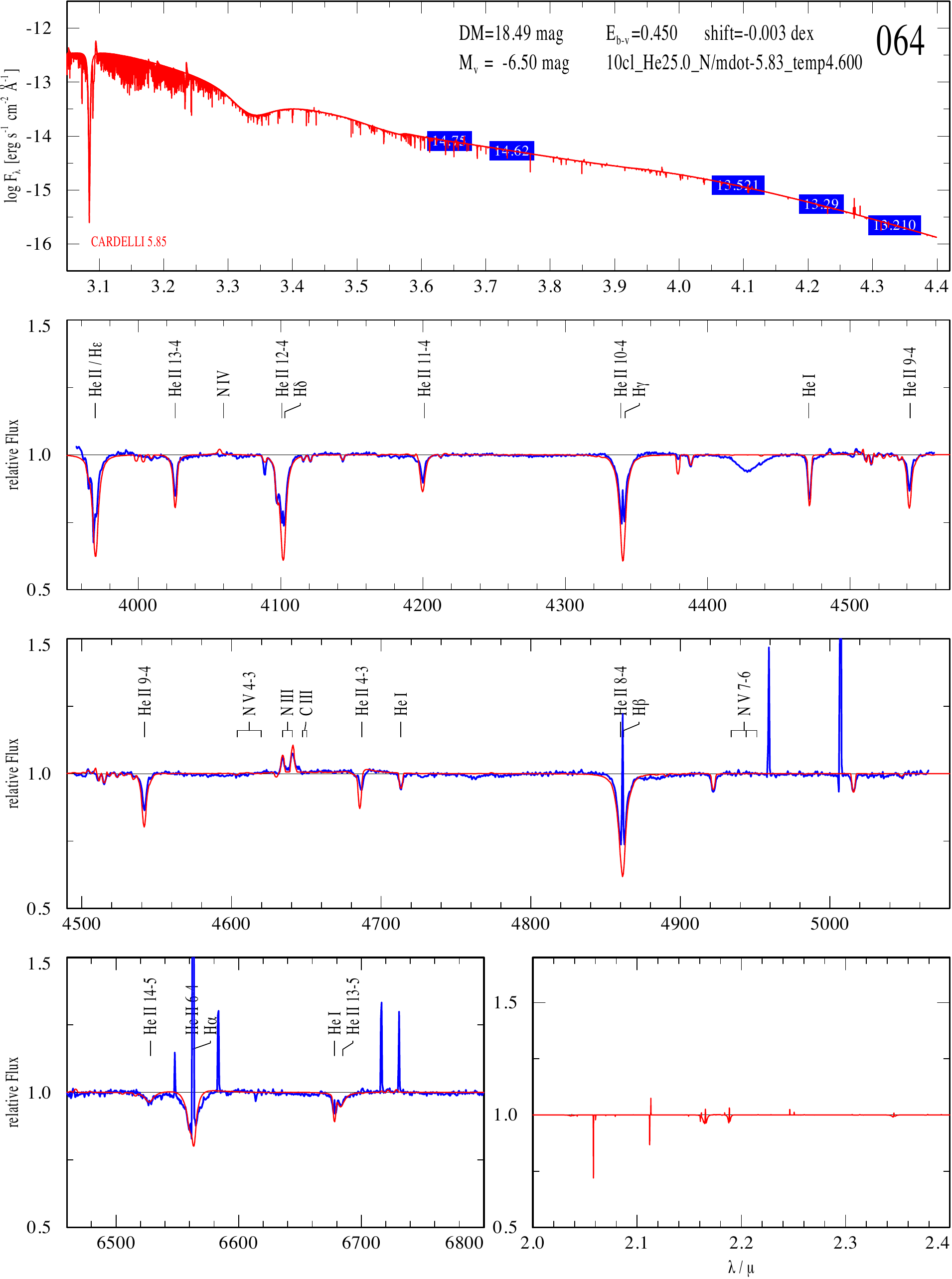}
\end{center}
\caption{The temperature of VFTS\,064 (O7.5 II(f)) is based on the He\,{\sc i}\,$\lambda 4471$ and N\,{\sc iii}\,$\lambda 4634/4640$ lines. $\dot{M}$ is based on the shape of the He\,{\sc ii}\,$\lambda 4686$ line. Nitrogen is enriched.}
\end{figure}
\clearpage
\begin{figure}
\begin{center}
\includegraphics[width=17cm]{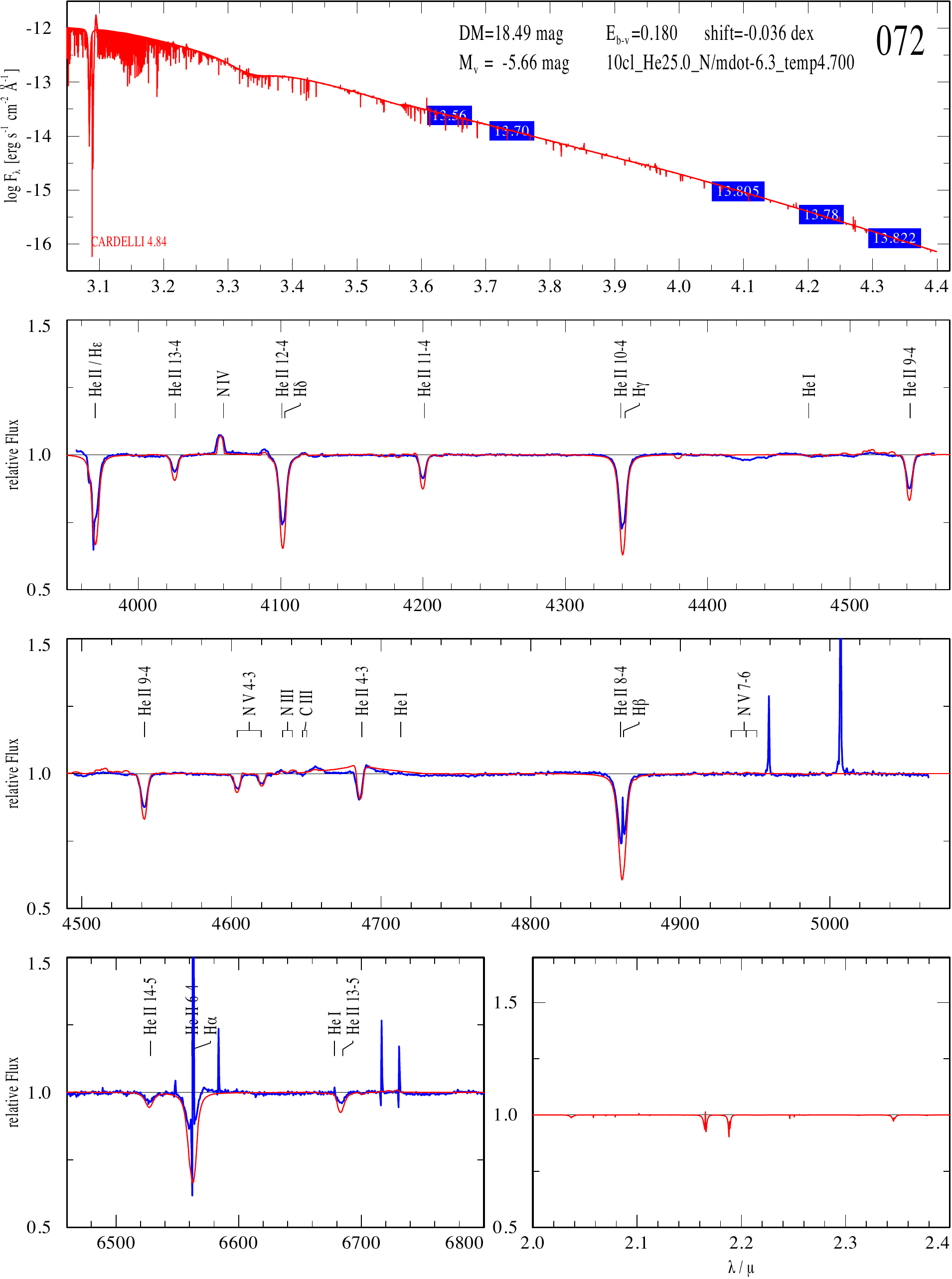}
\end{center}
\caption{The temperature of VFTS\,072 (O2 V-III(n)((f*))) is based on the lines N\,{\sc iii}\,$\lambda 4634/4640$, N\,{\sc iv}\,$\lambda 4058$,  N\,{\sc v} $\lambda4604/4620$, and on the absence of the He\,{\sc i}\,$\lambda 4471$ line. $\dot{M}$ is based on the line shape of He\,{\sc ii}\,$\lambda 4686$. Similar to Fig.\,\ref{f:016} the blue wing of He\,{\sc ii}\,$\lambda 4686$ and the red wing of $\mathrm{H_{\alpha}}$ are not properly reproduced by the model. The star is nitrogen enriched and fast rotating.}
\end{figure}
\clearpage
\begin{figure}
\begin{center}
\includegraphics[width=17cm]{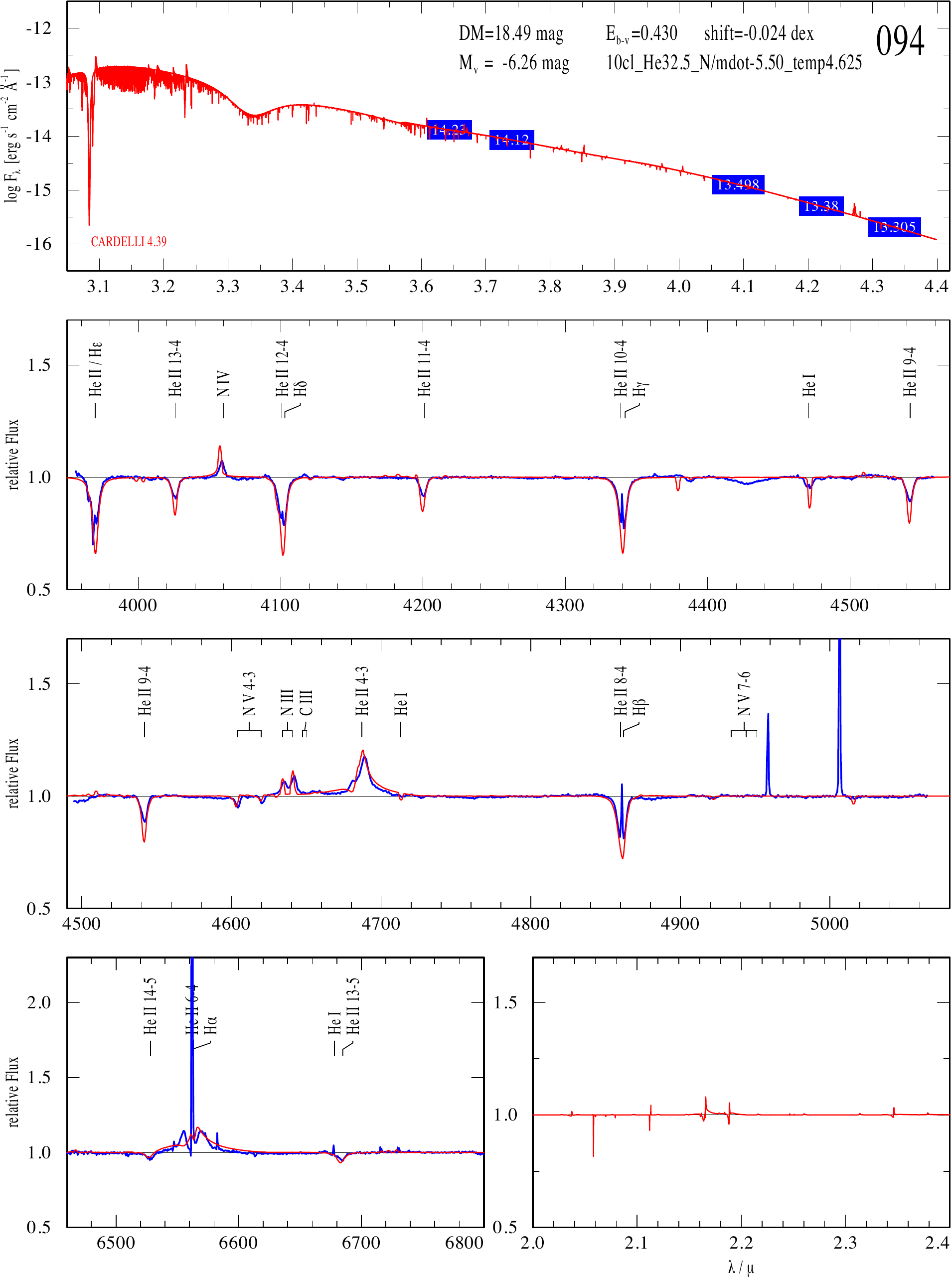}
\end{center}
\caption{He\,{\sc i}\,$\lambda 4471$ is too broad for VFTS\,094 (O3.5 Inf*p + sec) possibly as a result of its status as a SB2. The fit quality is reasonably good for a SB2, but the luminosity is uncertain as the contribution of the secondary is unknown.  N and He are enriched at the surface. The star is evolved with $\mathrm{H_{\alpha}}$ and He\,{\sc ii}\,$\lambda 4686$ in emission.}
\end{figure}
\clearpage
\clearpage
\begin{figure}
\begin{center}
\includegraphics[width=17cm]{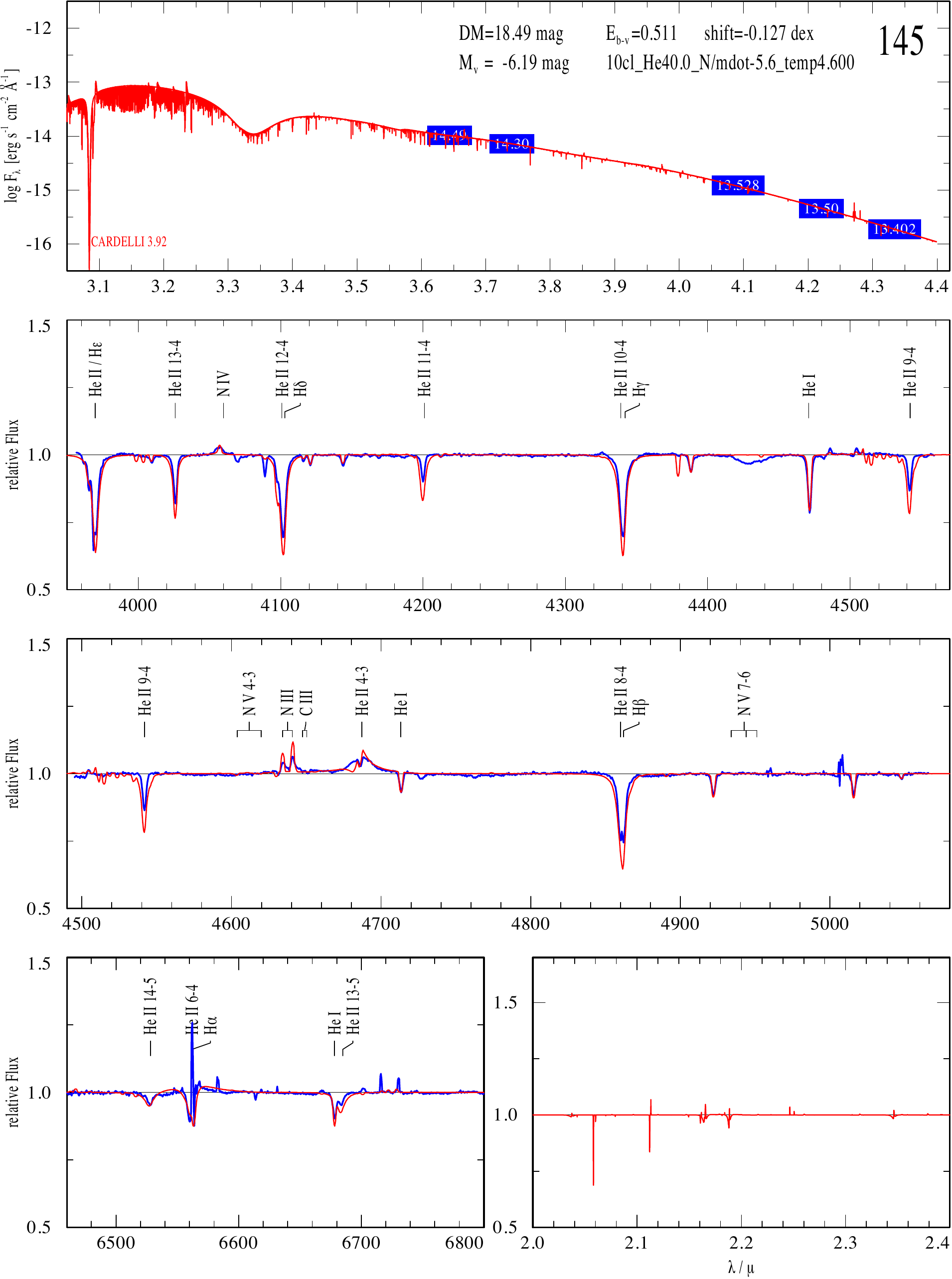}
\end{center}
\vspace{-0.3cm}
\caption{The temperature of VFTS\,145 (O8fp) is based on the lines He\,{\sc i}\,$\lambda 4471$, N\,{\sc iii}\,$\lambda 4634/4640$, and N\,{\sc iv}\,$\lambda 4058$. The He\,{\sc ii} absorption lines are narrower than the models which suggests a lower $\log g$ value than 4.0. The spectrum shows weak RV variations and LPVs. The star is multiple in the HST observations and therefore the luminosity might be overestimated. N is enriched. He\,{\sc ii}\,$\lambda 4686$ is quite strong relative to $\mathrm{H_{\alpha}}$. The only way to increase the He\,{\sc ii} emission line strength without increasing $\dot{M}$ is by increasing the He surface abundance.}
\end{figure}
\clearpage
\clearpage
\begin{figure}
\begin{center}
\includegraphics[width=17cm]{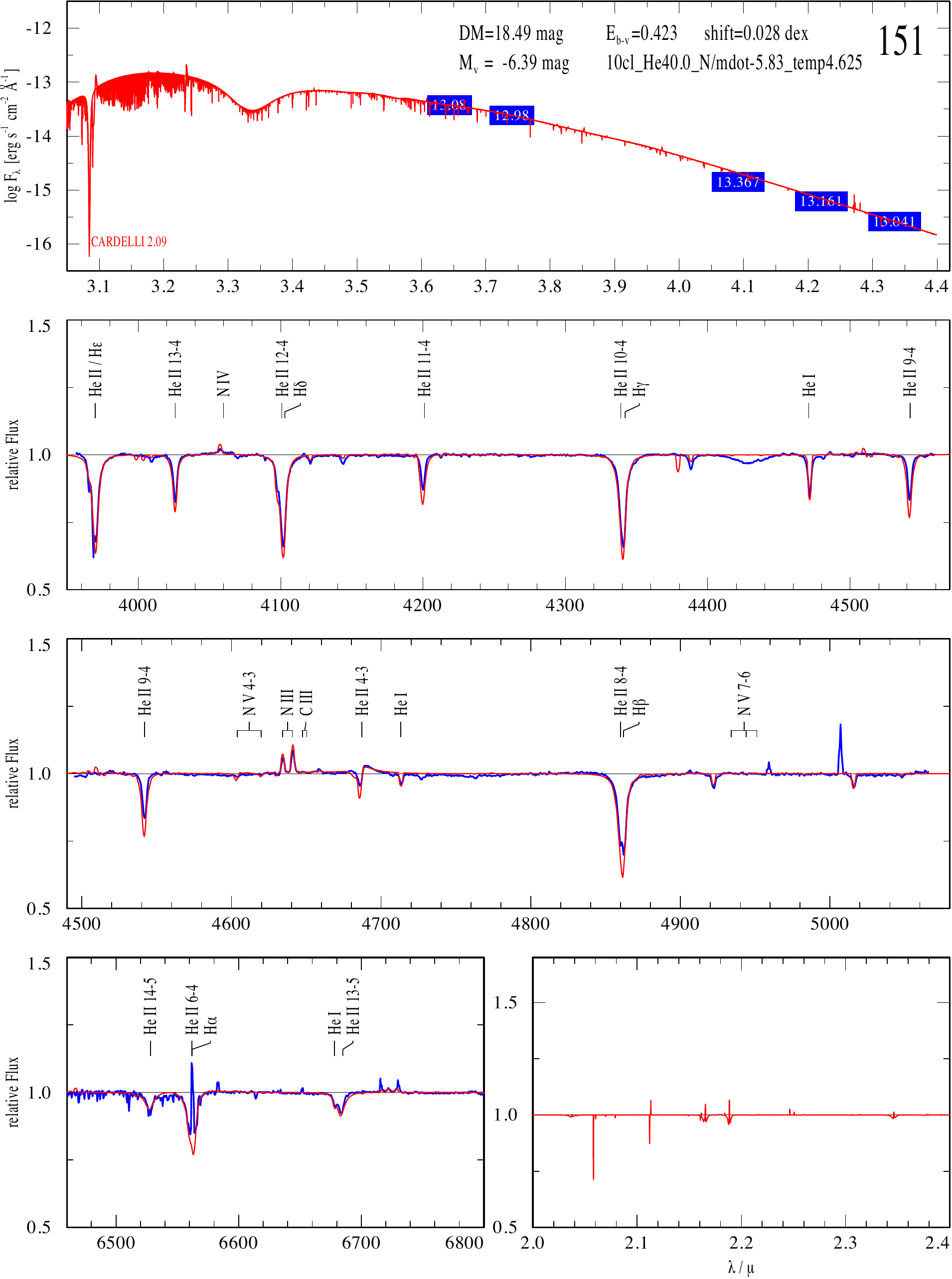}
\end{center}
\caption{The temperature of VFTS\,151 (O6.5 II(f)p) is based on the lines He\,{\sc i}\,$\lambda 4471$, N\,{\sc iii}\,$\lambda 4634/4640$, N\,{\sc iv}\,$\lambda 4058$, and N\,{\sc v} $\lambda4604/4620$. The He\,{\sc ii} absorption lines are slightly narrower than the models which suggests a $\log g$ value below 4.0. The spectrum shows weak RV variations and LPVs. The star is multiple in the HST observations and the luminosity might be overestimated. N is enriched. He\,{\sc ii}\,$\lambda 4686$ is in emission whilst $\mathrm{H_{\alpha}}$ is in absorption which suggests He enrichment at the surface.}
\end{figure}
\clearpage
\begin{figure}
\begin{center}
\includegraphics[width=17cm]{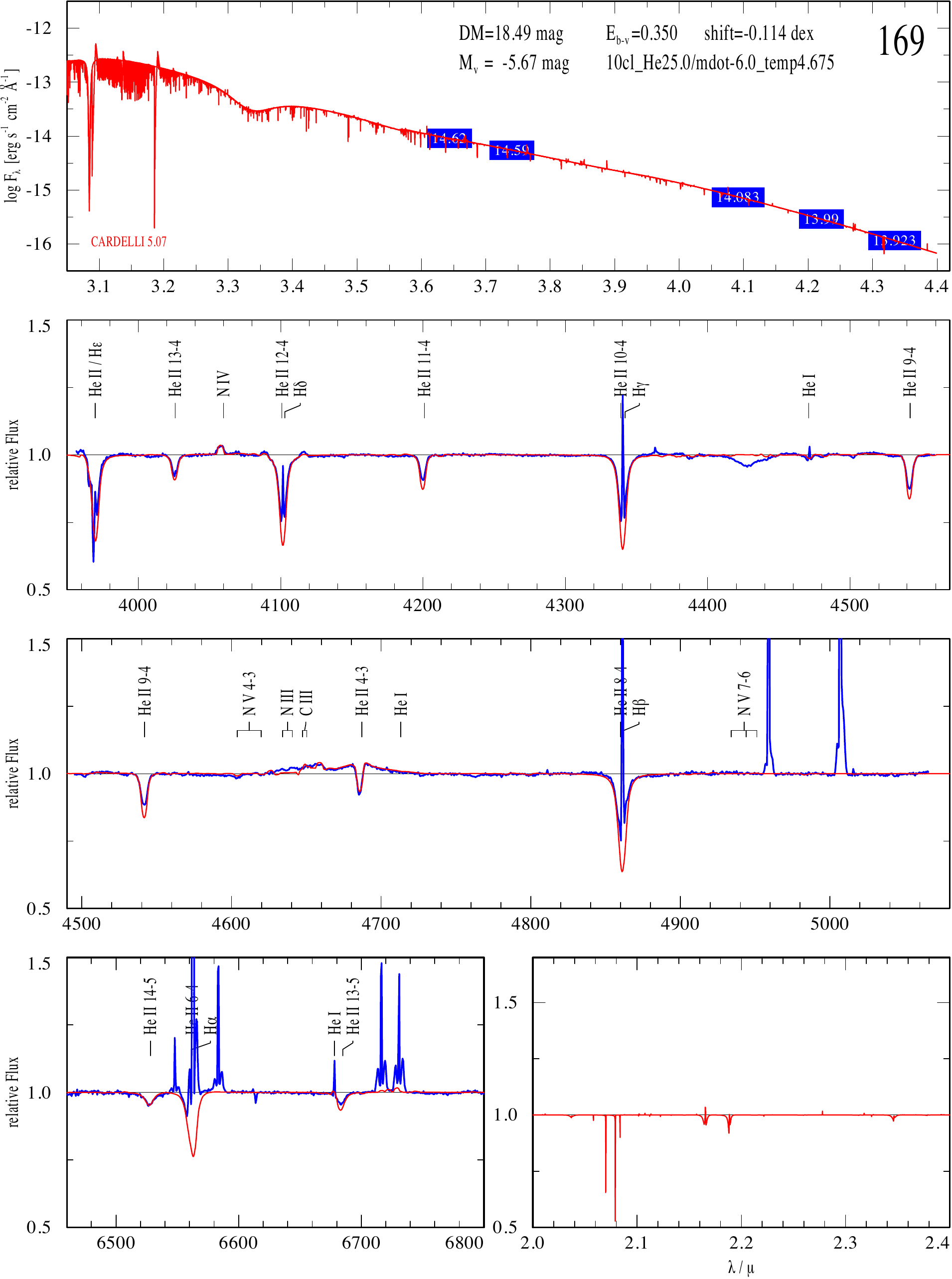}
\end{center}
\caption{The temperature of VFTS\,169 (O2.5 V(n)((f*))) is based on the lines He\,{\sc i}\,$\lambda 4471$, N\,{\sc iv}\,$\lambda 4058$, and N\,{\sc v} $\lambda4604/4620$. $\dot{M}$ is based on He\,{\sc ii}\,$\lambda 4686$. The stellar rotation is high and the model spectrum is convolved with a rotation profile. N-abundance is normal.}
\end{figure}
\clearpage
\begin{figure}
\begin{center}
\includegraphics[width=17cm]{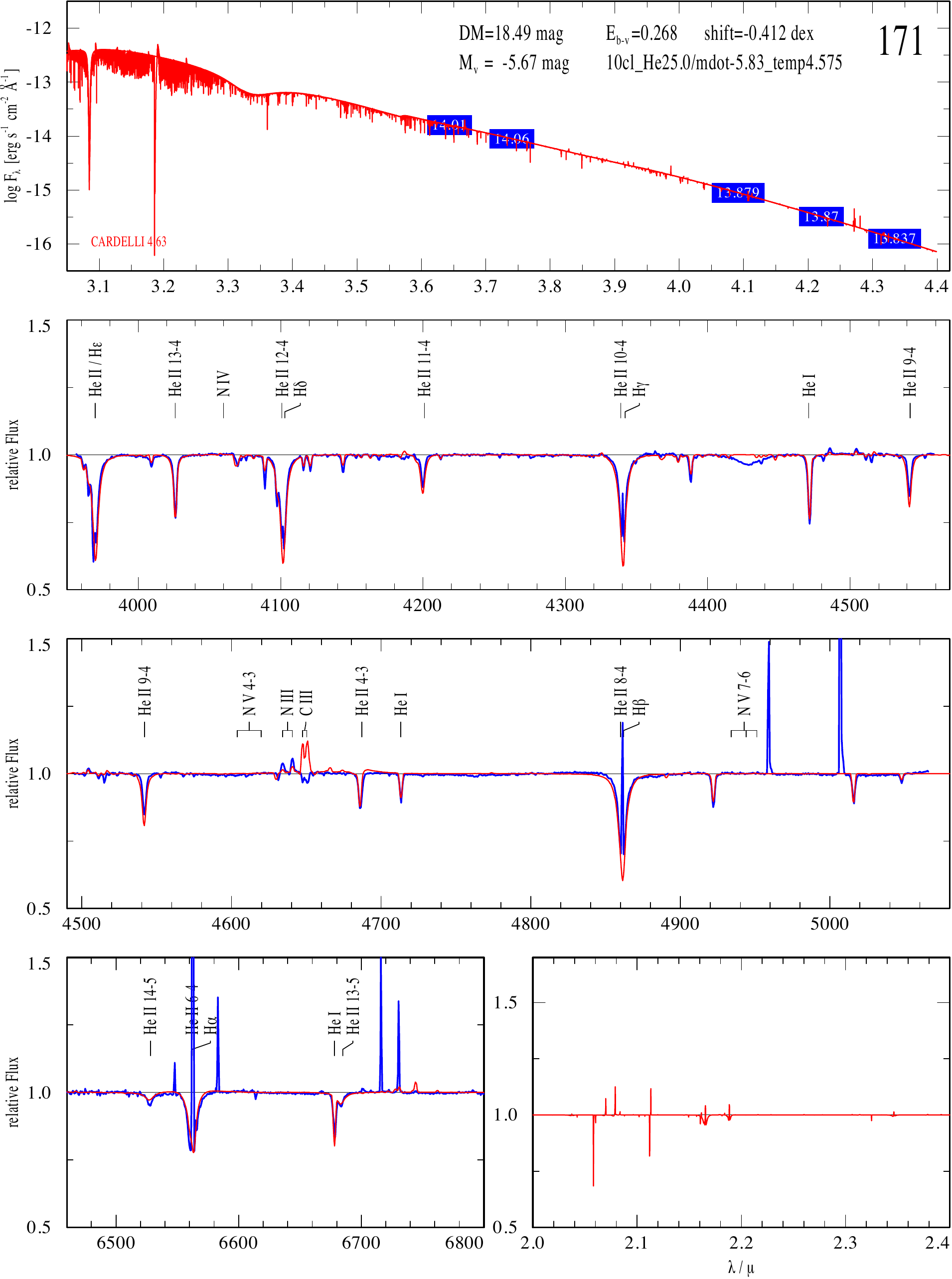}
\end{center}
\caption{The temperature of VFTS\,171 (O8 II-III(f)) is based on the lines He\,{\sc i}\,$\lambda 4471$ and N\,{\sc iii}\,$\lambda 4634/4640$. $\dot{M}$ is based on the line shape of He\,{\sc ii}\,$\lambda 4686$, but $\dot{M}$ of the best fitting model is a bit high. We note that C\,{\sc iii}\,$\lambda 4647/4650$ is in emission in the models, but in absorption in the observations (Appendix\,\ref{s:pot}). N-abundance is between normal and enriched.}
\end{figure}
\clearpage
\begin{figure}
\begin{center}
\includegraphics[width=17cm]{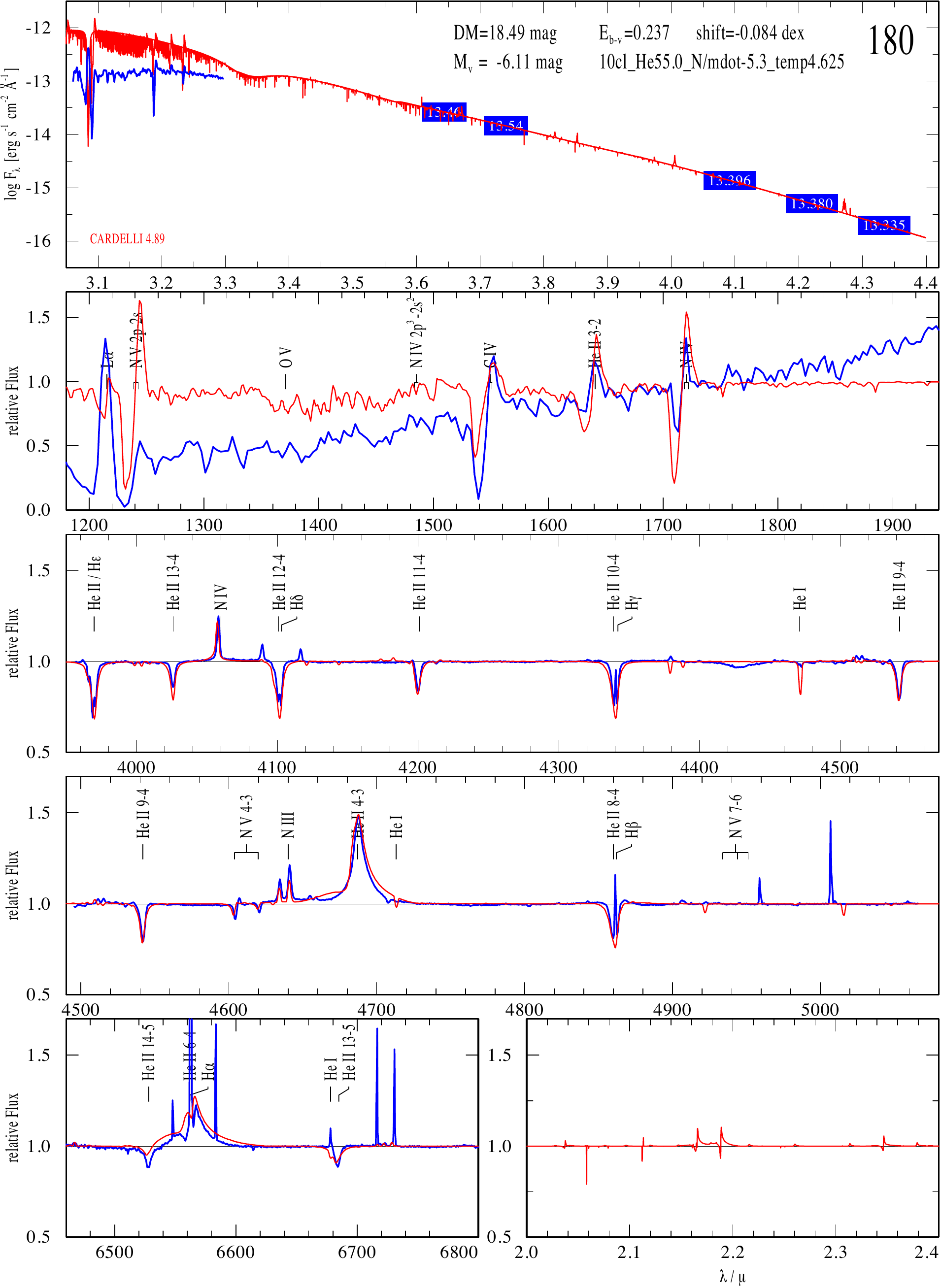}
\end{center}
\caption{The temperature of VFTS\,180 (O3 If*) is based on the lines N\,{\sc iii}\,$\lambda 4634/4640$, N\,{\sc iv}\,$\lambda 4058$, and N\,{\sc v} $\lambda4604/4620$. He\,{\sc i}\,$\lambda 4471$ is too strong in absorption, but it can be resolved by lowering $\log g$ which is suggested by the width of $\mathrm{H_{\gamma}}$ and $\mathrm{H_{\delta}}$. $\dot{M}$ is based on He\,{\sc ii}\,$\lambda 4686$ and $\mathrm{H_{\alpha}}$ and is slightly too high for the best fitting model.}
\end{figure}
\clearpage
\begin{figure}
\begin{center}
\includegraphics[width=17cm]{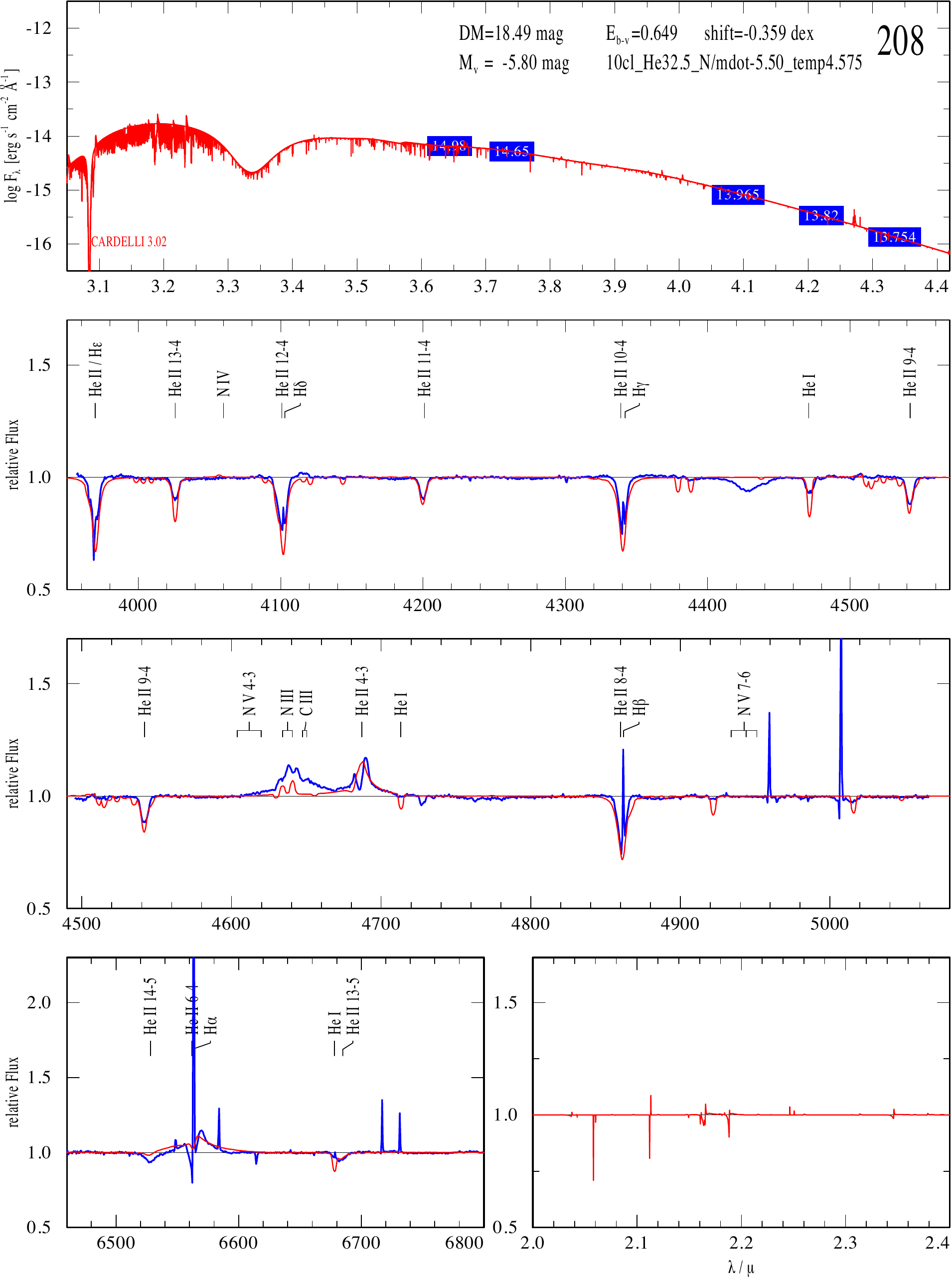}
\end{center}
\vspace{-0.3cm}
\caption{The temperature of VFTS\,208 (O6(n)fp) is based on the absence of the N\,{\sc iv}\,$\lambda 4058$ line. The line wings of the observations are narrower and He\,{\sc i}\,$\lambda 4471$ is too strong, which suggests a significantly lower $\log g$. The star is a fast rotator and the spectrum is peculiar (Appendix\,\ref{s:pot}).}
\end{figure}
\clearpage
\begin{figure}
\begin{center}
\includegraphics[width=17cm]{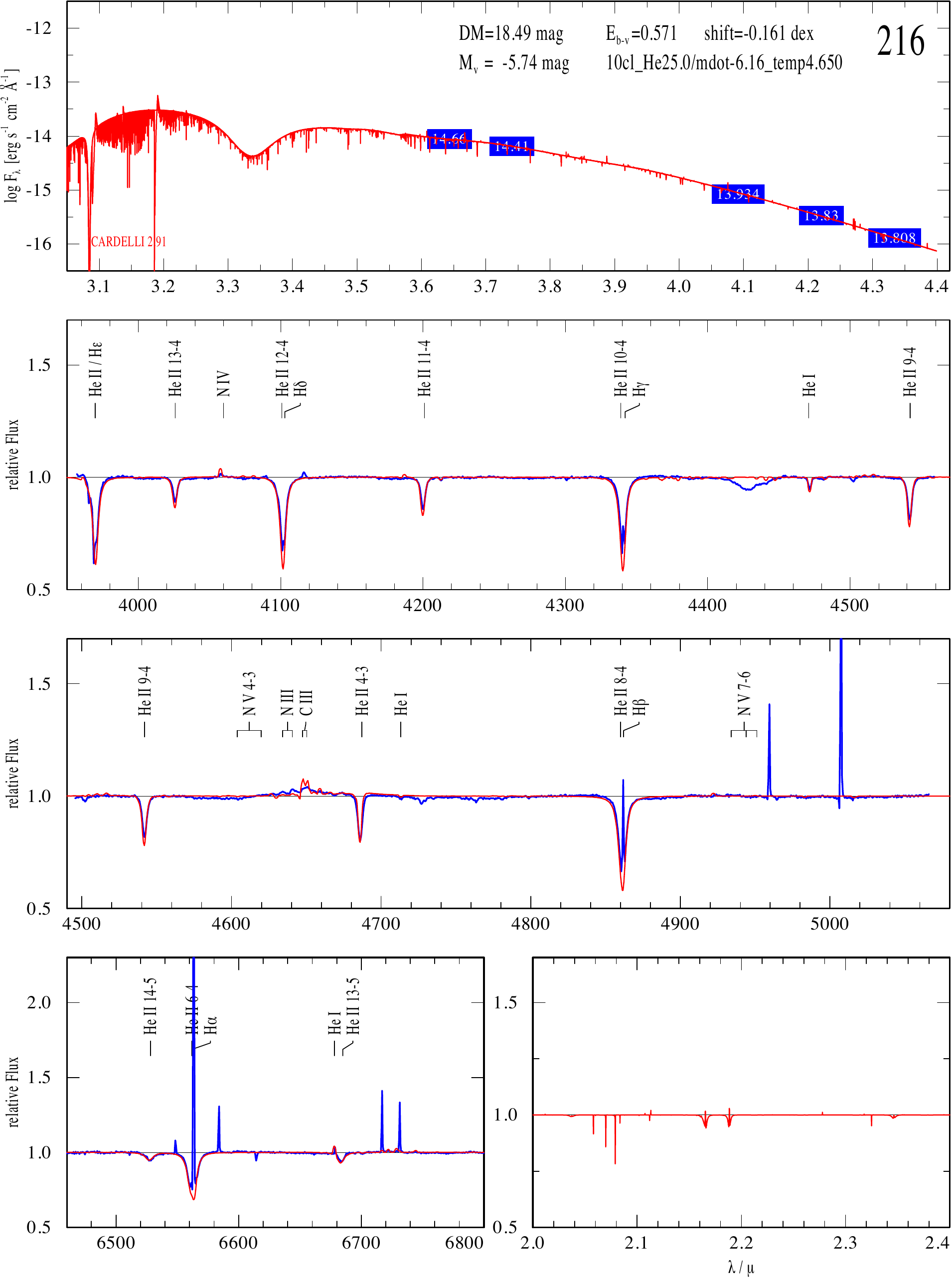}
\end{center}
\caption{The temperature of VFTS\,216 (O4 V((fc))) is based on the lines He\,{\sc i}\,$\lambda 4471$ and N\,{\sc iv}\,$\lambda 4058$. $\dot{M}$ is based on the line shape of He\,{\sc ii}\,$\lambda 4686$. N is not enriched at the surface.}
\end{figure}
\clearpage
\begin{figure}
\begin{center}
\vspace{-0.2cm}
\includegraphics[width=17cm]{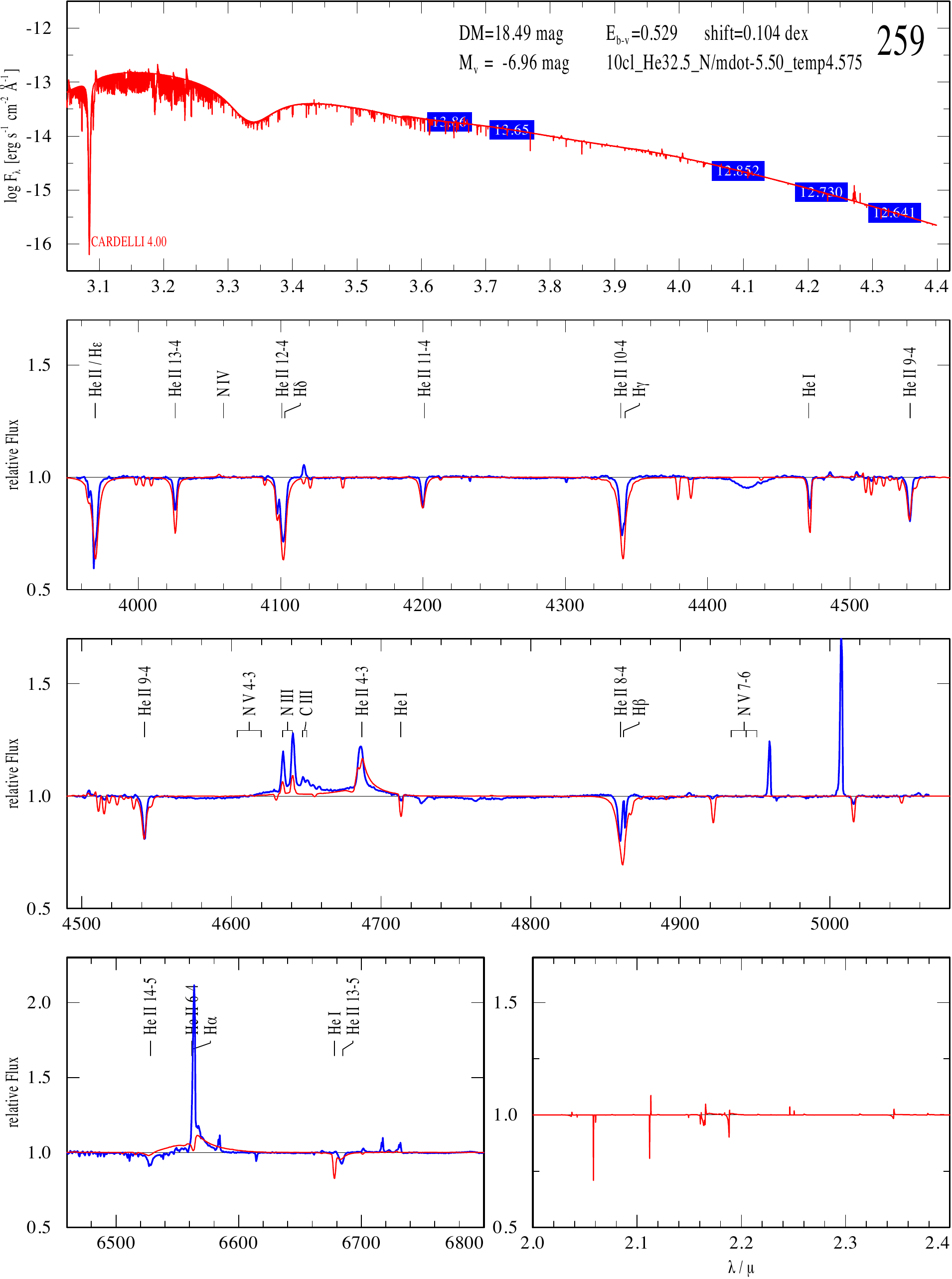}
\end{center}
\vspace{-0.6cm}
\caption{The temperature of VFTS\,259 (O6 Iaf) is based on the absence of the N\,{\sc iv}\,$\lambda 4058$ line. The line wings of the observations are narrower and He\,{\sc i}\,$\lambda 4471$ is too strong, which suggests a lower $\log g \approx 3.5$. Nitrogen is enriched, but N\,{\sc iii}\,$\lambda 4634/4640$ can only be reproduced by increasing the abundance additionally by a factor of two. C\,{\sc iii}\,$\lambda 4647/4650$ are present in the spectrum. Test models show that a reduction of $\log g$, enhancing N, and moving the starting point of clumping closer to the surface, improves the fit quality, but only slightly change the derived parameters.}
\end{figure}
\clearpage
\begin{figure}
\begin{center}
\includegraphics[width=17cm]{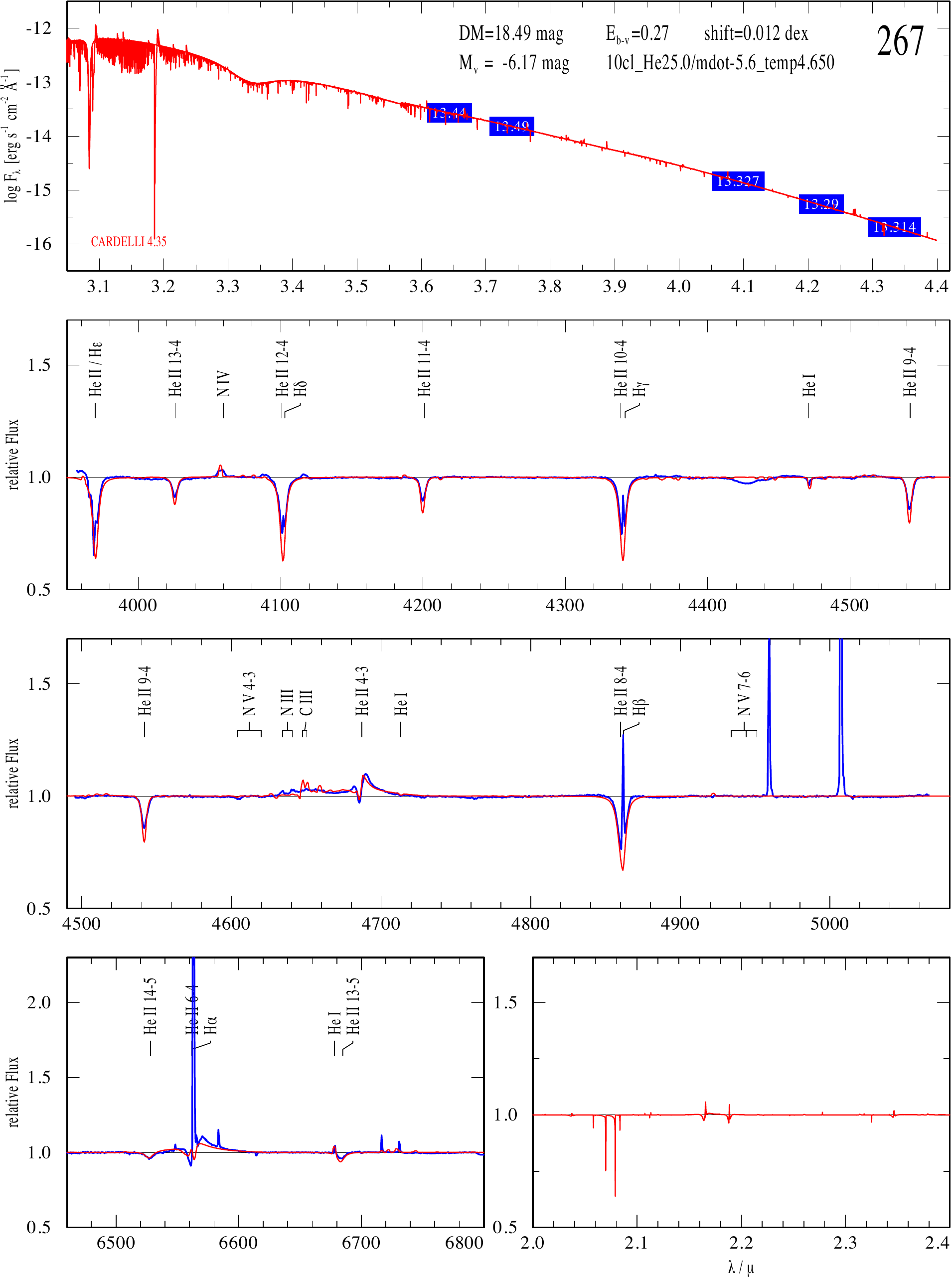}
\end{center}
\caption{The temperature of VFTS\,267 (O3 III-I(n)f*) is based on the lines He\,{\sc i}\,$\lambda 4471$ and N\,{\sc iv}\,$\lambda 4058$. $\dot{M}$ is based on the line shape of He\,{\sc ii}\,$\lambda 4686$. N-abundance is between normal and enriched.}
\end{figure}
\clearpage
\begin{figure}
\begin{center}
\includegraphics[width=17cm]{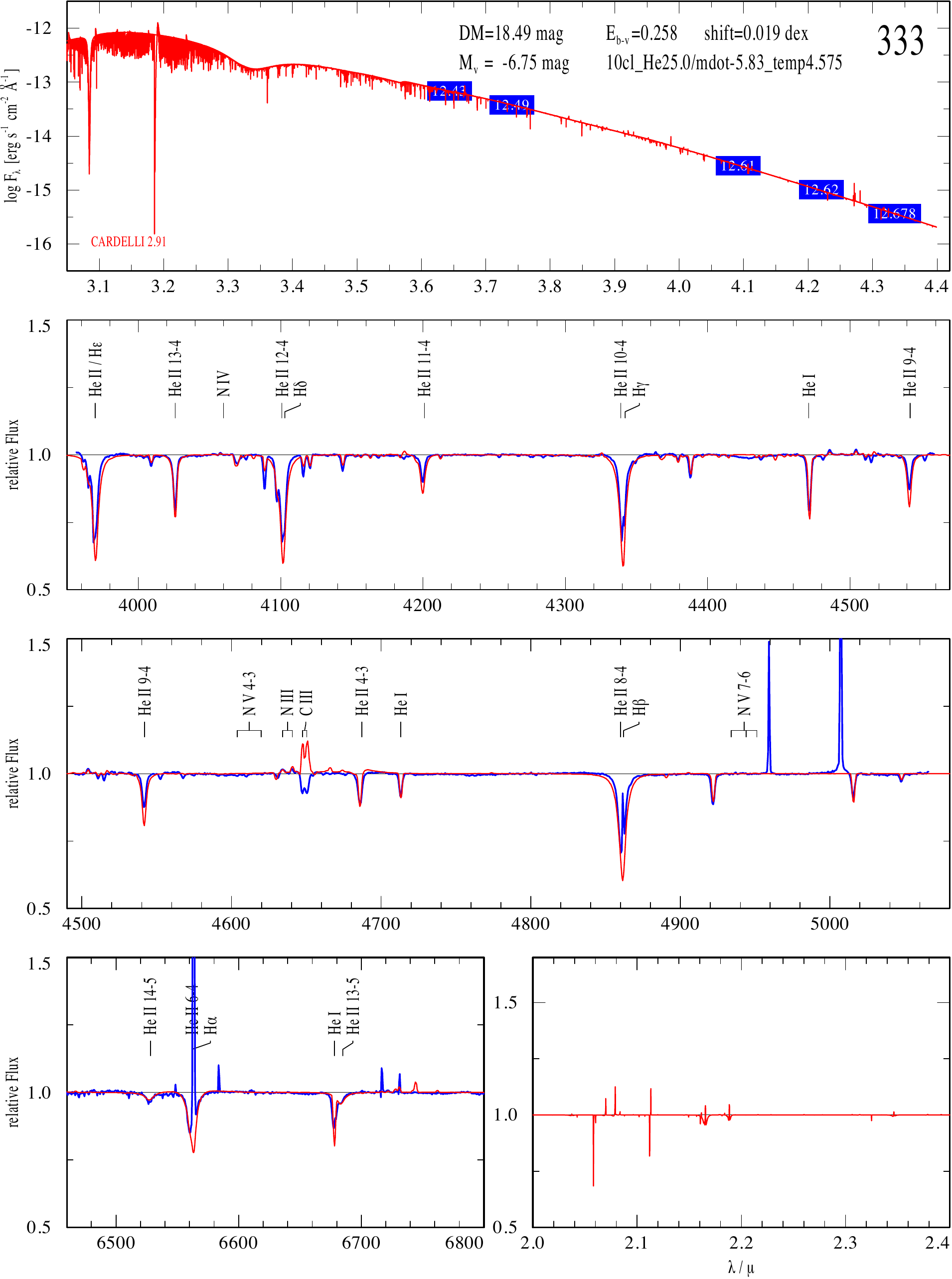}
\end{center}
\caption{The temperature of VFTS\,333 (O8 II-III((f))) is based on the lines He\,{\sc i}\,$\lambda 4471$ and N\,{\sc iii}\,$\lambda 4634/4640$. $\dot{M}$ is based on the line shape of He\,{\sc ii}\,$\lambda 4686$. We note that C\,{\sc iii}\,$\lambda 4647/4650$ is in emission in the models, but in absorption in the observations (Appendix\,\ref{s:pot}). N is not enriched at the surface.}
\end{figure}
\clearpage
\clearpage
\begin{figure}
\begin{center}
\includegraphics[width=17cm]{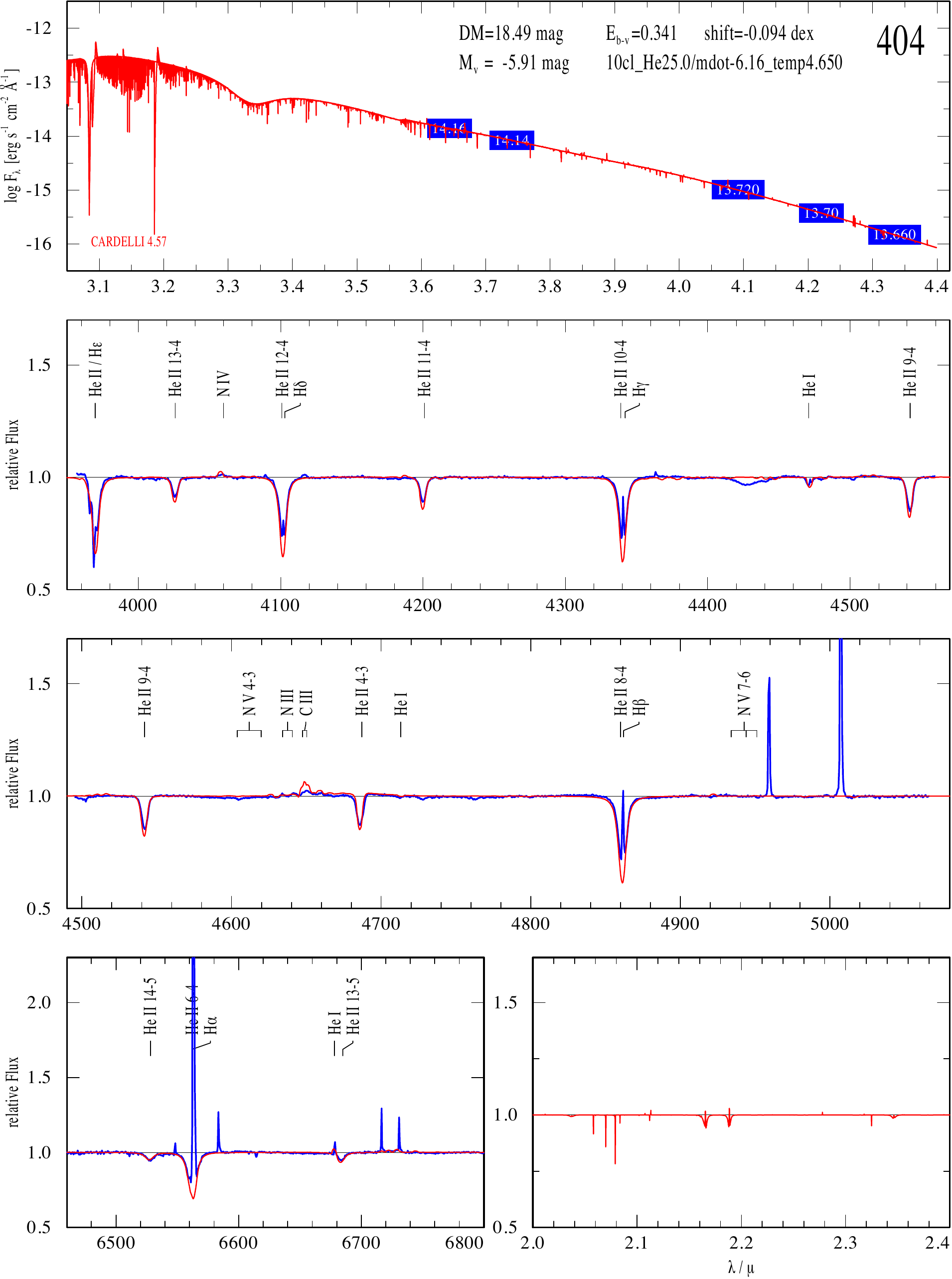}
\end{center}
\caption{The spectrum of VFTS\,404 (O3.5 V(n)((fc))) shows the characteristic of a SB1. The temperature is based on the lines He\,{\sc i}\,$\lambda 4471$, N\,{\sc iii}\,$\lambda 4634/4640$, and N\,{\sc iv}\,$\lambda 4058$. $\dot{M}$ is based on the line shape of He\,{\sc ii}\,$\lambda 4686$. N-abundance is normal.}
\end{figure}
\clearpage
\begin{figure}
\begin{center}
\vspace{-0.2cm}
\includegraphics[width=17cm]{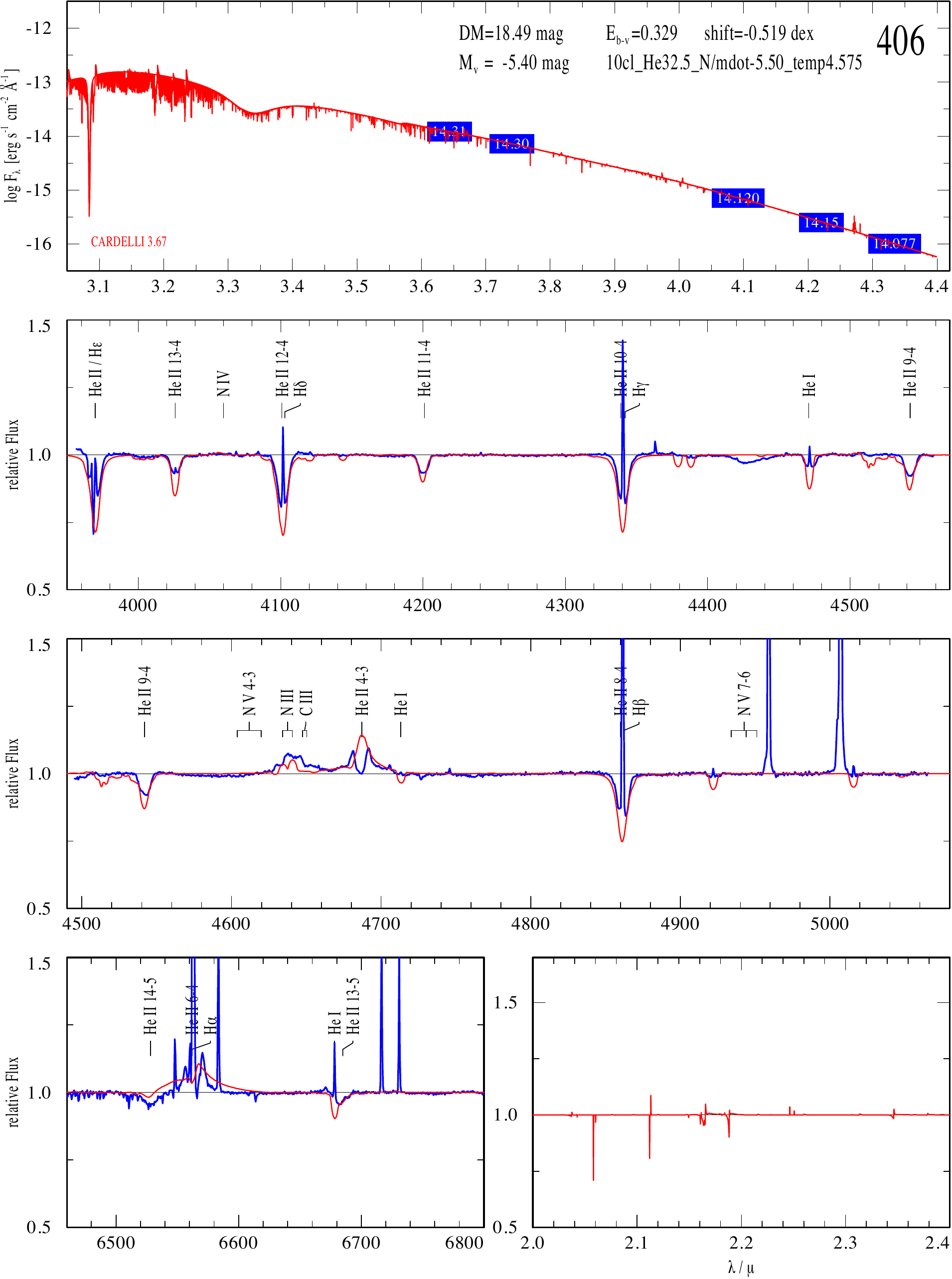}
\end{center}
\vspace{-0.6cm}
\caption{The temperature of VFTS\,406 (O6 nn(f)p) is based on the absence of the N\,{\sc iv}\,$\lambda 4058$ line. The line wings of the observations are narrower, which suggests a lower $\log g$. The star is a fast rotator and the spectrum is peculiar (Appendix\,\ref{s:pot}). Nitrogen is enriched.}
\end{figure}
\clearpage
\begin{figure}
\begin{center}
\includegraphics[width=17cm]{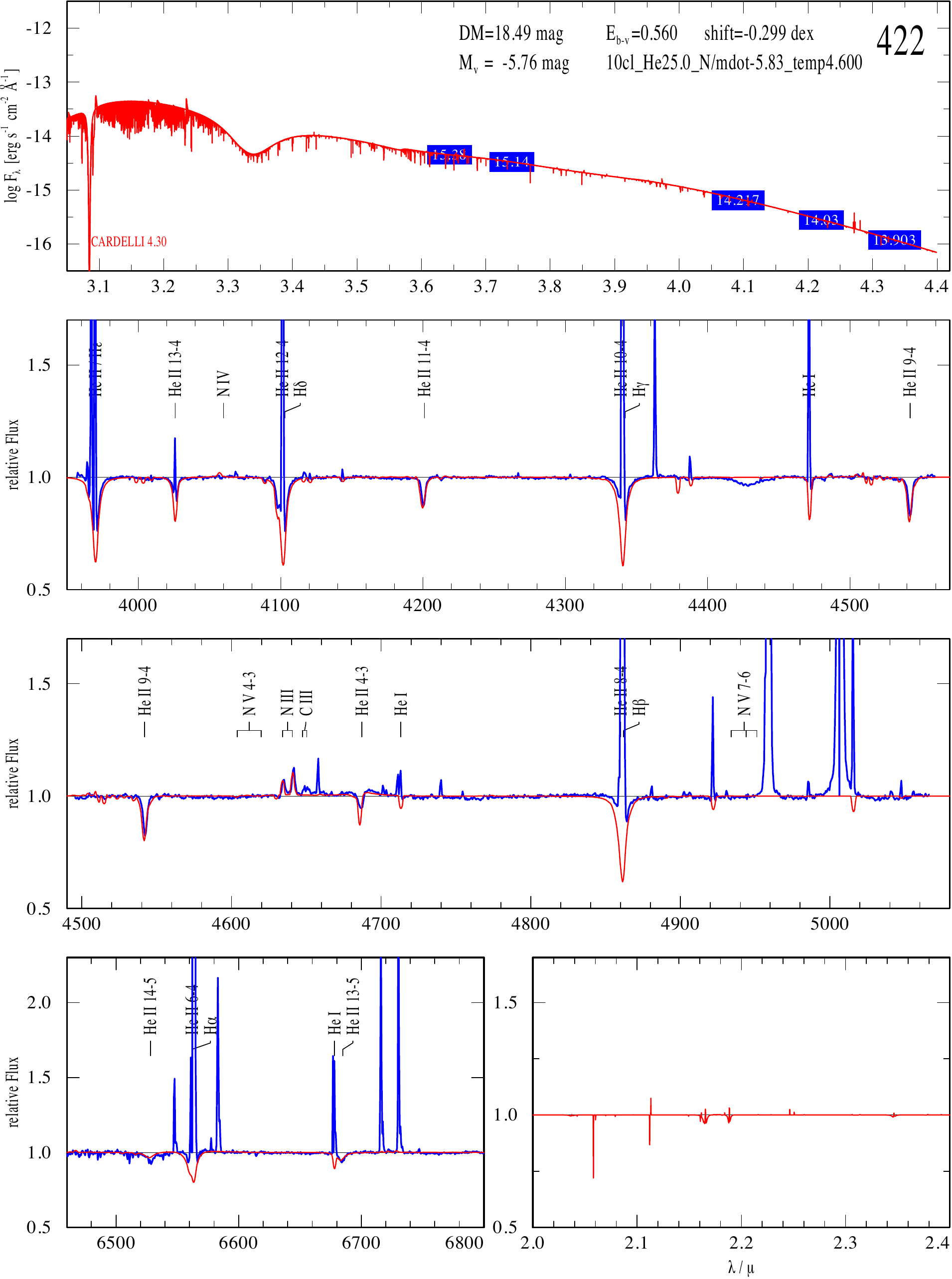}
\end{center}
\caption{The spectrum of VFTS\,422 (O4 III(f)) shows the characteristic of a SB1. The temperature is based on the lines N\,{\sc iii}\,$\lambda 4634/4640$ and N\,{\sc iv}\,$\lambda 4058$. The He\,{\sc i}\,$\lambda 4471$ line could not be used due to the strong nebular contamination. $\dot{M}$ is based on the line shape of He\,{\sc ii}\,$\lambda 4686$ and is a bit too high in the model. N-abundance is between normal and enriched.}
\end{figure}
\clearpage
\clearpage
\begin{figure}
\begin{center}
\includegraphics[width=17cm]{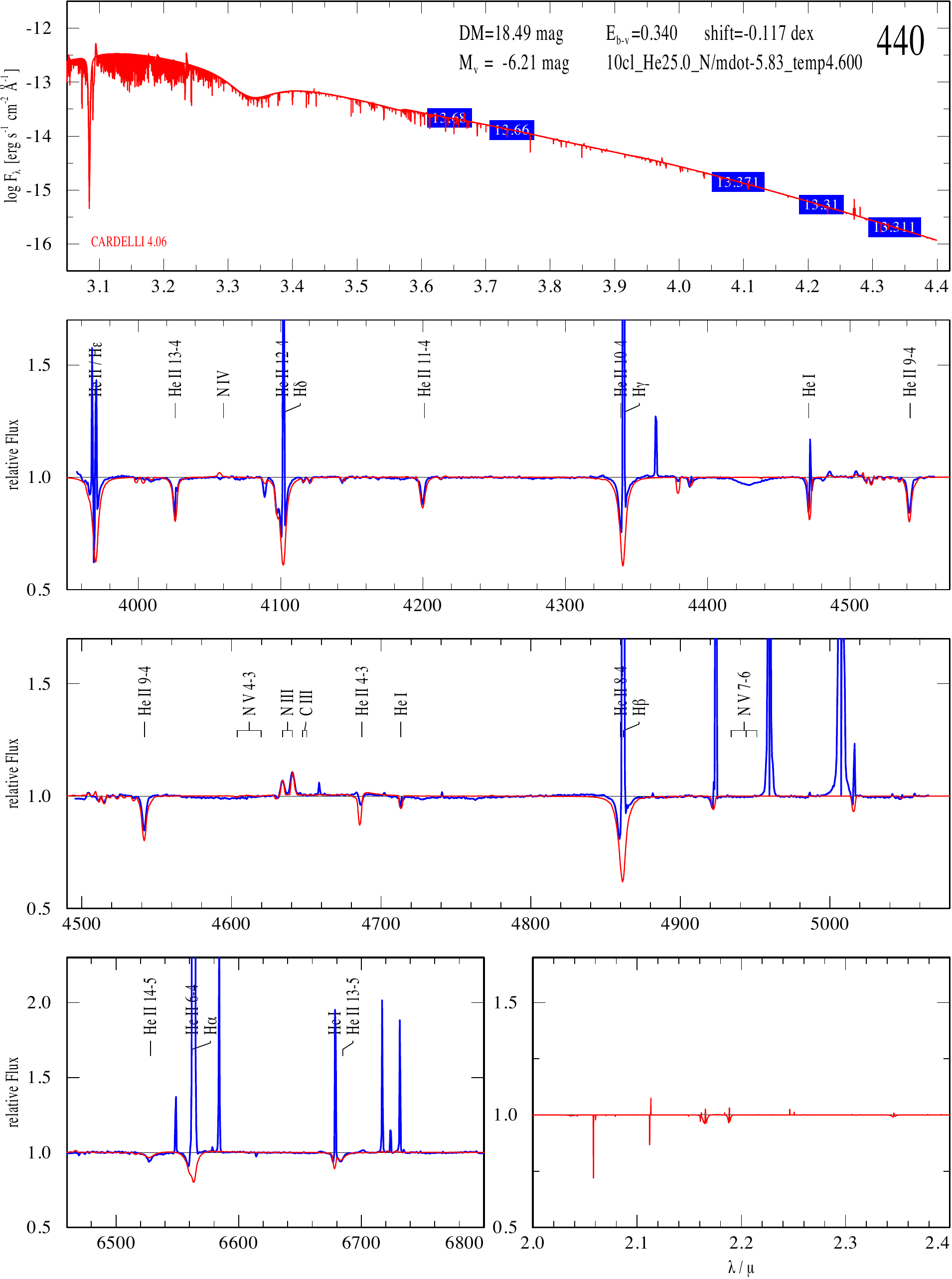}
\end{center}
\caption{The spectrum of VFTS\,440 (O6-6.5 II(f)) might be a SB1 or SB2. The temperature is based on the lines He\,{\sc i}\,$\lambda 4471$ and N\,{\sc iii}\,$\lambda 4634/4640$. $\dot{M}$ is based on the line shape of He\,{\sc ii}\,$\lambda 4686$ and is slightly too low in the model. Nitrogen is enriched at the stellar surface.}
\end{figure}
\clearpage
\begin{figure}
\begin{center}
\includegraphics[width=17cm]{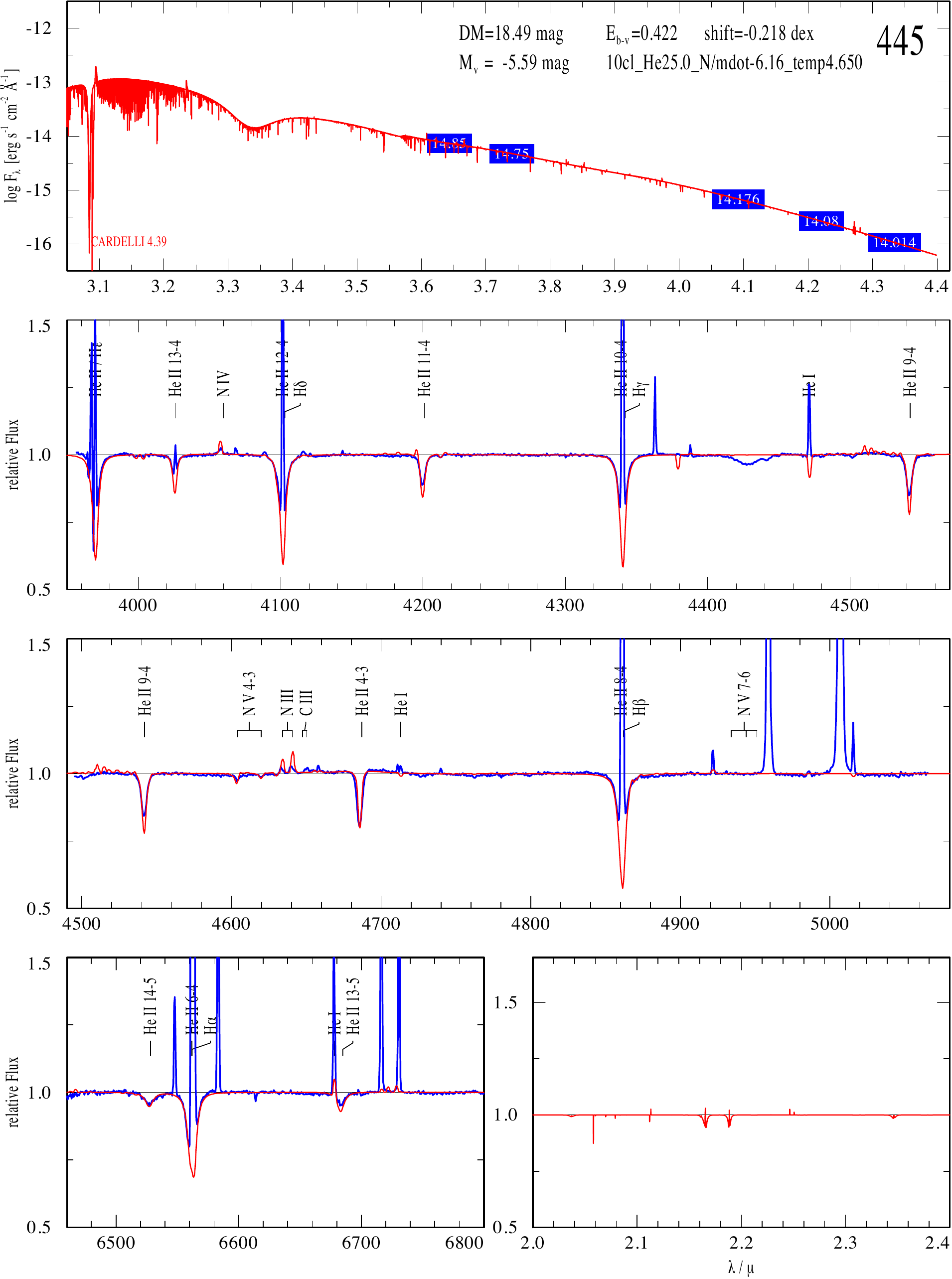}
\end{center}
\caption{The spectrum of VFTS\,445 (O3-4 V:((fc)): + O4-7 V:((fc)):) shows the characteristic of a SB2. The temperature is based on the lines N\,{\sc iii}\,$\lambda 4634/4640$, N\,{\sc iv}\,$\lambda 4058$, and N\,{\sc v} $\lambda4604/4620$. He\,{\sc i}\,$\lambda 4471$ could not be used due to the strong nebular contamination. $\dot{M}$ is based on the line shape of He\,{\sc ii}\,$\lambda 4686$. N-abundance is between normal and enriched.}
\end{figure}
\clearpage
\begin{figure}
\begin{center}
\includegraphics[width=17cm]{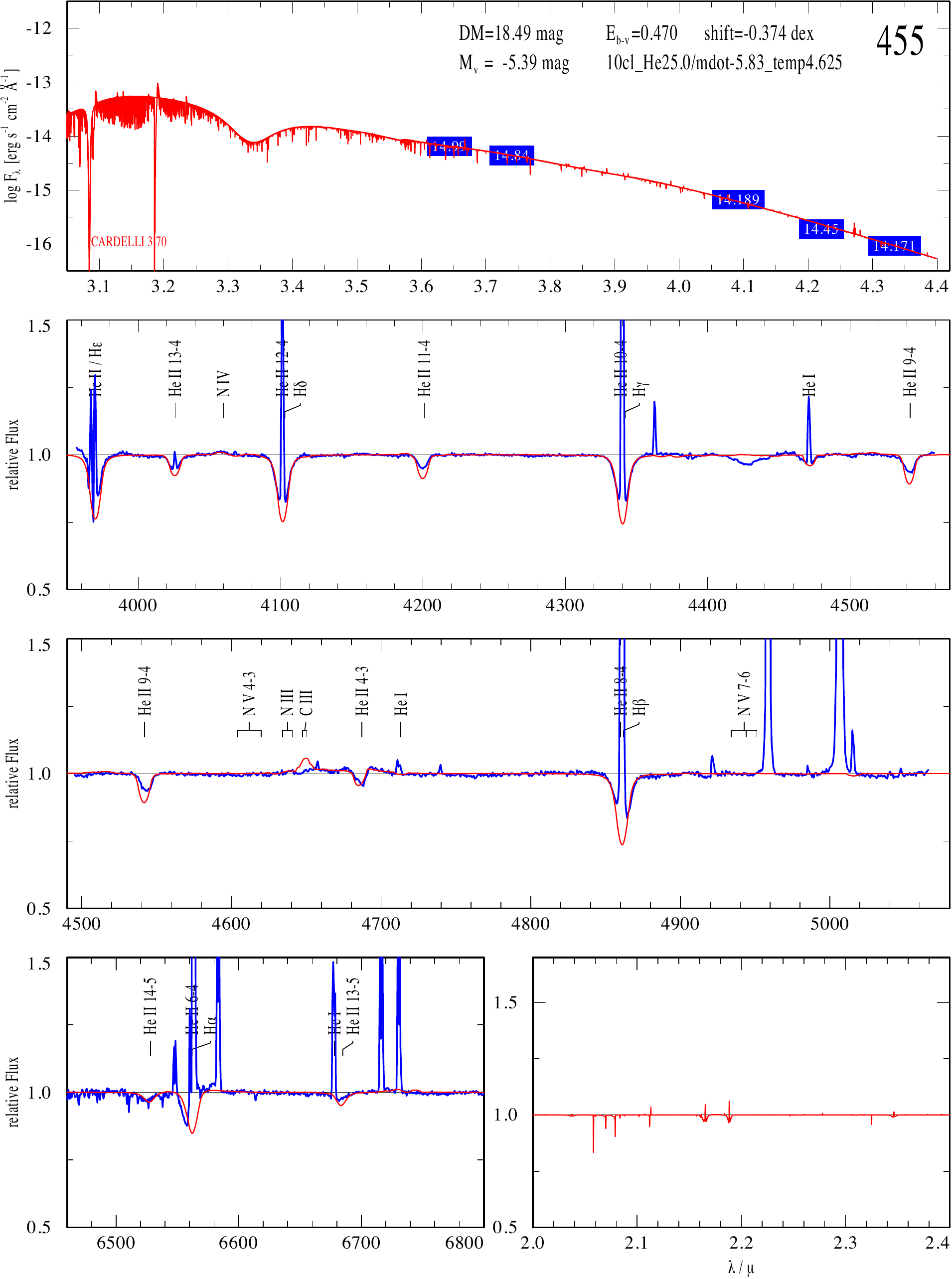}
\end{center}
\caption{The spectrum of VFTS\,455 (O5: V:n) shows the characteristic of a SB1. The temperature is based on the lines He\,{\sc i}\,$\lambda 4471$ and N\,{\sc iv}\,$\lambda 4058$. $\dot{M}$ is based on the line shape of He\,{\sc ii}\,$\lambda 4686$. The star is a fast rotator. N-abundance is between normal and enriched.}
\end{figure}
\clearpage
\clearpage
\clearpage
\begin{figure}
\begin{center}
\includegraphics[width=17cm]{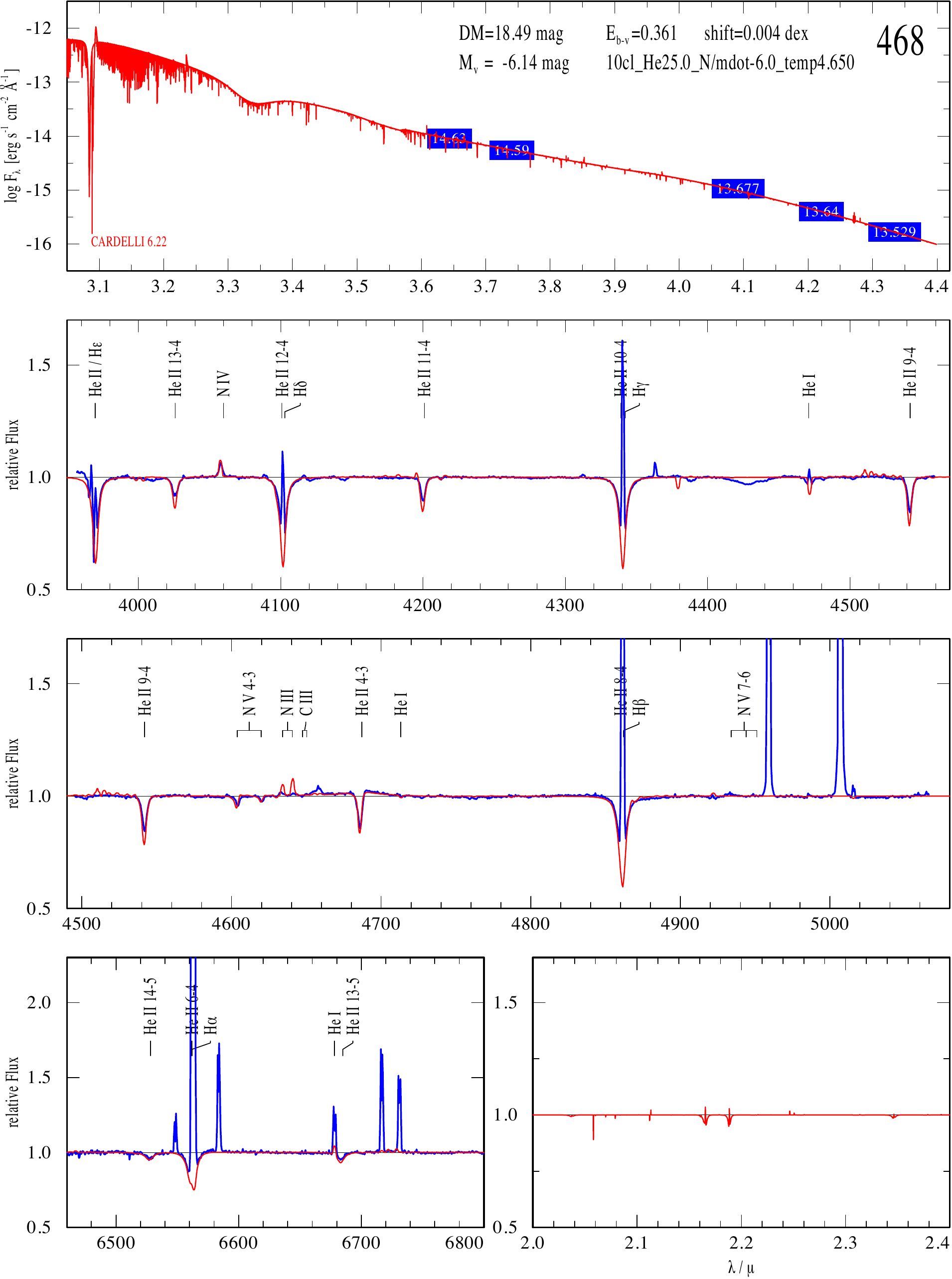}
\end{center}
\caption{VFTS\,468 (O2 V((f*)) + OB) has stars nearby, is multiple in HST images, and has a composite spectrum. The temperature is based on the lines He\,{\sc i}\,$\lambda 4471$, N\,{\sc iii}\,$\lambda 4634/4640$, N\,{\sc iv}\,$\lambda 4058$, and N\,{\sc v} $\lambda4604/4620$. $\dot{M}$ is based on the line shape of He\,{\sc ii}\,$\lambda 4686$. N is enriched.}
\end{figure}
\clearpage
\clearpage
\begin{figure}
\begin{center}
\includegraphics[width=17cm]{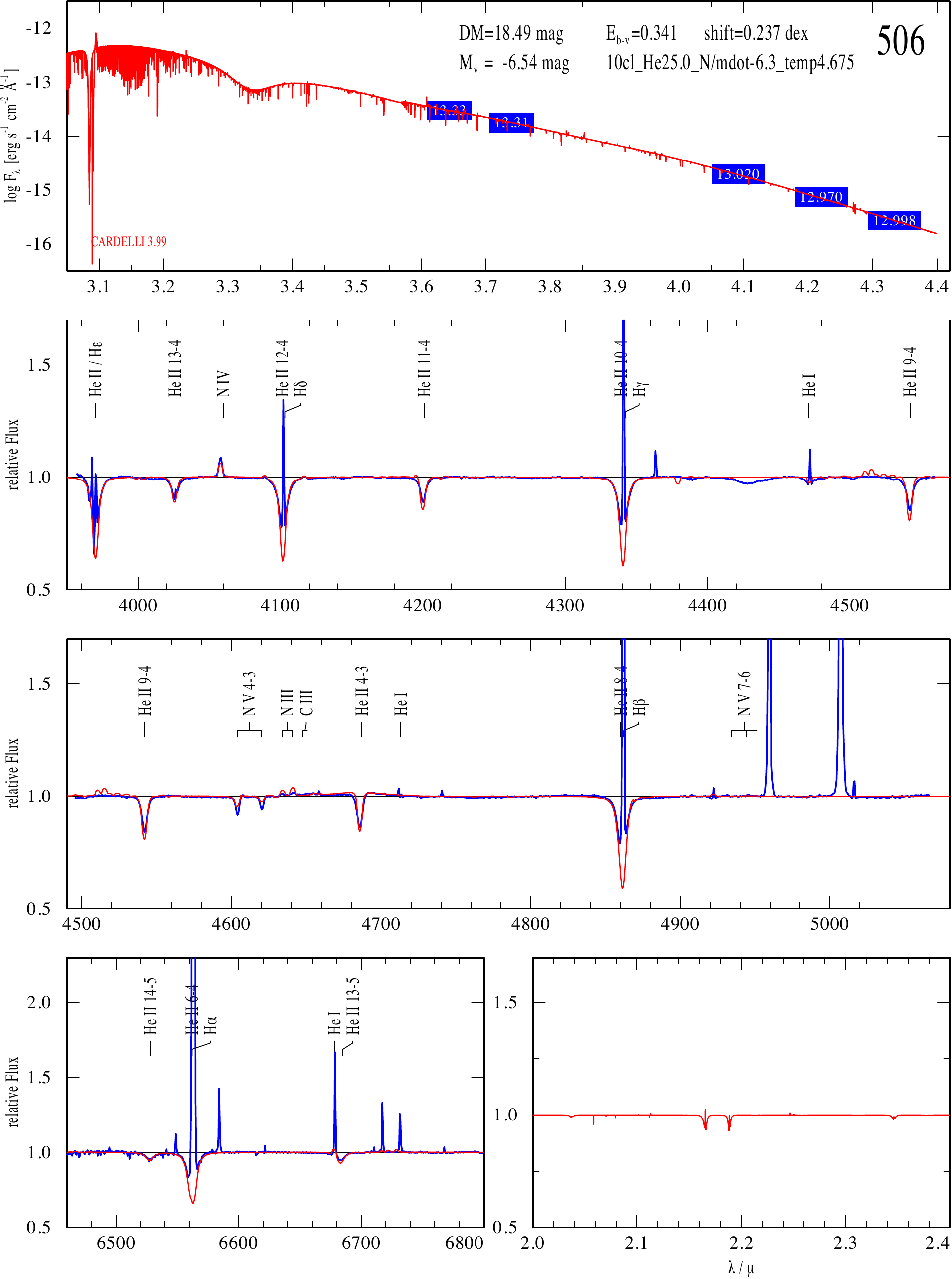}
\end{center}
\caption{The temperature of VFTS\,506 (ON2 V((n))((f*))) is based on the lines N\,{\sc iii}\,$\lambda 4634/4640$, N\,{\sc iv}\,$\lambda 4058$, and N\,{\sc v} $\lambda4604/4620$. The model temperature is slightly too low. $\dot{M}$ is based on the line shape of He\,{\sc ii}\,$\lambda 4686$. $\log g$ might be a bit higher, because the observations are a bit broader than the model. The star is fast rotating. N-abundance is between normal and enriched.}
\end{figure}
\clearpage
\clearpage
\begin{figure}
\begin{center}
\includegraphics[width=17cm]{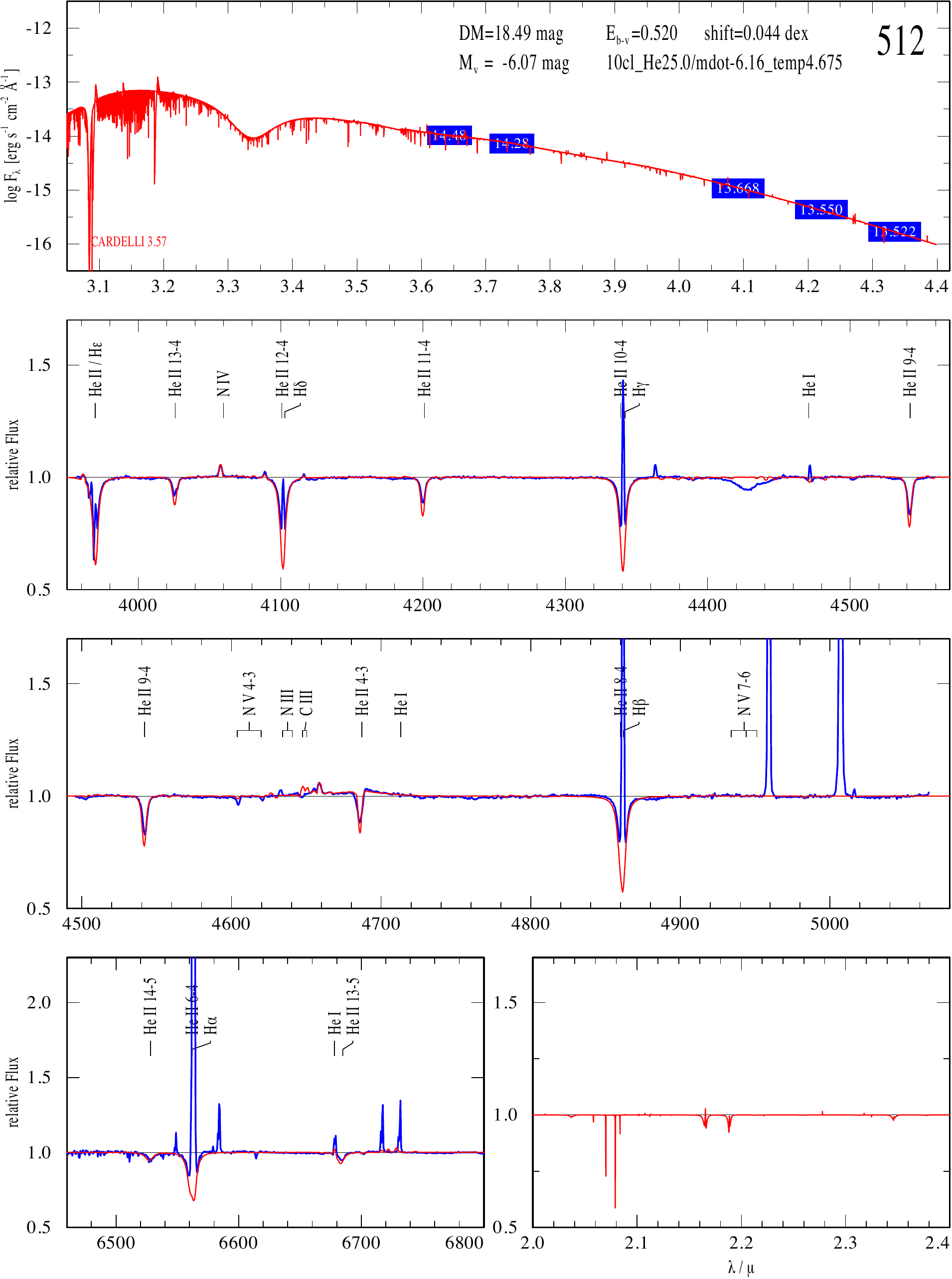}
\end{center}
\caption{The spectrum of VFTS\,512 (O2 V-III((f*))) shows the characteristic of a SB1. The temperature is based on the lines He\,{\sc i}\,$\lambda 4471$, N\,{\sc iv}\,$\lambda 4058$, and N\,{\sc v} $\lambda4604/4620$. $\dot{M}$ is based on the line shape of He\,{\sc ii}\,$\lambda 4686$. N-abundance is between normal and enriched.}
\end{figure}
\clearpage
\begin{figure}
\begin{center}
\includegraphics[width=17cm]{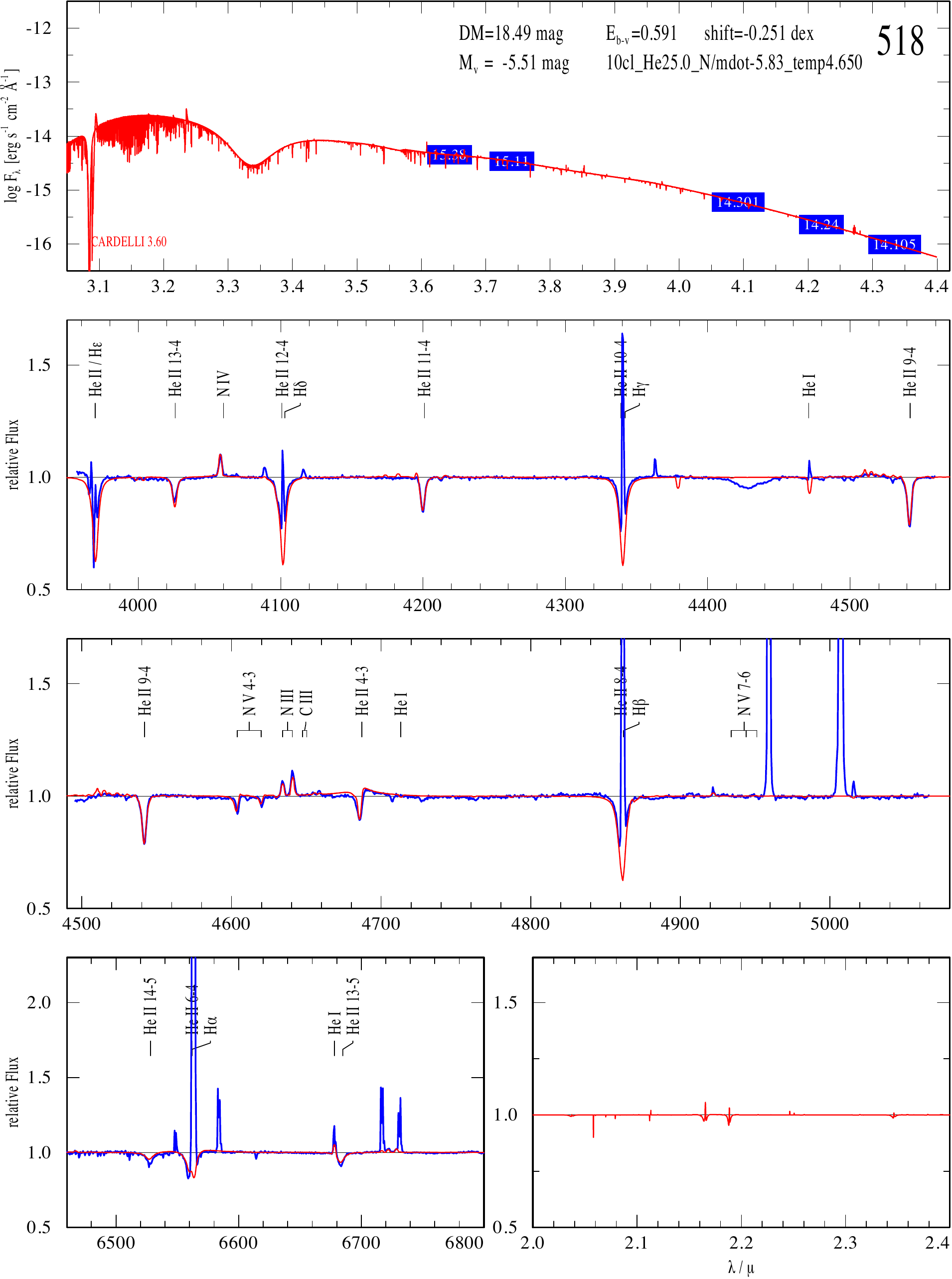}
\end{center}
\caption{The temperature of VFTS\,518 (O3.5 III(f*)) is based on the lines N\,{\sc iii}\,$\lambda 4634/4640$, N\,{\sc iv}\,$\lambda 4058$, and N\,{\sc v} $\lambda4604/4620$. $\dot{M}$ is based on the line shape of He\,{\sc ii}\,$\lambda 4686$. C is reduced and N is enriched.}
\end{figure}
\clearpage
\begin{figure}
\begin{center}
\includegraphics[width=17cm]{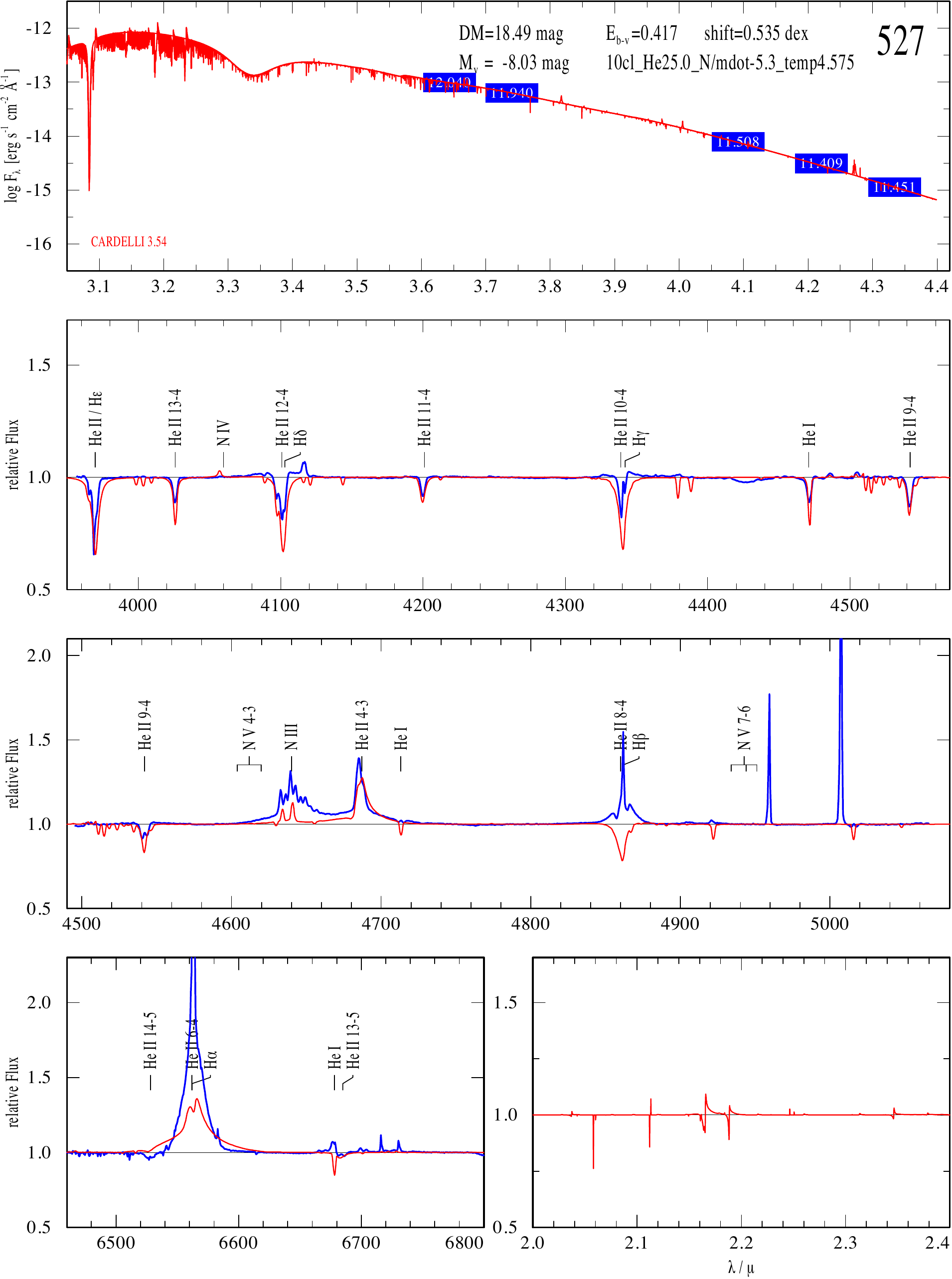}
\end{center}
\caption{The spectrum of VFTS\,527 (O6.5 Iafc + O6 Iaf) shows the characteristic of a SB2. The temperature and mass-loss rate are difficult to determine. We tried to fit the spectrum with a single model from our grid. The line width of the absorption lines suggests a lower $\log g$. The fit quality is poor as result of the SB2 characteristic and the derived $T_{\mathrm{eff}}$, $\dot{M}$ and the luminosity are quite uncertain. No He enrichment at the stellar surface. Combining two single star models would improve the fit quality.}
\end{figure}
\clearpage
\begin{figure}
\begin{center}
\includegraphics[width=17cm]{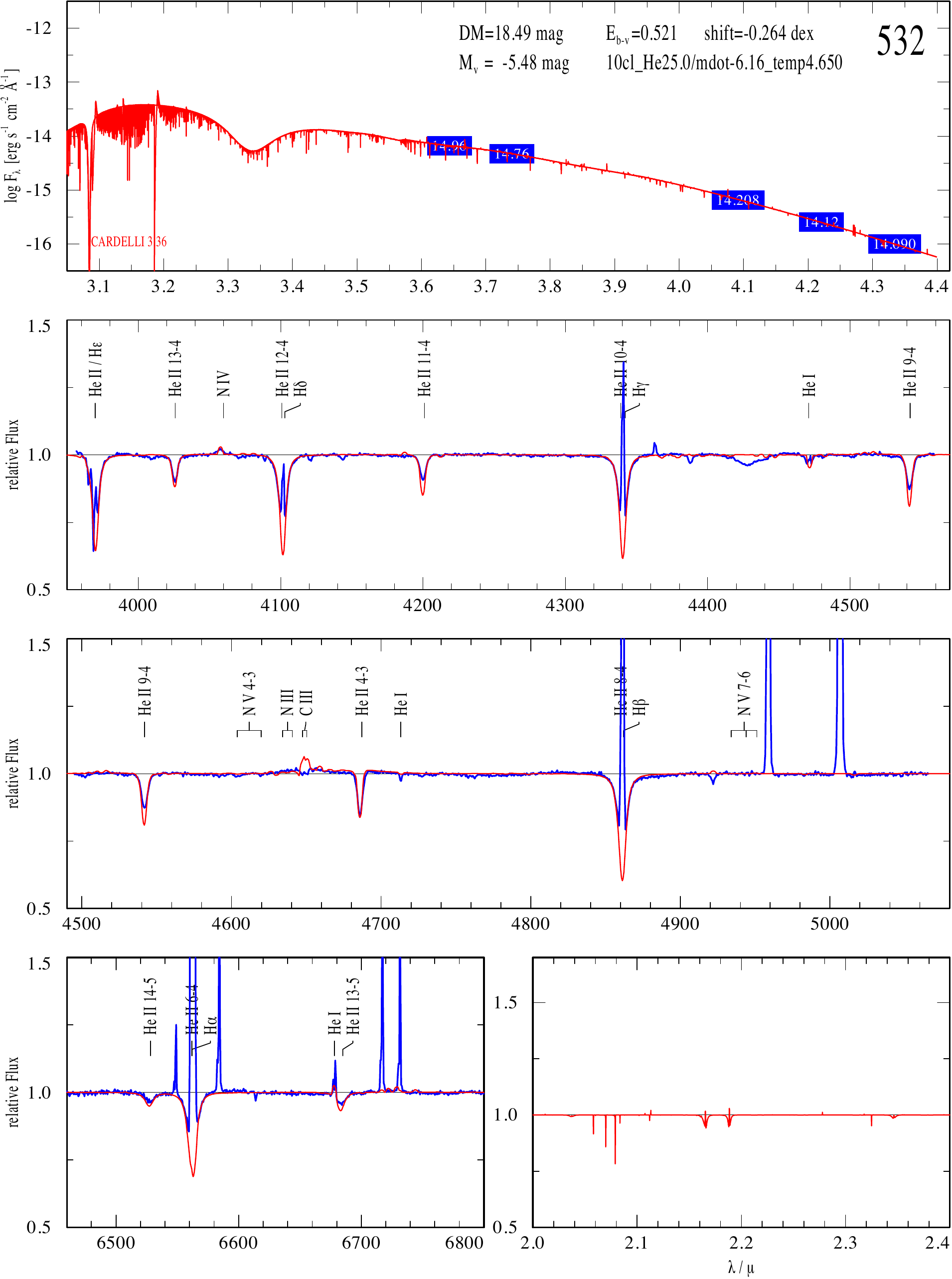}
\end{center}
\caption{The temperature of VFTS\,532 (O3 V(n)((f*))z + OB) is based on the lines He\,{\sc i}\,$\lambda 4471$ and N\,{\sc iv}\,$\lambda 4058$. $\dot{M}$ is based on the line shape of He\,{\sc ii}\,$\lambda 4686$. We note that C\,{\sc iii}\,$\lambda 4647/4650$ is in emission in the models, but in absorption in the observations (Appendix\,\ref{s:pot}). N is not enriched at the surface. The star is fast rotating.}
\end{figure}
\clearpage
\begin{figure}
\begin{center}
\includegraphics[width=17cm]{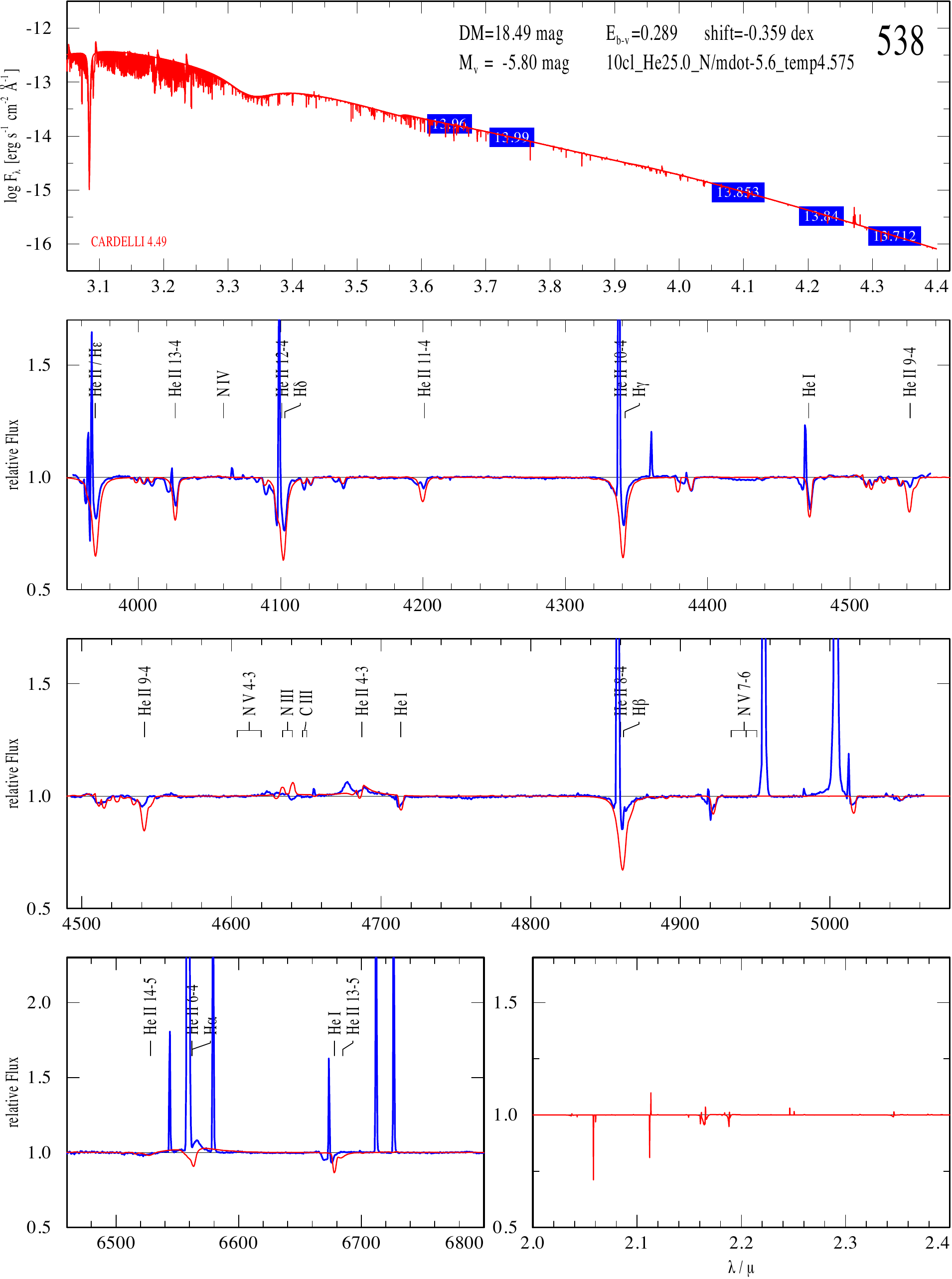}
\end{center}
\caption{The spectrum of VFTS\,538 (ON9 Ia + O7.5 Ia(f)) shows the characteristic of a SB2. The temperature is based on the He\,{\sc i}\,$\lambda 4471$ line. $\dot{M}$ is roughly based on the line shape of He\,{\sc ii}\,$\lambda 4686$.  The line width of the absorption lines suggests a lower $\log g$. The fit quality is rather poor and $T_{\mathrm{eff}}$ and $\dot{M}$ are quite uncertain.}
\end{figure}
\clearpage
\clearpage
\clearpage
\begin{figure}
\begin{center}
\includegraphics[width=17cm]{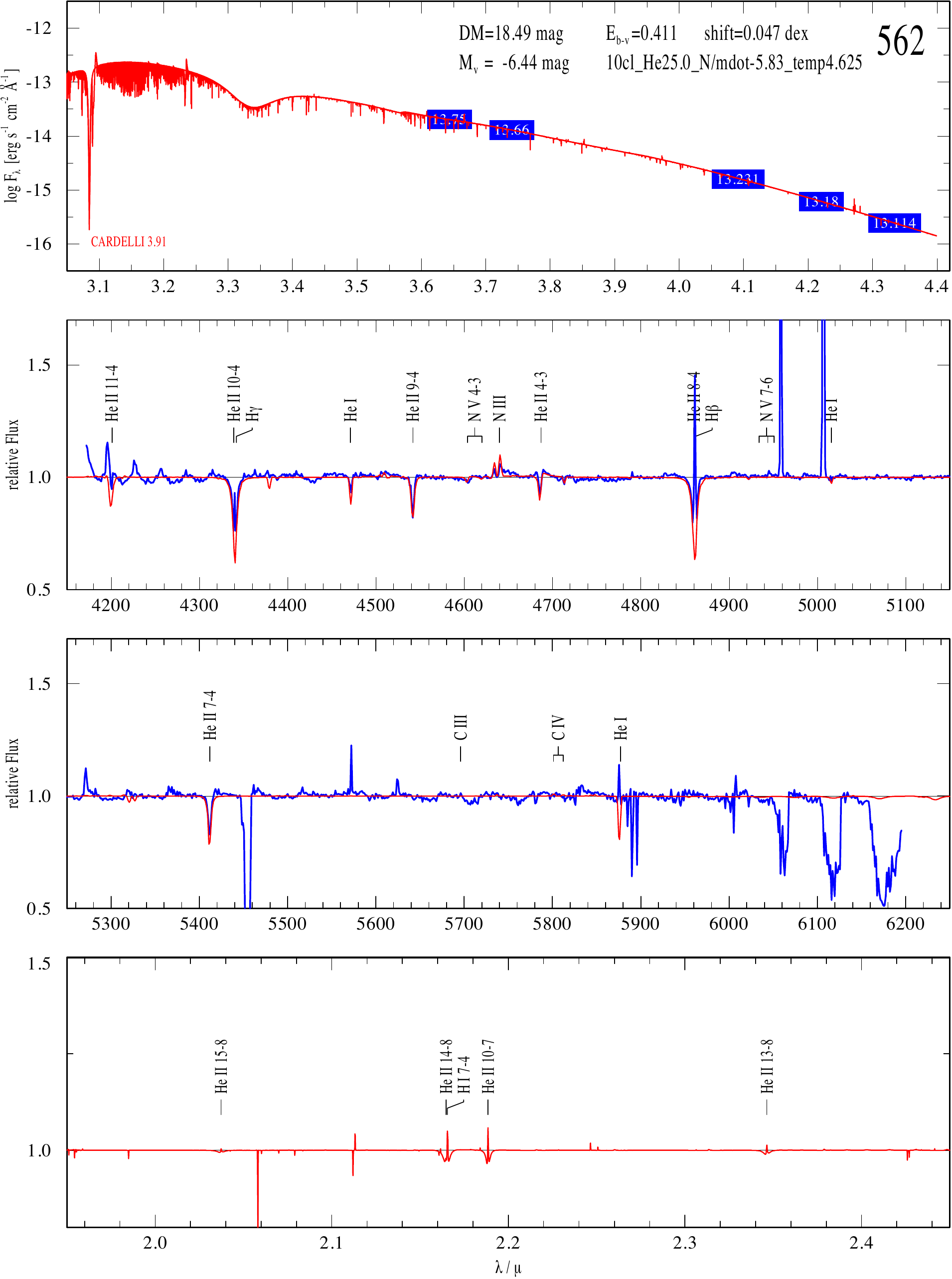}
\end{center}
\caption{VFTS\,562 (O4V): $\mathrm{H}_{\alpha}$ is not observed. Hence, the He abundance is uncertain. The temperature is based on the lines He\,{\sc i}\,$\lambda 4471$, N\,{\sc iii}\,$\lambda 4634/4640$, and N\,{\sc v} $\lambda4604/4620$. $\dot{M}$ is based on the line shape of He\,{\sc ii}\,$\lambda 4686$. The line width of the absorption lines suggests a lower $\log g$. The presence of C suggests a He-abundance of 25\% at the surface.}
\end{figure}
\clearpage
\begin{figure}
\begin{center}
\includegraphics[width=17cm]{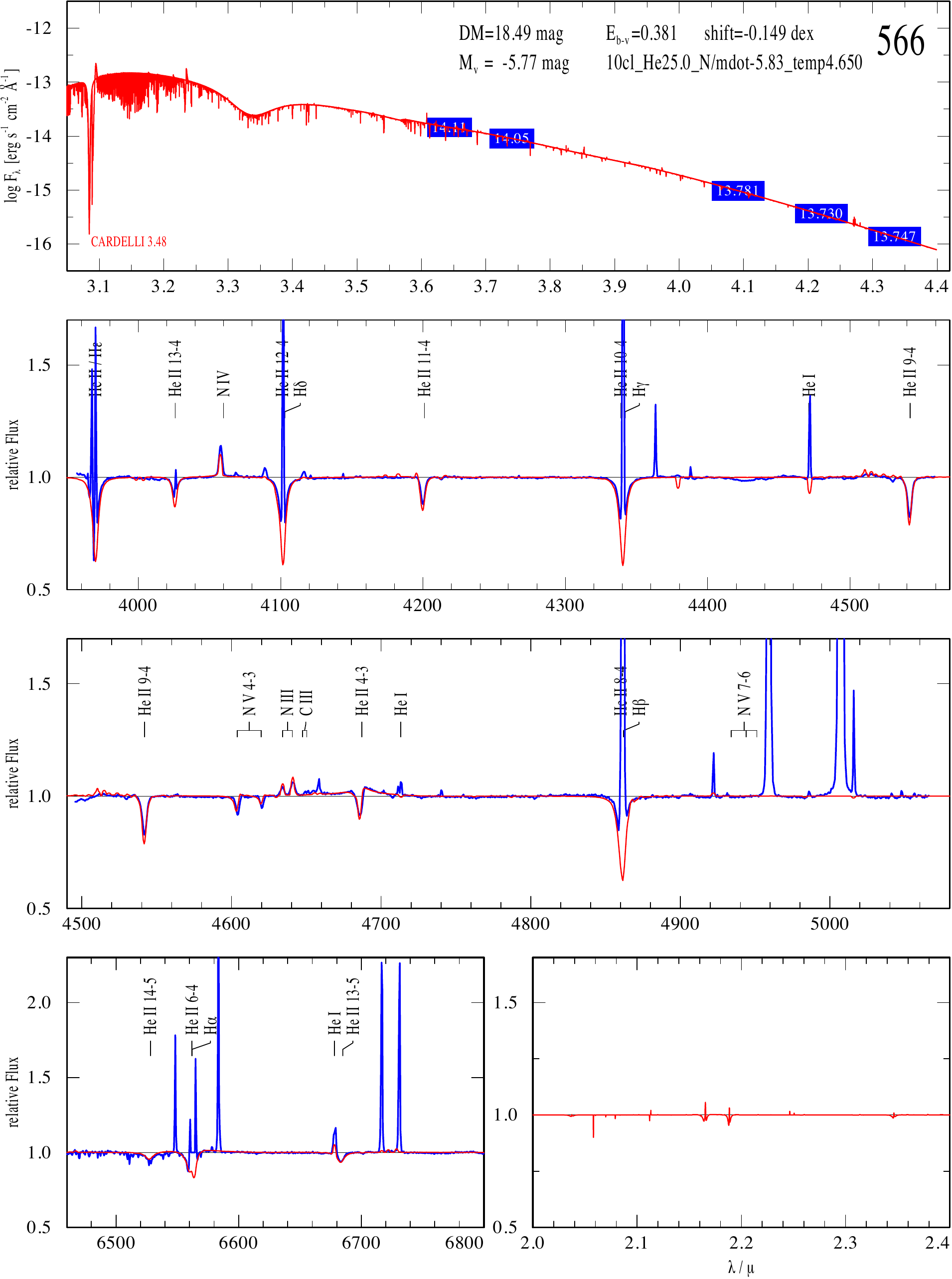}
\end{center}
\caption{The temperature of VFTS\,566 (O3 III(f*)) is based on the lines N\,{\sc iii}\,$\lambda 4634/4640$, N\,{\sc iv}\,$\lambda 4058$, and N\,{\sc v} $\lambda4604/4620$. $\dot{M}$ is based on the line shape of He\,{\sc ii}\,$\lambda 4686$. N is enriched.}
\end{figure}
\clearpage
\begin{figure}
\begin{center}
\includegraphics[width=17cm]{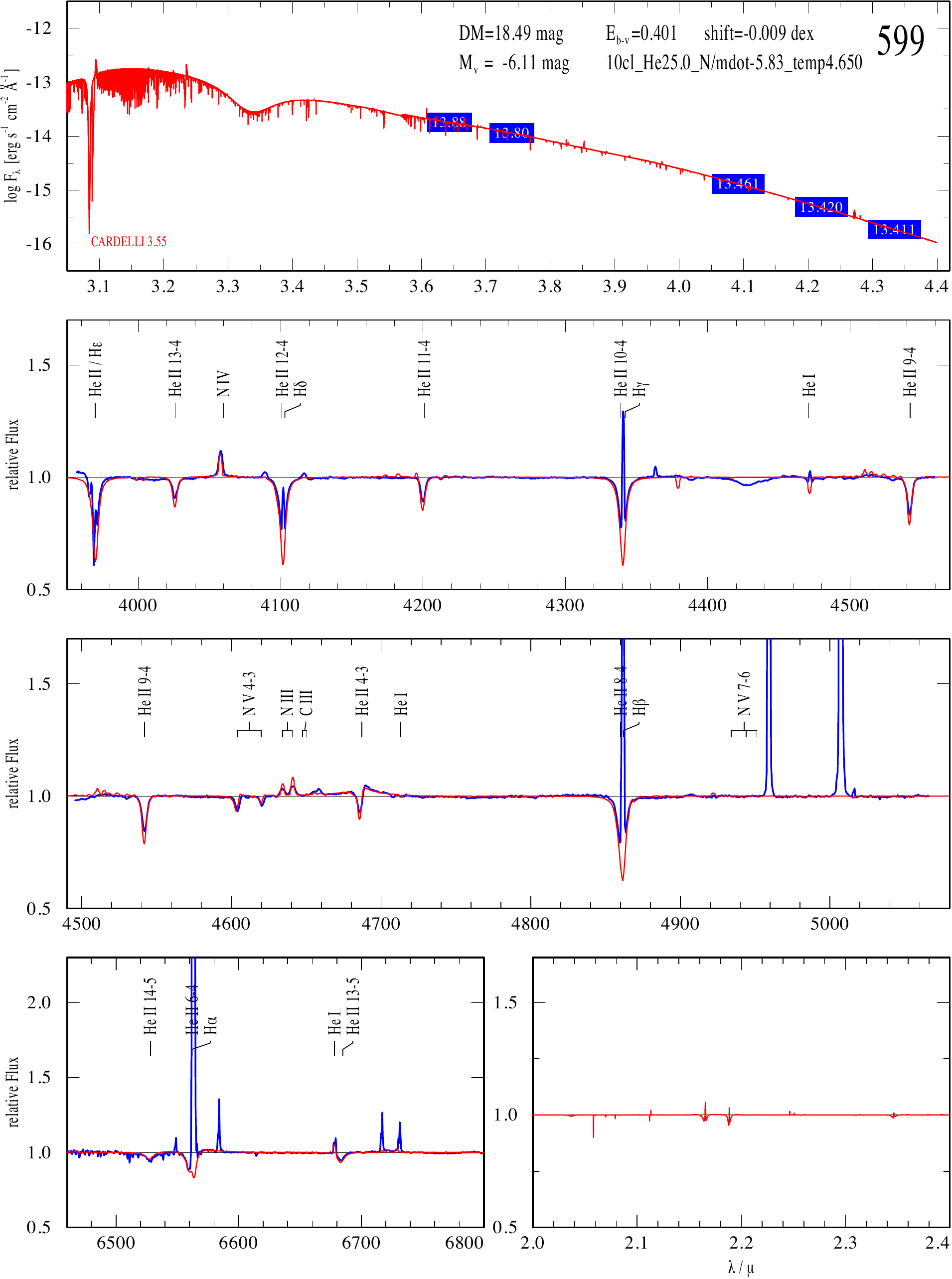}
\end{center}
\caption{The temperature of VFTS\,599 (O3 III(f*)) is based on the lines N\,{\sc iii}\,$\lambda 4634/4640$, N\,{\sc iv}\,$\lambda 4058$, and N\,{\sc v} $\lambda4604/4620$. $\dot{M}$ is based on the line shape of He\,{\sc ii}\,$\lambda 4686$. N is enriched.}
\end{figure}
\clearpage
\begin{figure}
\begin{center}
\includegraphics[width=17cm]{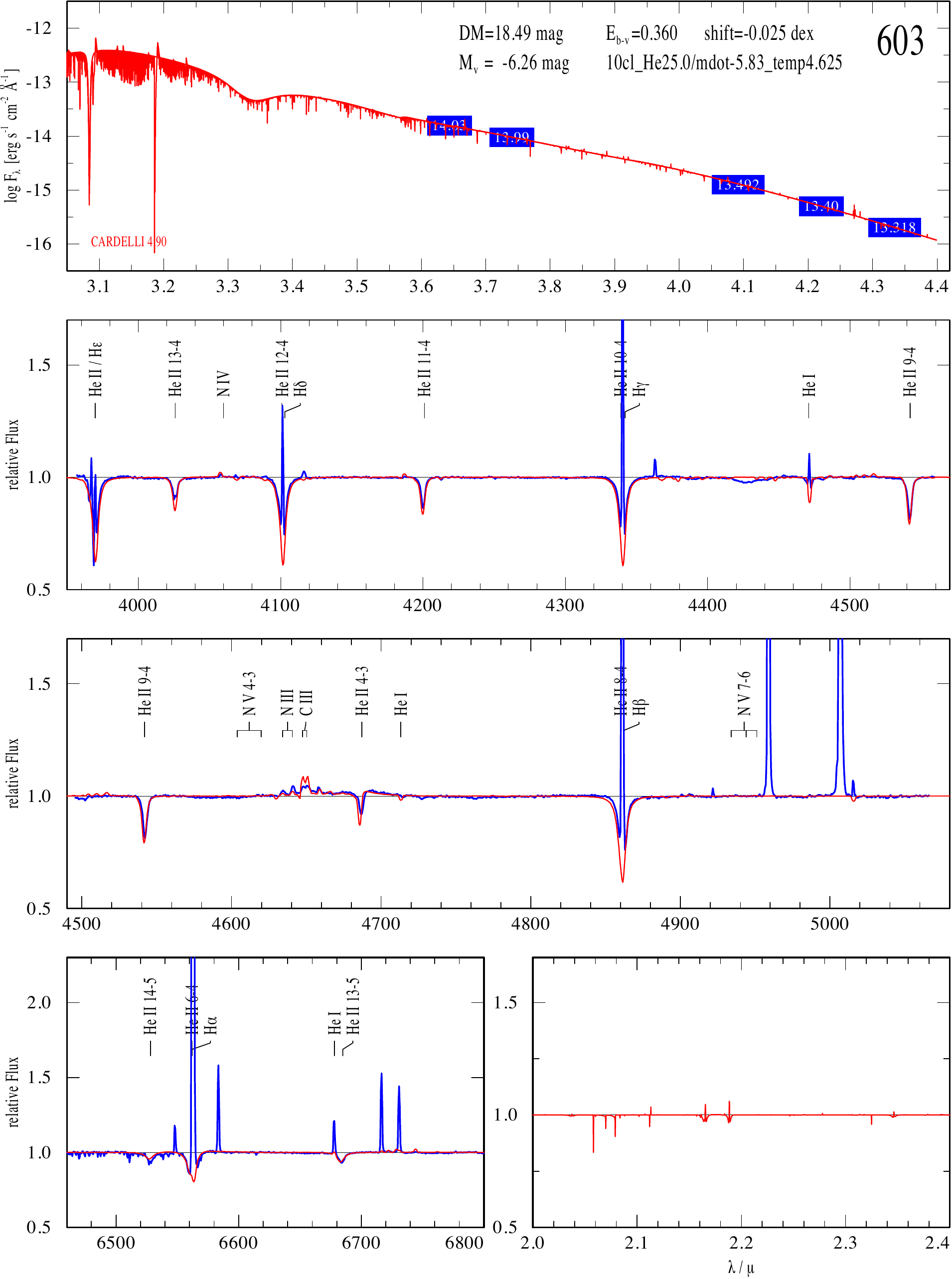}
\end{center}
\caption{The spectrum of VFTS\,603 (O4 III(fc)) shows the characteristic of a SB1. The temperature is based on the lines N\,{\sc iii}\,$\lambda 4634/4640$ and N\,{\sc iv}\,$\lambda 4058$. He\,{\sc i}\,$\lambda 4471$ is contaminated by nebular emission and is not used. $\dot{M}$ is based on the line shape of He\,{\sc ii}\,$\lambda 4686$.  N-abundance is between normal and enriched.}
\end{figure}
\clearpage
\begin{figure}
\begin{center}
\includegraphics[width=17cm]{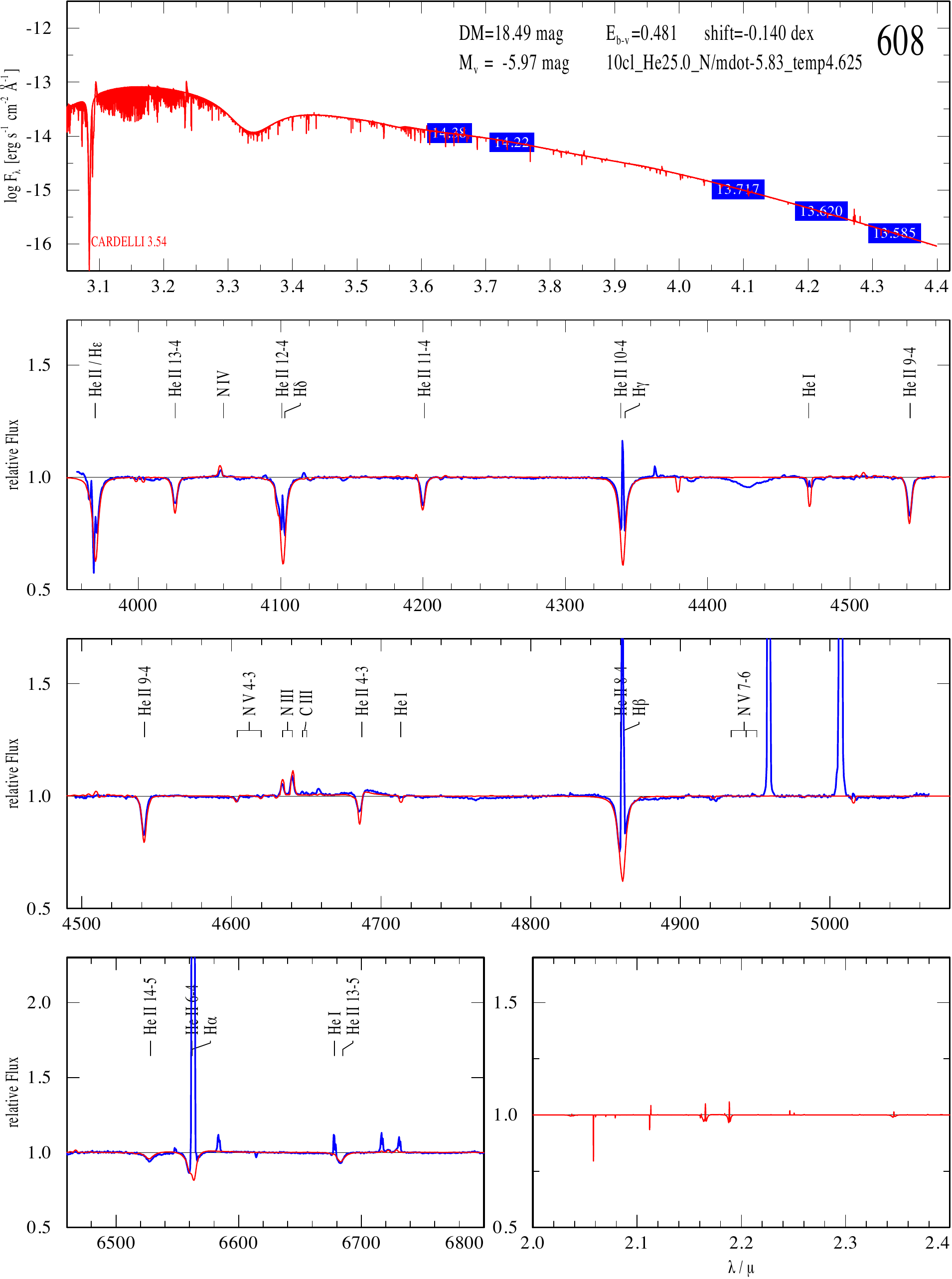}
\end{center}
\caption{The spectrum of VFTS\,608 (O4 III(f)) shows the characteristic of a SB1. The temperature is based on the lines N\,{\sc iii}\,$\lambda 4634/4640$, N\,{\sc iv}\,$\lambda 4058$, and N\,{\sc v} $\lambda4604/4620$. He\,{\sc i}\,$\lambda 4471$ is contaminated by nebular emission and is not used. $\dot{M}$ is based on the line shape of He\,{\sc ii}\,$\lambda 4686$. Nitrogen is enriched.}
\end{figure}
\clearpage
\clearpage
\begin{figure}
\begin{center}
\includegraphics[width=17cm]{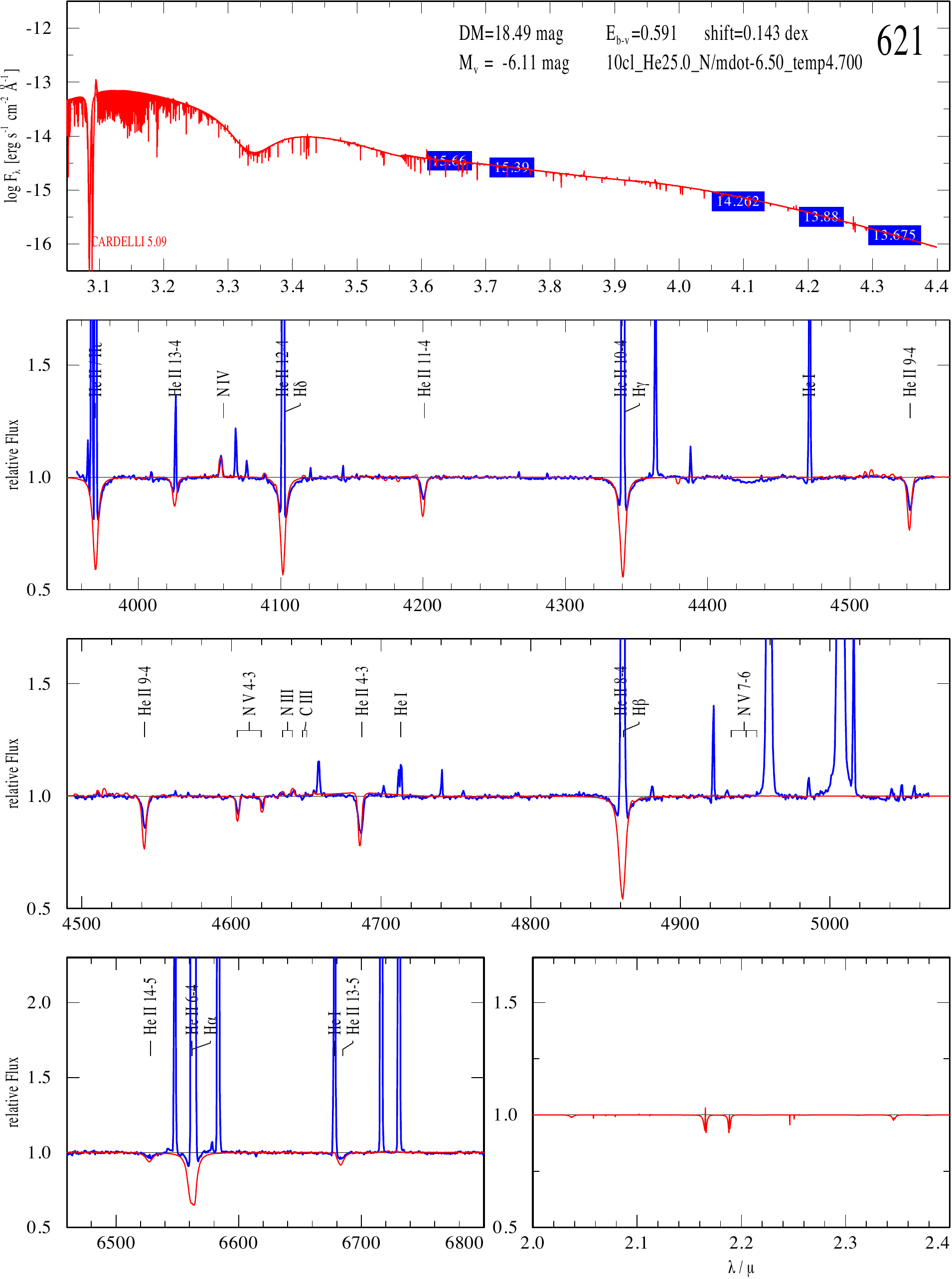}
\end{center}
\caption{The temperature of VFTS\,621 (O2 V((f*))z) is based on N\,{\sc iii}\,$\lambda 4634/4640$, N\,{\sc iv}\,$\lambda 4058$, and N\,{\sc v} $\lambda4604/4620$. $\dot{M}$ is estimated using the line shape of He\,{\sc ii}\,$\lambda 4686$ and is uncertain. N is enriched.}
\end{figure}
\clearpage
\begin{figure}
\begin{center}
\includegraphics[width=17cm]{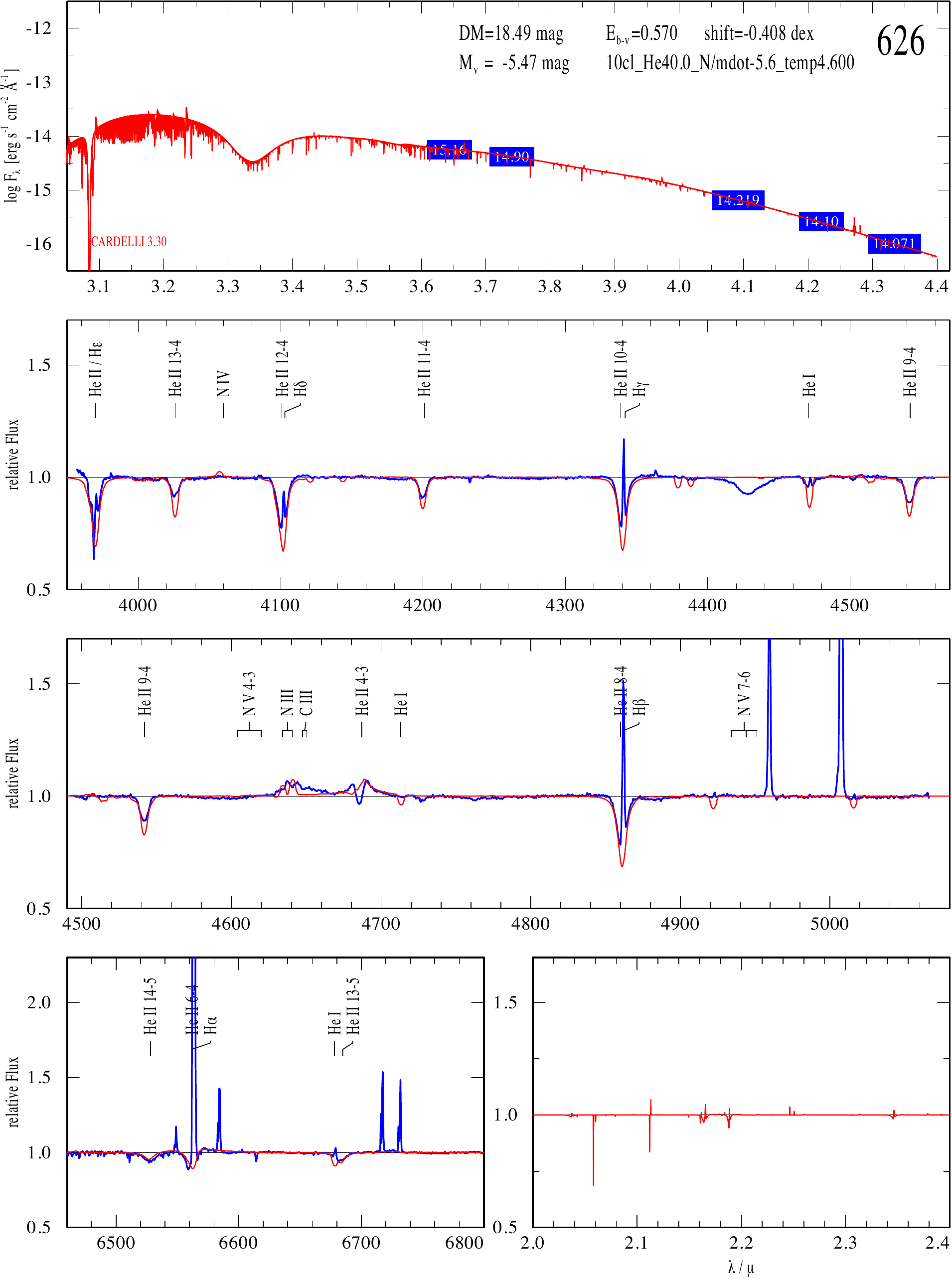}
\end{center}
\caption{The line profiles of VFTS\,626 (O5-6 n(f)p) are broadened as a result of rotation. The temperature is based on the lines He\,{\sc i}\,$\lambda 4471$ and N\,{\sc iii}\,$\lambda 4634/4640$. $\dot{M}$ is based on the line shape of He\,{\sc ii}\,$\lambda 4686$ and $\mathrm{H}_{\alpha}$. The best fit has a He-abundance of 40\%. N is enriched.}
\end{figure}
\clearpage
\begin{figure}
\begin{center}
\includegraphics[width=17cm]{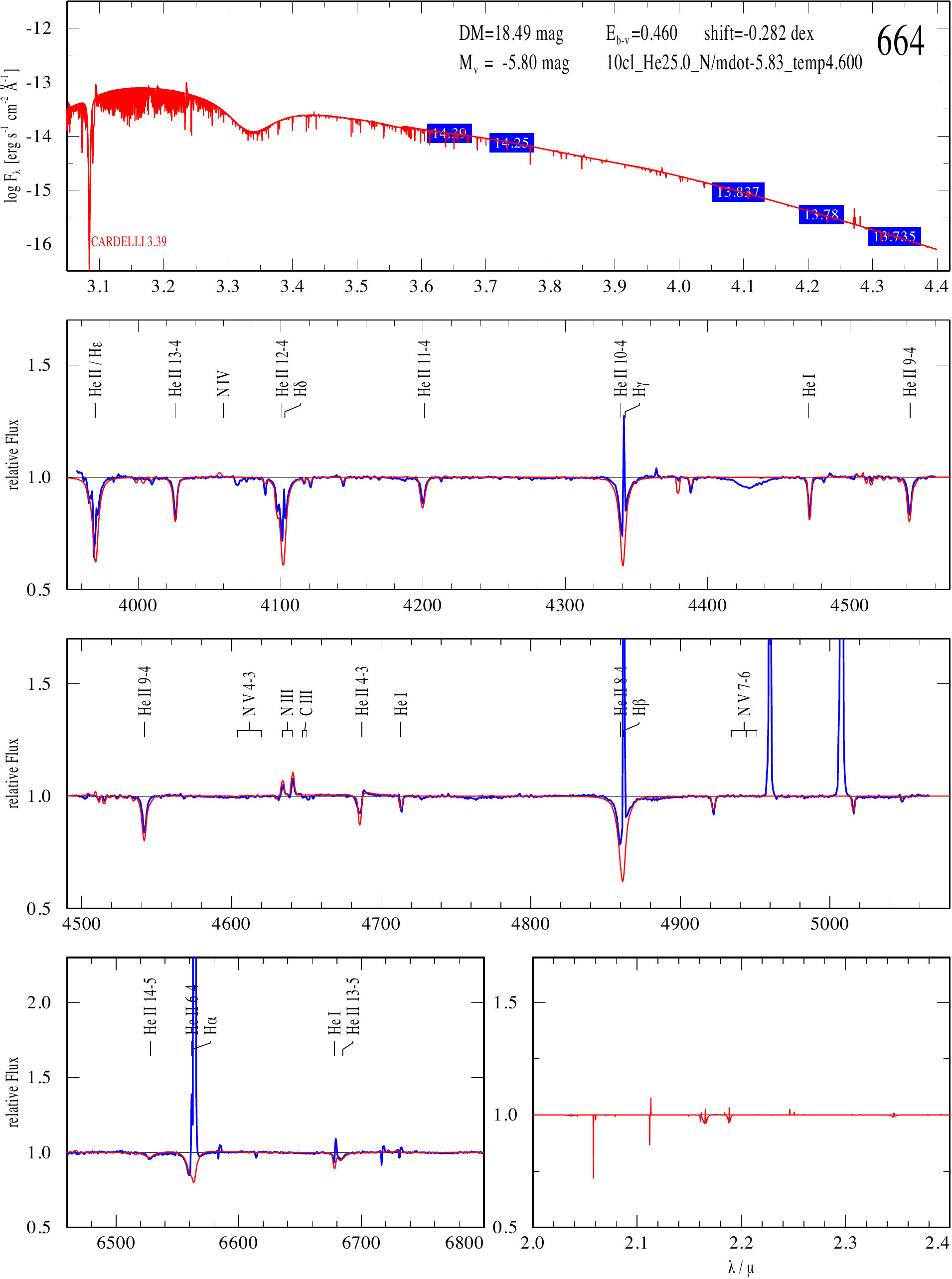}
\end{center}
\caption{The temperature of VFTS\,664 (O7 II(f)) is based on He\,{\sc i}\,$\lambda 4471$ and N\,{\sc iii}\,$\lambda 4634/4640$. $\dot{M}$ is based on the line shape of He\,{\sc ii}\,$\lambda 4686$. N is enriched.}
\end{figure}
\clearpage
\begin{figure}
\begin{center}
\includegraphics[width=17cm]{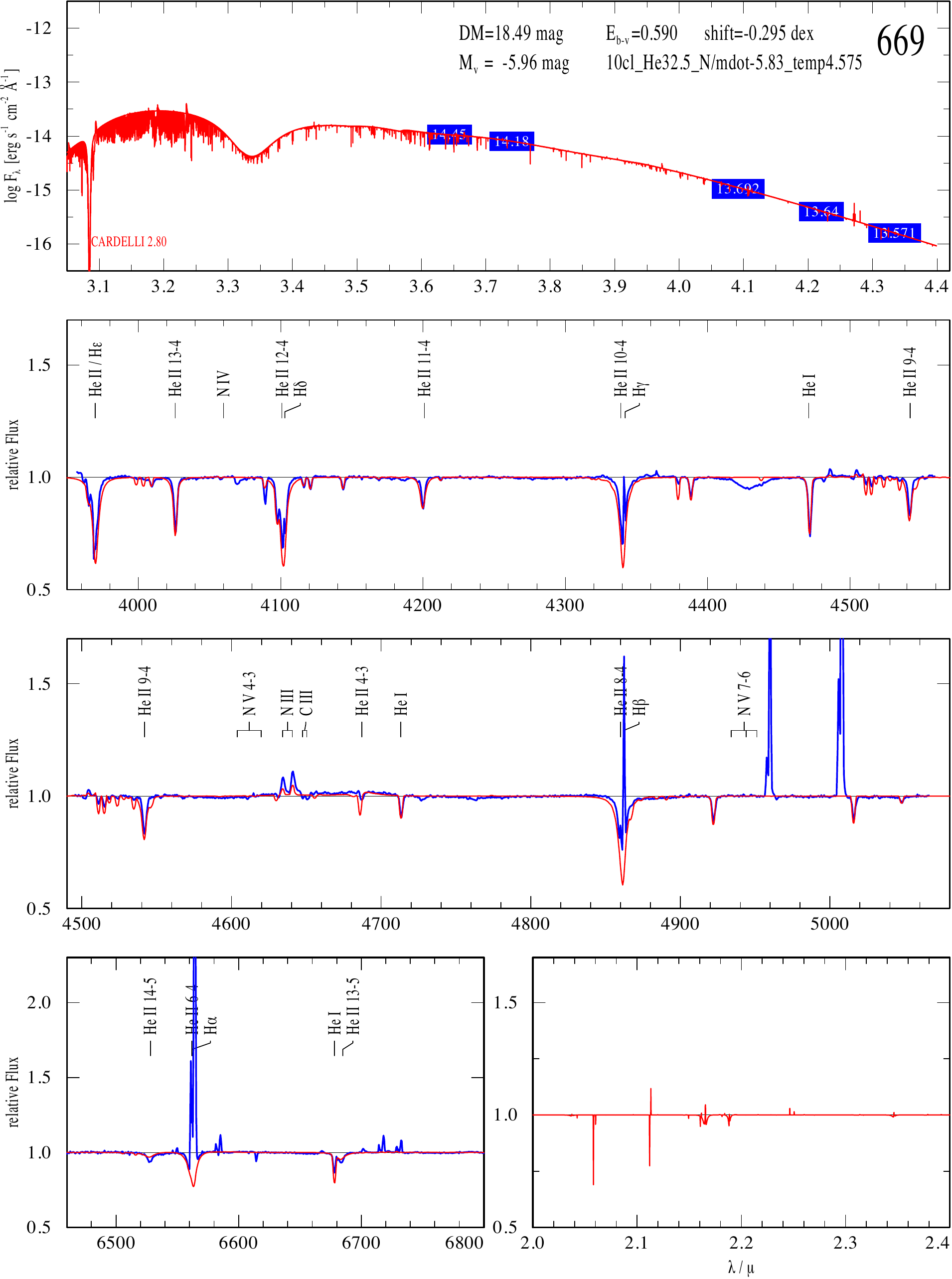}
\end{center}
\caption{The temperature of VFTS\,669 (O8 Ib(f)) is based on the lines He\,{\sc i}\,$\lambda 4471$ and N\,{\sc iii}\,$\lambda 4634/4640$. $\dot{M}$ is based on the line shape of He\,{\sc ii}\,$\lambda 4686$. N is enriched and He might be as well. The absorption lines are narrower than the model, which suggests a lower $\log g$. By lowering $\log g$ N\,{\sc iii}\,$\lambda 4634/4640$ would be stronger in emission and improve the fit (Fig.\,\ref{a:669_test}).}
\label{a:669}
\end{figure}
\clearpage
\begin{figure}
\begin{center}
\includegraphics[width=17cm]{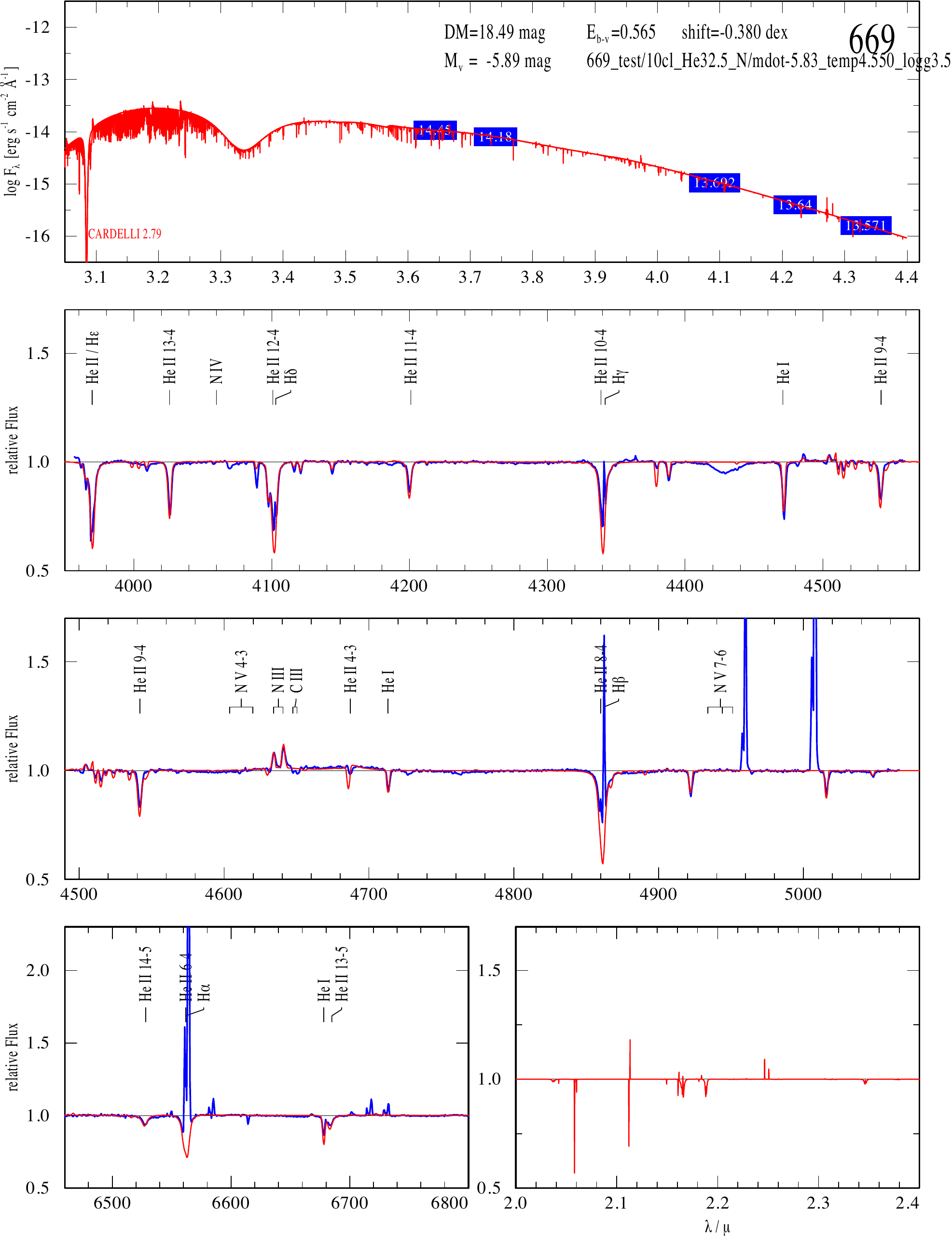}
\end{center}
\caption{Test model for VFTS\,669 (O8 Ib(f)) to investigate the predicted changes in temperature, luminosity and mass loss (see Figs.\,\ref{f:logg_lowT} and \ref{f:logg_highT}). As predicted the transformed mass-loss rate is still the same while the temperature ($-0.025$\,dex) and the luminosity ($-0.08$\,dex) are lower.}
\label{a:669_test}
\end{figure}
\clearpage
\clearpage
\clearpage
\begin{figure}
\begin{center}
\includegraphics[width=17cm]{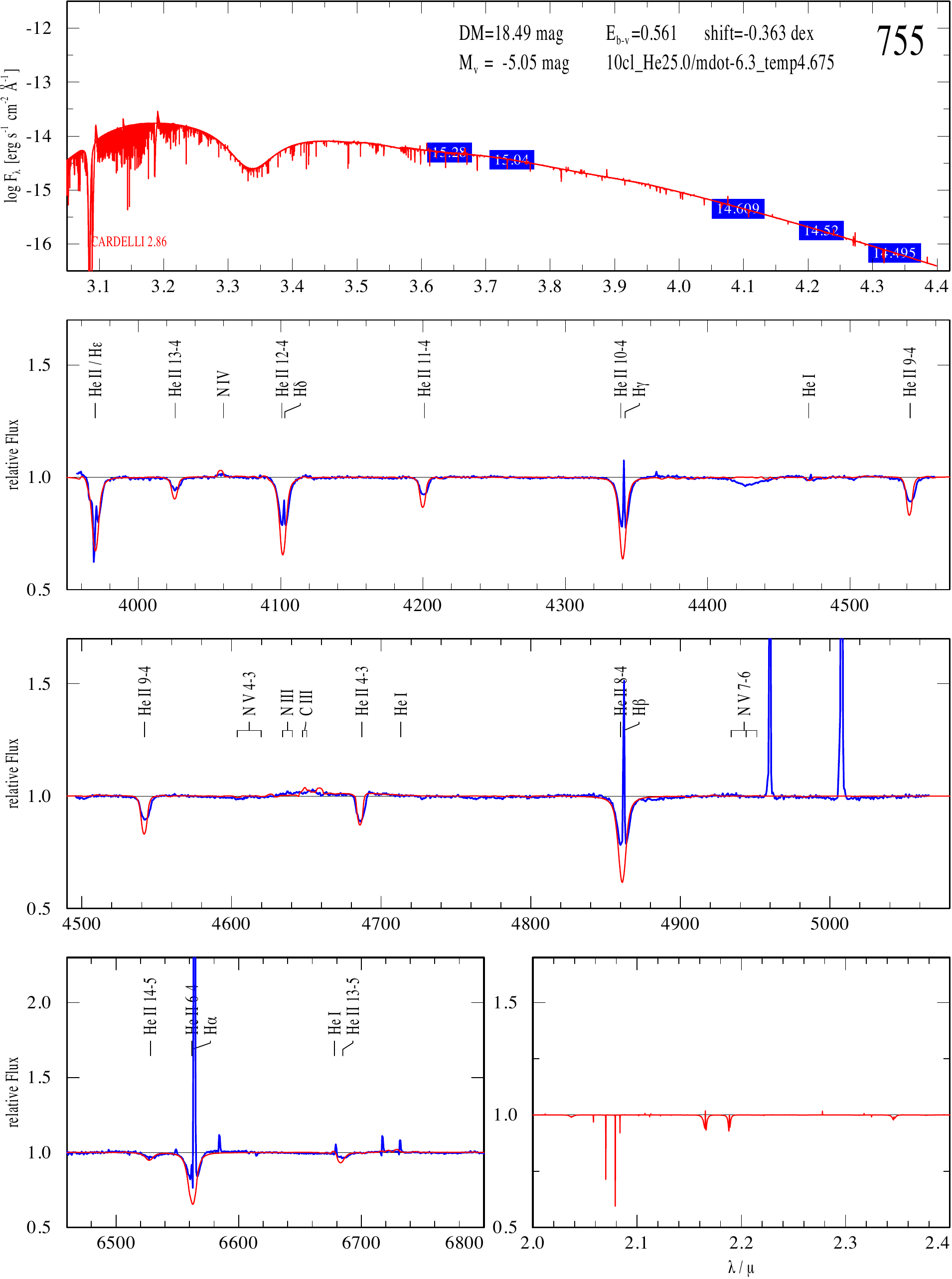}
\end{center}
\caption{The temperature of VFTS\,755 (O3 Vn((f*))) is based on the lines He\,{\sc i}\,$\lambda 4471$ and N\,{\sc iv}\,$\lambda 4058$. $\dot{M}$ is based on the line shape of He\,{\sc ii}\,$\lambda 4686$. N-abundance is normal.}
\end{figure}
\clearpage
\clearpage
\begin{figure}
\begin{center}
\includegraphics[width=17cm]{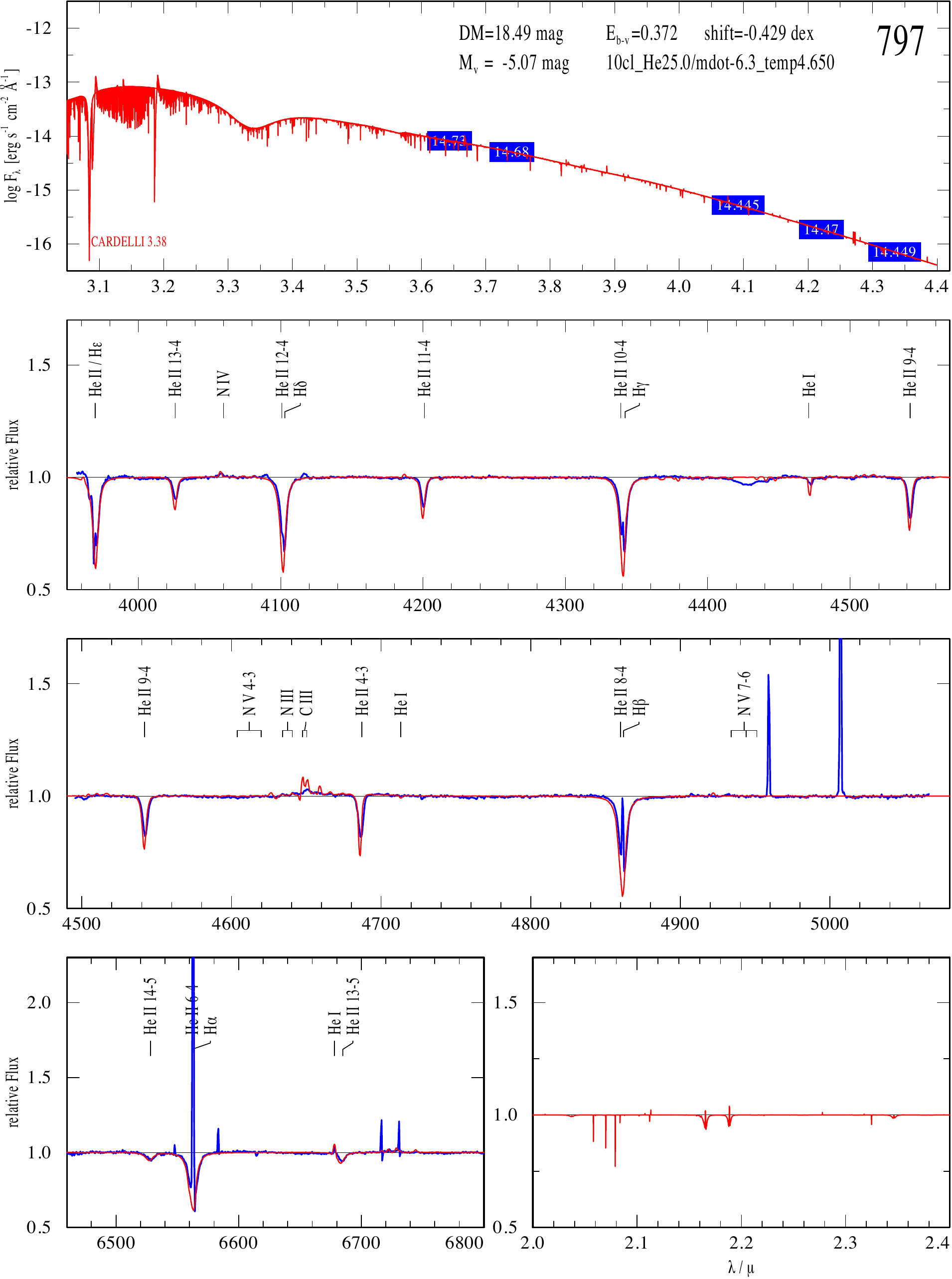}
\end{center}
\caption{The temperature of VFTS\,797 (O3.5 V((n))((fc))) is based on the lines He\,{\sc i}\,$\lambda 4471$ and N\,{\sc iv}\,$\lambda 4058$. $\dot{M}$ is based on the line shape of He\,{\sc ii}\,$\lambda 4686$. N is not enriched at the stellar surface.}
\end{figure}
\clearpage
\clearpage
\begin{figure}
\begin{center}
\includegraphics[width=17cm]{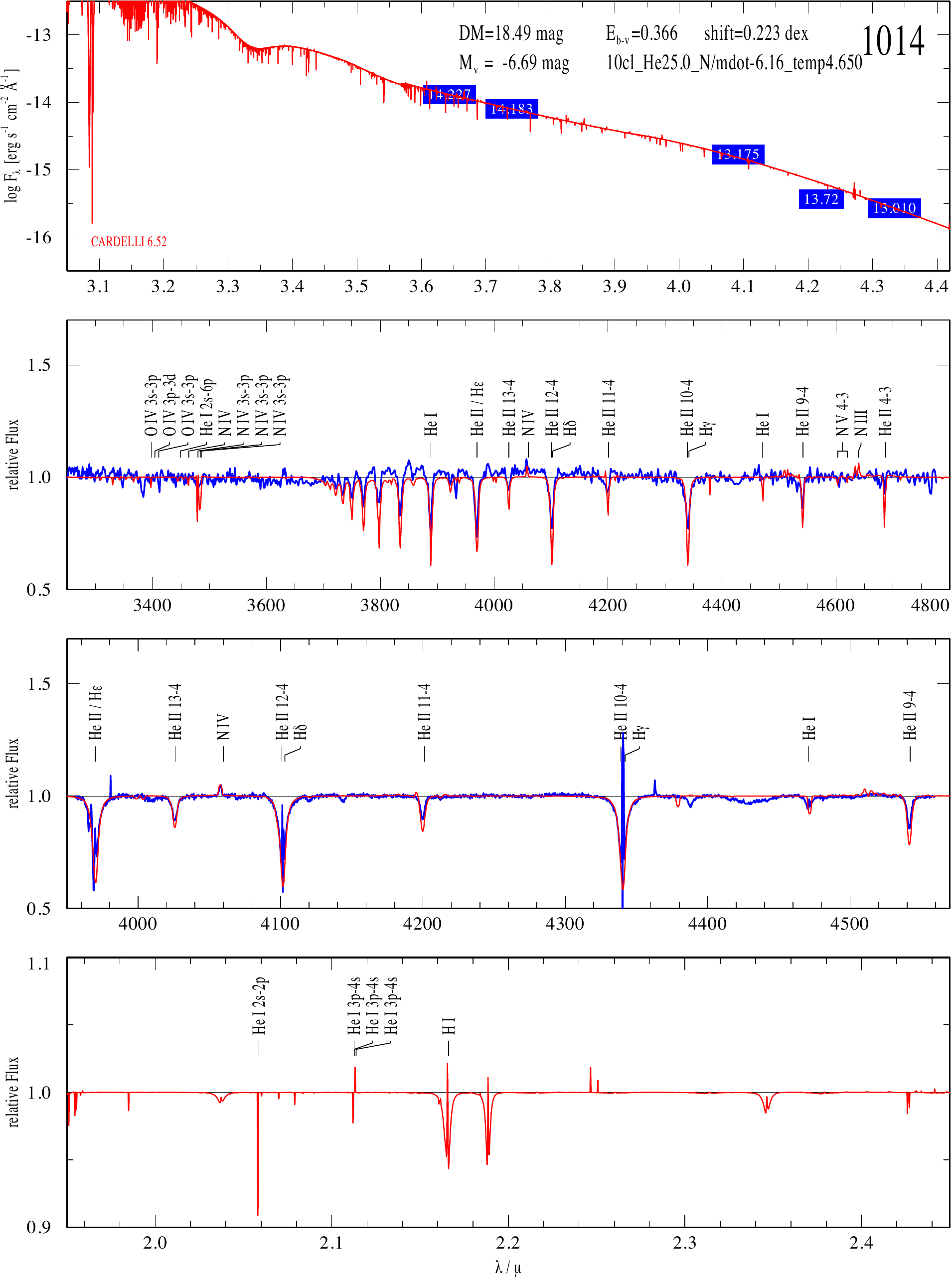}
\end{center}
\caption{The temperature of VFTS\,1014 (O3 V + mid/late O) is based on the lines N\,{\sc iv}\,$\lambda 4058$, N\,{\sc v} $\lambda4604/4620$, and on the He\,{\sc i}\,$\lambda 4471$ line. Unfortunately, the resolution and S/N of the HST observation is very low and $\dot{M}$ is based on N\,{\sc iv}\,$\lambda 4058$ instead of He\,{\sc ii}\,$\lambda 4686$. The $\log g$ is at least $\gtrsim$4 and the luminosity class is V, which suggests a young age and a He-abundance of $\approx$25\%. (Second panel: HST/FOS, third panel: ARGUS.)}
\end{figure}
\clearpage
\clearpage
\begin{figure}
\begin{center}
\includegraphics[width=17cm]{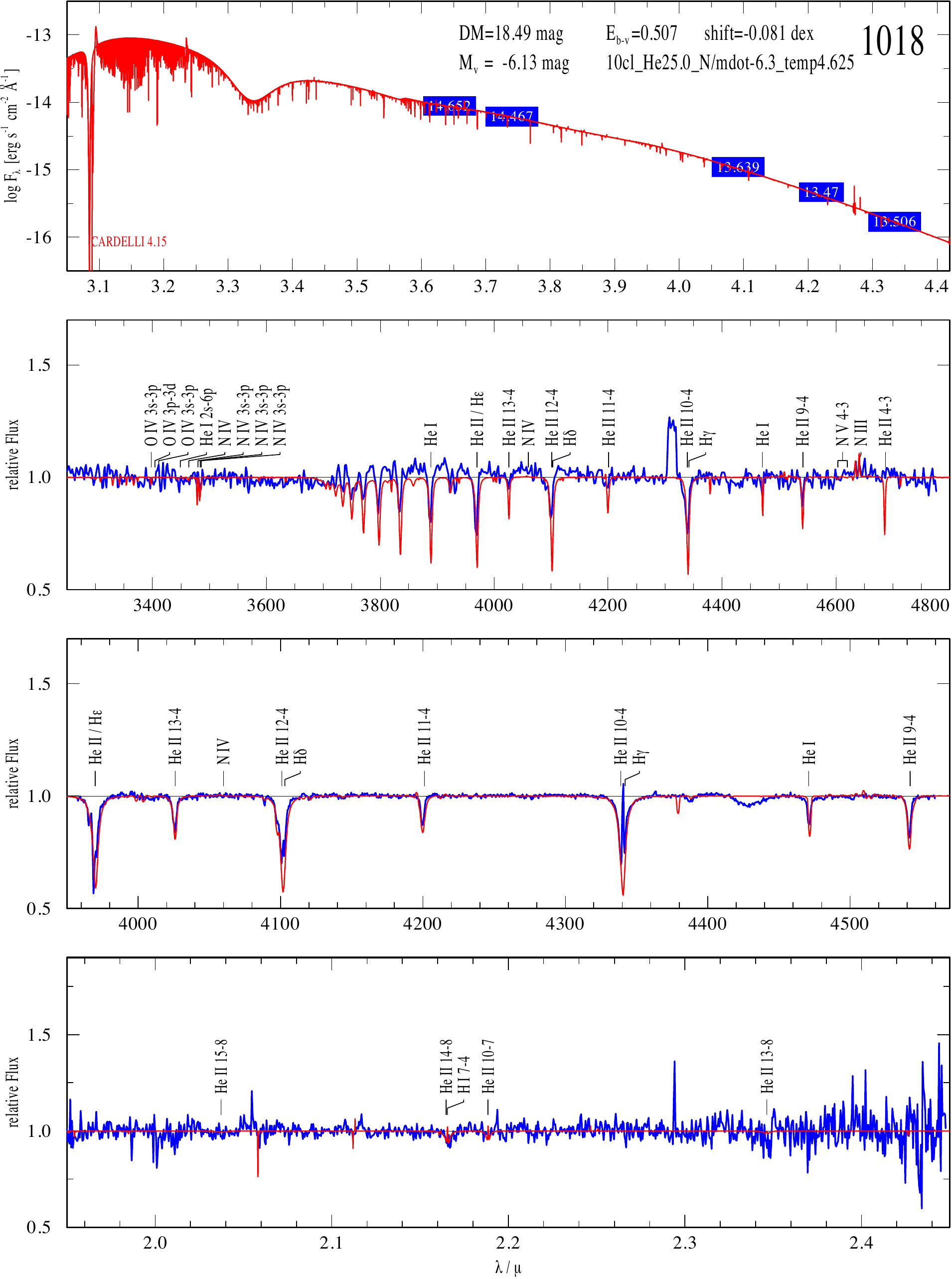}
\end{center}
\caption{The temperature of VFTS\,1018 (O3 III(f*) + mid/late O) is based on the lines N\,{\sc iii}\,$\lambda 4634/4640$, N\,{\sc iv}\,$\lambda 4058$, and He\,{\sc i}\,$\lambda 4471$. Unfortunately, the resolution and S/N of the HST observation is very low and $\dot{M}$ is based on the lines N\,{\sc iv}\,$\lambda 4058$, He\,{\sc ii}\,$2.19\mu m$, and $\mathrm{H}^{\rm Br}_{\gamma}$. The two near-IR lines are not in emission, which leads to an uncertainty in $\dot{M}$ and He-abundance. The star is N enriched. (Second panel: HST/FOS, third panel: ARGUS.)}
\end{figure}
\clearpage
\begin{figure}
\begin{center}
\includegraphics[width=17cm]{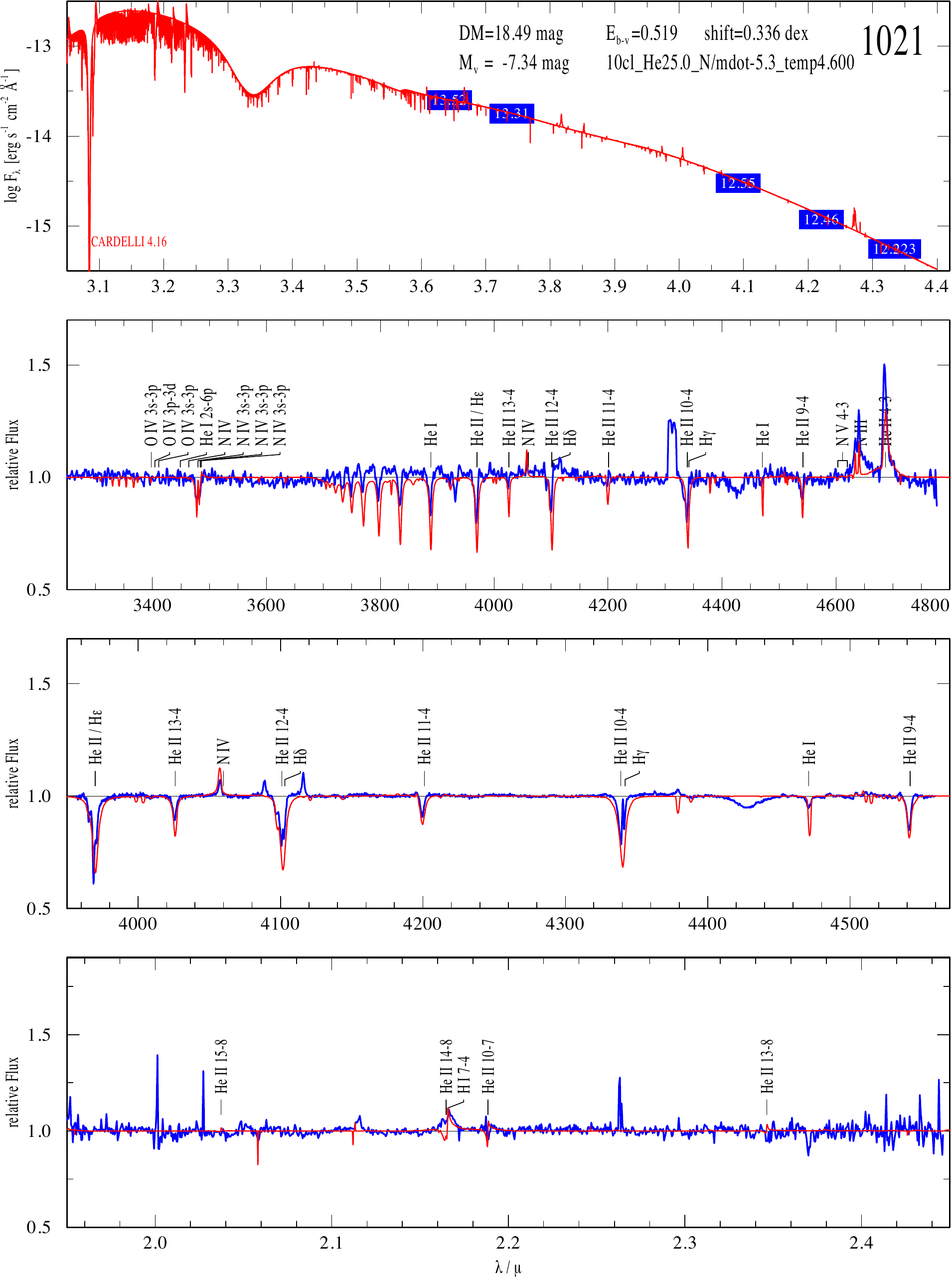}
\end{center}
\caption{The temperature of VFTS\,1021 (O4 If+) is based on the lines N\,{\sc iii}\,$\lambda 4634/4640$ and N\,{\sc iv}\,$\lambda 4058$, and He\,{\sc i}\,$\lambda 4471$. $\dot{M}$ and He-abundance are based on He\,{\sc ii}\,$2.19\mu m$ and $\mathrm{H}^{\rm Br}_{\gamma}$. The best fitting model $\dot{M}$ is too weak. The star suggests a unusually high N abundance. The fit quality is not good, but can be improved (see Fig.\,\ref{f:457_test}). The He-abundance is between 25\% and 32.5\%. (Second panel: HST/FOS, third panel: ARGUS.)}
\end{figure}
\clearpage
\clearpage
\begin{figure}
\begin{center}
\includegraphics[width=17cm]{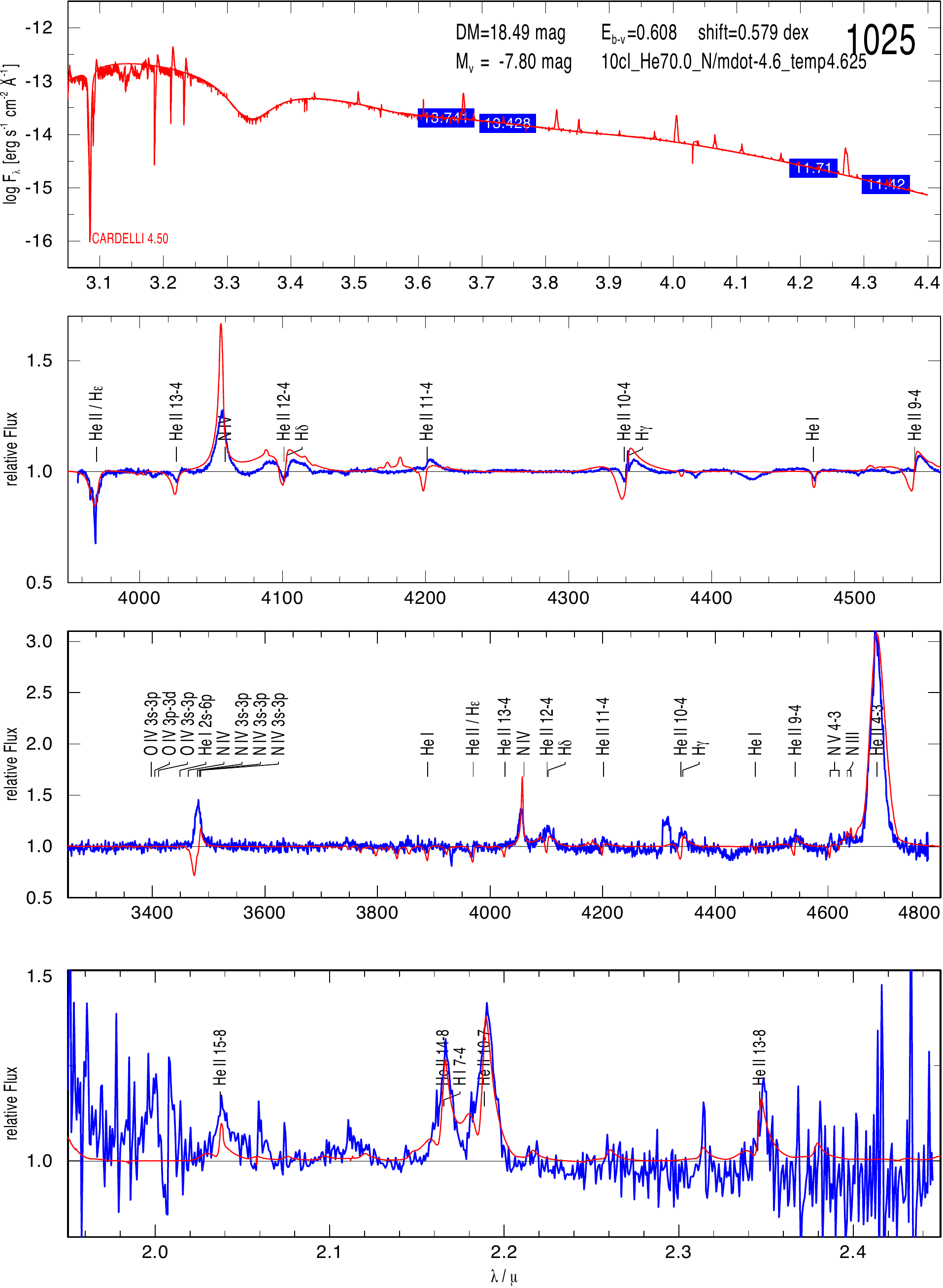}
\end{center}
\caption{The temperature of VFTS\,1025 (WN5h) is based on the lines N\,{\sc iii}\,$\lambda 4634/4640$ N\,{\sc iv}\,$\lambda 4058$, N\,{\sc v} $\lambda4604/4620$, and He\,{\sc i}\,$\lambda 4471$. $\dot{M}$ and He-abundance are based on the lines He\,{\sc ii}\,$\lambda 4686$, He\,{\sc ii}\,$2.19\mu m$, and $\mathrm{H}^{\rm Br}_{\gamma}$. N is enriched. (Second panel: ARGUS, third panel: HST/FOS. The ARGUS data were only approximately normalised as there prime use was the investigations of RV/LPV variations.)}
\label{a:1025}
\end{figure}
\clearpage
\begin{figure}
\begin{center}
\includegraphics[width=17cm]{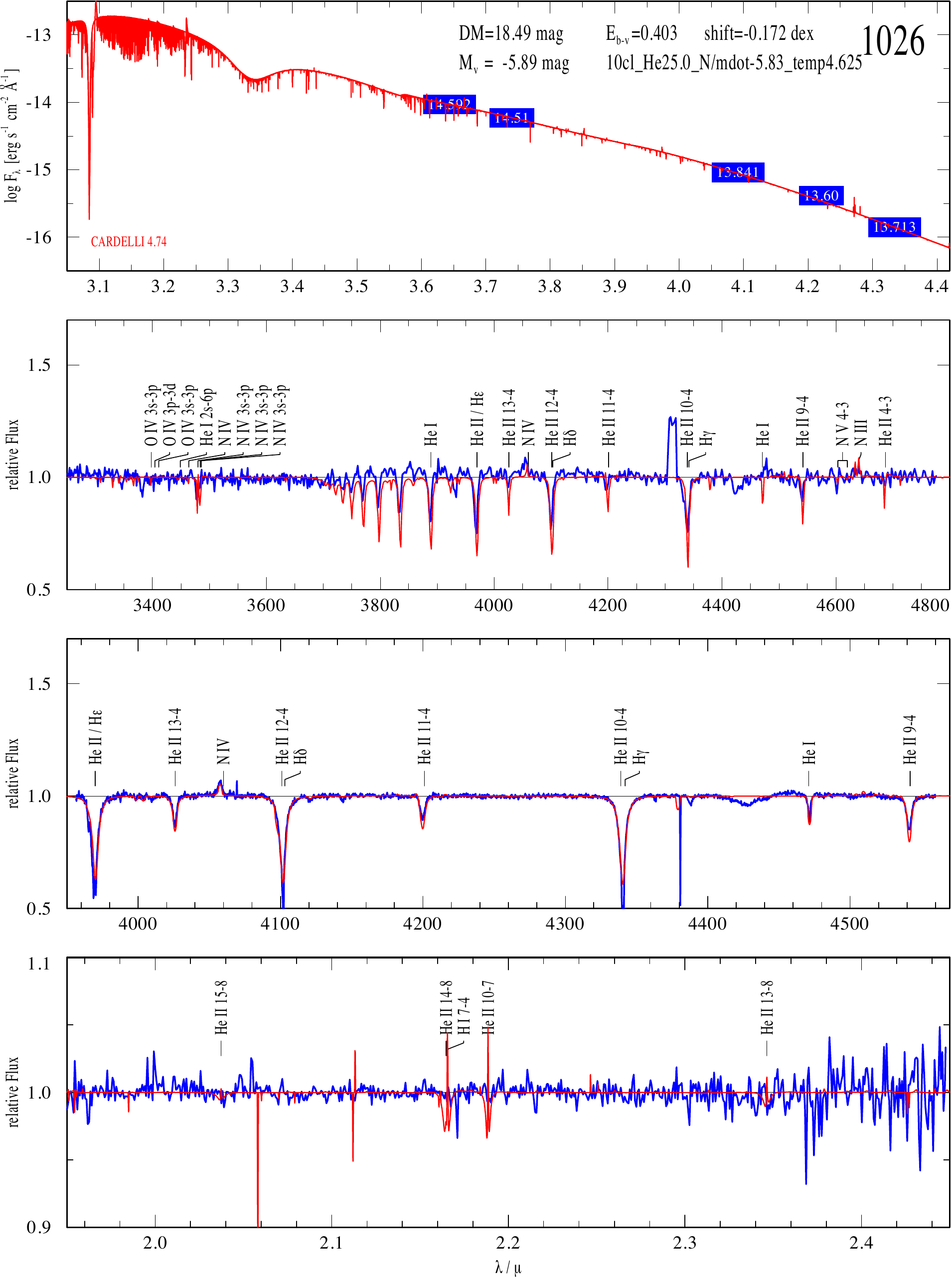}
\end{center}
\caption{The temperature of VFTS\,1026 (O3 III(f*) + mid/late O) is based on the lines N\,{\sc iv}\,$\lambda 4058$ and He\,{\sc i}\,$\lambda 4471$. $\dot{M}$ is estimated on the basis of the strength of the N\,{\sc iv}\,$\lambda 4058$ line and is highly uncertain. N is enriched. (Second panel: HST/FOS, third panel: ARGUS.)}
\end{figure}
\clearpage
\begin{figure}
\begin{center}
\includegraphics[width=17cm]{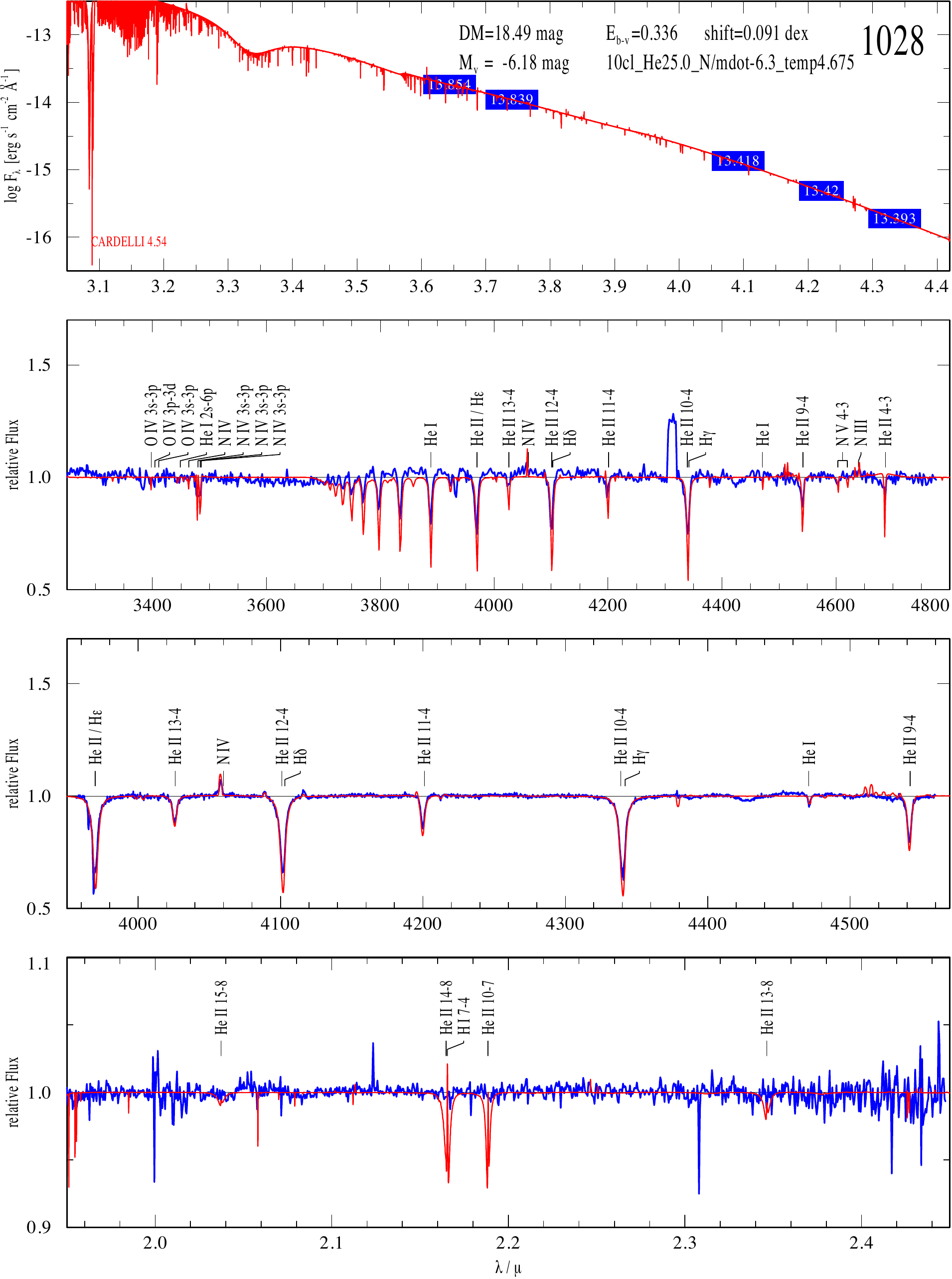}
\end{center}
\caption{The temperature of VFTS\,1028 (O3 III(f*) or O4-5V) is based on the lines N\,{\sc iv}\,$\lambda 4058$ and He\,{\sc i}\,$\lambda 4471$. $\dot{M}$ is estimated on the basis of the strength of the N\,{\sc iv}\,$\lambda 4058$ line and is highly uncertain. N is enriched. (Second panel: HST/FOS, third panel: ARGUS.)}
\end{figure}
\clearpage
\begin{figure}
\begin{center}
\includegraphics[width=17cm]{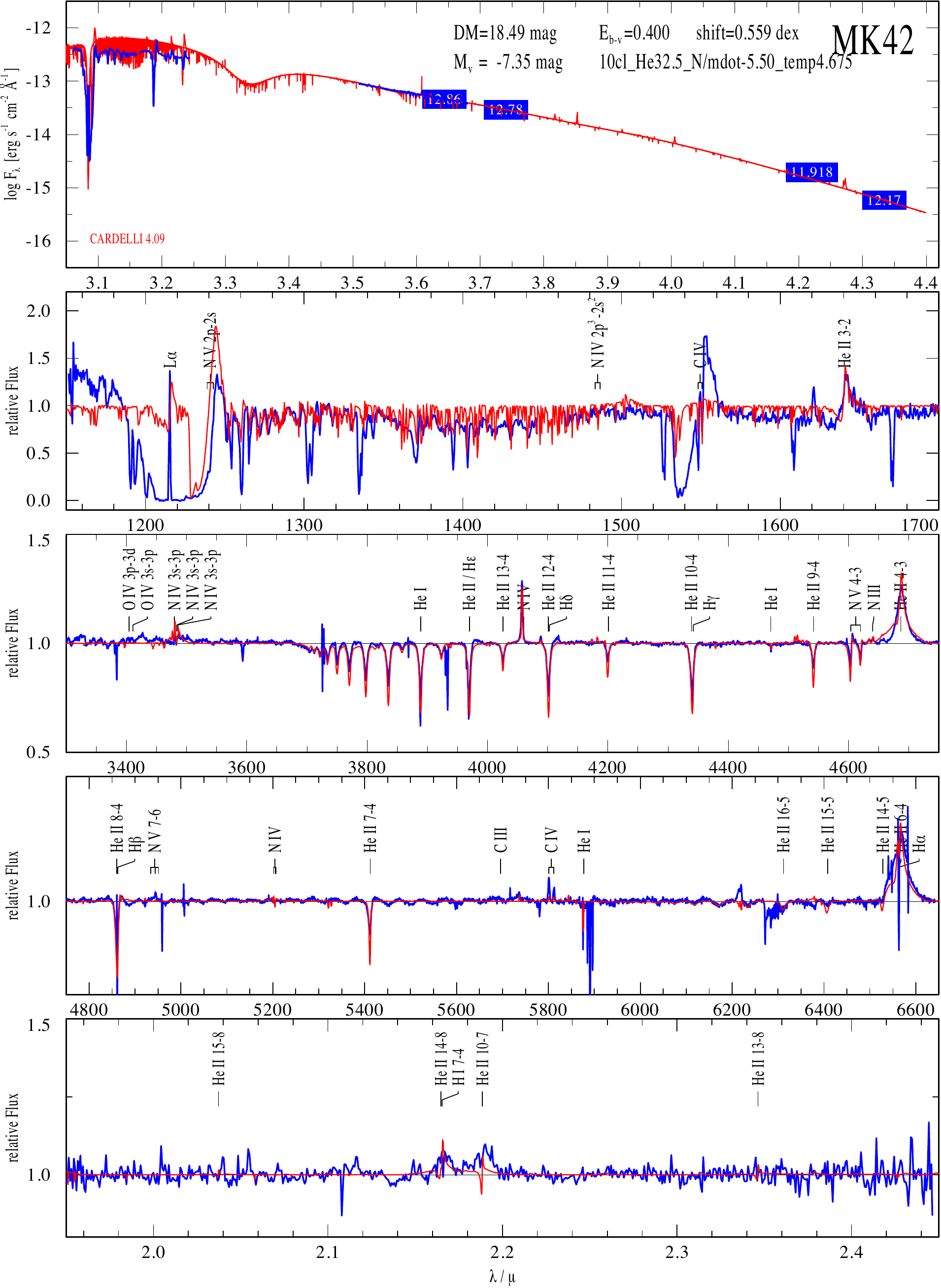}
\end{center}
\caption{The temperature of Mk42 (O2 If*) is based on the lines N\,{\sc iii}\,$\lambda 4634/4640$ N\,{\sc iv}\,$\lambda 4058$, N\,{\sc v} $\lambda4604/4620$, and He\,{\sc i}\,$\lambda 4471$. $\dot{M}$ and He-abundance are based on the lines He\,{\sc ii}\,$\lambda 4686$, $\mathrm{H}_{\alpha}$, He\,{\sc ii}\,$2.19\mu m$, and $\mathrm{H}^{\rm Br}_{\gamma}$. N is enriched. (Second panel: HST/GHRS, third and fourth panel: UVES.)}
\end{figure}

\end{appendix}
\end{document}